\documentclass[12pt,tightenlines]{tifrthesis}
\usepackage{hyperref, enumerate}
\usepackage[pagewise, mathlines]{lineno}
\usepackage[nottoc]{tocbibind}
\usepackage{booktabs}
\usepackage{setspace}
\usepackage{cite}
\usepackage{fancyhdr}
\usepackage{tocloft}
\usepackage{type1cm}
\usepackage{courier}
\usepackage{url}
\makeatletter
\def\url@leostyle{%
  \@ifundefined{selectfont}{\def\UrlFont{\sf}}{\def\UrlFont{\small\ttfamily}}}
\makeatother
\usepackage{graphicx}
\usepackage{relsize}
\usepackage[latin1]{inputenc}
\usepackage{amsfonts}
\usepackage{amsmath}
\usepackage{amssymb}
\usepackage{amsthm}
\usepackage{mathrsfs}
\usepackage{bm}
\usepackage{caption}
\usepackage{xcolor}
\usepackage{xspace}

\newcommand{\nc}{\newcommand}

\nc{\pt}{p_\mathrm{T}} \nc{\kt}{k_\mathrm{T}} \nc{\mt}{m_\mathrm{T}}
\nc{\Kt}{K_\mathrm{T}} \nc{\Mt}{M_\mathrm{T}} \nc{\pL}{p_\mathrm{L}}

\begin{document}
\author{Anjani Kumar Tiwari}
\regno{PHYS-159}
\title{Coherent random lasing in a \\[0.10in] disordered array of amplifying \\[0.10in] microresonators}
\dept{Department of Nuclear and Atomic Physics}
\gradtime{September, 2014}
\subject{Physics}
\maketitle

\advisorname{Prof. Sushil Mujumdar}
\disscopyright
\vspace*{0.6in}
\begin{center}
{\Large\bf DECLARATION}
\end{center}

\vskip 1cm

\noindent 
\startonehalfspace{\large This thesis is a presentation of my 
original research work and has not been submitted earlier as a whole 
or in part for a degree/diploma at this or any other 
Institution/University. Wherever contributions of others are 
involved, every effort is made to indicate this clearly, with due 
reference to the literature, and acknowledgement of collaborative 
research and discussions.}\\

\noindent 
\startonehalfspace{\large The work was done under the guidance of 
Prof. Sushil Mujumdar at the Tata Institute of Fundamental Research, 
Mumbai.}

\vskip 2.0cm

\begin{flushright}
\startonehalfspace{\large Anjani Kumar Tiwari}
\end{flushright}
\begin{flushright}
\startonehalfspace{\large {\bf [Candidate's name and signature]}}
\end{flushright}

\vskip 1.5cm

\noindent 
\startonehalfspace{\large In my capacity as supervisor of the 
candidate's thesis, I certify that the above statements are true to 
the best of my knowledge.}

\vskip 2.0cm

\begin{flushleft}
\startonehalfspace{\large Prof. Sushil Mujumdar}
\end{flushleft}
\begin{flushleft}
\startonehalfspace{\large {\bf [Supervisor's name and signature]}}
\end{flushleft}

\vskip 0.5cm
\noindent\startonehalfspace{\large Date: 21/04/2014}

\begin{acknowledgements}

\startonehalfspace{\large First and foremost, I am deeply grateful to my adviser, Prof. Sushil Mujumdar for his scientific guidance, valuable advice, support and encouragement in the journey towards my Ph.D. Working at the Nano-Optics and Mesoscopic Optics Laboratory (NOMOL) has been a great learning experience for me.}

\startonehalfspace{\large I have been extremely fortunate to have some very helpful labmates, and it was a great pleasure working together in a group. I sincerely thank Balu, Bikas, Girish, Randhir, Ravitej, Swapnesh, Shadak and Sreeman. I particularly acknowledge Ravitej for his valuable help in the Monte Carlo simulations (Chapter 2) and FDTD calculations (Chapter 8).
During this work, I have benefited from the discussion on various topics with Dr. Rajesh Nair and Dr. Shivakiran. I also thank Prof. Amol Dighe and Prof. R. Palit for having a keen interest in my research progress throughout my Ph.D.}

\startonehalfspace{\large I express my gratitude to the DNAP office, workshop, library, photography section and university cell for their help.}

\startonehalfspace{\large I thank my friends and colleagues Amaresh, Bhupender, Deepak, Pankaj, and Prashant for making my stay at TIFR a delightful and memorable experience.}

\startonehalfspace{\large Last but not least, I would like to thank my parents for their constant support over the years.}

\end{acknowledgements}

\begin{center}
\startcontentspace
\tableofcontents
\startdoublespace
\end{center}
\pagebreak

\begin{synopsis}
\vspace{-1cm}
\section*{\label{sec:level1}Introduction and motivation
}
A conventional laser consists of a gain medium and a carefully designed resonant cavity~\cite{syn_siegman86}. The most common configuration of a laser cavity is a Fabry-Perot (FP) resonator, made of two parallel mirrors. The consequent lasing intensity and frequency depend on the inversion and the physical parameters of the cavity.  If the gain overcomes all the cavity losses, the system crosses a threshold and lases. Any inhomogeneities causing scattering within the cavity introduce additional losses in the lasing process and hence are considered detrimental for lasing action. In 1968, Letokhov~\cite{syn_letokhov68} proposed the amplification of light in a random scattering environment instead of a well-defined set of mirrors. This proposal was motivated by the extension of the lifetime of a light wave in a multiply-scattering medium, similar to that created by a resonator cavity but without the wavelength-selectivity. Earlier experiments on this proposal were performed on laser powders, and the existence of stimulated emission was confirmed. A breakthrough result was reported by Lawandy et al.~\cite{syn_lawandy94}, who observed a significant collapse of emission linewidth of a dye-solvent medium with the addition of titanium dioxide nanoparticles. This system has been termed as a `random laser' since it exhibits laser-like characteristics of intensity enhancement and linewidth collapse after a pump threshold.

Random lasers are exciting optical sources where amplification is coupled with multiple scattering~\cite{syn_wiersma08}. Multiple scattering increases the interaction time of light with an inverted medium, and this results in enhanced stimulated emission. The structural disorder needed for scattering is realized by creating a random variation of refractive index in the gain medium. Since there are no prescribed restrictions on the nature of scatterers and the gain medium, a variety of different materials have been reported to show random lasing~\cite{syn_noginov05, syn_cao99, syn_fallert09, syn_sushil04, syn_gottardo08, syn_turitsyn10}. For example, Neodymium-doped powders, dye-doped microparticles, colloidal suspensions of titanium dioxide, zinc oxide, polystyrene in various gain media, etc. are used for this purpose.

Based on the feedback mechanism, random lasers are classified into two categories: (1) Diffusive random laser and (2) Coherent random laser. The former describes a system that emits a stable spectrum with a fixed number of modes (usually one or two) and a bandwidth that is narrowed by about an order of magnitude (typically reaching $4 - 5$~nm) compared to the homogeneous gain medium. In a diffusive random laser, the propagation of light is modeled as the diffusion of light with gain, and the phase information and the interference effects are neglected. This feedback mechanism involves the transport of intensity and is known as nonresonant or incoherent feedback. This kind of feedback is relevant in the disorder regime characterized by $L > \ell > \lambda$, where $L$ is the system size, $\ell$ is the mean free path of light of wavelength $\lambda$. In diffusive random lasers, the spectral peak position is predicted by the gain maximum. On the other hand, coherent random lasing differs significantly in emission characteristics. Under certain conditions of gain and pumping, the spectra from the samples exhibit several ultranarrow peaks (bandwidth~$\sim0.1$~nm) riding on an incoherent pedestal. These ultranarrow modes are temporally coherent, and strongly fluctuate in both intensity and frequency. Early studies of this phenomenon attributed these modes to Anderson-localized resonant modes of the disordered medium. Such a feedback, where the light wave follows a closed loop path over a randomly formed resonator, is termed resonant feedback. Clearly, this feedback is based on the electric field transport and requires the phase information of the scattered light wave. The phase shift requirement of $2\pi$ along the loop determines the lasing frequency~\cite{syn_cao99, syn_cao02}. In a disordered system, the interference effects become prominent when $\ell \sim \lambda$. In weaker samples, though,  the ultra-narrow modes can be understood without the requirement of light interference under conditions of high gain and small sample size~\cite{syn_sushil04}.

Theoretically, random lasing has been studied using various techniques. In the intensity transport regime, the diffusion equation was extensively used with concomitant gain~\cite{syn_wiersma96, syn_sajeevjohn96, syn_carminati07}. Since diffusion can be numerically simulated as random walks of particles, this technique was utilized by assuming a light pulse to comprise a packet of `photons' to understand the spectro-temporal features of random lasing~\cite{syn_genack97, syn_sushil04}. Resonant (field) feedback has been treated by solving Maxwell equations in a disordered environment using Finite-Difference Time-Domain (FDTD) calculations in one and two dimensional systems~\cite{syn_sebbah01, syn_soukoulis00}. The Transfer Matrix Method (TMM) has also been vastly used to study one-dimensional random lasers, with a particular focus on quasi-periodic systems with amplification~\cite{syn_soukoulis99, syn_soukoulis02}.

The important characteristics due to which random lasers particularly stand apart are their strong fluctuations in various physical parameters such as output intensity, lasing frequency, spatial mode extent etc.~\cite{syn_wiersma07, syn_sushil07, syn_lagendijk07}. Out of these, frequency fluctuations are the major impediment in practical applications of random lasers. This thesis investigates one roadmap towards minimization of frequency fluctuations. There exist a few mechanisms by which some control on random lasing frequency has been achieved. Of these, two prominently successful techniques involve (a) shaping the optical pump profile to select a particular mode~\cite{syn_sebbah01, syn_sebbah12} and (b) usage of monodisperse amplifying Mie scatterers~\cite{syn_gottardo08}. The former technique requires apriori knowledge of the spatial profile of a particular mode at the desired wavelength. Thus this technique, successfully demonstrated in coherent random lasers, achieves wavelength {\it selectivity} more than wavelength {\it control}. The latter technique, on the other hand, actually controls the location of modes in frequency. Hitherto, however, this technique has been demonstrated only in diffusive random lasers. In this thesis, we address the frequency behavior of coherent random lasing in a medium with monodisperse, resonant scatterers.

This synopsis is organized in the following manner. Section A  describes Monte Carlo simulations wherein we demonstrate efficient and frequency-controlled coherent random lasing in an ensemble of monodisperse, resonant scatterers with intra-scatterer gain, and propose a practical implementation of such a system. Next, experimental results are presented for polydisperse (Section B) and monodisperse (Section C) amplifying scatterers arranged in a disordered fashion in a linear array and the consequent frequency behavior. Motivated by the current literature on chains of resonators, the practical applicability of this system as a coupled-resonator optical waveguide (CROW) is examined (Section D). After that, we model our system using transfer matrices (Section E) and trace the origin of the modes to the gap states of an amplifying periodic-on-average random superlattice (aPARS). Next, we exploit spectral properties of such systems to achieve single-mode stabilized coherent random lasing (Section F). Finally, the statistics of the spatial extent of modes is further studied (Section G) where direct evidence of localization is obtained.

\section*{\label{sec:level2}Monte Carlo simulations of monodisperse random medium}
We employ Monte Carlo simulations to study the spectral features of a coherent random laser comprising monodisperse spherical scatterers. The algorithm essentially tracks the propagation of photons in a multiply scattering medium, coupled with inversion dynamics. The simulation has two parts: excitation and emission. During the excitation process, pump photons undergo three-dimensional random walks and are absorbed by ground state molecules which cause a local population inversion. In the next stage, spontaneously emitted fluorescent photons also perform random walks and interact with the excited medium and experience gain proportional to their dwell-time in the inverted region. Our system consists of randomly distributed monodisperse spherical scatterers, whose size ranges over $\sim 1 - 5~\mu$m. With these sizes, the particles are able to sustain resonances with appropriate quality factors. In this system, it is instructive to note the comparative behavior of inter-scatterer and intra-scatterer dwell-time ($\tau_d$) distribution, since the gain is proportional to the $\tau_d$, thereby influencing the spectral features. Figure~\ref{fig0p1}[A] shows two representative spectra from a conventional random laser, made of randomly shaped scatterers. The spectra show peaks at random frequencies which differ from pulse-to-pulse. Figure~\ref{fig0p1}[B] shows four spectra for a particle size of $1.09~\mu$m (refractive index contrast, $n = 1.8$) as a function of excitation pump energy for resonant scatterers with intra-scatterer gain. The spectra confirm persistent lasing with stable (well-defined frequency range), efficient (larger intensity in the coherent mode) lasing modes, the lasing frequency being determined by the Mie resonances of the scatterers.

\begin{figure}[h]
\begin{center}
\includegraphics[scale=.58]{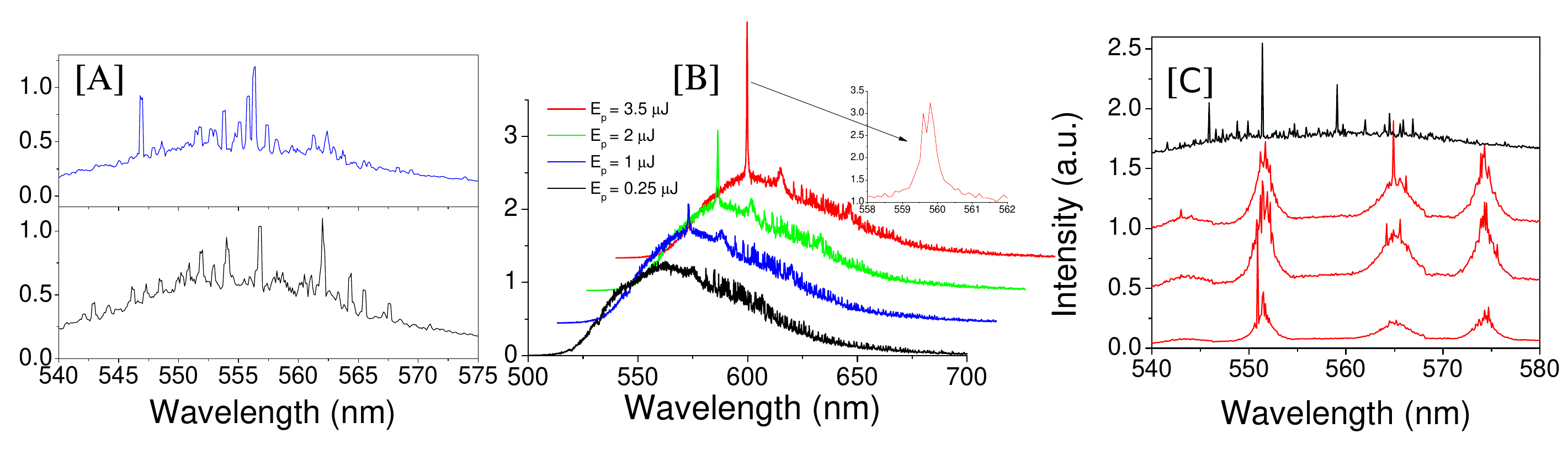}
\end{center}
\vspace{-20pt}
\caption{\label{fig:epsart}[A] Two numerically calculated spectra showing coherent modes from a conventional random laser [B] Emission spectra from a random laser with resonant scatterers with intra-scatterer gain. Inset: Mode exaggerated for clarity. [C] Emission spectra from amplifying Mie scatterers. Red curves: Monodisperse scatterer system. Black curve: With $12~\%$ polydispersity.}
\label{fig0p1}
\end{figure}

We, then, propose an experimentally realistic system capable of generating coherent random lasing, comprising monodisperse Mie scatterers with intra-scatterer gain. We conceive a high gain system that is made of a laser dye (gain cross-section $\sigma_{g}~\sim~10^{-16}$~cm$^{2}$). We assume that spherical microdroplets are created out of this dye, which have a refractive index contrast of $1.33$ (corresponding to the solvent for the dye) with the surrounding medium (air). Thus each microdroplet is an amplifying Mie scatterer. We study the spectral features of an ensemble of such microdroplets in the air or an aerosol. A polydisperse system is first simulated by assuming $12~\%$ spread in scatterer diameter (which was assumed to be $3.6~\mu$m). The black line in Figure~\ref{fig0p1}[C] depicts the spectrum from this system at a pump energy of $30~\mu$J. Lasing peaks appear at random positions. Next, we simulate spectra from the monodisperse system pumped at $7~\mu$J as shown in Figure~\ref{fig0p1}[C]. Multimode emission is observed at specific wavelength intervals. The spectra change from shot to shot indicating that fluctuations continue to exist, but the lasing peaks are restricted to the resonance band of the individual scatterer, which shows that the fluctuations are reduced. The measurements of the lasing peak distribution quantify the control achieved on the lasing wavelength.

\section*{\label{sec:citeref}Practical implementation of an aerosol random laser}
Motivated by the calculation, we created microdroplets using a device called the Vibrating Orifice Aerosol Generator (VOAG). The technique is based on the fracture of mechanical perturbation of an unstable liquid jet~\cite{syn_berglund73}. The schematic of the experimental setup is shown in Figure~\ref{fig0p2}[A]. In our case, the liquid is a solution of Rhodamine 6G dissolved in methanol. Using pressurized N$_{2}$  gas, this solution is forced through a micro-capillary having an in-built piezoelectric gate, the regulation of which creates polydispersity or monodispersity in the sizes and separations of the microdroplets. The microdroplet stream is irradiated by the second harmonic ($\lambda = 532$~nm) of an Nd: YAG laser to optically pump the thus-formed scatterers. Importantly, this system assumes the form of a linear array, and the longitudinal emission and the transverse emission have differing character. In the setup, the mirror M2 was placed at an angle of about $45^{\circ}$ from the vertical axis to redirect the longitudinal emission towards the spectrometer. The lens L1 was placed on a translation stage to collect and focus both the transverse and longitudinal emissions onto the entrance slit of the spectrometer, as depicted in the inset of Figure~\ref{fig0p2}[A]. A CCD camera (CCD2) was employed to image the microdroplet array.

\begin{figure}[h]
\begin{center}
\includegraphics[scale=1.35]{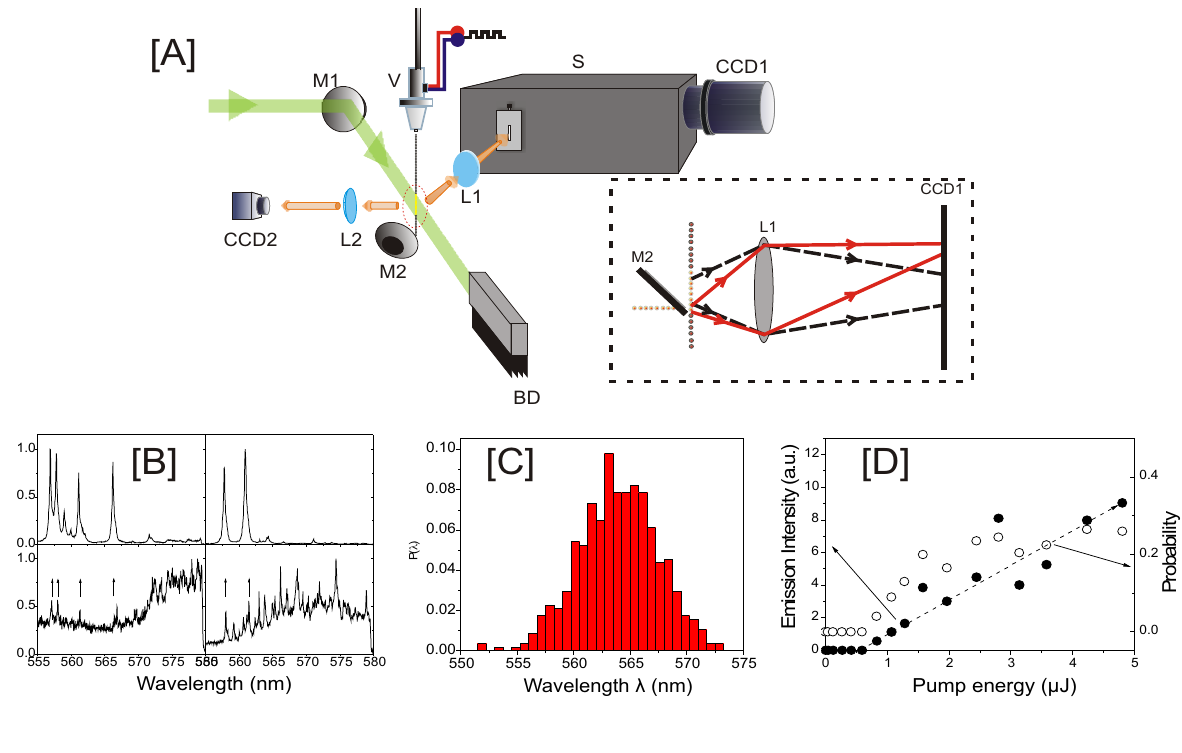}
\end{center}
\vspace{-40pt}
\caption{\label{fig:epsart}Coherent random lasing from polydisperse array of amplifying scatterers. [A] Experimental setup for the observation of coherent emission. Legend: V: Vibrating Orifice Aerosol Generator; M1, M2: Mirrors; L1, L2: Lenses; BD: Beam Dump, S: Spectrometer  [B] Representative emission spectra in longitudinal (top panels) and transverse (bottom panels) direction. [C] Continuous broadband distribution of lasing wavelength. [D] Observed lasing threshold behavior: Solid circles are the output coherent intensity, empty circles are the probability of random  lasing.}
\label{fig0p2}
\end{figure}

Figure~\ref{fig0p2}[B] shows two representative longitudinal spectra in the top panels and two transverse emission spectra in the bottom panels. Ultra narrowband coherent random modes are observed in the longitudinal direction. Every peak in the longitudinal direction has a corresponding weak mode in the transverse direction, as marked by the arrows. These peaks, when analyzed in the spectral images (not shown here), confirm the extent of the coherent modes over the entire array. The rest of the low-intensity peaks in the bottom panels can be traced to individual resonances. We irradiated a continuous microcolumn of the same laser dye with a comparable length and diameter and did not observe generation of coherent modes. This implies that discretization of refractive index plays a vital role in generating optical feedback. The lasing peaks fluctuate in both frequency and intensity. $500$ spectra were taken to analyze the frequency distribution of these lasing modes. As shown in Figure~\ref{fig0p2}[C], the coherent peaks occur at random wavelengths over a range of $\sim 10$~nm. This band is approximately determined by the width of the gain profile and the excitation energy. The variation of coherent intensity in the longitudinal direction as a function of pump energy is shown in Figure~\ref{fig0p2}[D]. Dash line is a linear fit to the coherent intensity, which is used to find the threshold energy. A threshold behavior is observed at $E_{p}=0.5~\mu$J, endorsing the lasing phenomenon.

\section*{\label{sec:citeref}A monodisperse aerosol random laser}
After that, we study the behavior of the above system in a resonant scattering environment. The monodisperse scatterers are generated by applying a periodic perturbation on the piezocrystal inside the capillary, which fractures the liquid jet into equal-sized microdroplets. The monodispersity is confirmed from the Whispering Gallery Modes (WGM's) of the individual microdroplets~\cite{syn_benner80}. One of the embodiments of the monodisperse droplet array is shown in Figure~\ref{fig0p3}[A]. In this case, the average size of the droplet is estimated to be $17.1~\mu$m. Across the array, we estimated a size variation of $\sim 100$~nm, which confirmed a strong monodispersity ($\sim 0.6~\%$). The separations were allowed to remain random as seen in the image.

\begin{figure}[h]
\begin{center}
\includegraphics[scale=1.35]{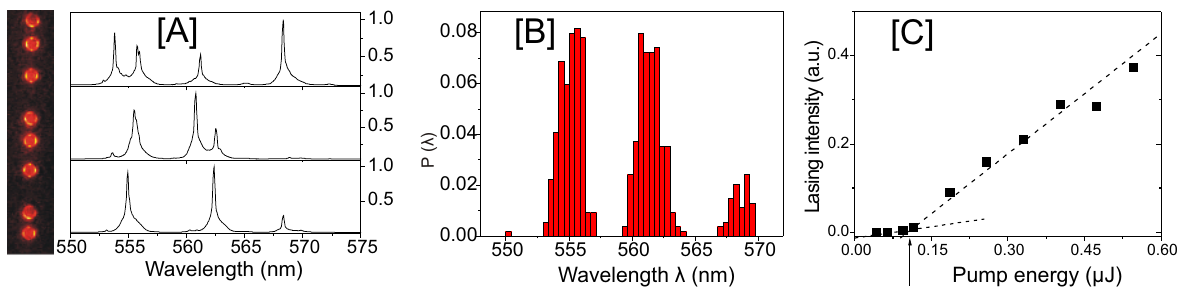}
\end{center}
\vspace{-30pt}
\caption{\label{fig:epsart}Coherent random lasing from monodisperse scatterers. [A] One manifestation of the microdroplet random array and three representative longitudinal emission spectra at a pump energy $E_{p} = 0.7~\mu$J. [B] Histogram of the lasing wavelengths shows discrete bunches. [C] Output intensity as a function of excitation pump energy, the collective mode shows a lasing threshold at $E_{p} = 0.12~\mu$J.}
\label{fig0p3}
\end{figure}

Three representative longitudinal spectra captured from the monodisperse array are shown in Figure~\ref{fig0p3}[A]. The peaks are strongly coherent as the incoherent pedestal is absent. The distribution of random lasing peaks is shown in Figure~ \ref{fig0p3}[B], which shows that these peaks appear only in discrete, specified intervals of wavelengths. Such bunching was observed whenever the scatterers were monodisperse. The width of each bunch is $\sim 2$~nm compared to $10$~nm of the polydisperse array, or of a conventional random laser. Thus, the monodisperse aerosol shows a reduction in frequency fluctuations. Notably, the bunch separation matches with the Fabry-Perot (FP) free spectral range of an individual scatterer and not the whispering gallery modes. Figure~\ref{fig0p3}[C] shows the threshold behavior in these monodisperse arrays, the threshold is almost an order of magnitude smaller than the polydisperse configuration.

\section*{\label{sec:citeref}Collective amplified modes in a bent array of spherical microresonators}

At this stage, we refer to the current literature on linear chains of microsphere resonators. In that regard, a linear array of microspherical resonators has been a system of interest in another field of integrated photonic circuits. Such arrays have been studied as candidates for waveguiding, as the so-called coupled-resonator optical waveguides (CROW's)~\cite{syn_yariv99}. The motivation here is to transmit an optical signal via coupling of successive resonators. Accordingly, linear chains of microspheres has been studied, wherein the coupling is mostly via evanescent modes. Radiative coupling occurs through nanojet-induced modes~\cite{syn_kapitonov07}. Such modes incur bending losses and hence are not considered candidates for bent CROW's~\cite{syn_chen06}. In this project, we considered our system as a potential candidate for active CROW's. The participating resonances are the Fabry-Perot modes and not the surface (whispering gallery) modes, and hence the coupling is direct far-field. We examine the ability of the gain to overcome bending, and other radiative, losses in bent arrays of the amplifying microdroplets. In the experiment, we employ a tapered nozzle which gently blows air on the the droplet stream to bend it in a regulated fashion. The pressure is maintained so as to just bend the stream, and not distort it completely. We achieve continuous bending up to $\sim~25^{\circ}$. Figure~\ref{fig0p4}[A] shows one such configuration where the array is bent at an angle of $\sim~10^{\circ}$.

\begin{figure}[h]
\begin{center}
\includegraphics[scale=0.90]{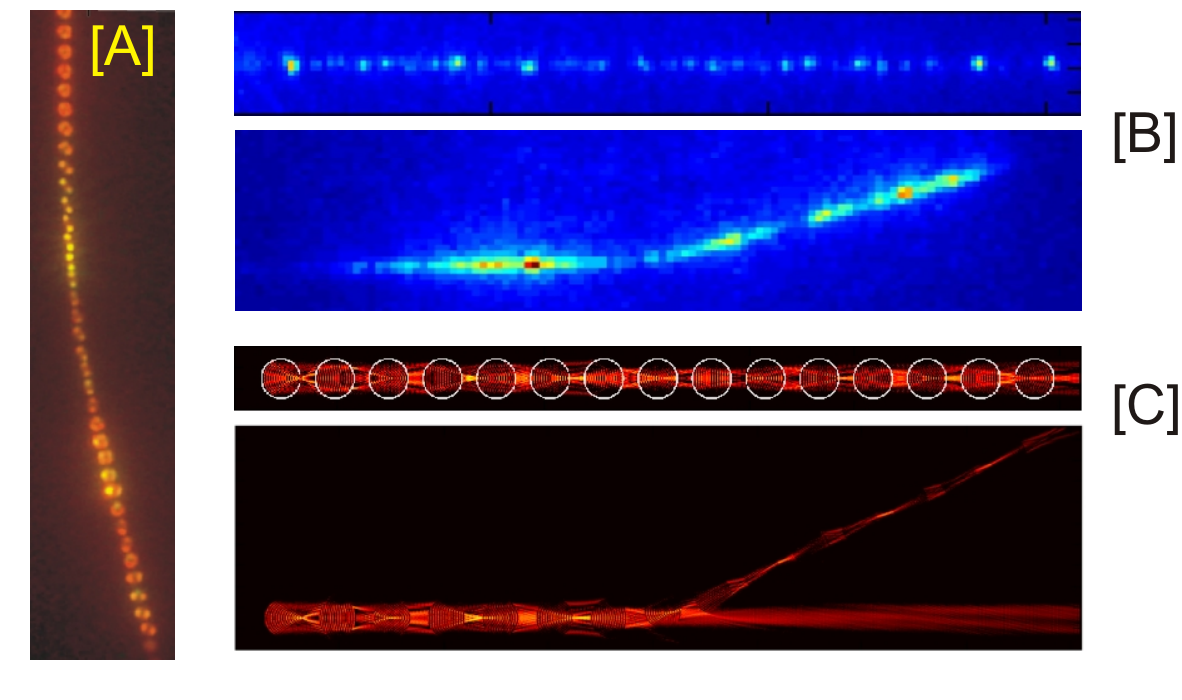}
\end{center}
\vspace{-20pt}
\caption{\label{fig:epsart}[A] An experimental realization of the bent array at an angle of  $\sim 10^{\circ}$. [B] Experimentally observed spatial profile of the extended mode using an ICCD attached to a spectrometer, each dot is a droplet and the array acts as a carrier of light. [C] Two mode profiles calculated using FDTD in a linear and bent arrangement.}
\label{fig0p4}
\end{figure}

Figure~\ref{fig0p4}[B] shows experimentally captured coherent modes for the straight and the bent array. It is clear from the image that all the resonators coherently participate and the bending does not prohibit the existence of the carrier mode. The dependence of the carrier mode on the bending angle was quantified by counting the number of modes at a given excitation energy, which roughly was halved when the bending angle was about $25^{\circ}$. Nonetheless, the gain in the system overcomes the far-field losses as well as bending losses and excites the mode across the entire array, which can even be upto $60$ microspheres long. Since it was difficult to experimentally measure the transmitted intensity and hence the bending losses, we carried out two-dimensional FDTD calculations to get better insights on the transmission. Figure~\ref{fig0p4}[C] shows a propagating mode excited at FP wavelength in the passive system. The bending loss is evident, and increases with angle. We have observed the reduction of these bending losses by adding gain in the simulations.
Further, we find that there is an advantage created by the spherical surface as against conventional plane FP resonators. As can be seen in the image, the mode is not necessarily centered on the axis of the array after the bend, but propagates via off-axis lensing from the microspheres. Such CROW's could be of interest in optofluidic devices.

However, the experimentally observed intensity variation along the array differs from the expected behavior. This difference can be traced to the origin of the collective longitudinal modes.

\section*{\label{sec:citeref}Origin of the longitudinal modes}
We then carried out theoretical activity to trace the origin of these random lasing modes. Since only the longitudinal emission possessed these unusual characteristics, and besides, since the frequency signatures in the emission were of a Fabry-Perot character, we only modeled the transport in one dimension. To that end, the droplet array was assumed to be a one-dimensional amplifying dielectric multilayered structure. The propagation of the electromagnetic field is then accurately described by the transfer matrix (TM) method. As described in the Introduction, the TM method has been extensively used to understand localizing and amplifying properties of one-dimensional random media~\cite{syn_soukoulis99}. In this method, successive multiplications of these transfer matrices provide transmission properties of the entire multilayer. Frequency characteristics of both passive and active systems can be analyzed using this method.
\begin{figure}[h]
\begin{center}
\includegraphics[scale=0.58]{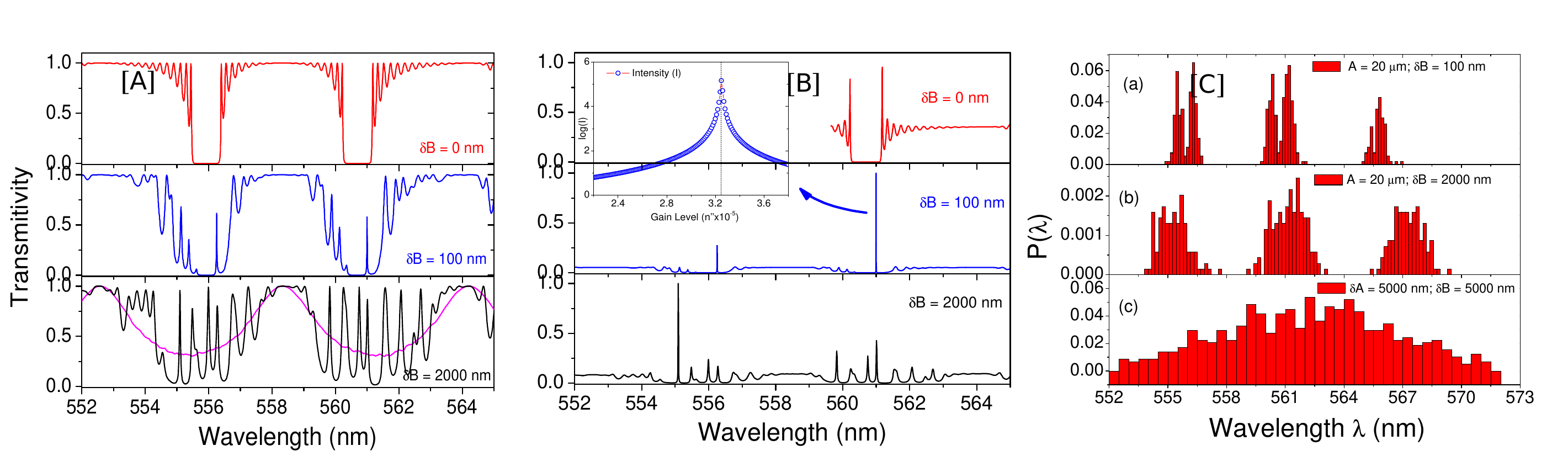}
\end{center}
\vspace{-30pt}
\caption{\label{fig:epsart}Transfer matrix calculations. [A] Spectra showing passive transmittance from an ordered and two random configurations. The magenta curve shows configurationally-averaged transmittance over $1000$ realization for the strongly random system. [B] Spectral profile for the same configuration under amplification, the inset shows the output intensity as a function of gain level $n^{\prime\prime}$ to identify the threshold. [C] Histogram of lasing wavelength for (a) a monodisperse system with weak disorder in spacing (b) a monodisperse system with strong disorder in spacing and (c) for polydisperse system with strong disorder in spacing.}
\label{fig0p5}
\end{figure}

Figure~\ref{fig0p5}[A] shows the transmission spectra from passive multilayers with a varying degree of randomness for a system of $20$ unit cells. Each unit cell has a dielectric layer of thickness $A = 20~\mu$m with an inter-layer spacing $B$ of $6~\mu$m. For a perfectly ordered configuration (red curve), two stop bands appear in the illustrated region of interest. Next, we introduce disorder in the interlayer spacing (quantified by $\delta B$, such that the separations are uniformly distributed over [$B-\delta B/2, B+\delta B/2$]) while keeping the thickness monodisperse. For weak randomness, the earlier band-edge modes migrate into the stopband region. At larger degree of positional disorder, several new modes are generated in the gap with randomly distributed quality ($Q$) factors. The frequency sensitivity of the passive system is revealed by the magenta curve which shows the configurationally averaged transmittance over $1000$ realizations. The separation between the two high transmission peaks is equal to the free spectral range (FSR) of an individual layer. Next, we introduced optical gain to the same system, and the spectra are shown in Figure~\ref{fig0p5}[B]. In the perfectly ordered configuration, the band edges have the highest $Q$ and hence the smallest threshold for lasing. In the weakly periodic configuration, the highest $Q$ modes are the perturbed band-edge modes. Under stronger disorder, multiple lasing modes are generated in the stopband region. We identified the frequency distribution of these lasing modes over $500$ realizations (Figure~\ref{fig0p5}[C]). A weakly disordered system shows a double-peaked bunch centered at the stopband of the underlying ordered configuration; these two bunches originate from the perturbed band-edges. When the sample is made slightly more disordered, the memory of the band-edges is washed out. Finally, when both the sizes and separations are strongly randomized, the frequency-sensitivity is lost, and the lasing peaks are distributed over a wide, continuous range. Comparison with experimental data shows that the separation between histogram bunches is equal to that seen in the experiments under identical layer thicknesses. Thus, the TM calculations reveal the origin of the experimentally observed lasing modes as arising from the gap states of these systems, known as amplifying periodic-on-average random systems.

\section*{\label{sec:citeref}Single-mode, stable coherent random lasing}
With the identification of the origin of the lasing modes, we explore the possibility to obtain single mode and stable coherent random lasing. In this scenario, stability implies a minimum fluctuation in lasing frequency, while single mode means that only one mode in the entire frequency band crosses the threshold and lases. In a gain environment, the lasing frequency depends on two main parameters, namely (i) Spectral position of high $Q$ modes and (ii) location of the gain maximum. The latter is a molecular property and cannot be altered, but for minor changes caused by the concentration of the lasing dye. Our study on aPARS shows that the highest Q modes appear at the minima of the underlying FP profile of the periodic system, and can hence be controlled by using appropriate system parameters.

\begin{figure}[h]
\begin{center}
\includegraphics[scale=.58]{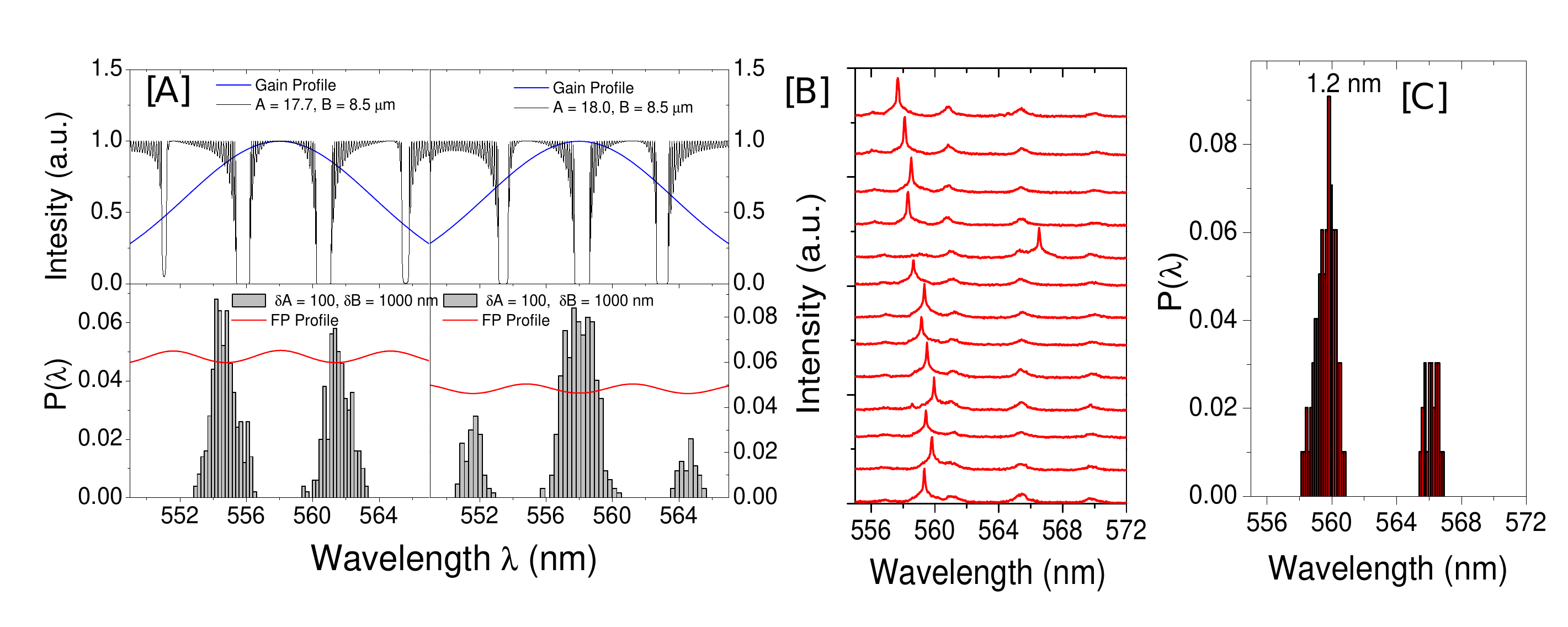}
\end{center}
\vspace{-30pt}
\caption{\label{fig:epsart}[A] Blue curves show the gain profile. Top left: Black curve is the spectrum from a periodic system with later thickness $17.7~\mu$m. Bottom Left: Red curve is the Fabry-Perot profile of a single dielectric layer; the histogram shows that the first lasing modes are equally populated in both bunches. Top right: Black curve shows the spectrum from a periodic system with layer thickness of $18.0~\mu$m. Bottom right: Red curve: Fabry-Perot profile of a single layer; the histogram shows that the lasing peaks are highly populated at the gain maximum. [B] Experimentally stabilized single mode coherent random lasing. [C] Histogram of the lasing wavelength, the fluctuations have been reduced to $\sim 1.2$~nm.}
\label{fig0p6}
\end{figure}

Figure~\ref{fig0p6} illustrates the idea to achieve stable coherent random lasing via mode matching. Mode matching refers to matching one of the minima of the FP profile with the maximum of the gain curve. (In conventional lasers with resonators, the cavity length is chosen so that one resonant mode matches with the gain maximum. In our case, the FP-profile of the single cell determines the stopband positions.) In Figure~\ref{fig0p6}[A], the blue lines show the frequency dependence of the gain profile. The black transmission spectrum (top left panel) arises from 30 monodisperse layers arranged in a perfectly periodic fashion. The size of each layer is $17.7~\mu$m with an inter-layer spacing of $8.5~\mu$m. The red line (bottom left) indicates the FP profile of a single layer. $500$ spectra are analyzed to study the frequency behavior of first lasing mode with $\delta A = 100$~nm and $\delta B = 1000$~nm. The distribution of lasing peaks (bottom left) shows two equally populated bunches, implying that the lasing emission randomly toggles from one frequency interval to another. Then we changed the slab width to $18~\mu$m to match the minimum of FP profile with the maximum of the gain curve (right panels). The resulting histogram shows the lasing distribution where $76~\%$ of the modes populate in the center bunch. Thus, a single mode is emitted per pulse within a range of $\sim2$~nm with a probability of $76~\%$. It should be noted that, since the gain exists only in the droplet and not outside, any randomness in the separations does not affect this behavior, apart from mildly widening the central bunch.

In the experiment, we changed the droplet diameter by varying the applied pressure and the perturbation frequency of the piezocrystal at the gate of the microcapillary. Figure~\ref{fig0p6}[B] shows a set of experimentally observed spectra which depict single mode lasing, with a very small variation in the lasing wavelength within a narrow band of $\sim 1.2$~nm. We found that over a small ensemble of about $20$ spectra, $\sim~90\%$ yield stable modes within the same band. In larger ensembles of over $100$ spectra (Figure~\ref{fig0p6}[C]), a stability of over $75~\%$ was obtained. This technique also requires the gain coefficient to be appropriately maintained, so as to avoid multimode emission over the lower quality resonators as well. Thus using a mode-matched aPARS, we have experimentally achieved stable, single mode random lasing over $1.2$~nm, compared to a $10$~nm band as seen in coherent random lasers.

\section*{\label{sec:citeref}Spatial extent and mode profile of random lasing modes in aPARS}
Transfer matrix calculations reveal that, in a periodic-on-average random superlattice (PARS) system, the localization length $\xi$ is sensitive to the frequency. In the vicinity of the band gap, $\xi < L$, and the system is capable of localizing the fields. The variation in the $\xi$ can be expected to appear in the spatial extent of the modes too. Figure~\ref{fig0p7}[A] shows the calculated transmission spectrum from a PARS consisting of $40$ unit cells. The right panel exaggerates the relevant spectral region. The magenta curve shows the band-structure of the underlying periodic system. The transmission spectrum of the random system has several peaks, out of which the three highest quality modes appear at $\lambda = 558.2, 558.9$ and $557.9$~nm. Figure~\ref{fig0p7}[B] shows the calculated normalized intensity distributions for these high $Q$ modes. The colour of the depicted modes corresponds with the marker arrows in the exaggerated spectrum.

\begin{figure}[h]
\begin{center}
\includegraphics[scale=.6]{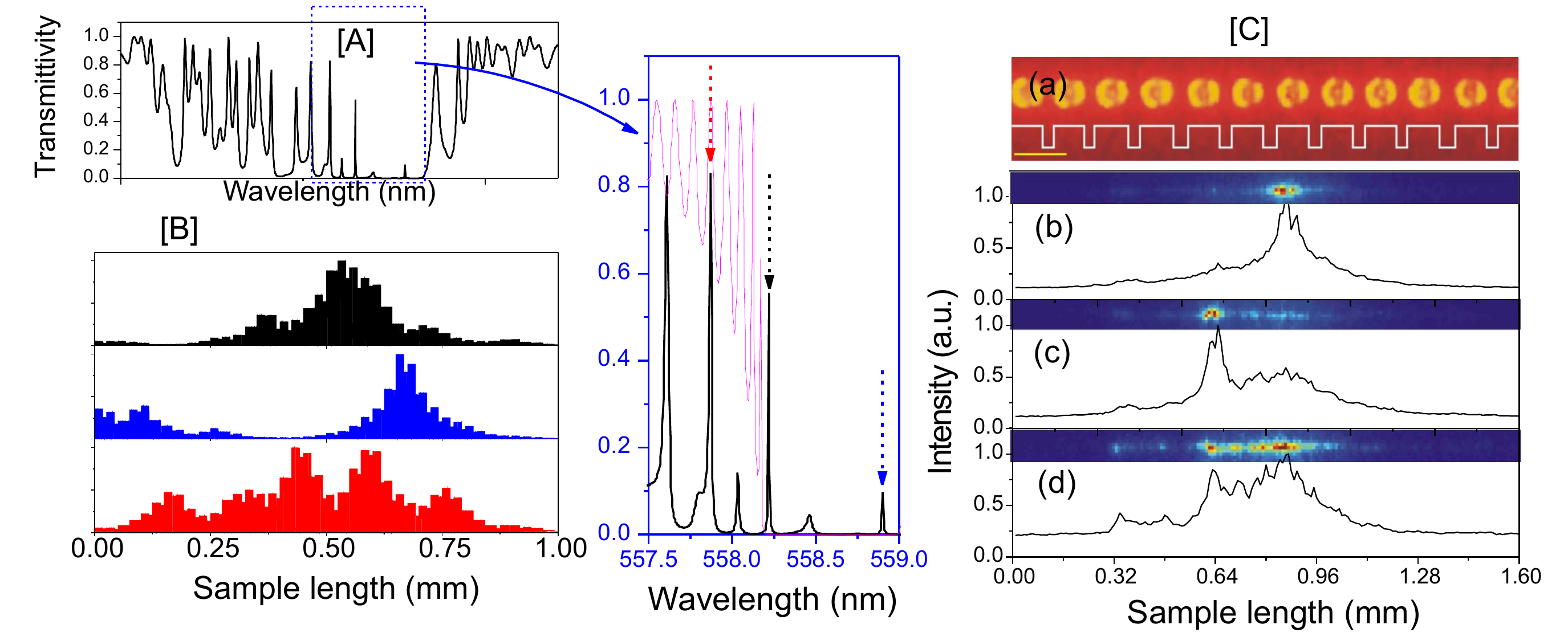}
\end{center}
\vspace{-30pt}
\caption{\label{fig:epsart} [A] Calculated spectrum from a passive PARS system. Right Panel:  Exaggeration of the region of interest, showing three highest quality lasing peaks. Purple plot  shows the underlying band-structure of the periodic system. [B] Spatial mode profiles (intensity) of these three lasing peaks, identified by the dotted arrows in same color. [C] Top: Image of the microdroplet array, and the axial refractive index profile. Bottom: Spectral images and intensity profiles of three lasing modes from the experimental aPARS system at a single disorder configuration.}
\label{fig0p7}
\end{figure}

We see that the tightly localized mode (blue profile) originates from the state deep inside the gap, while the weakly localized mode (black) is close to the passband region. A mode within the passband region is almost extended over the entire sample. Importantly, within one configuration, several lasing modes with strongly varying spatial extents can occur. Figure~\ref{fig0p7}[C](b)-(d) shows the experimentally observed mode profiles from a weakly random system, all observed from the same configuration under uniform pumping. Similar characteristics of varying spatial extent are seen, and some clear tightly-localized modes (b) are also observed. Thus, we conclude that the quasi-one-dimensional aPARS is capable of localizing light, and the observed lasing modes are obtained over amplification of localized states in a quasi-one-dimensional disordered medium. We, then, quantified the spatial extent of the modes by constructing the distribution of the Inverse Participation Ratios ($P_{2}'$s) of the localized modes, and we found that the weakly disordered system exhibited a broader distribution compared to the strongly disordered one. This behavior is consistent with the behavior of calculated localization lengths in one-dimensional random systems.

\section*{\label{sec:summary}Summary}
In summary, this thesis provides an experimental demonstration of a scheme to control the lasing frequency of a coherent random laser. This has been achieved by creating a linear disordered array of monodisperse microresonators. The co-operative longitudinal emission in the system obeys the effective quasi-one-dimensional multilayer formation along the axis. Given the control on sizes and separations of these resonators, we have successfully created an amplifying periodic-on-average random system. Such systems have been extensively studied in theory mostly in the passive domain, and experimental demonstrations are few, that too only in passive systems~\cite{syn_freilikher95, syn_deych98, syn_daozhong94}. Further, the experimental demonstrations in a multilayer do not allow for configurational averaging due to the non-availability of a large number of samples. Ours would be the first experimental demonstration of an amplifying PARS system. The inherent dynamic nature of our system allows for configurational averaging, and studying averaged quantities that are of prime interest in the physics of disordered systems. Thus, the experimental achievement of such a system, that is also easy to implement (not requiring sophisticated fabrication techniques), is a prime message of this thesis. Furthermore, a significant goal in the field of random lasers, namely, lasing over photon localized modes, has been conclusively achieved in these experiments. Novel physical results related to the distribution of the spatial extent of localized modes in amplifying disordered systems have been provided. We hope these results trigger interest in other groups worldwide to further study random lasing in such nearly periodic systems.
\end{synopsis}
\subsection*{Publications contributing to this thesis}

\begin{itemize}

\item {\bf Anjani Kumar Tiwari}, Ravitej Uppu and Sushil Mujumdar, {\it ``Frequency behavior of coherent random lasing in diffusive resonant media,''} {\bf Photonics and Nanostructures - Fundamentals and Applications}, {\bf 10}, Issue 4, pp. 416-422 (2012).

\item {\bf Anjani Kumar Tiwari}, Ravitej Uppu and Sushil Mujumdar, {\it ``Aerosol-based coherent random laser,''} {\bf Optics Letters}, {\bf 37}, 1053 (2012).

\item {\bf Anjani Kumar Tiwari}, Balu Chandra, Ravitej Uppu and Sushil Mujumdar, {\it ``Collective lasing from a linear array of dielectric microspheres with gain,''} {\bf Optics Express}, {\bf 20}, 6598 (2012).

\item {\bf Anjani Kumar Tiwari} and Sushil Mujumdar, {\it ``Random lasing over gap states from a quasi-one-dimensional amplifying periodic-on-average random superlattice,''} {\bf Physical Review Letters}, {\bf 111}, Issue 23, 233903 (2013).

\item {\bf Anjani Kumar Tiwari}, K. Shadak Alee, Ravitej Uppu and Sushil Mujumdar, {\it ``Single-mode, quasi-stable coherent random lasing in an amplifying periodic-on-average random system,''} {\bf Applied Physics Letters}, {\bf 104}, 131112 (2014).

\item {\bf Anjani Kumar Tiwari}, et al., {\it ``Direct observation of Anderson localization-induced random lasing,''} {\bf Manuscript to be prepared}.

\item {\bf Anjani Kumar Tiwari}, Ravitej Uppu, and Sushil Mujumdar, {\it ``Experimental demonstration of small-angle bending in an active direct-coupled chain of spherical microcavities,''} {\bf Applied Physics Letters}, {\bf 103}, 171108 (2013).

\end{itemize}

\subsection*{Other publications}

\begin{itemize}

\item Ravitej Uppu, {\bf Anjani Kumar Tiwari} and Sushil Mujumdar, {\it ``Identification of statistical regimes and crossovers in coherent random laser emission,''} {\bf Optics Letters}, {\bf 7}, 662 (2012).

\item Rajesh V. Nair, {\bf Anjani K. Tiwari}, Sushil Mujumdar and B. N. Jagatap, {\it Photonic-band-edge-induced lasing in self-assembled dye-activated photonic crystals,''} {\bf Physical Review A}, {\bf 85}, 023844 (2012).

\item Rajesh V. Nair, {\bf Anjani K. Tiwari}, Sushil Mujumdar and B. N. Jagatap, {\it ``Inhibition and enhancement of spontaneous emission using photonic band gap structures,''} {\bf Advanced Material Letters,} {\bf 4}, Issue 6, pp. 497-501 (2013).

\end{itemize} 

\chapter{Introduction}

\section{\label{sec:level1p1}A short history of the random laser}

Wave propagation in a random scattering environment is a subject with a long history~\cite{ishimaru78, hulst81, ping90}. Many natural media like the atmosphere, the ocean surface, clouds, fog, snow, etc. have inherent spatial inhomogeneities which vary randomly in space and time. Due to this inherent disorder, wave propagation becomes chaotic. The concept of disorder in optics was historically related to these media and related phenomena. However, in recent years, a field by the name mesoscopic optics has taken form, which relates more to the optical counterpart of disorder-oriented phenomena in condensed matter physics.
Perhaps the most exotic of these phenomena are Anderson Localization.
In 1958, Philip Anderson introduced the concept of electron localization in disordered systems and explained the disorder-induced metal-insulator transition~\cite{anderson58, lagendijk09}. He revealed that localization arises due to the quantum interference effect of electrons and the impurities assist coherent multiple scattering and subsequent interference of electron wave. As the localization phenomenon arises only due to the wave nature of electrons, it holds true for any kind of wave propagating in the random scattering environment. Accordingly, in the recent past, there have been extensive studies of disorder-induced localization of classical waves such as electromagnetic radiation, sound waves, elastic waves, etc.~\cite{ping90, john83, graham90}. Of these, the larger body of literature exists in the optical domain.

Over conventional electron localization, the study of light localization has additional advantages. For example, in the case of electrons, a sub-Kelvin temperature is required to produce interference effects, while optical experiments can be carried out at room temperature. Further, as photons are non-interacting unlike electrons, which interact through Coulombic forces, pure localization effects can be studied using optical waves. Moreover, the intensity and wavelength of photons can be controlled to very high finesse which is not possible for electrons in a metal. Another huge advantage with optical waves is the possibility to directly measure various properties like statistical distribution of intensity, temporal response, etc. Hence, the experimental study of localization with optical waves is much easier to perform than with electrons and has thus attracted considerable interest in the optics research community.
In 1984, S. John first introduced localization of optical waves~\cite{sajeev84}. At the same time, coherent backscattering (which is the signature of localization) was observed by Albada and Lagendijk~\cite{albada85} and Maret and Wolf~\cite{wolf85}. In 1987, S. John proposed strong localization of photons in the disordered dielectric superlattices~\cite{sajeev87, sajeev91}.

The entry of structured disorder in lasers is a rather intriguing story. At the heart of a conventional laser is an amplifying medium and a resonator. An optical resonator plays the most crucial role in lasing action and is responsible for coherence. The significance of resonators triggered a huge research activity in the 1960s~\cite{siegman99a, siegman99b}. It was believed for a long time that any inhomogeneity in the amplifying medium is detrimental to laser action as they induce scattering losses. However, in 1966, a group led by Letokhov demonstrated a new type of amplifying phenomenon where feedback was provided by a `bad' resonator~\cite{Ambartsumyan66}. Following that, in 1968, Letokhov combined the idea of multiple scattering and gain and theoretically predicted the amplification of light from a random scattering medium~\cite{letokhov68}. In such systems, the light was fed back into the amplifying region by multiple scattering via `nonresonant feedback,' a term which will be clarified later. This idea received a massive boost in 1994 when Lawandy et al. reported laser-like emission and linewidth narrowing from an alcoholic solution of Rhodamine 6G by adding TiO$_2$ microparticles~\cite{lawandy94}. They observed that upon increasing the pump energy, the emission intensity diverged and the bandwidth collapsed to $\sim 10$~nm. It was during this time that the system acquired the name random laser. This work stimulated the area of random lasers and similar observations were made by several groups in various dye-scatterer and other systems like laser crystal powders, etc.~\cite{gouedard93, noginov95, siddique96}. A variety of new phenomena in these systems were discovered, making the field of random lasers an integral component of mesoscopic optics.

\section{\label{sec:level1p2}Conventional lasers and resonant feedback}

Conventional lasers consist of a pumping source, a gain medium, and a well-defined resonator cavity~\cite{siegman86}. The pumping source is required to establish the population inversion condition in the gain medium. Amplification occurs via stimulated emission process when a photon interacts with the inverted gain medium. A photon, while propagating in the inverted gain medium, gathers an intensity exponentially proportional to its propagation length $I \propto \text{exp}(L/l_{g}$). Here, $l_{g}$ is the gain length and is defined as the path length over which the initial intensity is amplified by a factor of $e$.

\begin{figure}[h]
\begin{center}
\includegraphics[scale=1.2]{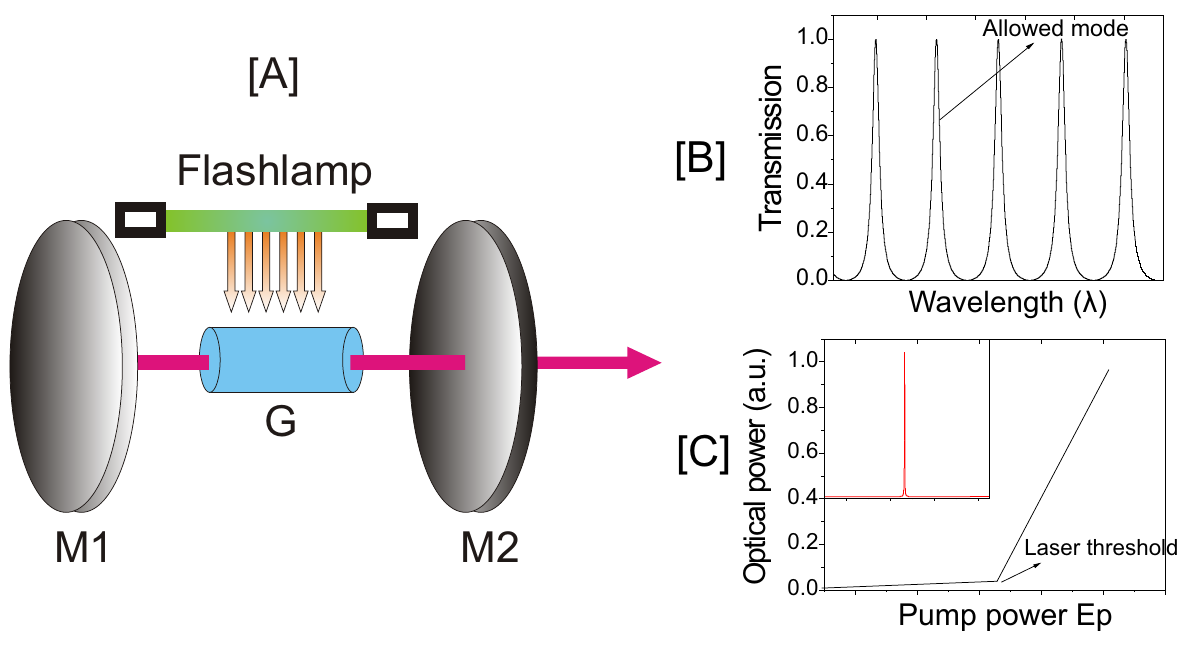}
\end{center}
\vspace{-35pt}
\color{blue}\caption{[A]: Schematic representation of a laser. Legend: G: gain medium. Flashlamp: pumping source to excite the gain medium. M1 and M2: two mirrors to form a Fabry-Perot cavity. [B]: Allowed modes of the cavity. [C]: The threshold behavior of a typical laser, inset shows the intense peak in the spectrum.}
\label{fig1p1}
\end{figure}

Understandably, the optical resonator plays the most important role in the generation of lasing action. It increases the output intensity by feeding the light back into the gain medium and thereby increasing the propagation length $L$. As depicted in Figure~\ref{fig1p1}[A], two parallel mirrors M1 and M2 form a cavity. The operation frequency of the laser depends on the physical parameters of the cavity such as its dimensions, reflectivity of the mirrors, etc. The constructive interference condition is given by
\begin{equation}
2L_\text{cav} =n\lambda
\label{eq1p1}
\end{equation}
where, $L_\text{cav}$ is the length of the cavity, $n$ is an integer multiple and $\lambda$ is the wavelength of light. Clearly, the resonant wavelength gets selectively chosen by the cavity as shown in Figure~\ref{fig1p1}[B]. The cavity fed back light at these wavelengths into the gain medium. If the total gain overcomes the resonator losses, the system crosses threshold and begins to lase. At threshold, the output intensity shows a sudden growth, as shown by the arrow in Figure~\ref{fig1p1}[C] which marks the threshold. It is known that any inhomogeneities within the cavity introduce additional scattering losses and hence are considered detrimental for lasing action.

\section{\label{sec:level1p3}Laser-like emission from disordered media}

\begin{figure}[h]
\begin{center}
\includegraphics[scale=1]{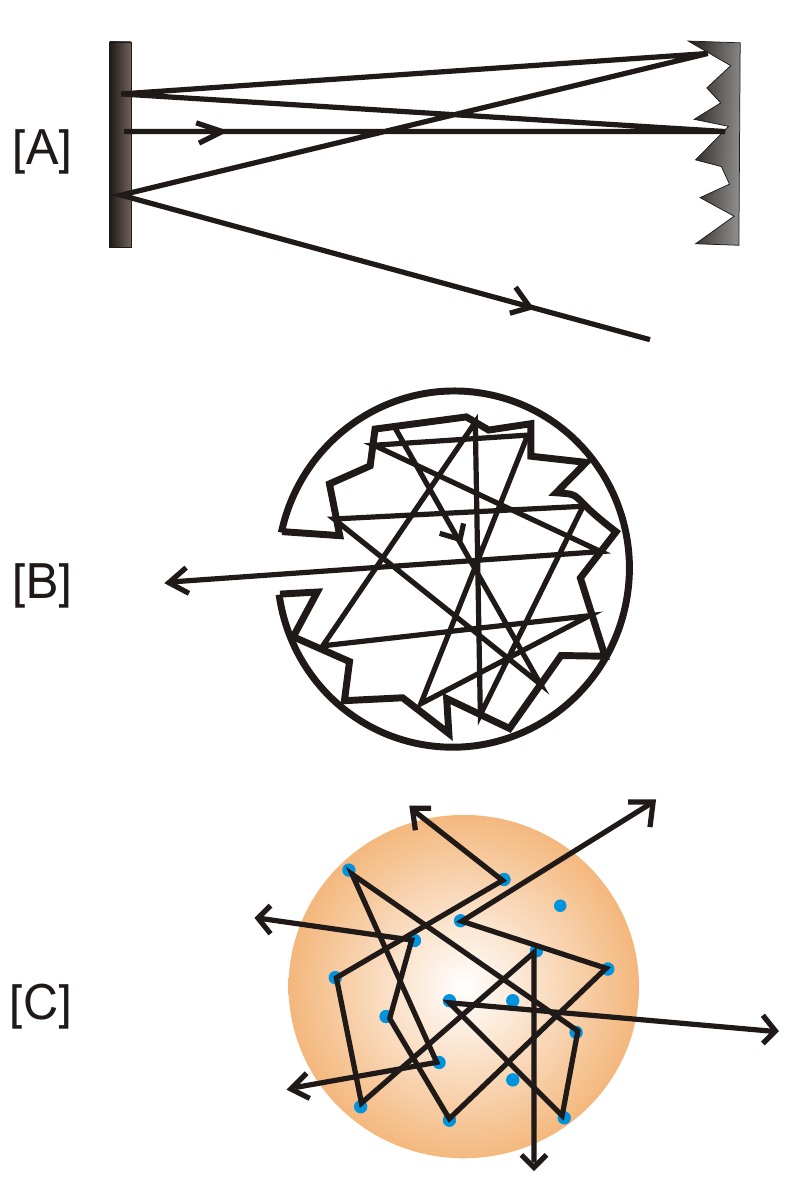}
\end{center}
\vspace{-20pt}
\color{blue}\caption{Design of different cavities with non resonant feedback [A]: A Fabry-Perot cavity formed by a mirror and a scattering surface. [B] A circular cavity with scattering inner wall and a small outlet. [C] Randomly arranged scatterers in an amplifying medium, the gain region is indicated by the yellow color.}
\label{fig1p2}
\end{figure}

Clearly, in a conventional laser, the length of the cavity determines the lasing frequency and hence it is sensitive to the operational conditions like mechanical vibration, thermal expansion and so on. The group led by Letokhov proposed a new type of laser which was based on a unique feedback mechanism devoid of resonances~\cite{Ambartsumyan66,letokhov68}. Figure~\ref{fig1p2}[A] depicts the idea where one of the FP mirrors is replaced by a diffused scattering surface. This configuration forbids the possibility of forming a stable resonance. In fact, the rough surface is responsible for many low-quality modes which overlap in frequency and thus wash out the possibility of any specific resonance. Interestingly, even in this situation, under strong pumping, the emission intensity gets enhanced at the gain maximum. Figure~\ref{fig1p2}[B] depicts another schematic where the inner wall of an inhomogeneous circular cavity provides similar feedback.

Letokhov proposed the amplification of light in a random scattering environment with gain (see Figure~\ref{fig1p2}[C]). He considered an amplifying random scattering environment where the spontaneously emitted light experiences both multiple scattering and amplification. The relevant length scales to quantify the scattering phenomenon are the scattering mean free path and the transport mean free path. As these length scales are frequently used in this thesis, we first define these quantities. The scattering mean free path $l_{s}$ determines the average distance between two scattering events and is defined as
\begin{equation}
l_s =\frac{1}{N\sigma_{s}}
\label{eq1p2}
\end{equation}
where $N$ is the scatterer density and $\sigma_{s}$ is the scattering cross-section of the individual scatterer. Since $l_{s}$ does not contain the information of scattering anisotropy a more relevant parameter in the multiple scattering environment is transport mean free path ($l_{t}$). It is defined as the average distance traveled by the photons before their direction of propagation gets completely randomized and is given by

\begin{equation}
l_{t} =\frac{l_{s}}{1-g}
\label{eq1p3}
\end{equation}
where $g$ is the angular anisotropy parameter in scattering and given by $g$ = $\langle cos\theta \rangle$, where $\theta$ is the scattering angle. In a random scattering medium, the actual path length traveled by the photon is much larger than the chord distance between the starting and the ending points. The total distance between the starting and the ending points of photons in the gain region determine amplification. This frames the concept of amplification length ($l_\text{amp}$). If the gain length of the photon is $l_{g}$, then the amplification length is given by
\begin{equation}
l_\text{amp} =\sqrt{\frac{l_{t}l_{g}}{3}}
\label{eq1p4}
\end{equation}

In the diffusive scattering regime ($\lambda<l_{t}<L_\text{sys}$, where $L_\text{sys}$ is the length of the system), the propagation of light can be described by the random walks of photons, and the dynamics is governed by the diffusion equation

\begin{equation}
\frac{\partial^{2} I(r,t)}{\partial t} = D\nabla^{2} I(r,t) + {\frac{v}{l_\text{amp}}I(r,t)} + S(r,t)
\label{eq1p5}
\end{equation}

where $D = vl_{t}/3$, is the Boltzmann diffusion coefficient and $v$ is the transport velocity of light in the medium. The second term on the right side describes the amplification and $S(r, t)$ represents the source term. Letokhov found that the solution of the above equation diverges at some value of gain. He showed that, in a diffusive amplifying medium, if the volume gain is larger than the surface losses, the system crosses the threshold of random lasing~\cite{letokhov68}. In a slab geometry, the critical thickness $L_{cr}$ above which the system starts to lase is given by~\cite{wiersma96}

\begin{equation}
L_{cr} =\pi l_\text{amp} = \pi\sqrt{\frac{l_{t}l_{g}}{3}}
\label{eq1p6}
\end{equation}
This innovative idea of mirrorless resonator thus opened a new possibility of lasing with nonresonant feedback. Early experiments on this proposal were performed on laser powders~ \cite{varsanyi71, fork74, fork79} and the existence of scattering-enhanced stimulated emission was confirmed. The first breakthrough result of significant gain narrowing at threshold was reported by Lawandy et al.~\cite{lawandy94}. They observed a substantial spectral and temporal narrowing in a dye solution with the addition of titanium dioxide nanoparticles. This system eventually has been termed as a `random laser', since, after a certain pumping threshold, it exhibits laser-like properties of intensity enhancement and bandwidth narrowing assisted by randomness. The feedback generated by the randomness is termed `nonresonant' feedback. 

\section{\label{sec:level1p4}The coherent random laser}

\begin{figure}[h]
\begin{center}
\includegraphics[scale=1]{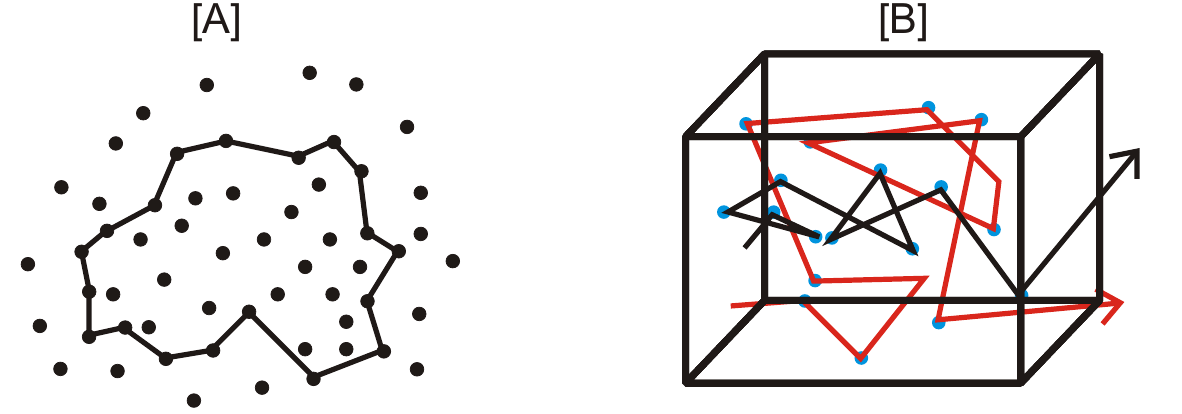}
\end{center}
\vspace{-20pt}
\color{blue}\caption{[A] A schematic diagram of the formation of a closed loop path through the recurrent scattering of light in ZnO powder. [B]: Trajectories of photons under diffusive propagation modeled as random walks.}
\label{fig1p3}
\end{figure}

While the above-mentioned system was termed a laser, it still lacked a high degree of spatial or temporal coherence. The existence of temporal coherence was first reported by the group led by H. Cao. They used ZnO semiconductor powder to realize the essential conditions of high gain and strong scattering~\cite{cao99, cao02}. They observed ultranarrow lasing
peaks (bandwidth $\sim 0.5$~nm) on top of an incoherent background. The origin of coherent random lasing was attributed to the recurrent scattering of light (see Figure~\ref{fig1p3}[A]), motivated by the fact that the scattering mean free path in the system was less than the emission wavelength. The recurrent scattering arises due to the coherent feedback of light between the scatterers which are the key factor for photon localization in the Anderson model. Such a feedback, where the light wave follows a closed loop path over a randomly formed resonator, is termed resonant feedback. Clearly, this feedback is based on the electric field transport and requires the phase information of the scattered light wave. The phase shift requirement of 2$\pi$ along the loop determines the lasing frequency, essentially, involving resonant feedback in the random medium.

However, subsequent experiments with dye-scatterer systems showed that such sharp lasing peaks can also be observed in very weakly scattering systems that fall in the diffusive transport regime~\cite{cao00}. The explanation for this behavior was provided by Mujumdar and Wiersma et al.~\cite{sushil04}. They showed that diffusion can enhance the lifetime of spontaneous emission photons to an excessive degree. They modeled diffusive propagation as random walks of fluorescent photons that gather gain during propagation. The diffusion process in the finite-sized system, realizes a few long and rare paths(Figure~\ref{fig1p3}[B]), which realize an excessive gain on the virtue of their path length. Since these are statistically rare, the peaks created by the paths do not get averaged out during the complete relaxation of the inverted medium. Thus, they create an intense peak in the spectrum. This model did not involve interferences and explained the coherent peaks in a diffusive scenario, thus involving nonresonant feedback. The frequency of the coherent modes is decided by the spontaneous emission events that excite the amplified extended modes. Nonetheless, further discussions based on resonant feedback persisted in the field, wherein the origin of the modes was attributed to the Fabry-Perot resonances that form between two nanoparticles resulting in equispaced modes~\cite{wu06}. Theoretically, resonant (field) feedback has been implemented by solving Maxwell equations through Finite-Difference Time-Domain (FDTD) techniques in one and two dimensional random systems~\cite{sebbah01, soukoulis00}. The Transfer Matrix Method (TMM) has also been vastly used to study one-dimensional random lasers, with a particular focus on quasi-periodic systems with amplification~\cite{soukoulis99,soukoulis02}.

Over the years, a variety of materials were identified to show random coherent lasing~\cite{polson04, tulek10, turistsyn10, strangi06, noginov05}. In most cases, the scatterers are separately added to the amplifying medium. For example, when nanoparticles like polystyrene or TiO$_2$ are added in a laser dye solution, the nanoparticles act as scatterers, and the dye solution serves as the amplifying medium. On the other hand, in some samples, both scattering and amplification can be provided by the same entity. For example, Nd-doped powders, dye-doped microparticles, etc. serve this purpose. Regardless of the constituent materials, the emission spectra from such systems exhibit strong, temporally coherent peaks that fluctuate in frequency, intensity, spatial mode extent, etc.~\cite{zaitsev07, zaitsev06, hackenbroich05, hcao01, beenakker96, wiersma07, sushil07, lagendijk07, uppu10}. Out of these, frequency fluctuations are the major impediment in practical applications of random lasers. Hence, there is considerable interest in the community in developing techniques to overcome them.

\section{\label{sec:level1p5}Attempts of frequency control in random lasers}

One of the important properties of a laser device is good stability in the intensity and frequency of the lasing modes. The resonator design exactly determines the lasing frequency. Accordingly, a device can be fabricated at any desired wavelength, assuming the gain profile covers it. In contradiction, in the random laser, the inherent spatial disorder rules out this possibility. In general, the emission spectrum of a coherent random laser has distinct multimode lasing peaks which fluctuate over a range of $10-15$~nm determined by the gain profile. For most of the real-life applications, it is desirable for the random laser to have a better control on the number of lasing modes and their wavelength. In that regard, various techniques such as, usage of monodisperse Mie scatterers, shifting the effective gain curve of the amplifying material, shaping the optical pump profile and deliberately creating a point defect, etc. have been demonstrated to control the lasing frequency~\cite{gottardo08, lagendijk11, sebbah12, sebbah13, shiva12, leonetti13, fujiwara13}. Here, we briefly review these techniques.

In one of the earliest attempts of frequency control, Gottardo et al. used self-assembled amplifying Mie scatterers and exploited their resonant features~\cite{gottardo08}. In these experiments, Polystyrene microspheres of diameters larger than the wavelength of light such as $0.9~\mu$m, $1.0~\mu$m and $1.2~\mu$m were used. The inherent polydispersity in the diameters was $\sim 2~\%$. The amplification was incorporated by infiltrating DCM dye (dissolved in ethanol) into the sample. The packing fraction of the sample was $\sim 55~\%$, and the typical thickness of a sample was nearly $100~\mu$m. Multiple scattering events were ensured as the thickness of the sample was much larger than the scattering mean path ($2-3~\mu$m). Under optical pumping, samples with particle diameters of $0.9~\mu$m, $1.0~\mu$m and $1.2~\mu$m exhibited lasing at $602.9$~nm, $637.5$~nm and $616.5$~nm respectively which were all away from the gain maximum. These lasing peaks appeared close to the transmission minimum of the samples. Thus, the lasing wavelength was successfully tuned by changing the diameter of the spherical scatterers showing wavelength control of diffusive random lasers.

Lagendijk et al. presented a way of tuning the diffusive lasing wavelength by absorption engineering~\cite{lagendijk11}. In their experiment, they used a solution of Rhodamine 640P dye in methanol as the gain medium and Titania particles as elastic scatterers. A non-fluorescent dye (Quinaldine Blue, quantum efficiency $0.1~\%$) was added into the system to shift the lasing peak towards the blue wavelength. By changing the concentration of the non-fluorescent dye, they showed a shift of $14$~nm in the peak emission wavelength from $606$~nm to $592$~nm.

Sebbah et al. used gain engineering to achieve frequency control. First, they numerically demonstrated that any desired mode of a coherent random laser can be extracted from the emission spectrum~\cite{sebbah12}. This involved choosing a correct spatial profile of the pump. To this end, they used an iterative optimization algorithm. The basic idea behind this technique
was to feed the gain into a preselected mode. They successfully showed that the emission of a coherent random laser can be brought into the single mode regime by this selective excitation process. Furthermore, the threshold of the selected modes was lower when the system was excited with the optimized pump, as compared to a uniform pump profile. This method was shown to work also for weak scattering media, where the modes were spatially extended and therefore strongly overlapping. Later on, they carried out experiments to demonstrate this technique~\cite{sebbah13}. The active control on the pump was achieved by a spatial light modulator (SLM). Single mode operation was shown in an optofluidic random laser of 3~mm long structured PDMS~\cite{shiva12}. Clearly, this method is applicable only for static systems and requires {\it a priori} knowledge of the spatial profile of a particular mode. Furthermore, the basic principle behind this technique is wavelength {\it selectivity} rather than wavelength {\it control}.

In another technique, Fujiwara et al. proposed that the random lasing properties can be controlled by intentionally creating a defect site in the random sample of monodisperse scatterers~\cite{fujiwara13}. Their random sample contained submicrometer sized ZnO nanoparticles which act as scatterer and gain medium. To incorporate the point defects, a small amount of green fluorescent polystyrene particles ($900$~nm) were added into the solution. The fluorescence of polystyrene allows locating the point defects. Upon optical excitation, the system showed unique lasing properties like suppression of the number of lasing modes, lowering of the threshold, and control of lasing wavelength. The wavelength control in the random laser was attributed to the average resonant properties of the ZnO particle film, whereas, the location of the defect determined the spatial position of the lasing sites.

\section{\label{sec:level1p6}Light localization}

Anderson et al. pointed out that the dimensionality of the disordered system is a crucial aspect in the physics of localization~\cite{abrahams79}. Based on the scaling theory, they suggested that, if the dimension of the system is less than or equal to two, all the states can be localized, provided that the system is sufficiently large. On the other hand, the 3D system requires certain critical disorder strength. The condition on the critical disorder is known as Ioffe-Regel criterion ($kl_{s} \leq 1$, where $k$ is the wave vector of photon and $l_{s}$ is the scattering mean free path)~\cite{iofee60}.

In the recent years, there have been many attempts to observe Anderson localization of light. Here, we briefly discuss some of the important results. To observe strong localization in 3D, Wiersma et al. prepared a strongly scattering system from micrometer-sized GaAs semiconductor particles (refractive index = $3.48$)~\cite{wiersma97}. By grinding semiconductor powders, samples with different particle sizes ($10~\mu$m, $1~\mu$m and $300$~nm) were prepared, which offered a range of scattering strengths. In the experiment, the transmission coefficients as a function of sample thickness were measured in diffusive as well as in the localized regime. It was observed that, in the weaker sample, the transmission coefficient decreased linearly implying diffusion. On the other hand, in the stronger samples, it followed an exponential decay, a hallmark of Anderson localization. While this was the first experimental report on Anderson localization in the near visible regime, it also generated a discussion regarding the contribution of absorption in the sample, which also realizes an exponential decay of transmittance~\cite{maret99}. To overcome the issues related to absorption, the group led by Genack used a different approach to identify localization. They reported that the variance of the normalized total transmission (var(s$_{a}$)) can serve as a tag for localization. They claimed that, regardless of absorption, if g$^{'} \leq  1$, the system achieves localization, where, g$^{'}$ was a new localization parameter and given by g$^{'} \equiv 2/3$ var(s$_{a}$)~\cite{chabanov00}. These results were observed in the microwave domain. In the microwave frequencies, it is easier to achieve localization because the large scatterer sizes allows the fabrication of scatterers with a good control on randomness. Besides, with larger $\lambda$, the Ioffe-Regel criterion is more accessible. Indeed, the same group had already demonstrated photon localization in the microwave regime~\cite{garcia91, genack91}, wherein they had studied transmission and diffusion coefficients and intensity correlations in a random
mixture of aluminum and Teflon spheres. Recently, the group of Maret reported visible light transport in three-dimensional bulk samples comprising of TiO2 nanopowders~\cite{sperling13}. They
measured the time-dependent broadening of an incident laser pulse in the transverse direction of the sample. The evolution of transverse width with time provided a direct measure of the localization which was independent of absorption effects. However, these results need to clarify issues relating to inelastic scattering of light realized by nonlinear interactions~\cite{schffold13}. This inelastically scattered light arrives late and can create an impression of long-time delay of scattered light. 

Lower-dimensional systems have already been investigated for light localization. In the microwave regime, localization effects in a two-dimensional system have been reported~\cite{rachida91} where the scattering was provided by the collection of dielectric cylinders randomly placed between two parallel aluminum plates on a square lattice. Sharp peaks in the transmission spectrum were observed. These sharp peaks had high electric field regions and were attributed to the localized states. More recently, strongly confined Anderson-localized cavity modes were reported in disordered photonic crystal waveguides. Essentially, the system consisted of one-dimensional waveguides fabricated in two-dimensional lattices. By engineering the disorder in the photonic lattice in the vicinity of the waveguide, high quality-Anderson-localized modes (Q$\sim$10$^5$) have been demonstrated~\cite{vollmer07, lodahl10}.

In the current times, a different kind of localization, called transverse localization, is being vigorously investigated. This concept, introduced by Lagendijk et al., describes a situation
where the wave propagates in one direction but is confined in the transverse direction by disorder~\cite{lagendijk89}. This is relatively easier to observe and elucidates a lot of physics of Anderson Localization. Lahini et al. demonstrated transverse localization in a one-dimensional lattice of coupled optical waveguides patterned on an AlGaAs substrate ($5$~mm long disordered photonic lattice)~\cite{lahini08}. In their experiment, a light beam was injected at the input end of a disordered photonic lattice. The beam then evolved as it propagated along the structure and coherently tunneled between the neighboring waveguides. They showed that, in an ordered structure, the light beam follows normal ballistic propagation properties. However, as the disorder strength increases, light tends to localize around the input position. At even higher disorder, they observed an exponentially decaying transverse profile. Segev et al. demonstrated Anderson localization of light in two-dimensional disordered nonlinear photonic lattices~\cite{segev07}. The disordered photonic lattice was constructed from a $10$~mm-long SBN:60 crystal using optical induction technique. By varying the disorder strength of the sample, they showed the transition from the ballistic regime to diffusive transport regime. They observed that, as the disorder strength was increased, the inverse participation ratio (a quantity that indicates the mode spatial extent, also defined later in this thesis) and its relative fluctuations increased. Further, when Anderson localization occurred, the transverse field acquired an exponentially decaying tail from the center of the incident beam.

In the scenario of amplification, the occurrence of light localization had led to the concept of mirrorless lasing via random resonators~\cite{pradhan94}. The earliest claim to the same was made by Cao et al. when they observed ultranarrow coherent lasing modes from ZnO nanopowders. The origin of the ultranarrow modes was attributed to the recurrent scattering of light and Anderson localization, though the claim of Anderson localization was not explicitly supported~\cite{cao99}. In another experiment, V. Milner and A. Z. Genack demonstrated localization induced low-threshold random lasing~\cite {genack05}. The sample consisted of a stack of partially reflecting cover slides of random thickness ($80~\mu$m - $120~\mu$m). The cover slides were arranged in a one-dimensional fashion, and the amplifying layers (Rhodamine 6G solution) were placed between them. When the stack was pumped with a pulse laser, the emitted light gets localized due to multiple scattering between the interfaces. They observed low threshold ultranarrow lasing when the system size was much larger than the localization length.

\section{\label{sec:level1p7}Thesis}

This thesis describes the experimental study of coherent random lasing from a unique system consisting of an amplifying aerosol. The aerosol is achieved in a quasi-one-dimensional fashion with random spacing between the individual liquid microdroplets. We implemented a roadmap towards minimization of frequency fluctuations in this coherent random laser. We show that by using monodisperse scatterers, the fluctuations can be reduced into a single band of $\sim 1.2$~nm. Transfer matrix analysis and finite difference time domain techniques are used to explain the experimental data. By studying the spatial profile of the random lasing modes, we attribute the observed random lasing to the Anderson localization states of the periodic-on-average random system (PARS). 

This thesis is organized as follows.

{\bf Chapter 2:} In this chapter, Monte Carlo simulations based on the propagation of photons were performed, which showed that efficient and intense lasing modes can be obtained with a system of intra-scatterer gain. This is studied in the diffusive scattering domain. We observed frequency-controlled coherent random lasing in such a system. Based on these simulations, we propose a practical implementation of an aerosol, comprising liquid microdroplets in air. Monodispersity in the microdroplets can be expected to induce frequency control.

{\bf Chapter 3:} In this chapter, we describe the generation process of the aerosol and its characterization. The aerosol is in the form of arbitrarily shaped dye-doped microdroplets. We show that, upon optical excitation, the longitudinal emission spectrum consists of coherent random lasing (collective) modes. Using direct spatiospectral imaging of the modes, we confirm that they are extended over several microdroplets which coherently participate in the emission.

{\bf Chapter 4:} In this chapter, we implement monodispersity in the above system and carry out the experimental study of random lasing modes from a linear array of monodisperse microspheres. We confirm that the wavelength of the collective modes is sensitive to the size of microdroplets. Our analysis of lasing bands indicates that the Fabry-Perot resonances of individual microdroplets participate in the collective emission. Furthermore, we discuss various properties of the collective lasing modes like threshold behavior, the effect of system size, angular distribution and polarization anisotropy in this system.

{\bf Chapter 5:} This chapter contains a theoretical discussion of plane wave propagation in a one-dimensional structure comprising a dielectric multilayer. We model the microdroplet array as an amplifying periodic-on-average random superlattice (aPARS) system. We perform transfer matrix calculations with gain to analyze the origin of the lasing modes and their properties. Based on our calculations, we conclude that the observed random lasing modes originate from the high-quality gap states of the underlying periodic system. The ensemble-averaged analysis reveals that the gap states are restricted in the specific frequency range governed by the size of the individual layer.

{\bf Chapter 6:} Next, we exploit spectral properties of PARS systems to achieve single-mode, stable coherent random lasing. To that end, we use the concept of spectral mode-matching which refers to matching the gain maximum with the spectral realm of gap states. In the experiment, we used the mode matching technique and reduce the frequency fluctuations of the random lasing modes into a single band of $\sim 1.2$~nm.

{\bf Chapter 7:} In this chapter, we study the spatial properties of the modes using transfer matrix calculations. We then present direct evidence of Anderson localization of light in the
microdroplet array. To quantify the spatial extent of the modes, we calculate localization length and the Inverse Participation Ratio (IPR) and discuss the statistics of the IPR. This chapter conclusively proves the occurrence of Anderson localization in the amplifying PARS system.

{\bf Chapter 8:} Motivated by the current literature on chains of resonators, we examine the practical applicability of this system as a coupled-resonator optical waveguide (CROW). To that end, we experimentally demonstrate collective amplified modes along a bent chain consisting of microdroplet resonators. Furthermore, to get a better insight of the system, we present finite-difference time domain calculations in a bent array of infinite cylinders. We observe that the light propagation is realized by the filaments of the nanojet-induced modes (NIMs) which were formed at the edge of the cylinders due to the lensing effect induced by the curved surfaces. This chapter emphasizes the efficacy of the system as a CROW.
	
\chapter[Frequency fluctuations in coherent random laser, inter-scatterer vs~.~.~.] {Frequency fluctuations in coherent random laser, inter-scatterer vs intra-scatterer gain}

In this chapter, we address the issue of frequency fluctuations in coherent random lasers and show that the fluctuations can be reduced by using resonant scatterers. Our calculation, based on Monte Carlo simulations, shows that an intra-scatterer gain system provides more efficient and cleaner lasing spectra. The chapter is organized as follows: In section \ref{sec:level2.1}, we discuss the details of the photon transport Monte Carlo technique and parameters involved in studying the dye-scatterer random lasing system using this technique. In section \ref{sec:level2.2}, we consider a sparse system of monodisperse scatterers and compare the emission properties of an inter-scatterer gain system (the gain being present between the scatterers) versus an intra-scatterer gain system (gain inside the scatterers). In section \ref{sec:level2.3}, we consider a dense scattering system and show that stable and nearly thresholdless lasing can be achieved for intra-scatterer gain system. Finally, in section \ref{sec:level2.4}, we propose a practical system made of an amplifying aerosol to control the frequency fluctuations and discuss its emission properties.

\section{\label{sec:level2.1}Photon transport Monte Carlo simulations}

To study the spectral behavior of coherent random laser, we perform the same Monte Carlo simulation implemented in references~\cite{mujumdar00, sushil04, mujumdar10, ruppuaip, ruppu11, anjpnfa}. The simulation space is a box of dimensions $10$~mm~$\times~10$~mm~$\times~10~$mm which contains a uniform distribution of scatterers in an amplifying environment, akin to the dye scatterer random system. We assume that the box contains N dye molecules, decided by the concentration of the amplifying medium. The system was discretized into $10~\mu$m~$\times~10~\mu$m$\times~10~\mu$m boxes and the number of excited ($N_{1}$) and ground ($N_{0}$) state molecules in each box was registered to accurately calculate the local gain. We model light propagation as random walk of photons inside the computational box. Interference effects are not accounted for in the simulation. This assumption is motivated by the fact that the scattering mean free path of photons is much larger than their wavelength. Thus, the photons are in the diffusive transport regime where we can neglect the possibility of recurrent scattering between the scatterers. In the simulation, we incorporate the exact scattering parameters and calculate total dwell-time of light due to multiple scattering. The simulation consists of two major parts, the first part handles the excitation process and the second part deals with the subsequent emission. 

In the first process, a pump beam is incident on the front surface of the computational box. The pump beam is assumed to consist of an ensemble of photons or energy packets traveling along the longitudinal direction. The total number of such packets determines the pump energy. We set the wavelength of the incident photon bunch at $532$~nm, corresponding
to a frequency-doubled Nd: YAG laser. Upon entering the front surface of the sample, the photon bunch suffers multiple scattering from various scattering sites and undergoes a three-dimensional random walk. The random walk was implemented by translating the position of the photon packet after the $i^{th}$ scattering event as 

\begin{equation}
x_{i+1} = x_i+p_i
\label{eq2p1}
\end{equation}
where,
\begin{equation}
|p_i| = -\ell_s \ln\Sigma,
\label{eq2p1}
\end{equation}

$\Sigma$ is a uniformly distributed random number between ($0$, $1$) and $\ell_{s}$ is the scattering mean free path. Thus, $|p_{i}|$ belongs to an exponential distribution of path lengths between successive scattering events, with the mean $\ell_{s}$. The angular coordinates ($\theta_{0}$, $\phi_{0}$) of all photon packets before the first event were taken to be ($0$, $0$) to simulate a collimated beam. For subsequent scattering events, the angular coordinates ($\theta_{i}$, $\phi_{i}$) were picked from a uniform distribution over $4\pi$ steradian. During the random walk, the photons interact with the ground state molecules and experience absorption as determined by its absorption cross-section $\sigma_{abs}$. The ground state ($N_{0}$) and excited state ($N_{1}$) populations in each unit cell of the discretized volume are accordingly updated and recorded. This phenomenon terminates when either the photon bunch is completely absorbed or scattered out from the sample. At the end of excitation stage, the simulation records the final distribution of ground and excited state molecules. 
In the second stage, the fluorescence is initiated. Spontaneous emission events generate energy packets with unit intensity, which corresponds to one photon. The spatial distribution of the spontaneous events is determined by the spatial excitation profile. The frequency of the spontaneously emitted photons is picked from a cumulative distribution of wavelengths constructed from the emission cross-section profile $\sigma_{em}$. In our simulation, we maintained the spectral resolution at $0.1$~nm. Similar to the pump light, the spontaneously emitted photons also perform three dimensional random walks through the active random medium. During the random walk, each spontaneously emitted photon experiences amplification or absorption depending on the local population inversion. The random walks in the amplifying medium are terminated when either the photons exit the sample or are completely reabsorbed by the ground state molecules. After the complete de-excitation process, total number of emitted photons is recorded as a function of wavelength to construct the emission spectra.

\section{\label{sec:level2.2}Sparse system: Resonant amplifying scatterers}

\begin{figure}[h]
\begin{center}
\includegraphics[scale=0.8]{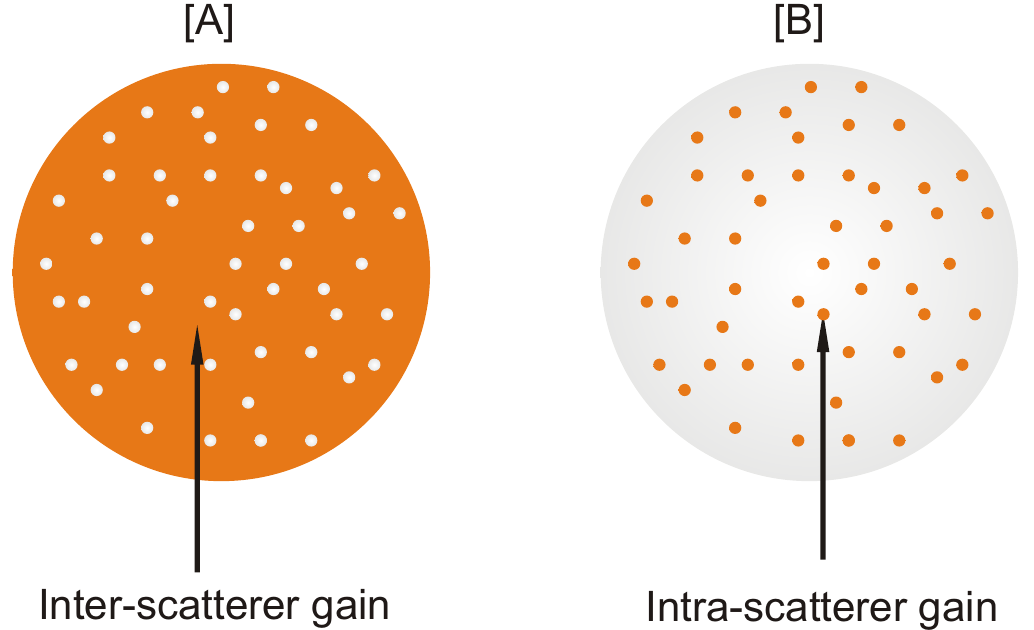}
\end{center}
\vspace{-30pt}
\color{blue}\caption {Schematic representation of the inter-scatterer [A] and intra-scatterer [B] gain systems. Orange color represents the gain region, the volume fraction of the scatterers is $1~\%$.}
\label{fig2p1}
\end{figure}

\begin{figure}[h]
\begin{center}
\includegraphics[scale=0.55]{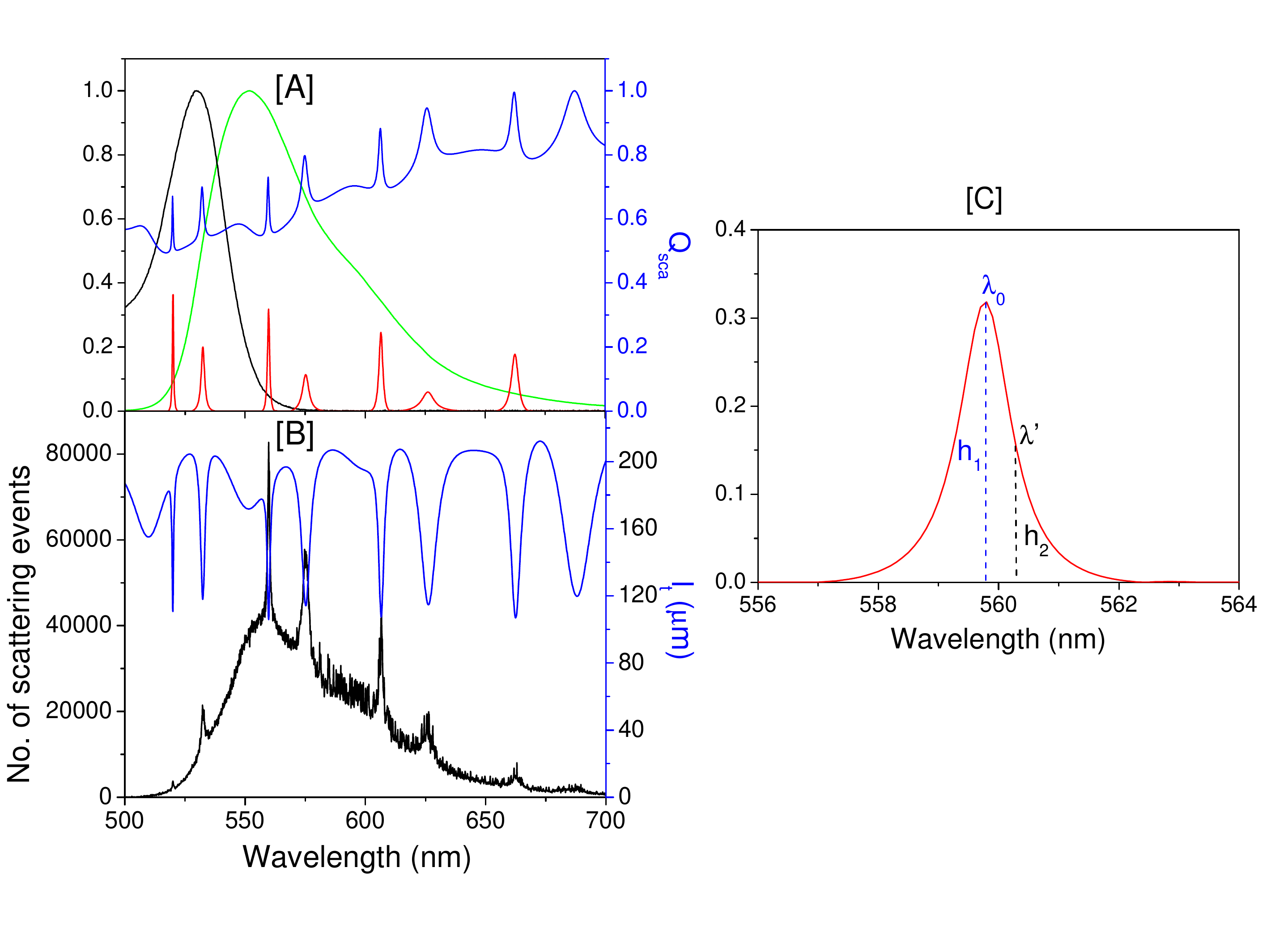}
\end{center}
\vspace{-50pt}
\color{blue}\caption{[A] Black and green curves: absorption and fluorescence spectra of the gain medium. Blue curve: Mie scattering efficiency for a particle of $1.09~\mu$m with a refractive index contrast of $1.8$ with ambient medium (here, alcohol, $n = 1.329$). Red curve: Extracted high quality modes from the scattering profile. [B] Blue curve: Variation of scattering mean free path as a function of wavelength. Black curve: Number of scattering event before the photons exit from the three dimensional computational box. [C] Extracted mode centered at $559.8$~nm.}
\label{fig2p2}
\end{figure}

Figure~\ref{fig2p1} shows the schematic representation of the inter-scatterer and intra-scatterer gain systems with monodisperse resonant scatterers. An inter-scatterer gain system (Figure~\ref{fig2p1}[A]) has the amplification between the scatterers while in the intra-scatterer gain system (Figure~ \ref{fig2p1}[B]), the gain is present inside each scatterer. We first consider a system with a volume fraction of $1~\%$, which implies that the scatterers are sparsely distributed. Various parameters of the resonant scatterer system are shown in Figure~ \ref{fig2p2}. These numbers were calculated for a system comprising TiO$_2$ (refractive index $n = 2.4$) spheres of diameter $1.09~\mu$m with a refractive index contrast with respect to the ambient medium (here, methanol) of $n = 1.8$. The black and the green curves in the top panel are the absorption and emission spectra of the Rhodamine dye. The blue curve in the top panel depicts the calculated Mie scattering efficiency ($Q_{scat}$) within the spectral band of interest~\cite{matzler02}. The amplitude of these resonance profiles determines the scattering mean free path, while the width determines the quality factor. We first extract these resonances by fitting a Lorentzian profile $L$($\lambda$) which is defined by 

\begin{equation}
L(\lambda) =\frac{1}{\pi}[\frac{\Delta\lambda/2}{{(\lambda - {\lambda}_0)}^2+({\Delta\lambda/2})^2}]
\label{eq2p1}
\end{equation}

where $\Delta$ $\lambda$ is the full width at half maximum (FWHM) of the Lorentzian profile and $\lambda_{0}$ specifies the location of the peak. The red curve shows these extracted high quality modes where the dwell-time of the photons is relatively large. The $Q_{scat}$ profile also has various low quality modes, but the quality factor of these modes is few orders of magnitude smaller and hence do not contribute strongly to the overall transport.

The blue curve in Figure~\ref{fig2p2}[B] is the variation of transport mean free path ($l_{t}$) with wavelength. By definition, $l_{t}$ is inversely proportional to the $Q_{scat}$ and is given by

\begin{equation}
l_{t} =\frac{2d}{3(1-g)\phi Q_{scat}}
\label{eq2p3}
\end{equation}

where, $d$ is the particle diameter, $\phi$ is volume fraction of the scatterers and $g = \langle \text{cos}\theta \rangle$, where $\theta$ is the scattering polar angle [This equation is consistent with the definition of $l_{t}$ given in section \ref{sec:level1p3}]. Thus, in a monodisperse scattering environment, the transport mean free path $l_{t}$ is smaller at the positions of the Mie resonances indicating stronger diffusion. Furthermore, at each resonance, stronger scattering enhances the number of scattering events encountered by the fluorescent photons. The black curve shown in the bottom panel is the total number of scattering events encountered by the fluorescent photons before they exit from the computational box as registered by the simulation. At each scattering event, the fluorescent photon spends some resident time inside the scatterer, which is determined from the quality factor of the corresponding resonance. In a multiple scattering environment, $\tau_{intra}$ is calculated by the cumulative resident times within the multiple scatterers, while $\tau_{inter}$ is obtained from the total path-lengths of the photons. Figure~\ref{fig2p2}[C] shows the parameters of the extracted resonance, which will be discussed later regarding calculations of the intra-scatterer dwell-times.

\begin{figure}[h]
\begin{center}
\includegraphics[scale=0.55]{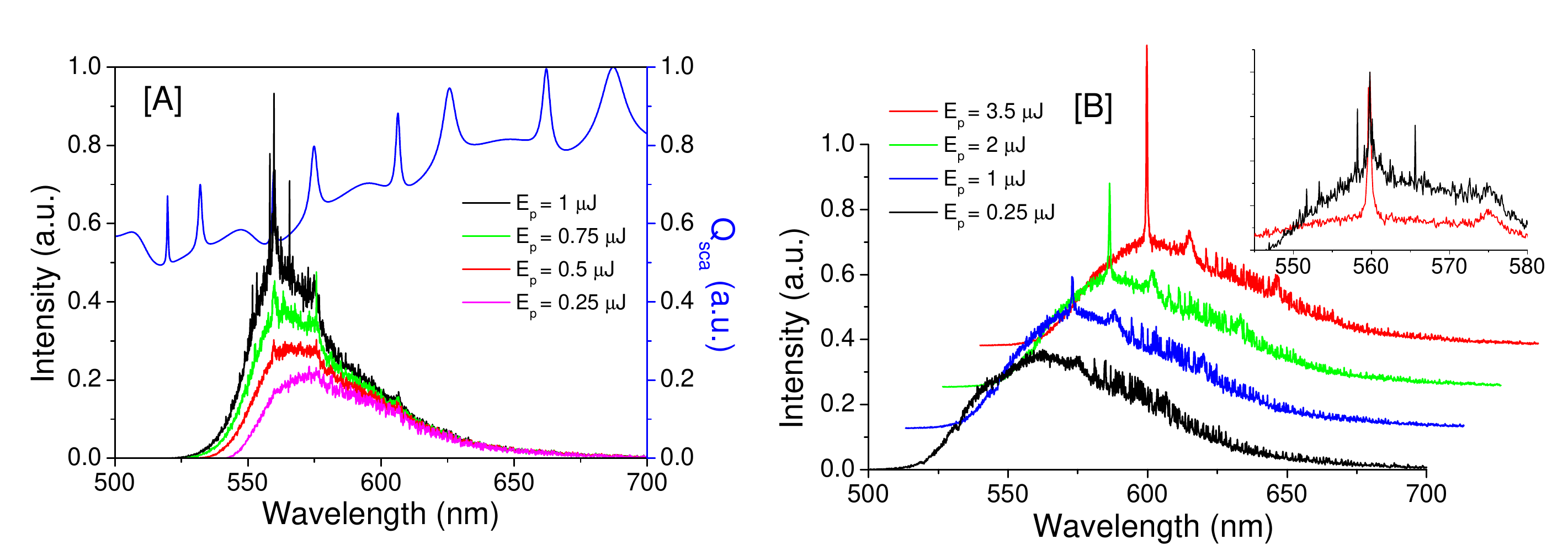}
\end{center}
\vspace{-30pt}
\color{blue}\caption{[A] Inter scatterer gain system: Blue curve is the Mie scattering efficiency curve. Four typical emission spectra at different pump energies are shown. The lasing peaks appear near one resonance of an individual resonator. Further, multiple modes are generated at the resonance. [B] Intra-scatterer gain system: Here, the spectra are cleaner and more stable. The inset is the exaggerated region for clarity. Black curve: inter-scatterer gain system. Red curve: intra-scatterer gain system.}
\label{fig2p3}
\end{figure}

Figure~\ref{fig2p3} shows the consequence of resonance on the emission spectrum. Figure~\ref{fig2p3}[A] depicts the emission spectrum for inter-scatter gain system at the said pump energies. The blue curve is the variation of $Q_{scat}$ with wavelength. At low pump energy ($E_{p}$), the system yields the typical fluorescence spectrum of Rhodamine dye. As we increase the pump excitation energy, lasing peaks develop in the spectrum. The peak positions are determined by the Mie resonances and the spectrum shows maximum enhancement at the highest quality mode. In this system, we observed that, in addition to the coherent modes in the vicinity of $\lambda_{res}$, multiple low intensity modes also appeared across the spectral range. The overall emission spectrum shows gain narrowing because of multiple amplified modes. Figure~\ref{fig2p3}[B] shows the representative spectra for intra-scatterer gain system. Here, the high intensity coherent modes appear only at the Mie resonance. Furthermore, in this system, the spectra are much cleaner and more stable\footnote{By `stable', we mean that the coherent lasing peaks have less frequency fluctuations under optical pumping.} compared to the inter-scatterer gain system as seen in the inset.

To understand the emission in the inter-scatterer and intra-scatterer gain systems, we estimated $\tau_{inter}$ and $\tau_{intra}$, which are the dwell-times between and within scatterers, respectively. The $\tau_{intra}$ is extracted from the Mie scattering profile by estimating the quality factor of each resonance, defined as $Q=\frac{{\lambda}_{0}}{\Delta\lambda}$, where, $\Delta\lambda$ is the FWHM of the resonance and $\lambda_{0}$ specifies the center wavelength. The lifetime of the photon inside a scatterer is proportional to the quality factor and is given as $\tau=\frac{Q{\lambda}_0}{2\pi c}$. At the central resonant wavelength, the dwell-times in the scatterers are exponentially distributed with a time constant $\tau$. The total intra-scatterer dwell-time in the sample was calculated by the cumulative sum of all these exponentially distributed variables. At another wavelength in the resonant band, the dwell-time in the scatterers will be exponentially distributed with a time constant of $\tau^{'}$. Intuitively, this decay will be faster than that at the central wavelength, implying thereby that the $\tau^{'}$ is smaller, the best scaling of which is given by $\tau^{'} = (h_{2}/h_{1})\times \tau$, where $h_{1}$ is the height of the extracted Lorentzian at the central wavelength (Figure~\ref{fig2p2}[C]) and $h_{2}$ is the height of the Lorentzian at the other wavelength.

\begin{figure}[h]
\begin{center}
\includegraphics[scale=0.62]{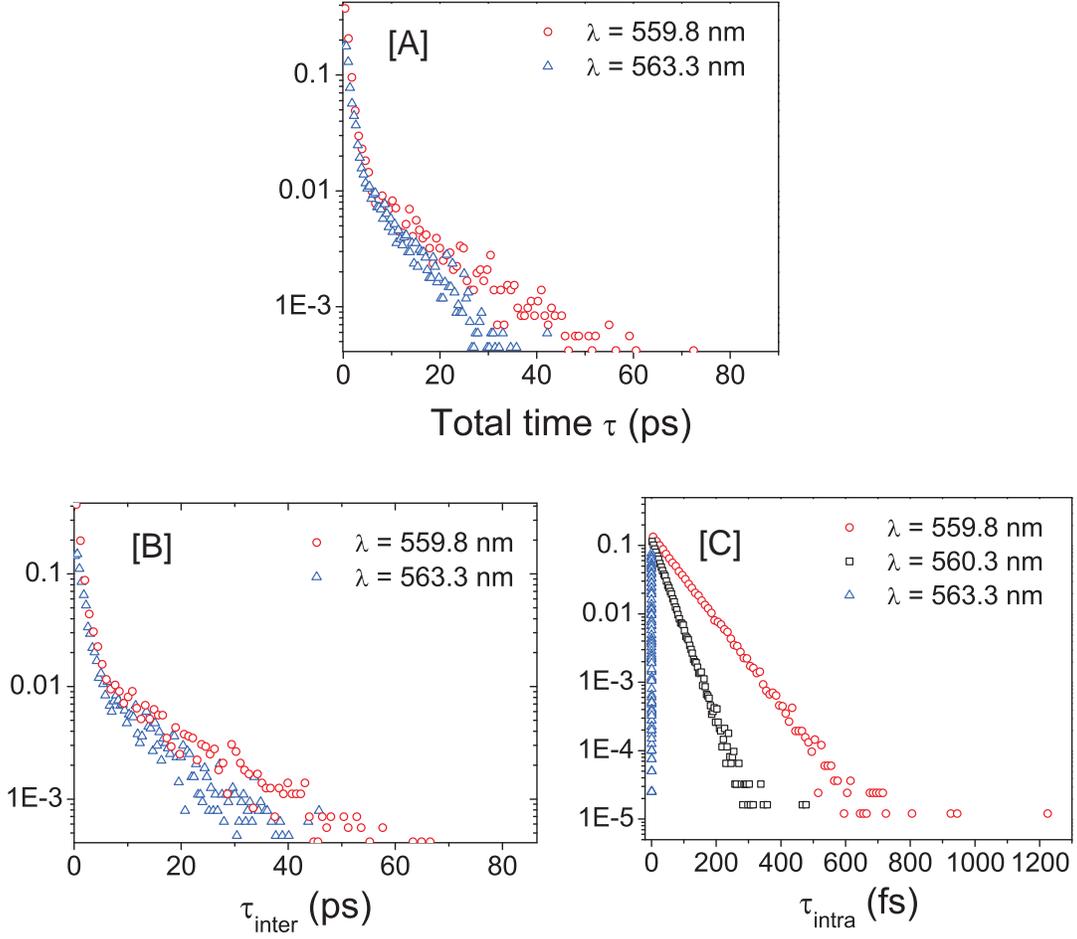}
\end{center}
\vspace{-30pt}
\color{blue}\caption{[A] Distribution of total dwell-time of photons at resonant $\lambda_{res} = 559.8$~nm (red) and at nonresonant $\lambda_{nres} = 563.3$~nm (blue) wavelengths. [B] Distribution of inter-scatterer dwell-time at the same wavelengths. [C] Distribution of intra-scatterer dwell-time at three wavelengths $\lambda_{res} = 559.8$~nm, $\lambda^{'} = 560.3$~nm and $\lambda_{nres} = 563.3$~nm.}
\label{fig2p4}
\end{figure}

By incorporating these relations, we estimate the dwell-time distribution of photons before they exit the computational box. Two wavelengths are chosen to estimate the dwell-time distributions, the resonant $\lambda_{res} = 559.8$~nm, and the nonresonant $\lambda_{nres} = 563.3$~nm. Figure~\ref{fig2p4}[A] shows the total dwell-time distribution at these wavelengths. From the total dwell-time, we separate the distribution of $\tau_{inter}$ and $\tau_{intra}$. As shown in Figure~\ref{fig2p4}[B], $\tau_{inter}$ has a slight increase at the resonant wavelength. Note that the difference is only modest, because the large number of scattering events encountered by the photons are compensated by the small $l_{t}$. This leads to a smaller disparity in the dwell-time distribution at $\lambda_{res}$ and $\lambda_{nres}$.

Figure~\ref{fig2p4}[C] discusses $\tau_{intra}$ at three wavelengths, namely, resonant $\lambda_{res} = 559.8$~nm, nonresonant $\lambda_{nres} = 563.3$~nm, and an intermediate wavelength at the half max $\lambda^{'} = 560.3$~nm of the resonance. Here, obviously the disparity between $\tau_{inter}$ and $\tau_{intra}$ is large, clearly because the nonresonant light is not sustained in the scatterer. Importantly, in this sparse system, the total time distribution is dominated by the $\tau_{inter}$ (note the difference in the X axis scale between \ref{fig2p4}[B] and \ref{fig2p4}[C]). The cleaner and more stable spectra in the intra-scatterer gain system can now be attributed to a larger disparity in the $\tau_{intra}$ at $\lambda_{res}$ and $\lambda_{nres}$. The modes at $\lambda_{nres}$ experience only a marginal amplification. Hence, only those modes which lie at the resonant frequencies are preferentially amplified. On the other hand, in an inter-scatterer gain system, as the disparity between $\lambda_{res}$ and $\lambda_{nres}$ is small, occasionally the extended modes at $\lambda_{nres}$ can also experience gain and contribute in lasing.

\section{\label{sec:level2.3}Dense scattering system}

In the previous section, we discussed a sparse system with a volume fraction ($\phi$) of 1$\%$. Small $\phi$ results in larger $\tau_{inter}$ and hence, the emission favors the inter-scatterer gain configuration. In this section, we consider a system with denser packing, which results in reducing the scattering mean free path. Thus, naturally the $\tau_{inter}$ and $\tau_{intra}$ distributions are altered. Here, we assume a volume fraction of $10~\%$ which corresponds to a mean free path of about $5-7~\mu$m. Clearly, the diffusion approximation still holds in the system as the mean free path is several times larger than the wavelength. 
\begin{figure}[h]
\begin{center}
\includegraphics[scale=0.35]{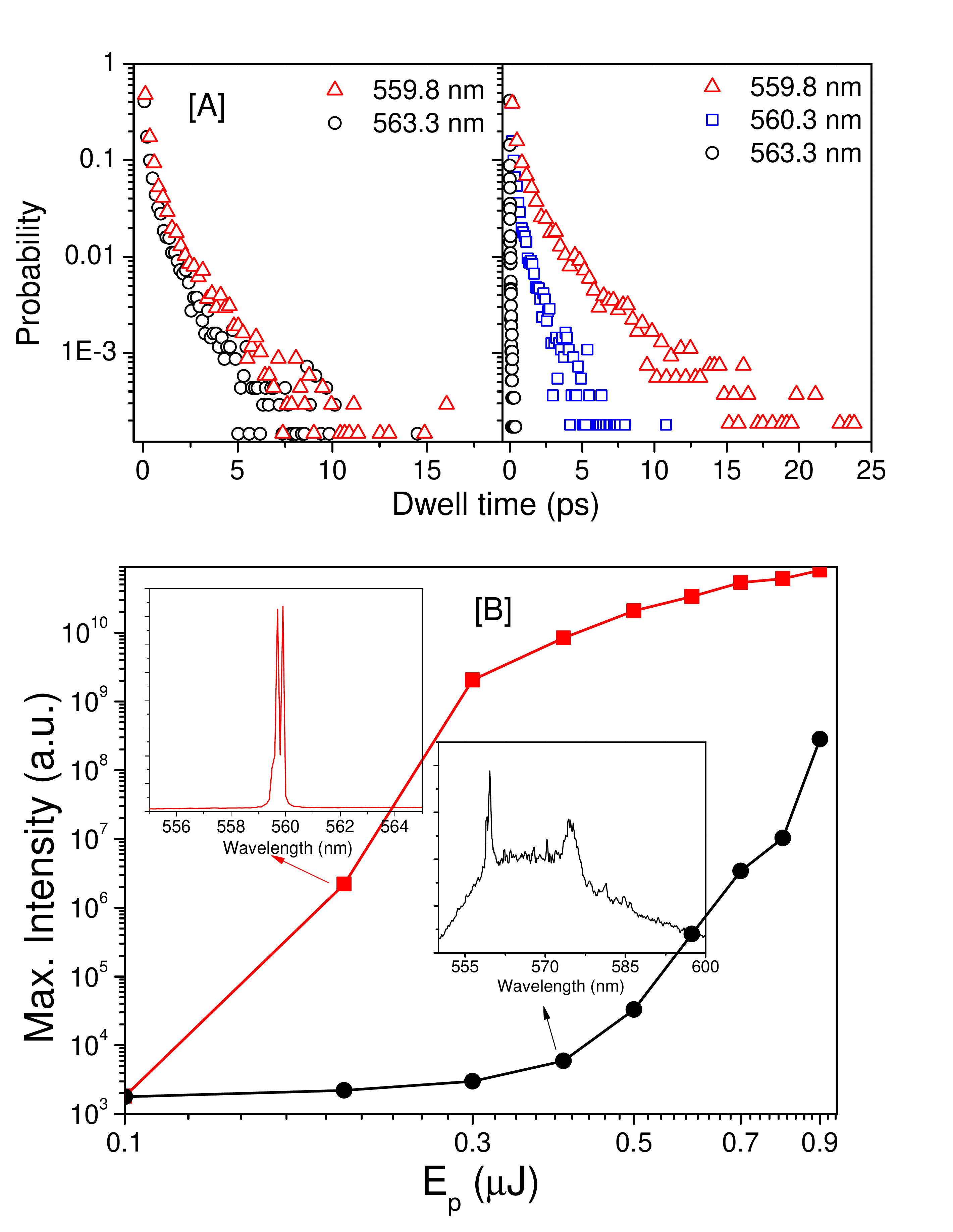}
\end{center}
\vspace{-35pt}
\color{blue}\caption{[A] Left panel: Distribution of inter-scatterer dwell-time of photons at $\lambda_{res} = 559.8$~nm (red) and at $\lambda_{nres} = 563.3$~nm (black). Right panel: Distribution of intra-scatterer dwell-time at $\lambda_{res} = 559.8$~nm, $\lambda_{\text{FWHM}} = 560.3$~nm and $\lambda_{nres} = 563.3$~nm. [B] Variation of lasing peak power with the excitation energy for inter-scatterer (black) and intra-scatterer (red) gain system with two representative spectra in the inset at the energies marked by the arrows.}
\label{fig2p5}
\end{figure}

Fig. \ref{fig2p5}[A] shows the distribution of both $\tau_{inter}$ (left panel) and $\tau_{intra}$ (right panel) at resonant and at nonresonant wavelengths for the dense system. Here, $\tau_{intra}$ shoots up (few picoseconds) due to the increased number of scattering events, and the inter-scatterer and intra-scatterer dwell-times are now comparable. Fig. \ref{fig2p5}[B] further depicts the behavior of the lasing peak intensity as a function of excitation intensity. The black curve corresponds to the inter-scatterer gain system. A clear threshold is observed at $E_{p}~\sim~0.3~\mu$J. The bottom inset shows a representative spectrum at $E_{p} = 0.4~\mu$J. The spectra consist of coherent random lasing peaks riding on a broadband pedestal. With higher excitation energy, multiple coherent modes start to grow. On the other hand, in the intra-scatterer gain system, the lasing intensity enhances rapidly even with a very small excitation energy as shown by the red curve. This is due to the excessive residence times of the photons inside the amplifying scatterers. At $\lambda_{res}$, the absorbed energy is rapidly channeled into the resonant modes and thus the incoherent pedestal is suppressed. Consequently, very clean, ultra-narrowband coherent random lasing is observed in this system. A representative spectrum at $E_{p} = 0.2~\mu$J is shown in the top inset. Upon further excitation, gain saturation occurs and the maximum intensity of the coherent peaks shows a saturation behavior. Thus, based on these analyses, we conclude that very low or almost thresholdless random lasing can be achieved from a dense ensemble of amplifying resonant scatterers.

\section{\label{sec:level2.4}A practical proposal}

A random ensemble of resonant scatterers has already been used in the context of random lasers~\cite{gottardo08}. In those experiments, photonic glasses consisting of monodisperse spherical scatterers (diameter $\sim~1~\mu$m, polydispersity $< 2~\%$) were used. The gain was incorporated by adsorbing the dye molecules on the surface of the spheres. Thus the gain was present on the surface of spheres. Under optical pumping, diffusive random lasing was observed that was sensitive to the resonances of the system. In this system, coherent random lasing was not reported. This could be because the gain was not present inside the spheres. On the other hand, active scatterers like ZnO powders can serve the requirement of high gain. However, the polydispersity of $5-8~\%$ inherently present~\footnote{This is the expected polydispersity created in the synthesis process for a particle size of about $1~\mu$m.} in the microparticles washes the overall resonant effect of the system. Thus, despite their high gain, ZnO powders are not suitable for frequency sensitive coherent random lasing.

\begin{figure}[h]
\begin{center}
\includegraphics[scale=0.6]{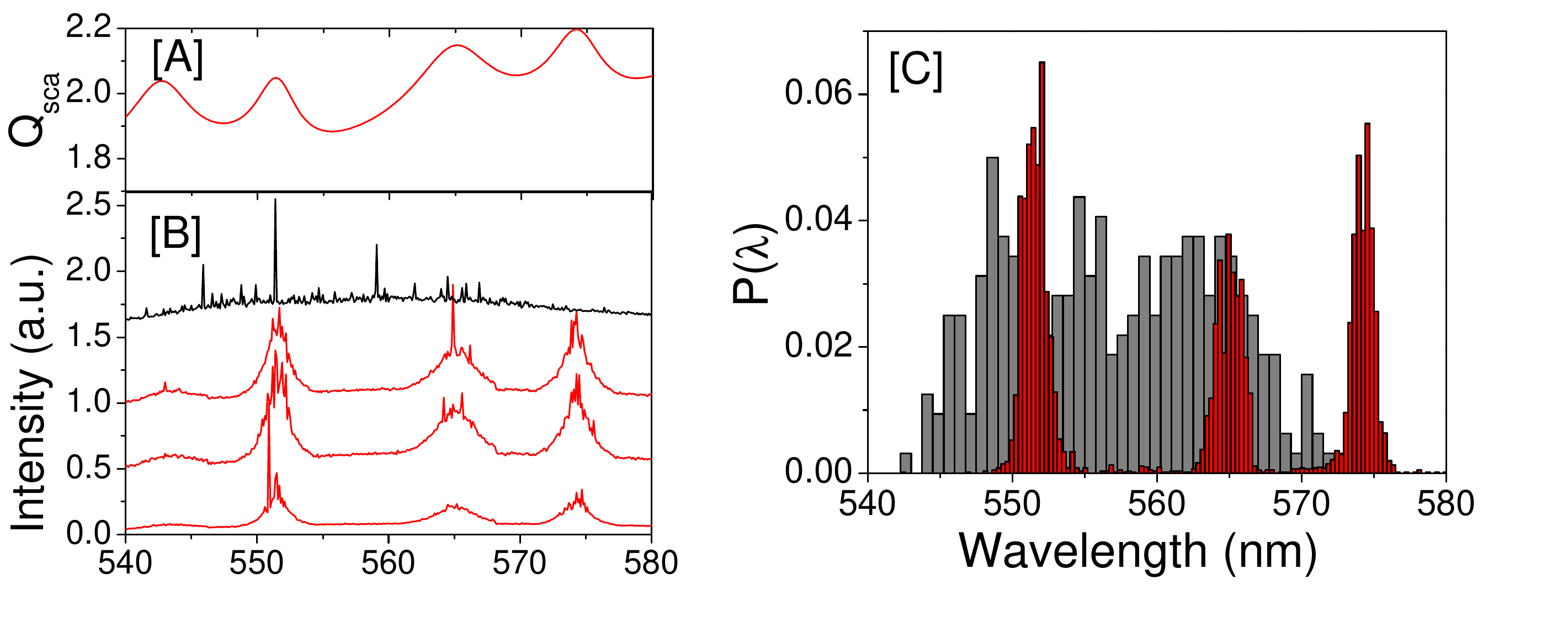}
\end{center}
\vspace{-30pt}
\color{blue}\caption{[A] Scattering efficiency curve ($Q_{sca}$) for a methanol microdroplet. [B] Red curve: Three calculated spectra from a mondisperse aerosol for a volume fraction of $10~\%$. Black curve: Calculated spectrum of the aerosol with a polydispersity of $12~\%$. [C] Gray and red histograms are the distribution of the lasing wavelengths from the aerosol system.}
\label{fig2p6}
\end{figure}

To observe frequency control in a coherent random laser, we propose a potentially realistic system comprised of Mie scatterers. The fact that the dye molecules have high quantum efficiency in alcohol ($\sigma_{g} \sim 10^{-16}$~cm$^{2}$) makes them the right material to fulfil the requirement of high gain.
A possible mechanism to create the amplifying scatterers is by creating microdroplets from the said dye solution. Such microdroplets will have a refractive index contrast of $1.329$ (of the solvent, with respect to air) and will offer a high gain for the light resident inside it. A random ensemble of such microdroplets, technically called an aerosol, will act as an amplifying disordered medium. We simulated optical transport in such an aerosol system, and the results are illustrated Figure~ \ref{fig2p6}. The red curve in Figure~\ref{fig2p6}[A] shows the Mie scattering profile ($Q_{sca}$) of a spherical microdroplet with $d = 3.6~\mu$m and $n = 1.329$. Compared to the earlier system, the resonances are broader due to the lower refractive index of these scatterers. Three representative simulated spectra from a monodisperse system pumped at $7~\mu$J are shown by the red curves in Figure~\ref{fig2p6}[B]. The effect of the broadband nature of the resonance is evident. Multimode emission is observed at each resonance location. Furthermore, these modes fluctuate within the broad resonance band. A summary of frequency fluctuations is obtained by creating a histogram of the lasing frequencies over $75$ spectra. The red columns are the frequency distribution of lasing modes from the monodisperse aerosol system. It is clear that the distribution is discretized and the lasing peaks appear at each resonance (red histogram).

In our calculation, we also studied the effect of polydispersity. The black line in Figure~\ref{fig2p6}[B] is a representative spectrum from the polydisperse system by assuming a $12~\%$ spread in the scatterer diameter (at $E_{p} = 30~\mu$J). Note that large excitation energy is required to trigger the coherent modes, which is also a consequence of the polydispersity. The gray columns in Figure~\ref{fig2p6}[B] are the probability distribution of lasing frequencies from the polydisperse system. Here, the lasing peaks appear at random positions, and the distribution is broadband and continuous. This histogram exemplifies the normal frequency fluctuations seen in coherent random lasers. In contrast, a monodisperse amplifying scatterer system emits modes only in discrete frequency intervals. Thus, it can serve to control frequency fluctuations in a coherent random laser.

{\bf Summary:}
We have simulated coherent random laser emission from monodisperse amplifying scatterers and presented a comparative behavior of two systems, namely, inter-scatterer and intra-scatterer gain systems. We find that the intra-scatterer gain system provides intense, much cleaner and stable random lasing in a very narrow band of wavelengths, determined by the underlying Mie resonances. In a system of dense scatterers, the lasing phenomenon starts at very low excitation energies and the system shows almost thresholdless lasing. Based on these conclusions, we discuss the possibility of an experimentally realistic system comprising an amplifying aerosol.	
\chapter{Practical implementation of an aerosol random laser}

In this chapter, we follow up our previous simulation work with experimental demonstration of coherent random lasing from an amplifying aerosol. The aerosol consists of a linear array of polydisperse, arbitrarily shaped and randomly placed dye-doped microdroplets. Under optical excitation, ultranarrow, coherent random lasing is observed in the longitudinal direction of the array. Spectral imaging and threshold behavior confirms the coherent participation of each microdroplet.
This chapter is organized in the following way: In section \ref{sec:level3.1}, we discuss the experimental setup and characterize the aerosol. We, then, discuss the frequency fluctuations and threshold behavior of coherent random lasing in section \ref{sec:level3.2}. Finally, in section \ref{sec:level3.3}, we compare the emission properties of the aerosol system with that of Rhodamine dye filled in a microcolumn.

\section{\label{sec:level3.1}Experimental setup and characterization of the amplifying aerosol}
To generate an aerosol of dye-doped microdroplets, we use a glass microcapillary attached to a Teflon tube. The other end of the tube is connected to a liquid chamber, filled with a $1$~mM Rhodamine dye solution (dissolved in methanol). Using a non-interacting gas ($N_{2}$) supply, the liquid chamber is pressurized to about $15$ atmospheres which forces the liquid into the capillary at a constant rate. Under this pressure, the exiting liquid jet, which is mechanically unstable, breaks into an array of fine microdroplets.

\begin{figure}[h]
\begin{center}
\includegraphics[scale=1.35]{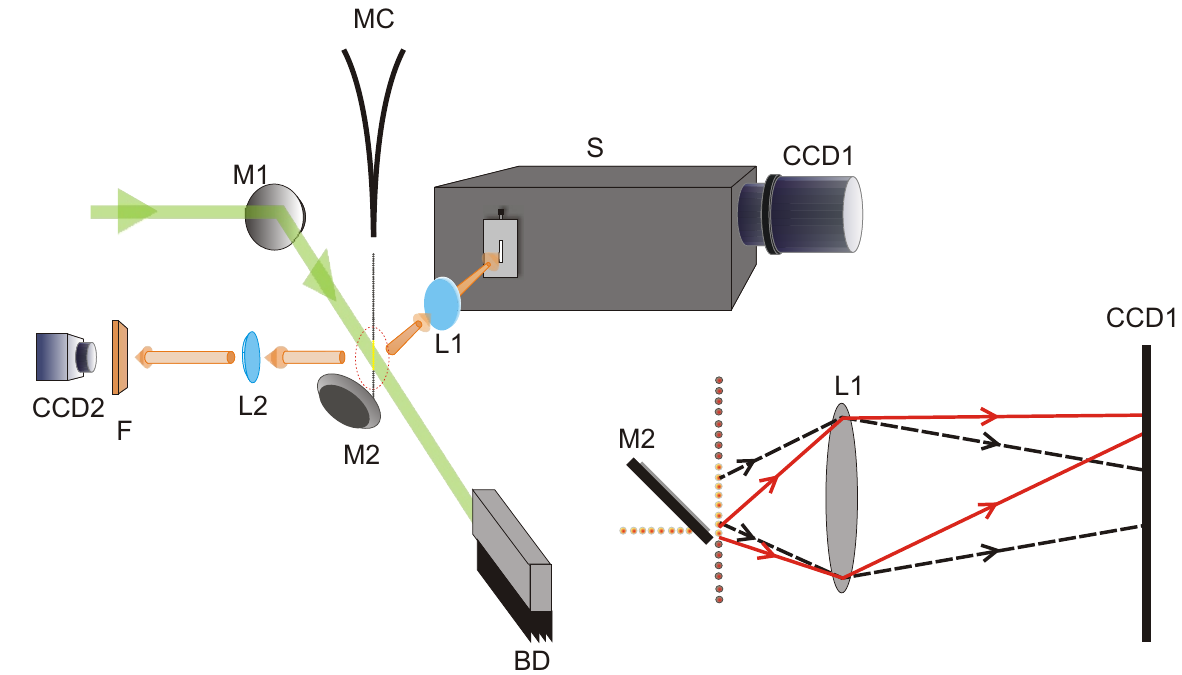}
\end{center}
\vspace{-25pt}
\color{blue}\caption{Schematic representation of experimental setup. Legend: MC: microcapillary, F: filter, S: spectrometer, M1, M2: mirrors, BD: beam dump, CCD1: CCD for spectral measurement, CCD2: CCD for imaging the scatterer, L1, L2: lens. Inset shows the schematic to obtain simultaneous spectra from the transverse (black ray) and longitudinal (red ray) direction.}
\label{fig3p1}
\end{figure}

The schematic of the setup to observe coherent random lasing from the aerosol is shown in Figure~\ref{fig3p1}. The droplet array is excited by the second harmonic of a Nd: YAG laser ($\lambda = 532$~nm, pulsewidth $\sim 25$~ps) using the steering mirror M1. The beam is focused to a focal spot of $\sim 1$~mm to pump a sufficient number of droplets ($\sim 20$). Lens L2 (focal length $f = 50$~mm) is used to image the array on an imaging CCD (CCD2). The filter F is used to attenuate the scattered green light from the droplets. Mirror M2 is placed at an angle of about $42^{\circ}$ with respect to the axis of the array. This allows us to redirect the longitudinal emission towards the entrance slit of an imaging spectrometer (S). Lens L1 ($f = 150$~mm) is mounted on an XYZ translation stage to simultaneously collect the emission in the longitudinal (along the array axis) and transverse direction (perpendicular to the array axis) as shown in the inset.

\begin{figure}[h]
\begin{center}
\includegraphics[scale=1.4]{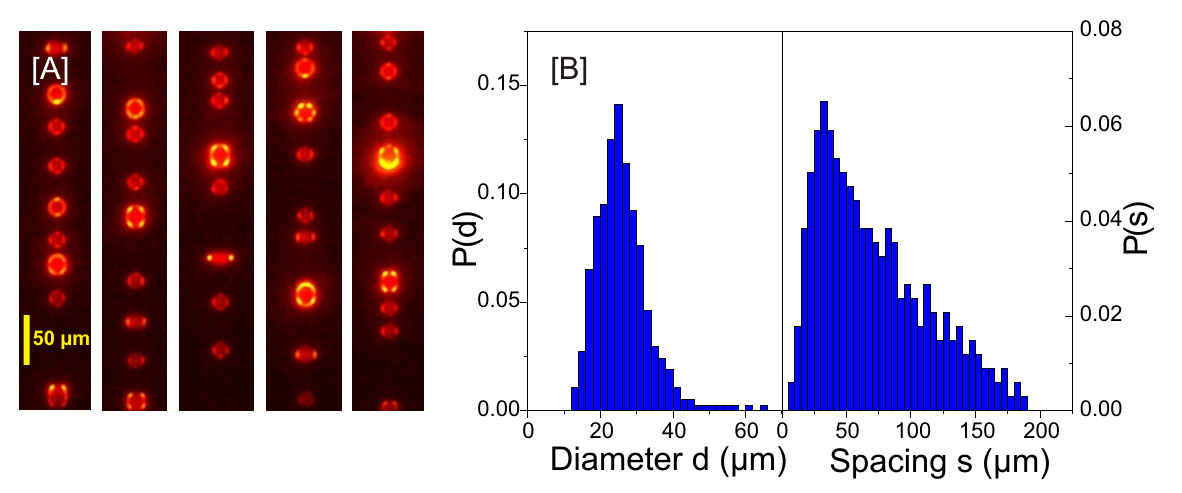}
\end{center}
\vspace{-30pt}
\color{blue}\caption{[A] Random manifestations of the microdroplet array, the arrays are uncorrelated to each other. [B] Probability distribution of the longitudinal size (left panel) and the spacing (right panel) over $500$ realizations of the array; the most probable size and spacing are $25~\mu$m and $40~\mu$m respectively.}
\label{fig3p2}
\end{figure}

Figure~\ref{fig3p2}[A], shows a few realizations of the highly polydisperse microdroplet array. It is clear that, every configuration is uncorrelated to each other. The size and spacing between the droplets were estimated using the imaging CCD which has a rectangular $1024 \times 768$ pixel sensor, the size of each pixel being $4.65~\mu$m $\times~4.65~\mu$m~\footnote{As mentioned in the camera specifications.}. In the experiment, the droplet array was magnified by a factor of about $10$. By counting the total number of pixels covered by the droplet, we estimated its size. Note that, the number of pixels occupied by the image depends on its magnification. For the said magnification, $1~\text{pixel} = 4.65/10~\mu$m $= 465$~nm. Furthermore, as the surface of the microdroplets in the image does not have a sharp boundary, there is an uncertainty of about $1~\mu$m in the measurement~\footnote{Although this characterization technique gives only approximate estimates, we note that this chapter does not require exact parameters for analysis. A more rigorous technique will be described in the next chapter.}.

Figure~\ref{fig3p2}[B], shows the distribution of the longitudinal size of droplets and the inter-droplet spacing (surface to surface) over $500$ configurations. The droplet size varied from $10~\mu$m to $50~\mu$m and the spacing varied from $5~\mu$m to $180~\mu$m. The size and spacing depended on various physical parameters like viscosity of the liquid, flow rate of the solution and pressure applied on the chamber etc. 

\section{\label{sec:level3.2}Coherent random lasing from the polydisperse droplet array}
\subsection{\label{sec:level3.2a}Longitudinal and transverse emission spectra}

\begin{figure}[h]
\begin{center}
\includegraphics[scale=0.5]{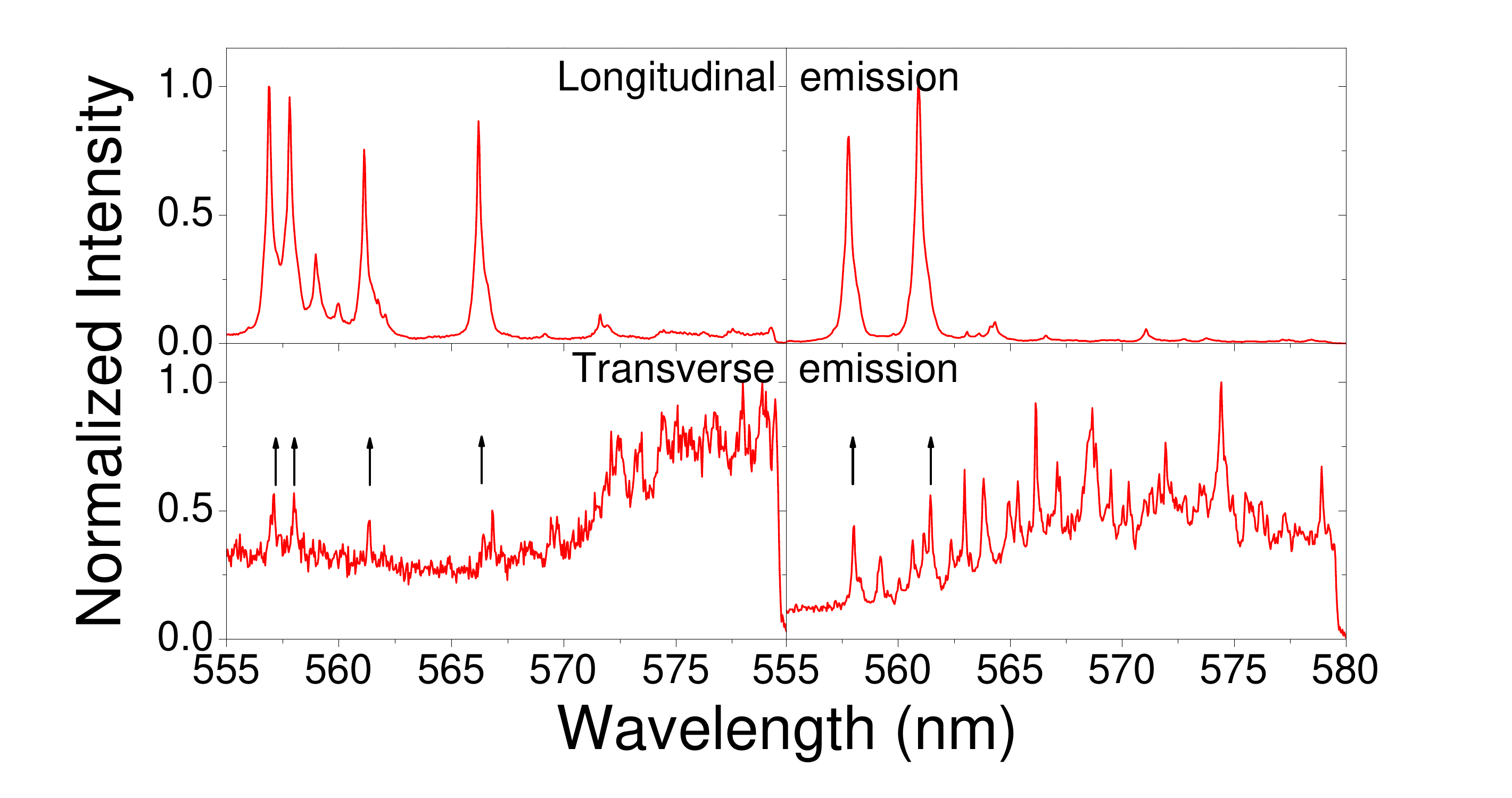}
\end{center}
\vspace{-35pt}
\color{blue}\caption{Simultaneously captured longitudinal and transverse emission spectra. Whenever there is a peak in the longitudinal direction, a corresponding low intensity peak is also observed in the transverse direction. Arrows identify the correlated transverse and longitudinal peaks.}
\label{fig3p3}
\end{figure}

Two representative spectra of the longitudinal emission at $E_{p} = 4.2~\mu$J are shown in the top panel of Figure~\ref{fig3p3}. The spectra are multimode and no periodicity is observed in the coherent peaks. At higher pump energies, the coherent peaks exhibited a self-averaging behavior. The bottom panel illustrates the simultaneously measured transverse emission spectra, which show a multitude of random, low-intense coherent peaks on a broad fluorescence background. The intensity in the longitudinal direction is approximately $15$ times stronger as compared to the transverse direction. As indicated by the arrows in Figure~\ref{fig3p3}, whenever there is a coherent peak in the longitudinal direction, there is a corresponding low intensity peak in the transverse direction. The linewidth of the intense peaks in the longitudinal direction was measured to be resolution limited to $\sim 0.25$~nm, which corresponds to a minimum coherence length of $\sim 1.6$~mm. Due to the changing configuration of the array, the coherent random lasing peaks show shot-to-shot fluctuations in both frequency and intensity.

\subsection{\label{sec:level3.2b}Spectral image, threshold and frequency fluctuations}

\begin{figure}[h]
\begin{center}
\includegraphics[scale=.8]{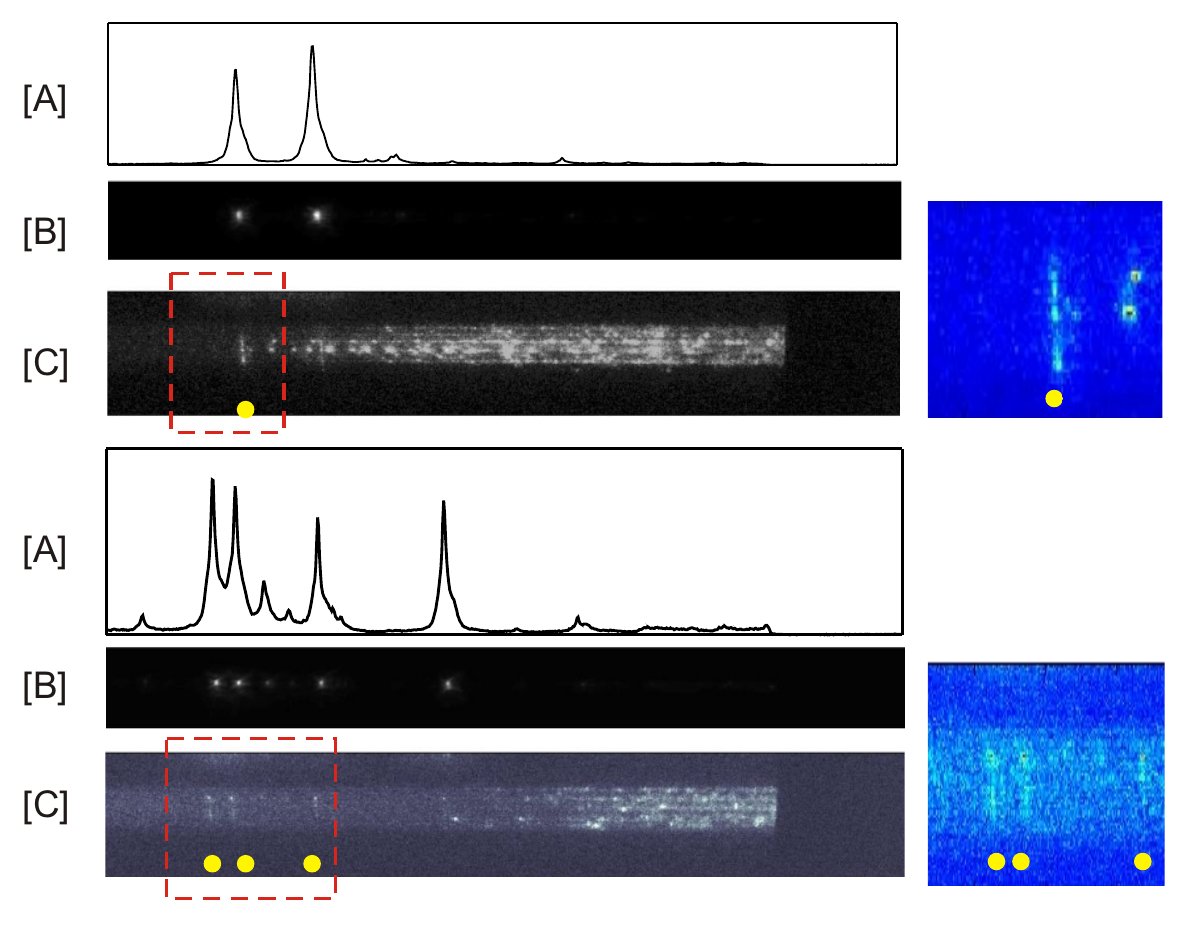}
\end{center}
\vspace{-30pt}
\color{blue}\caption{Two shot-to-shot spectra from the droplet array, [A] Longitudinal emission spectra [B] Corresponding spectral image [C] Simultaneously captured transverse spectral image. Every bright spot in the transverse direction has a vertical bright streak, which confirms the coherent participation of droplets. Side panel: magnified image of the area enclosed by the dotted rectangle.}
\label{fig3p4}
\end{figure}

We extract further details of the random lasing modes from spatiospectral imaging as shown in Figure~\ref{fig3p4}. The top panel [A] illustrates the longitudinal spectrum. Middle panel [B] shows the corresponding longitudinal spectral image from which the spectrum in [A] is extracted. Two intense spots are observed in [B] which corresponds to the two coherent peaks in the emission. The bottom panel [C] is the simultaneously captured transverse spectral image. The transverse image also consists of many bright spots, which represent the internal resonances of the microdroplets. Since the droplets in the array are polydisperse, these intra-particle resonances do not occur at the same wavelengths. The low-intensity peaks seen in transverse direction (Figure~\ref{fig3p3}) are the result of these intra-particle resonances. We found that the longitudinal emission does not show any correlation with the intra-particle resonances, which confirm that the coherent peaks do not originate from the single microdroplets. 

Importantly, every intense spot in the longitudinal direction (seen in [B])has a corresponding low intensity vertical streak in the transverse emission (seen in [C]). These are labeled by yellow dots underneath. For clarity, the right panel shows the magnified image of the region marked by the dotted rectangle. It is clear that, each vertical streak is extended over the entire pumped region and there is a perfect positive correlation between the transverse extended mode and the high intensity longitudinal peak. This observation confirms that the lasing peak observed in the longitudinal direction originates from the cooperative behavior of all droplets present in the pumped array.

\begin{figure}[h]
\begin{center}
\includegraphics[scale=0.60]{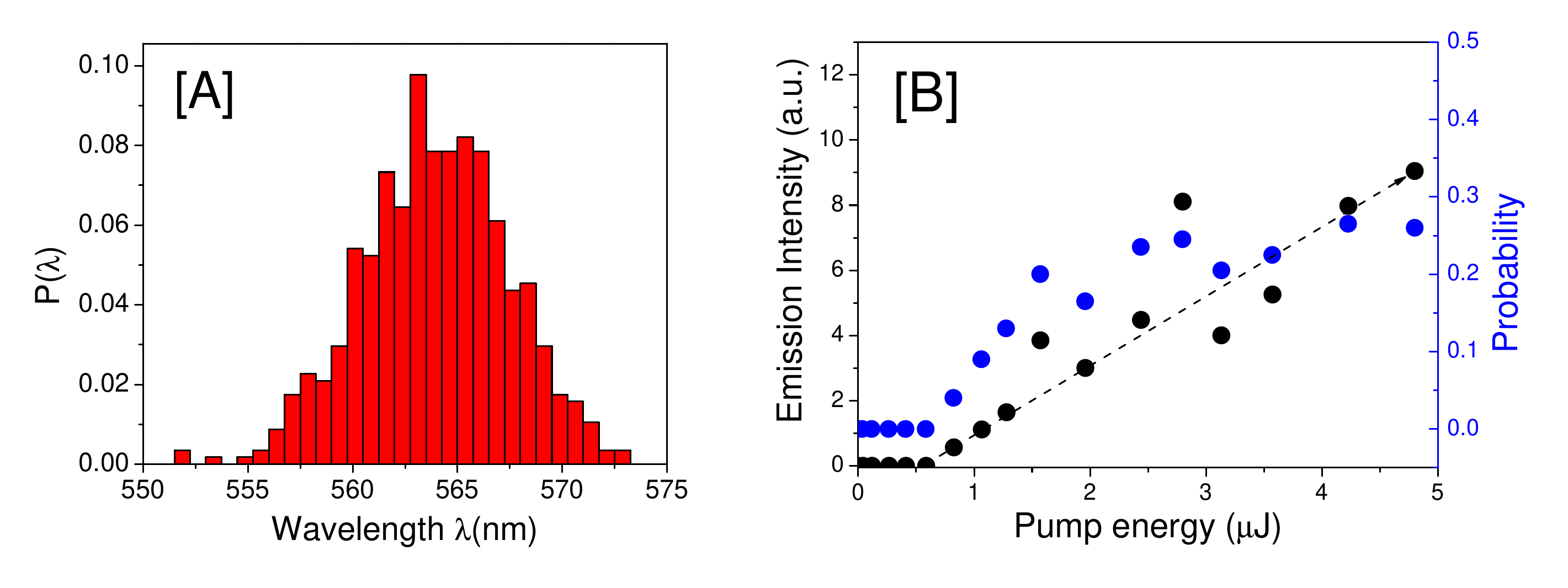}
\end{center}
\vspace{-25pt}
\color{blue}\caption{[A] Frequency behavior of coherent random lasing peaks. The peaks appear over a range of $10$~nm. [B] Threshold behavior of lasing emission. The black circles show the longitudinal emission intensity as a function of pump energy. The blue circles show the probability of lasing. The emission intensity shows a lasing threshold at $E_{p} = 0.6~\mu$J.}
\label{fig3p5}
\end{figure}

To study the frequency fluctuations of the random lasing modes, we collected $500$ spectra. The frequency distribution of the coherent peaks is shown in Figure~\ref{fig3p5}[A]. Like a conventional random laser, the distribution is continuous and centered at the gain maximum of Rhodamine dye. The width of the distribution is about $10$~nm indicating that the lasing frequency fluctuates over $10$~nm. Figure~\ref{fig3p5}[B] shows the threshold behavior exhibited by the aerosol system under excitation. The longitudinal emission intensity (black circles) has a clear lasing threshold at $E_{p} = 0.6~\mu$J, after which the intensity grows linearly on the average, with strong fluctuations. It is known that, in a coherent random laser, the system does not emit random lasing modes at each excitation pulse and a probability of random lasing can be measured~\cite{fallert09}. This probability quantifies the ratio of spectra exhibiting coherent modes to the total number of spectra. In our system, the lasing probability also shows this threshold behavior at the same energy (blue circle). We find that at threshold, about $4~\%$ of the spectra show lasing. At a higher pump power ($4~\mu$J) the lasing probability rises to about $25~\%$, and beyond that it shows a saturation behavior.

\section[Emission properties of a dye-scatterer random laser in a microchannel~.~.~.~.]{\label{sec:level3.3}Emission properties of a dye-scatterer random laser in a microchannel geometry} 

In this section, we compare the emission properties of the aerosol with the same dye solution, with suspended scatterers, in a microchannel geometry. We show that the discretization of the liquid jet in the form of microdroplets plays a vital role in the coherent random lasing. First, we briefly discuss the fabrication procedure of a microchannel geometry. To that end, we used a commercial micropipette puller (Sutter Instruments, P$-2000$). It consists of a $20$~W CO$_{2}$ laser with a beam diameter of $3.5$~mm. The CO$_{2}$ laser beam is internally redirected by a retro-reflective mirror to provide relatively uniform intensity on a mounted capillary. A quartz glass capillary (inner diameter: $700~\mu$m and outer diameter: $1000~\mu$m) is loaded in the puller and the laser light scans a specified longitudinal area. During the scan, the CO$_{2}$ laser heats and melts the exposed area of the capillary. At the same time, an internal tension device pulls the capillary along its major axis which results in the filament formation. The filamentation speed of the capillary can be controlled by adjusting the velocity of initial tension. Next, the laser is deactivated and after a preset delay, a hard pull is applied. Due to the hard pull, the capillary breaks into two separate pieces. The parameters affecting the size and shape of the pulled capillary include heating temperature, scanned area, pulling strength, etc.

\begin{figure}[h]
\begin{center}
\includegraphics[scale=0.6]{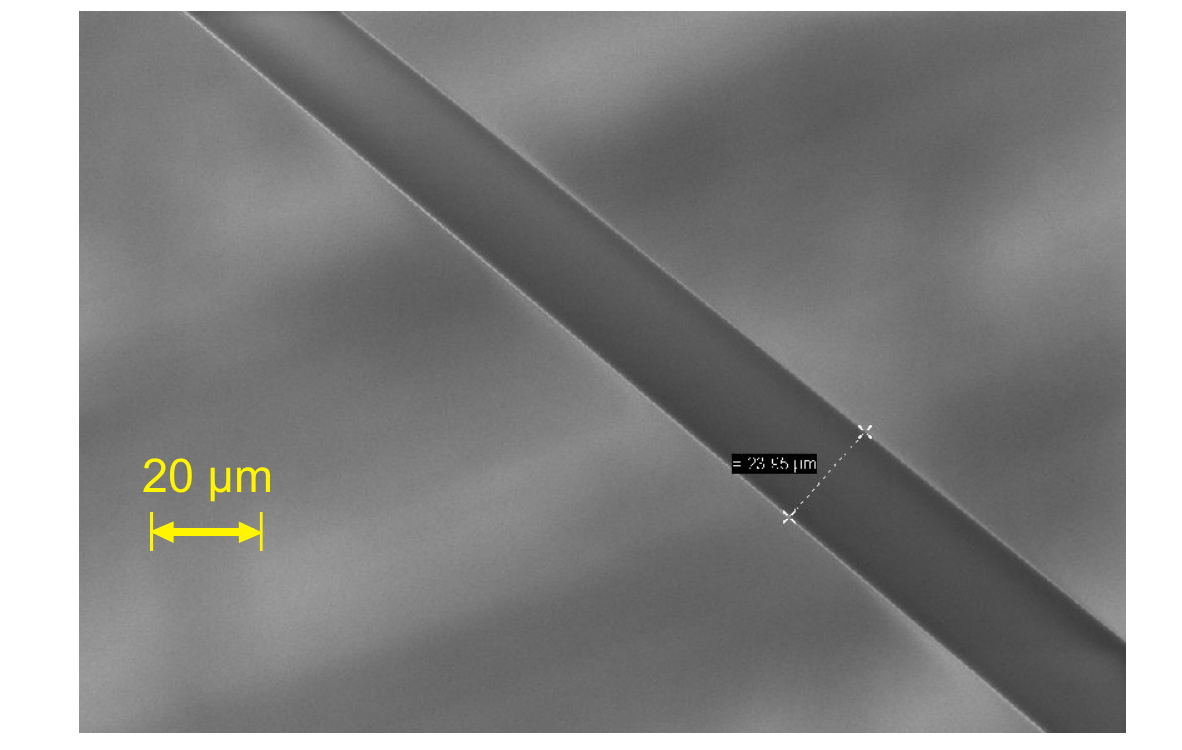}
\end{center}
\vspace{-25pt}
\color{blue}\caption {Image of the pulled capillary. The internal diameter of the capillary was measured to be $24~\mu$m.}
\label{fig3p6}
\end{figure}

\begin{figure}[h]
\begin{center}
\includegraphics[scale=1.4]{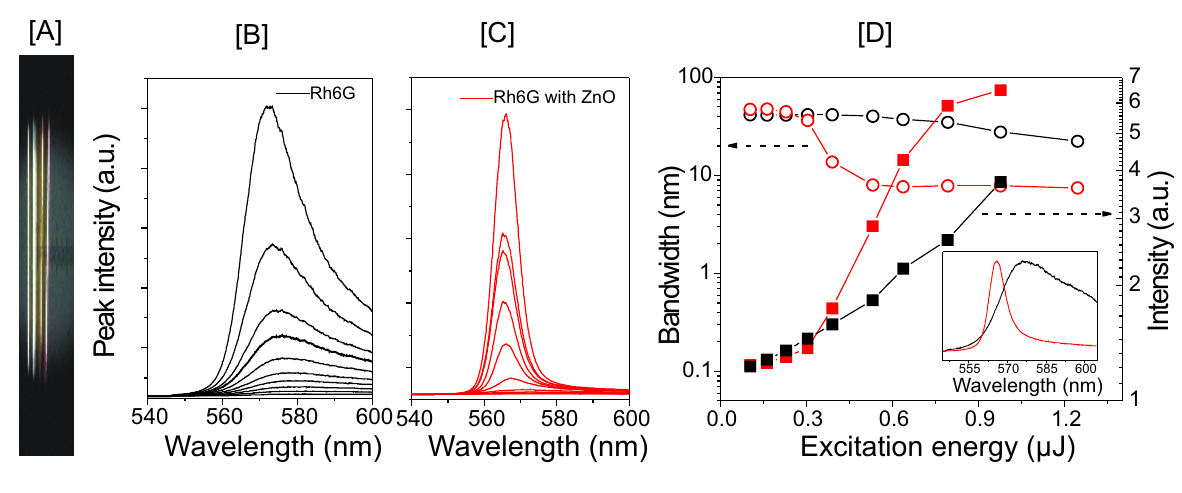}
\end{center}
\vspace{-20pt}
\color{blue}\caption{Longitudinal emission spectra from the dye solution filled in the microchannel geometry [A] Image of the microchannel [B] Emission spectra of the pure dye at various pump excitation energy. [C] Emission spectra of the dye solution with ZnO nanopowder. [D] Spectral features of emission: square markers- from pure Rh6G/methanol solution, circle markers- with ZnO nanopowder. Inset shows the corresponding spectrum at $E_{p}$ = $0.53~\mu$J.}
\label{fig3p7}
\end{figure}

The image of a fabricated microcapillary is shown in Figure~ \ref{fig3p6}. The uniform region of the capillary ($\sim 1$~mm in length) has an internal diameter of about $24~\mu$m and the aperture formed at the tip was measured to be about $10~\mu$m (not shown in the figure). The dye solution was filled by capillary action. The image of the filled capillary is shown in Figure~\ref{fig3p7}[A]. The uniform region of the microcapillary was pumped by the same laser with a focal spot of about 1~mm. The longitudinal emission spectra at various pump energies are shown as the black curves in Figure~\ref{fig3p7}[B]. The neat dye solution exhibited normal fluorescence. Next, we randomized the dye solution by adding ZnO nanoparticles (refractive index $2.0$, size $\sim 10$~nm, concentration $7.5 \times 10^{14}$/cc). The spectra at the same excitation energy are shown by the red curves in Figure~\ref{fig3p7}[C]. Spectral narrowing due to the randomness is immediately evident from the figure.

The longitudinal spectral response of the dye in the microchannel geometry is quantified in Figure~\ref{fig3p7}[D]. The black and the red markers show the respective response of pure and disordered dye solution as a function of pump excitation intensity. The pure dye solution has normal fluorescence properties like steady emission enhancement (see the black squares) and gradual bandwidth narrowing. At higher power ($1.2~\mu$J), the bandwidth was observed to be $\sim 20$~nm. On the other hand, the disordered dye solution exhibits the signature of diffusive random lasing. Here, the output intensity shows a rapid growth and the bandwidth collapses above the threshold energy of $E_{p} = 0.3~\mu$J. Importantly, in the microchannel geometry, coherent random lasing was not observed.This confirms that merely a microcolumnar random lasing geometry does not realize coherent modes. In the aerosol system, the breakage of the continuous solution into tiny droplets is responsible for the coherent lasing modes. Adequate optical feedback assisted by the high gain of the dye realizes coherent random lasing in the microdroplet array.

{\bf Summary:}\\ In summary, we have demonstrated coherent random lasing from an amplifying aerosol. The emission spectra from the aerosol array consist of ultra-narrowband random lasing modes. The transverse spectral image confirms that the random lasing emission arises due to coherent participation of the droplets. These modes are not the resonances of individual droplet as the optical feedback is extended over the entire illuminated array. The lasing emission is strongly directional along the array and exhibits a clear threshold behavior. Furthermore, the lasing peak fluctuates in both frequency and intensity due to fluctuating configuration of the aerosol array.	
\chapter{A monodisperse aerosol random laser}

In the previous chapter, we studied coherent random lasing from a random array of polydisperse microdroplets. The lasing action was the consequence of cooperative participation of all microdroplets present in the pumped region. We discussed the properties of collective emission like ultranarrow bandwidth, frequency fluctuations and threshold behavior. In this chapter, we experimentally study the optical emission behavior of the array when the amplifying microdroplets are monodisperse, well-separated and randomly arranged. We find that the collective lasing phenomenon happens at specific wavelength intervals. The analysis of lasing frequency indicates the participation of Fabry-Perot resonances of the individual microdroplets. We further report on the angular distribution and polarization anisotropy of the collective emission. This chapter is organized as follows: Section \ref{sec:level4.1} briefly describes the
generation process of monodisperse microdroplets. In section \ref{sec:level4.2}, we discuss the well-known Whispering Gallery Resonances of an individual spherical scatterer and their utility in sizing the droplets. Then in section \ref{sec:level4.3}, we discuss the frequency quantization in the collective emission. Finally in section \ref{sec:level4.4}, we present the general properties of the cooperative emission like angular distribution, polarization anisotropy and threshold behavior. 

\section{\label{sec:level4.1}Generation of monodisperse amplifying aerosol}

\begin{figure}[h]
\begin{center}
\includegraphics[scale=1.3]{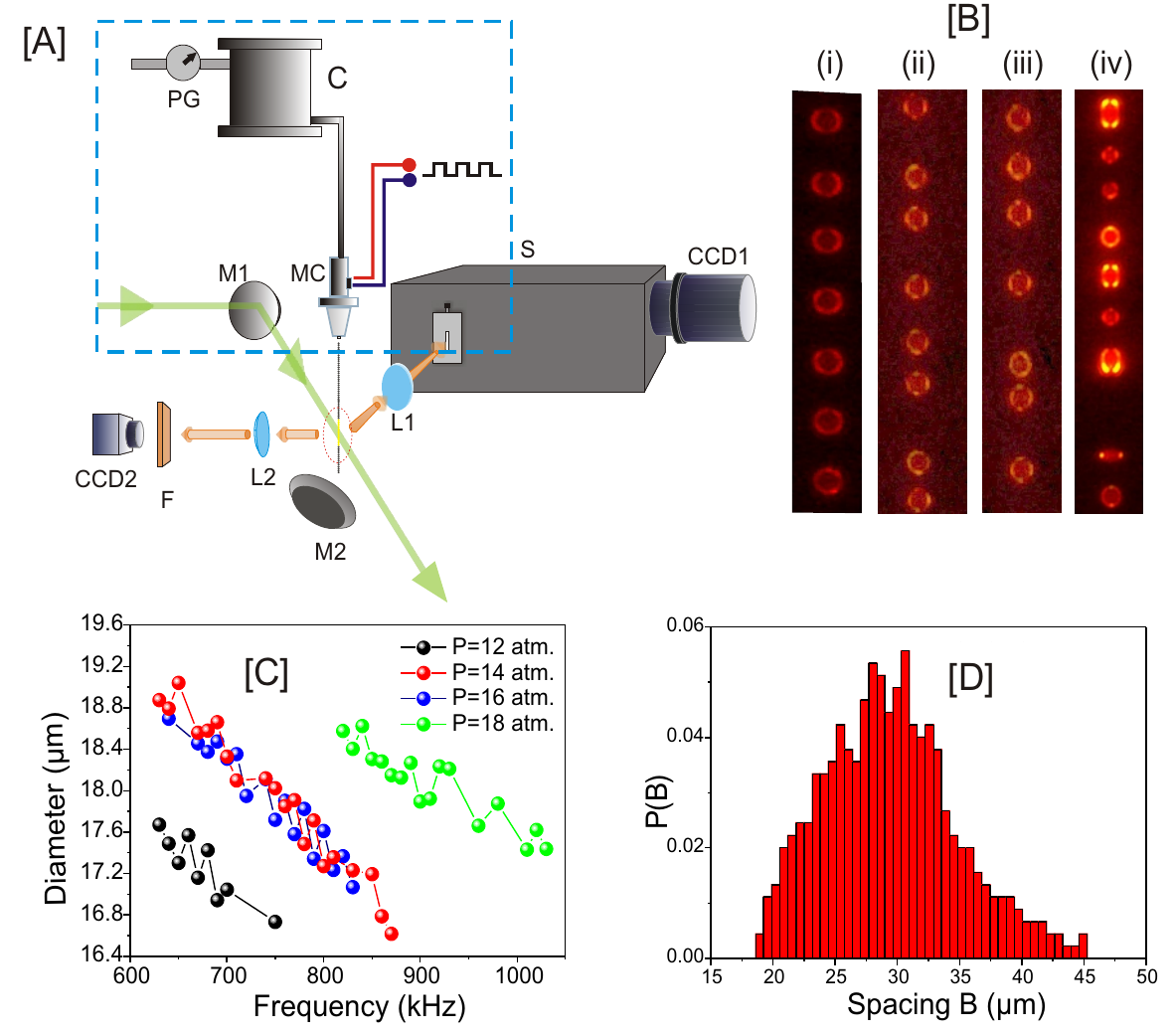}
\end{center}
\vspace{-30pt}
\color{blue}\caption {[A] Experimental setup to observe random lasing from the droplet array; dotted rectangle highlights the piezo driven micro capillary. [B] Four manifestations of the array. [C] Variation of droplet sizes as a function of piezo frequency at various pressures. [D] Distribution of spacing between nearest-neighbor droplets.}
\label{fig4p1}
\end{figure}

The basic process of microdroplet formation using a narrow capillary was described in the previous chapter. When a liquid is forced out of a narrow aperture under pressure, the exiting liquid jet, which is mechanically unstable, breaks spontaneously into an aerosol of fine microdroplets. Without an external influence, the jet disintegrates into a stream of polydisperse,
irregularly shaped microdroplets. To break the liquid jet into equal sized droplets, we perturb the liquid stream with an appropriate periodic perturbation. The drop formation process can be controlled by the frequency and amplitude of the perturbation. In our microcapillaries, the perturbation on the stream is done by using an embedded piezo crystal. An electric square wave signal of $3$~V peak-to-peak at a few hundred kHz applied to the piezo generates monodisperse microdroplets. In this configuration, the microcapillary functions as a vibrating orifice aerosol generator (VOAG), a device which has been traditionally used to study Mie lasers~\cite{lin90, berguland73, qian86, chang96}. The schematic of the experimental setup is shown in Figure~\ref{fig4p1}. The highlighted dotted rectangle marks the VOAG assembly. PG is the pressure gauge to monitor the pressure applied on the liquid chamber (C) and MC is the microcapillary having an embedded piezo crystal. The right panel shows the representative embodiments of the array. The size of the microdroplets depends on various physical parameters such as viscosity of the liquid, diameter of the orifice, pressure applied on the liquid, etc. The droplet diameter ($2a$) can be estimated using the following relation~\cite{berguland73}

\begin{equation}
2a =(\frac{6Q}{\pi f})^{1/3}
\label{eq4p1}
\end{equation}
where $Q$ is the volumetric flow rate of liquid through the orifice and $f$ is the driving frequency. Various monodisperse configurations of the array of droplets with diameter $16~\mu$m to $25~\mu$m can be achieved by scanning the frequency ($500 - 1000$~kHz) and pressure ($10 - 20$~atmospheres). Figure~\ref{fig4p1}[B] shows four representative embodiments of the array which we identify as weakly periodic (i), aperiodic (ii and iii) and polydisperse (iv) array with random spacings. Figure~\ref{fig4p1}[C] shows the size variation of monodisperse microdroplets as a function of perturbation frequency at four different pressures. The size and monodispersity of the droplets are estimated using the information of Whispering Gallery Modes (WGMs) which is discussed in the next section. Importantly, at a given frequency and pressure, the array maintains its monodispersity for a sufficiently long time ($30 - 40$~minutes), which allows us to perform our experiments on various configurations with the same degree of disorder. Figure~\ref{fig4p1}[D] shows the distribution of interdroplet spacing (surface to surface) over $500$ configurations. It is clear from the histogram that the spacing between the droplets has a continuous distribution with a spread of over $20~\mu$m~\footnote{The actual process of characterization is explained in section\ref{sec:level4.2}}. It appears that smaller spacing is more probable than larger, which could be a consequence of the fluid dynamic generation process. But this variation is not the focus of our investigation. The size and the spacing distributions can be changed by varying the perturbation frequency.

\section{\label{sec:level4.2}Whispering gallery modes for characterization}

\begin{figure}[h]
\begin{center}
\includegraphics[scale=1.3]{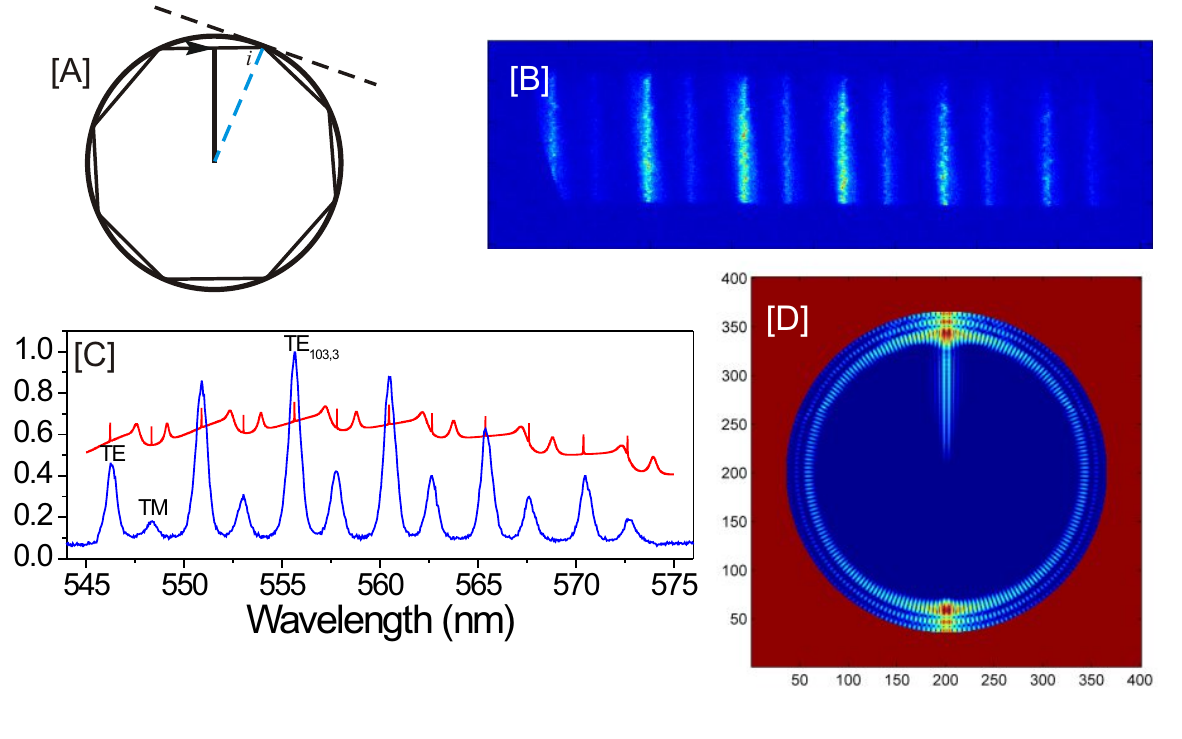}
\end{center}
\vspace{-35pt}
\color{blue}\caption{[A] Schematic representation of the formation of the WGMs [B] Spectral image of the droplet array, the showing TE and the TM modes [C] Blue curve: Experimentally observed spectrum from the array; Red curve: Mie scattering efficiency curve from a single scatterer of $16.5~\mu$m with a refractive index contrast of $1.34$. [D] Field distribution of the TE$_{103, 3}$ mode at the equatorial plane.}
\label{fig4p2}
\end{figure}

In an ideal condition, the VOAG generates highly monodisperse microdroplets with a very narrow size distribution. By using simple imaging techniques, this distribution cannot be resolved. In the current study, unlike in chapter 3, we need a much more accurate method of measuring sizes and spacing of the microdroplets. To this end, we use the transverse emission spectrum which consists of Whispering Gallery Modes (WGMs) of the individual microdroplets. The basic principle of the WGM formation is shown in Figure~\ref{fig4p2}[A]. Let us consider a ray of light (pointed by the arrow) propagating inside a sphere of radius a with a refractive index of m relative to the ambient medium. If the ray hits the surface at an angle of incidence $i$ $>$ $i_c$ = $sin^{-1}$($1/m$), the phenomenon of total internal reflection takes place. A resonance is hit if the totally internally reflected ray undergoes multiple internal reflections and closes in on itself. In this eventuality, light gets trapped inside the sphere. For large spheres (diameter $2a\gg\lambda$), the trapped ray propagates close to the surface and travels a distance of $\sim 2\pi a$ in each round trip. Internal electric field within the sphere is expected to be large whenever the ray satisfies the resonance condition, $\pi a = n \lambda$, where $n$ is an integer multiple. Thus, the spectral position of these resonances depends on the sphere size and its refractive index contrast with the ambient medium.

Figure~\ref{fig4p2}[B] shows the spectral image of a representative monodisperse array, captured by the spectroscopic ICCD. The image consists of several vertical bright stripes of alternating high and low intensity. These are the Transverse Electric (TE) and Transverse Magnetic (TM) WGM modes, depending on whether the electric or the magnetic field is parallel to the spherical surface. Whenever the droplets in the array achieve monodispersity, the WGMs of all the droplets occur at the same wavelength. The broadband emission spectrum of the dye ensures that several resonant modes are excited. The spacing between two consecutive TE or TM modes is given by~\cite{campillo88}

\begin{equation}
\Delta\lambda =\frac{\lambda^2}{2\pi a}\Delta x
\label{eq4p2}
\end{equation}

where, 
\begin{equation}
\Delta x =  \frac{tan^{-1}\sqrt{(m^2-1)}} {\sqrt{(m^2-1)}} 
\label{eq4p3}
\end{equation}

The blue curve in Figure~\ref{fig4p2}[C] is the spectrum of the droplet array which is obtained by the vertical addition of the image. Here, the spacing between two TE or TM WG modes is $4.81$~nm at a central wavelength of $555.63$~nm. Using equation \ref{eq4p2}, this corresponds to a droplet size of $16.66~\mu$m. The equation \ref{eq4p2} is valid for large sphere diameters, and it is of interest to validate its usage by explicit Mie theoretical calculations.

For a spherical scatterer, the scattering efficiency $Q_{sca}$ is given by the Mie theory~\cite{bohren83}.

\begin{equation}
Q_{sca} = \frac{2}{x^2} \displaystyle\sum\limits_{n=0}^\infty (2n+1)(|a_n|^2+|b_n|^2)
\label{eq4p4}
\end{equation}
where, $x$ is the size parameter defined by $x = ka$, $k$ being the wave number ($2\pi$/$\lambda$) in the ambient medium. The infinite summation in the series can be truncated after $n_{max}$ terms which is given by 

\begin{equation}
n_{max} =x+4x^{1/3}+2
\label{eq4p5} 
\end{equation} 
The parameters $a_{n}$ and $b_{n}$ are the Mie coefficients and are given by 

\begin{equation}
a_n =\frac{m^2j_n(mx)[xj_n(x)]'-\mu_1j_n(x)[mxj_n(mx)]'}{m^2j_n(mx)[xh_n^{(1)}(x)]'-\mu_1h_n^{(1)}(x)[mxj_n(mx)]'}
\label{eq4p6}
\end{equation}

\begin{equation}
b_n =\frac{\mu_1j_n(mx)[xj_n(x)]'-j_n(x)[mxj_n(mx)]'}{\mu_1j_n(mx)[xh_n^{(1)}(x)]'-h_n^{(1)}(x)[mxj_n(mx)]'}
\label{eq4p7}
\end{equation}

where, $\mu_{1}$ is the relative magnetic permeability of the sphere. For nonmagnetic spheres, $\mu_{1} = 0$. The function $j_{n}$ and $h_{n}$ are the spherical Bessel functions of order $n$ and primes represent the derivatives with respect to the arguments. To calculate the Mie scattering coefficients and $Q_{sca}$, we used the Matlab functions by Matzler~\cite{matzler02}. The red curve in Figure~\ref{fig4p2}[C] is the plot of $Q_{sca}$ versus wavelength. To get the best match, we first calculate the scattering efficiency curve for the experimentally observed size parameters and then shift the curve by tweaking the particle size. The WGM peak positions show an excellent agreement with the experimental observation for a particle size of $16.5~\mu$m. This value is within $\sim 100$~nm from the size calculated using equation~\ref{eq4p2}, which suggests the validity of the equation. 

As discussed in section \ref{sec:level3.1}, we used an imaging CCD (CCD2) to estimate the spacing between the droplet. The precise measurement of the droplet size also helps us in this regard. Once we know the exact diameter of the droplets, we can back calculate the effective pixel size of the imaging CCD more accurately and hence the spacing. Figure~\ref{fig4p2}[D] shows the field distribution at a representative resonance wavelength at $\lambda = 555.63$~nm (red color corresponds to highest field density). From the Mie scattering coefficients, we identify this WGM as the third order TE$_{103, 3}$ mode. It is clear that the eigenmode is confined at the surface. The field exists in the form of a ribbon about the equatorial plane. Roughly, this volume can be estimated to be $\sim 175~\mu$m$^3$, assuming the radius vector spans $20^{\circ}$ about the equatorial plane~\footnote{This angle is roughly estimated from FDTD calculation.}. Since this is a very small volume, the lasing in the WGM modes is initiated at very low threshold energies.
 
\section{\label{sec:level4.3} Collective lasing from the monodisperse microdroplet array}

\begin{figure}[h]
\begin{center}
\includegraphics[scale=0.7]{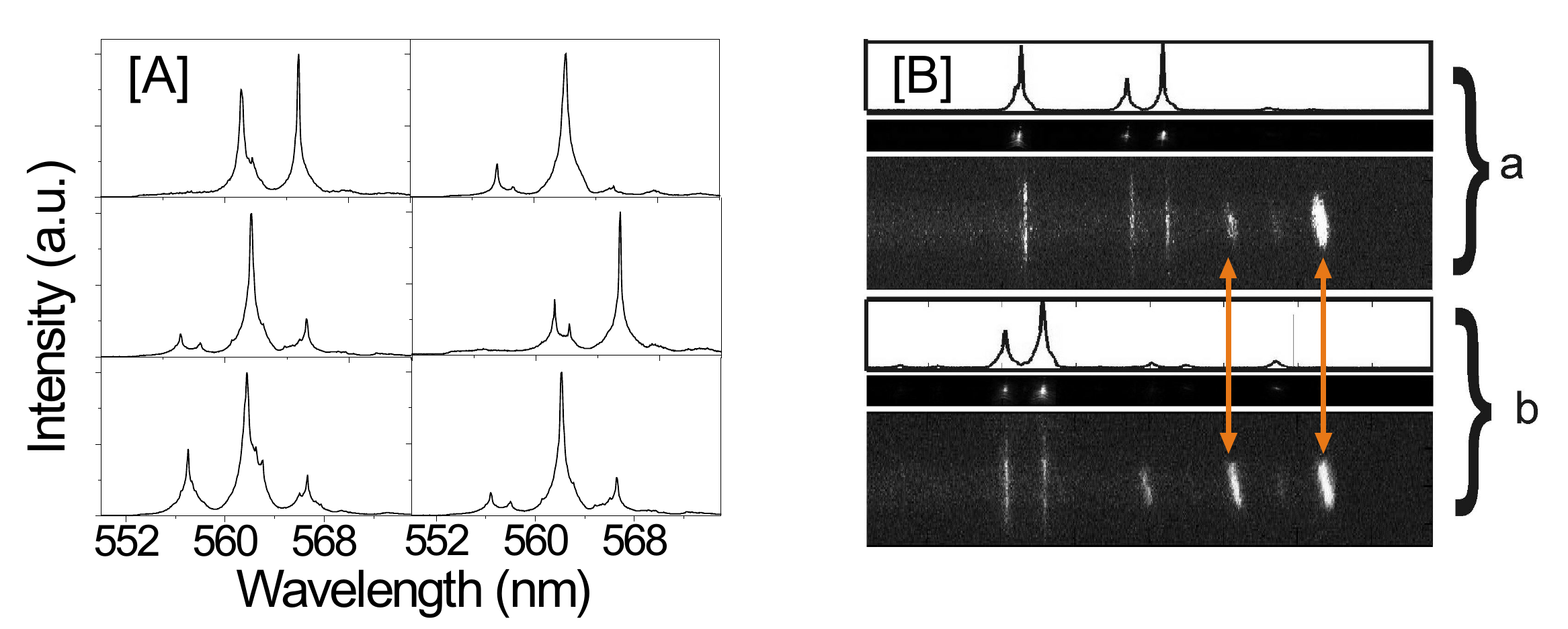}
\end{center}
\vspace{-25pt}
\color{blue}\caption{Spectra from the monodisperse array, [A] Longitudinal spectra [B] Two sets `a' and `b' of longitudinal (middle panel) and corresponding transverse (bottom panel) spectral images, arrows indicate the WGMs. See text for details.}
\label{fig4p3}
\end{figure}

The emission properties of the monodisperse aerosol in the longitudinal direction are summarized in Figure~\ref{fig4p3}, which presents similar analysis as in Figure~\ref{fig4p3}, but for this monodisperse configuration.

Figure~\ref{fig4p3}[A] illustrates six shot-to-shot emission spectra at a pump energy $E_{p} = 0.55~\mu$J. As discussed in section \ref{sec:level3.2} both transverse and longitudinal emission can simultaneously be collected by the lens L1. The spectra consist of multimode coherent peaks with a resolutionlimited bandwidth of $\sim 0.2$~nm. The number of modes increases with increasing excitation energy. Figure~\ref{fig4p3}[B] illustrates simultaneously captured spectral images in both the longitudinal and transverse directions for two different excitation pulses, labeled as `a' and `b'. Within each (`a' or `b') data set, the top panel shows the longitudinal spectrum, the middle panel shows the longitudinal spectroscopic image, while the bottom panel depicts the transverse spectroscopic image. The transverse spectral image consists of WGMs (labeled by the red arrows) and the scattered fraction of collective modes. The WGMs of all the droplets occur at the same wavelength and remain fixed as long as the array maintains its monodispersity. Small inclination of the WGMs observed in the spectral images is due to a minor monotonic variation in microdroplet sizes. Heterogeneity in the droplet sizes is estimated by the width of the WGM patches. In our experiments, we find that even in the best monodisperse configurations, the droplets have a size variation of about $50$~nm. Similarly, the spacings between consecutive spheres is not exactly equal and fluctuations occur there too. We believe, the heterogeneity in the array arises due to the operational limits of VOAG, fluctuations in the applied pressure and the air current, etc. It is clear from Figure~\ref{fig4p3}[B] that, corresponding to every high intense spot in the middle panel, there is a vertical bright streak in the transverse direction. Similar to that observed in the polydisperse array (see section \ref{sec:level3.2}), these streaks are extended over the entire pumped region and comprise of scattered light from the collective mode~\cite{anj12ol}.

\subsection{\label{sec:level4p3a}Discretization of frequency distribution}
\begin{figure}[h]
\begin{center}
\includegraphics[scale=0.5]{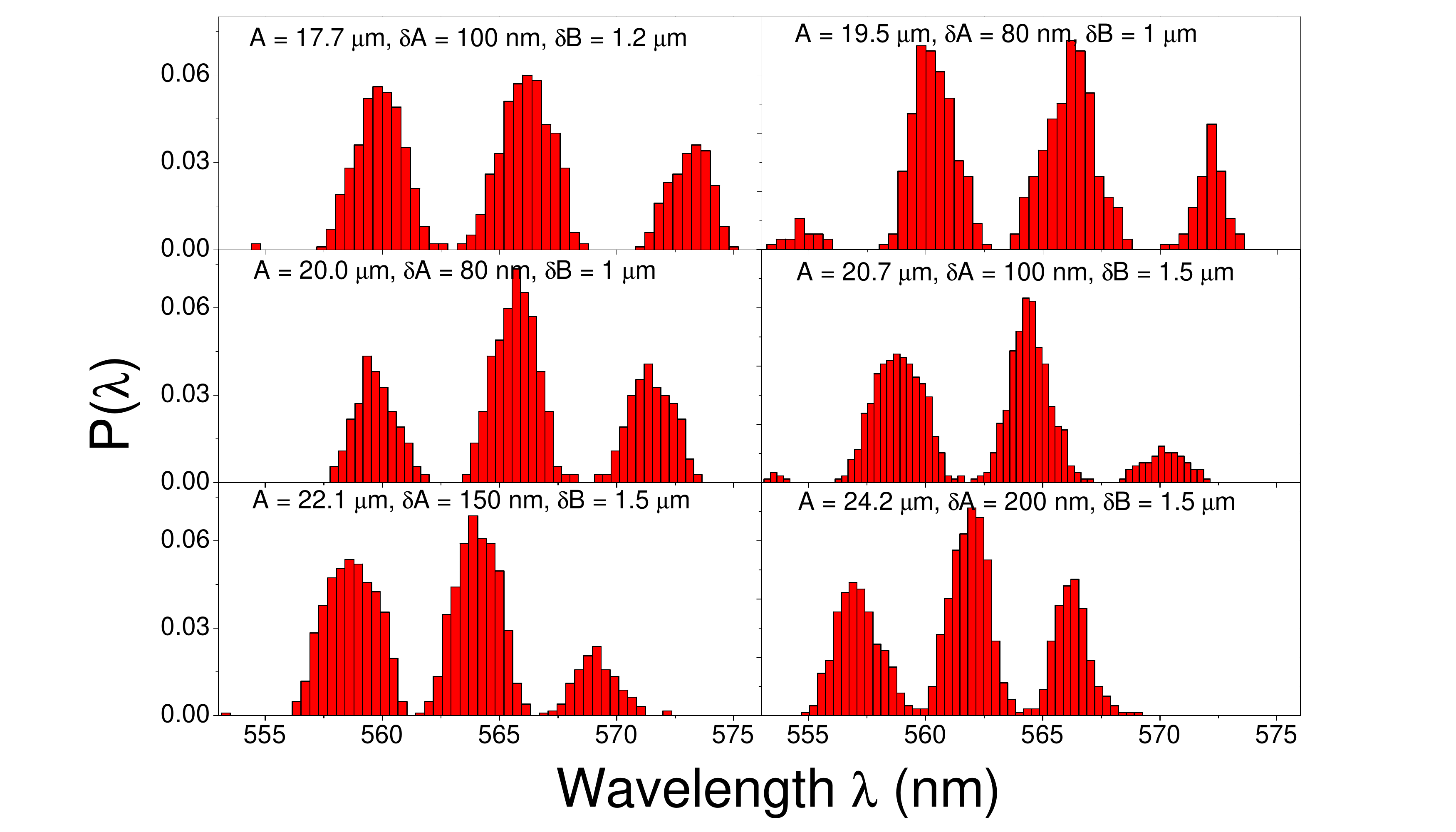}
\end{center}
\vspace{-30pt}
\color{blue}\caption{Frequency distribution of the lasing modes at various droplet sizes.}
\label{fig4p4}
\end{figure}

In section \ref{sec:level3.2b}, we showed that in a polydisperse aerosol, the collective modes appear at random wavelengths spanning a range of about $10$~nm. In the present monodisperse case also, the spectra fluctuate from pulse-to-pulse and the frequency sensitivity is not evident in an individual spectrum. To see the effect of monodispersity, we observe the lasing frequency over $500$ spectra. Figure~\ref{fig4p4} shows the distributions of the mode frequency in monodisperse droplet arrays for various microdroplet diameters ranging from about $18~\mu$m to $24~\mu$m. Clearly, the histograms show distinct bunches, implying that the collective lasing peaks appear in certain intervals and are forbidden at other wavelengths. The continuous distribution of the polydisperse array is thus discretized in the monodisperse system. We observed that, the number of lasing bunches in a given frequency range depends on the droplet diameter, gain profile and the excitation energy.

\begin{figure}[h]
\begin{center}
\includegraphics[scale=0.4]{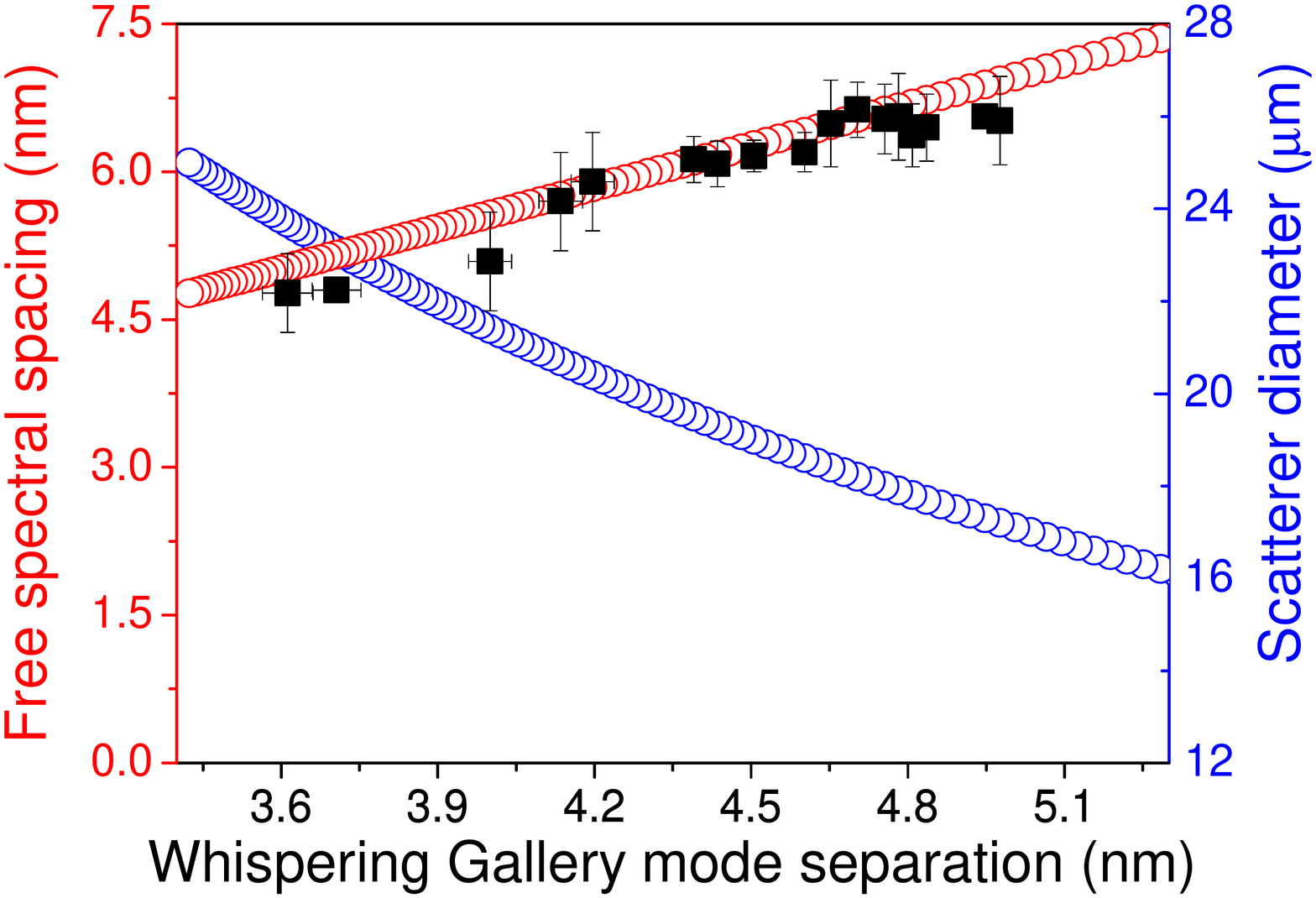}
\end{center}
\vspace{-30pt}
\color{blue}\caption{Frequency characteristics of the histogram bunches. X axis shows the WGM mode separation. The curve marked by the blue circles shows the calculated droplet diameter, as labeled on the right Y axis. The red circles show the Fabry-Perot (FP) free spectral range for a FP resonator with the corresponding droplet diameter. The Left Y axis shows the experimentally measured separation in the histogram bunches marked as the black squares.}
\label{fig4p5}
\end{figure}

The origin of the varying separations between the bunches is revealed in Figure~\ref{fig4p5}. On the X axis, we show the WG mode separation, which is the measurable quantity in our experiment. The right Y axis represents the diameter of microdroplets calculated by WGM separation using equation \ref{eq4p2}. For these diameters, we calculate the free spectral spacing of the Fabry-Perot resonances formed between the diametrically opposite surfaces of a single microdroplet ($\delta\lambda = \lambda^{2}/2md$). This FSR is shown by the red circles, scaled on the left Y-axis. The black squares indicate the experimentally measured separations in the histogram bunches. Horizontal error bars are the standard deviations in particle size, estimated by the spread in WG modes. Vertical error bars indicate variations in bunch separation. As can be seen from Figure~\ref{fig4p4}, the interbunch separation between two neighboring bunches in a histogram is not the same. The vertical error-bar characterizes this variation, with the black square being the average separation. Clearly, the experimental data show an excellent agreement with the Fabry-Perot free spectral range measured over the entire range of diameters achievable in our system. This analysis implies that the Fabry-Perot resonances of the droplet participate in longitudinal emission and the distributed feedback along the array triggers the collective lasing modes.

\section{\label{sec:level4.4}Characteristics of the collective modes}
\subsection{\label{sec:level4.4a}Threshold behavior}

\begin{figure}[h]
\begin{center}
\includegraphics[scale=0.55]{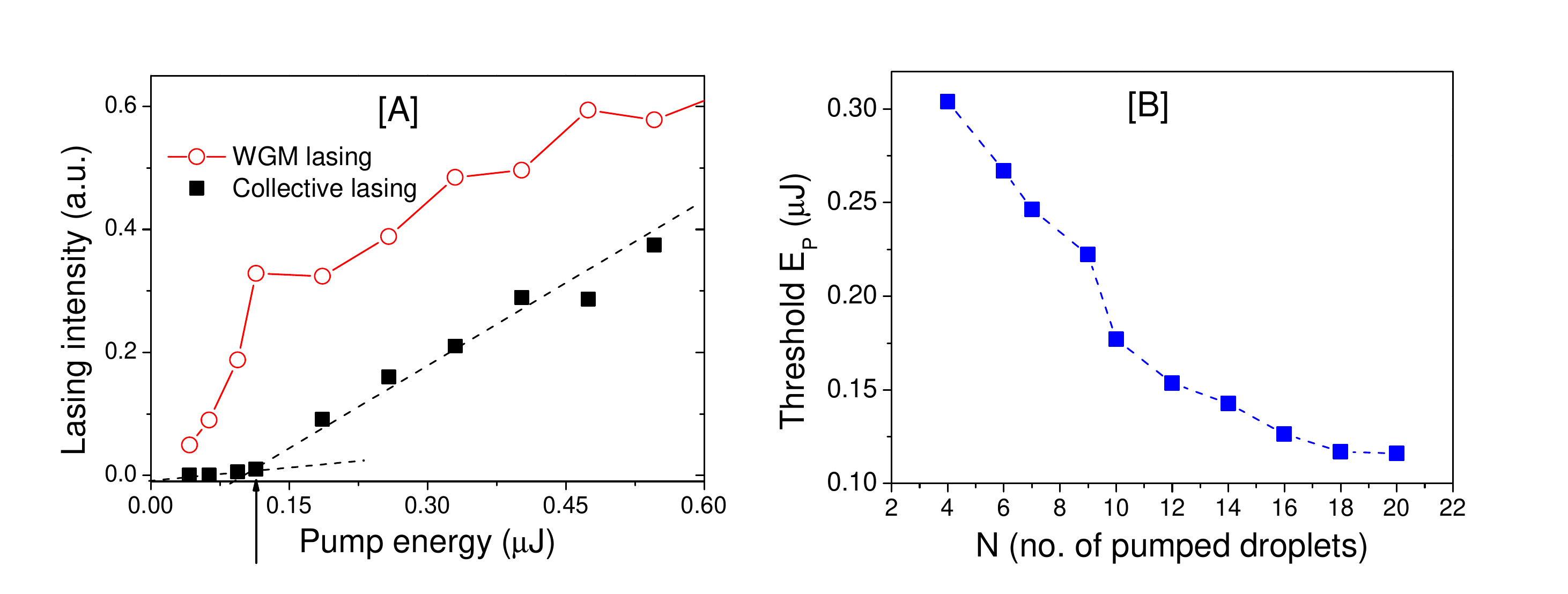}
\end{center}
\vspace{-25pt}
\color{blue}\caption{[A]: Red circles: output intensity of the WGMs as a function of input energy; black squares: output intensity of the collective modes. Dashed line: linear fit to the experimental data. Output intensity shows a lasing threshold at $E_{p} = 0.12~\mu$J for the collective modes. [B]: Lasing threshold versus number of excited microdroplets.}
\label{fig4p6}
\end{figure}

Whether an optical signal constitutes laser emission is ascertained by the existence of a threshold in its Input/Output characteristics, as discussed in section\ref{sec:level1p2} regarding a conventional laser. Figure~\ref{fig4p6} depicts the threshold behavior in the longitudinal emission when $\sim 20$ microdroplets in the array were excited. The black squares show the variation of the spectrally averaged output intensity as a function of the excitation energy. Every data point is averaged over $100$ spectra. The black dashed line is a linear fit to the experimental data. As indicated by the arrow, the intensity variation shows a clear change in the slope at $E_{p} = 0.12~\mu$J. Below this threshold energy, collective lasing is not observed. This threshold is an order of magnitude smaller than that for the polydisperse configuration (see section\ref{sec:level3.3}). The red circles show the output intensity of the WGMs. Since the WGMs have a higher quality factor and a smaller mode volume, the inversion for the WGM modes is reached rapidly and hence they have a smaller lasing threshold. Although we could not capture the exact lasing threshold of the WGM lasing, it can be inferred that the quality factor of the collective modes is lower than that of WGM's. On the other hand, the threshold of collective lasing can be lowered by increasing the size of the pumped array as shown in the right panel in Figure~\ref{fig4p6}. It depicts the variation of the lasing threshold with the number of pumped microdroplets. In the experiment, the number of amplifying droplets was controlled by changing the size of the pump spot. It is clear that the threshold energy monotonically decreases as we increase the size of pump spot. When the number of pumped microdroplets was greater than $16$, the threshold energy decreases too slowly to be measured in our setup. We observed that when the number of pumped droplets was less than $4$, the system did not lase, indicating a very high lasing threshold~\footnote{Pumping with very intense light leads to damage of the droplets, and hence was not pursued.}.

\subsection{\label{sec:level4.4b}Angular distribution of the collective emission}

\begin{figure}[h]
\begin{center}
\includegraphics[scale=0.55]{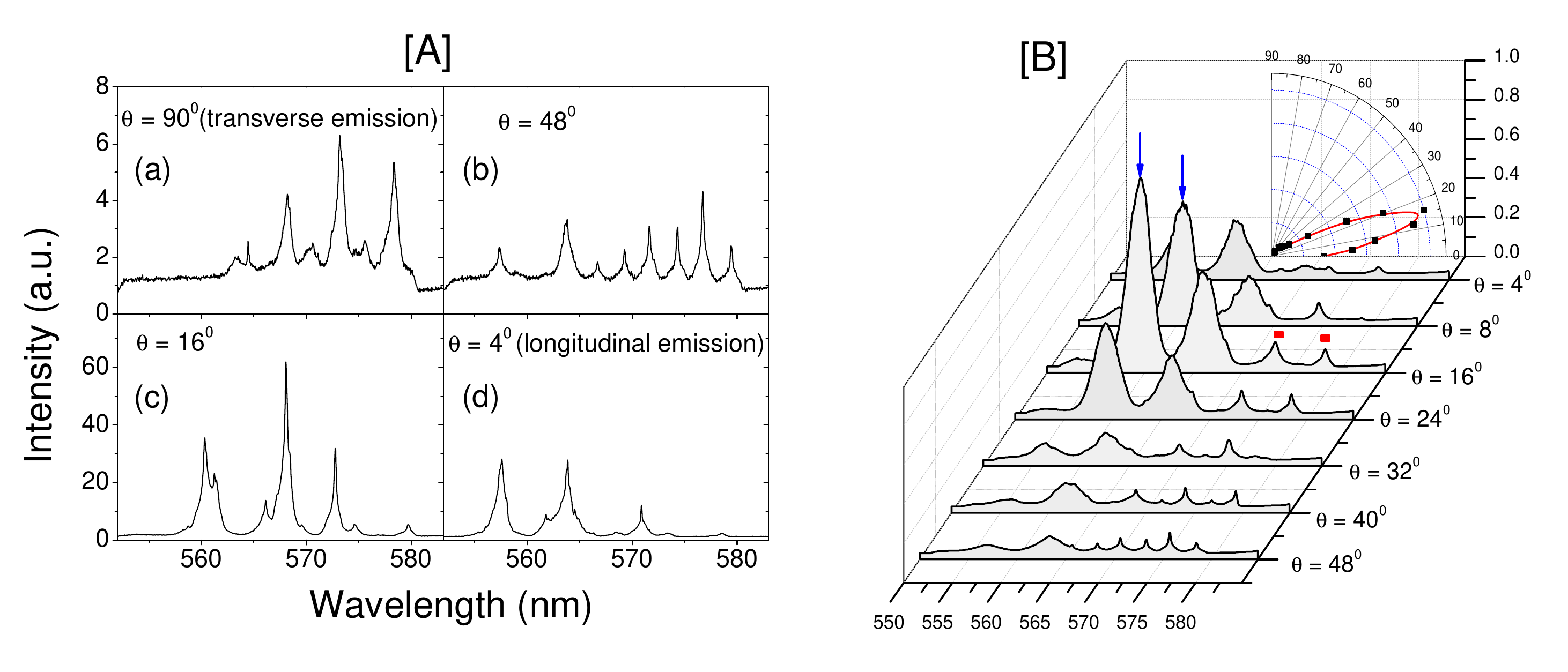}
\end{center}
\vspace{-30pt}
\color{blue}\caption{Angular distribution: [A] Emission spectra from the droplet array, (a) transverse, (b) $48^{\circ}$, (c) $16^{\circ}$ and (d) longitudinal direction at an excitation energy of $10~\mu$J. [B] Ensemble average spectra over $200$ pulses. Inset is the angular distribution of spectrally averaged coherent emission intensity. The maximum intensity is observed at $16^{\circ}$.} 
\label{fig4p7}
\end{figure}

A significant feature of any novel optical source is the angular distribution of its emission. To measure the angular distribution of lasing intensity from the array, the mirror M2 was mounted on a rotational stage. The mirror was rotated in a steps of $2^{\circ}$ and $200$ spectra were collected at every step. Figure~\ref{fig4p7} [A] shows representative single shot emission spectra at four different angles. Here, $\theta$ represents the polar angle with respect to the axis of the microdroplet array. Thus $\theta = 90^{\circ}$ corresponds to the transverse emission and $\theta = 0^{\circ}$ corresponds to the longitudinal emission~\footnote{$\theta = 0^{\circ}$ could not be measured because the droplet stream would fall on the mirror M2.}. 

Figure~\ref{fig4p7}[B] depicts the ensemble averaged spectra at various angles. Here, the coherent intensity is distributed into two bunches as marked by the blue arrows. We observed that, in the transverse direction, the WGM lasing (marked by the red rectangles) dominates over the collective lasing, while in the longitudinal direction, the emission intensity is dominated by the collective lasing modes. We separated out the contribution of the collective modes from the ensemble averaged spectra. The inset in Figure~\ref{fig4p7}[B] is the polar plot of the spectrally averaged intensity in the collective modes. Intuitively, one can expect that the intensity of the lasing modes should maximize in the longitudinal direction as the gain is maximum along the droplet chain. However, the collective modes show a minimum at $\sim 16^{\circ}$. Apparently, the system emit minimal light in the direction of maximal gain. This is a rather intriguing observation since it is known that the gain anisotropy in random lasers maximises the output in the direction of gain. It can be surmised that the feedback from the surfaces of the microdroplet plays a crucial part and is responsible for this effect. We will comment about this outcome in the next chapter.  

\subsection{\label{sec:level4.4c}Polarization anisotropy}

\begin{figure}[h]
\begin{center}
\includegraphics[scale=1.1]{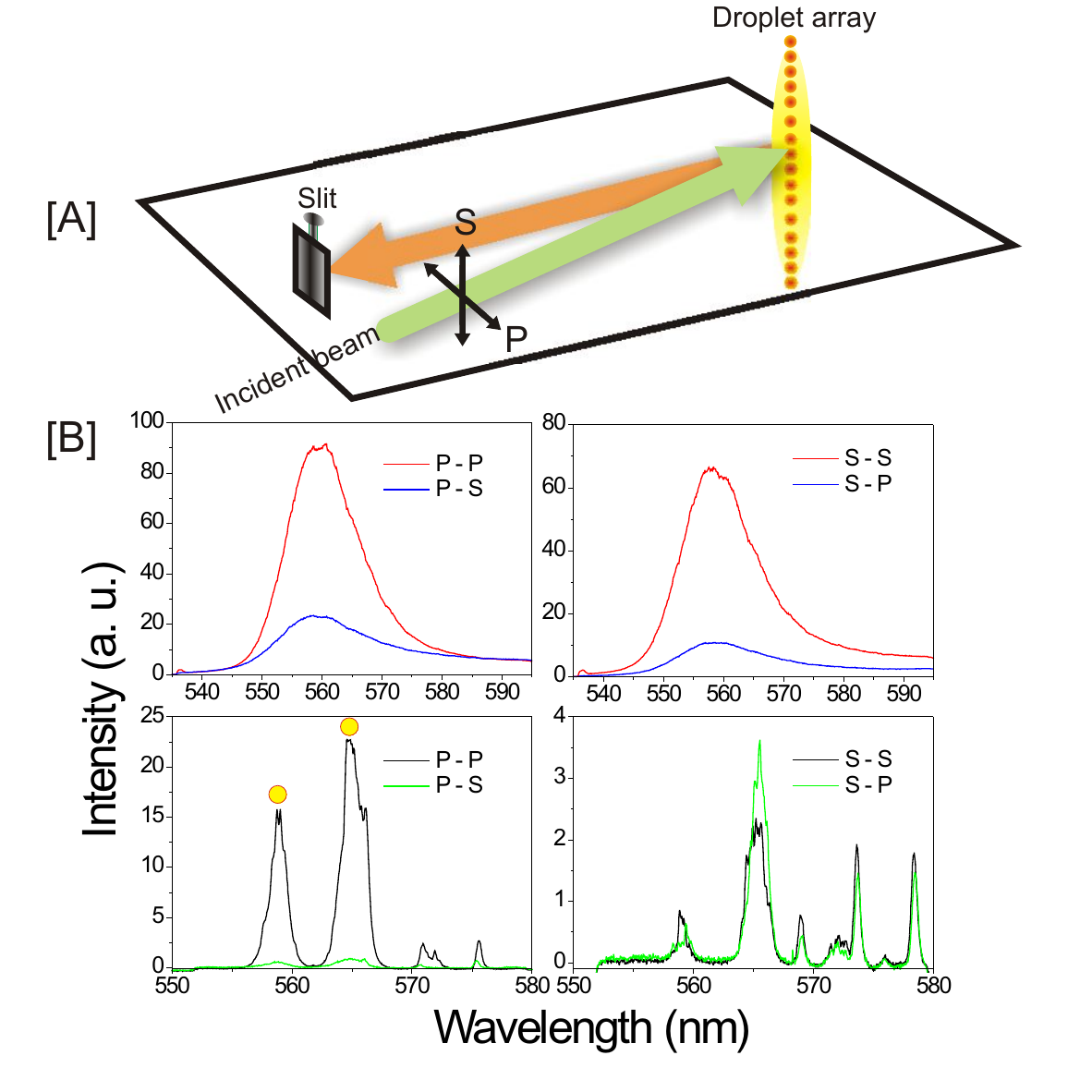}
\end{center}
\vspace{-35pt}
\color{blue}\caption{Polarization anisotropy: [A] Sketch of the experimental setup. [B] Top panel: Ensemble averaged spectra from the Rhodamine dye filled in a cuvette. Bottom panel: Ensemble averaged spectra from the droplet array. The angle between the pump laser and spectrometer was kept at $\sim 10^{\circ}$.} 
\label{fig4p8}
\end{figure} 

In this section we discuss the polarization properties of collective emission in the microdroplet array. We first study the fluorescence signal from the pure dye solution filled in a cuvette at $2$~mM concentration. As shown in Figure~\ref{fig4p8} [A], a linear polarized beam (P or S)~\footnote{P and S represent the electric field oscillations $\parallel$ and $\perp$ to the plane determined by the directions of excitation and observation.} was incident on the sample. The fluorescence signal from the sample was collected at about $100$ from the pump direction and an analyzer was placed before the entrance slit of spectrometer to separate the polarization state.

The degree of polarization is estimated by quantifying the polarization anisotropy (r), which is defined by~\cite{fleming76}
\begin{equation}
r =\frac{I_{\parallel} - I_{\perp}}{I_{\parallel} + 2 I_{\perp}}
\label{eq4p8}
\end{equation}
where $I_{\parallel}$ and $I_{\perp}$ denote the emission intensities parallel and perpendicular to the polarization
state of the excitation beam. The absorption of the linearly polarized beam is high if the electric field vector of the incident beam and the dipole moment of the molecules are oriented parallel to each other. On the other hand the absorption will be zero if the beam and the dipole moments are oriented perpendicular to each other. Therefore, the molecules with their dipole moment oriented along the field of the incident light get preferentially excited. This cause an polarization anisotropy in the subsequent emission. Any decay in the anisotropy occurs due to the orientational relaxation of the dipoles which depends on the size and shape of the rotating molecules. The time dependence of the anisotropy is expressed by~\cite{eichler79}

\begin{equation}
r(t) =r_{0} exp (\frac{-3t}{\rho})
\label{eq4p9}
\end{equation}

where r$_{0}$ and $\rho$ are the limiting anisotropy and rotational relaxation time. The polarization in the emission from the system depends on the relative timescales of fluorescence ($\tau_{F}$) and the reorientation ($\tau_{R}$) of the molecules. If $\tau_{F}$ $\gg$ $\tau_{R}$, the dipole distribution induced by the pump laser will completely relax to the equilibrium before the effective fluorescence takes place, which results in a minimal anisotropy. However if $\tau_{F}$ $\ll$ $\tau_{R}$, the fluorescent molecules emit before the reorientation takes place and molecules have no time to relax in the equilibrium. Thus, the emission spectra are expected to have large anisotropy.

Typically, the rotation time of Rhodamine 6G molecules in methanol solvent is $\tau_{R}\sim 100$~ps~\cite{porter77} and $\tau_{F}$ is $\sim 3$~ns. Top panel in Figure~\ref{fig4p8}[B] shows the experimentally observed polarization spectra pumped by the P and the S polarized beam at $E_{p} = 6~\mu$J. As expected, the bulk dye emission showed partial polarization and fluorescence preferentially followed the polarization state of the pump beam. The anisotropy in the emission was observed to be $\sim 0.35$ and $0.55$ for the P and S polarized pump respectively. Clearly the FWHM of the emission is $\sim 15$~nm which is an order of $2$ less than the typical fluorescence of the Rhodamine dye in methanol. This confirms that the ps pump laser had initiated the stimulated emission process in the bulk dye solution. This stimulated emission contributes and enhance the polarization anisotropy. It should be noted that the difference in the polarization anisotropy for P and S pump ($0.35$ and $0.55$ respectively), arises due to the dissimilar response of spectrometer grating for P and S polarized light.

We next study the collective lasing spectra obtained from the droplet array under similar pump energy. As the lasing spectra fluctuate in both frequency and intensity we analyzed the ensemble averaged intensity over $1000$ pulses. The experimental observation from the array is summarized in the bottom panel of Figure~\ref{fig4p8}[B]. The first two peaks (marked by the yellow circles) correspond to the collective modes while other low intense peaks are the WGMs of the droplet. When the array was pumped by the P polarized beam, the output emission intensity became highly P polarized ($r \sim 0.85$). Furthermore, when the droplet was pumped with S polarized beam, the anisotropy was lost ($r \sim 0.01$).

The above observations can be explained using equation \ref{eq4p9}. The dye molecules whose dipole moment are oriented parallel to the P polarized pump gets preferentially excited. As the lasing mode build-up time by definition is fast, the molecules do not get enough time to attain the reorientation. Furthermore, as the lasing modes start to build up, the stimulated process triggers the molecules to emit in the same polarization state. On the other hand, when the array was pumped by the S polarized beam, we observed that contribution of P and S polarized components was almost equal, and the intensity of the modes were relatively smaller. This behavior can be understood as follows. Clearly, S polarized beam preferentially excite the molecules having their dipole in the S plane. Since the lasing modes also exist along the array, the oriented molecules cannot contribute in the collective lasing mode (in this situation both wavevector {\bf k} and electric field {\bf e} will be parallel to each other). Next, when the dipoles reorient, they settle equally in the other two orthogonal directions. Thus, the lasing spectra in this situation does not have preferential polarization.

{\bf Summary:}\\ In summary, we have demonstrated coherent random lasing from a monodisperse array of amplifying microdroplets. The spectra consist of coherent lasing peaks with smaller incoherent pedestal in comparison to the polydisperse array. The frequency distribution of the lasing modes was restricted to a relatively narrow wavelength interval. The spacing between bunches show the participation of Fabry-Perot-resonances of individual resonator. The monodisperse array has an order of magnitude smaller lasing threshold than the polydisperse droplet configuration. We discussed the angular distribution and polarization anisotropy of the random lasing modes.	
\chapter[Lasing from a quasi-one-dimensional, amplifying period-on-average~.~.~.~]{Lasing from a quasi-one-dimensional, amplifying period-on-average random superlattice}

In the previous chapters (3 and 4), we discussed the experimental results on coherent random lasing from an array of amplifying microdroplets where the droplet diameter was larger than the optical wavelength by about a factor of $40$. We showed that, the random lasing modes originate from the co-operative behavior and coherent participation of the microdroplets present in the pumped region. In a monodisperse array, the collective lasing modes exist in the longitudinal direction and the frequency behavior of the modes reflects the Fabry-Perot profile of an individual resonator. Motivated by these observations, now we perform transfer matrix calculations with gain to understand the origin of the random lasing modes and their frequency behavior. To this end, we model the microdroplet array as a one-dimensional periodic-on-average random superlattice (PARS) system with gain. PARS refers to a weakly randomized system derived from an underlying periodic system. Theoretically, PARS systems have been extensively examined in the past~\cite{sajeev87, mcgurn93, freilikhr95, deych98}. For example, issues such as variation of the localization length~($\xi$) in the gap~\cite{mcgurn93}, enhanced transmission coefficient in the gap~\cite{freilikhr95}, and statistics of Lyapunov exponent in PARS systems have been addressed~\cite{deych98}. In this chapter, based on our calculations, we conclude that the experimentally observed random lasing modes originate from the high quality modes arising in the stopgap region of the underlying periodic system. This chapter is organized as follows: In section \ref{sec:level5p1} we discuss light  propagation in a one-dimensional disordered system using the Transfer Matrix Method (TMM). In section \ref{sec:level5p2}, we discuss the transmission properties of monodisperse and polydisperse multilayer systems. Then, in section \ref{sec:level5p3}, we report on the frequency fluctuations in these systems. Finally, in section \ref{sec:level5p4}, we discuss the effect of disorder and system size on the lasing threshold.

\section{\label{sec:level5p1} The transfer matrix method}

The properties of light propagation in a dielectric multilayer can be obtained by solving Maxwell's equations with appropriate boundary conditions. In a one-dimensional multilayer system, it is convenient to use the transfer matrix method to address the transmission and reflection properties of light~\cite{born80, chang03}. This technique is briefly described below.

\subsection{\label{sec:level5p1a} Scattering and phase matrices}

\begin{figure}[h]
\begin{center}
\includegraphics[scale=0.7]{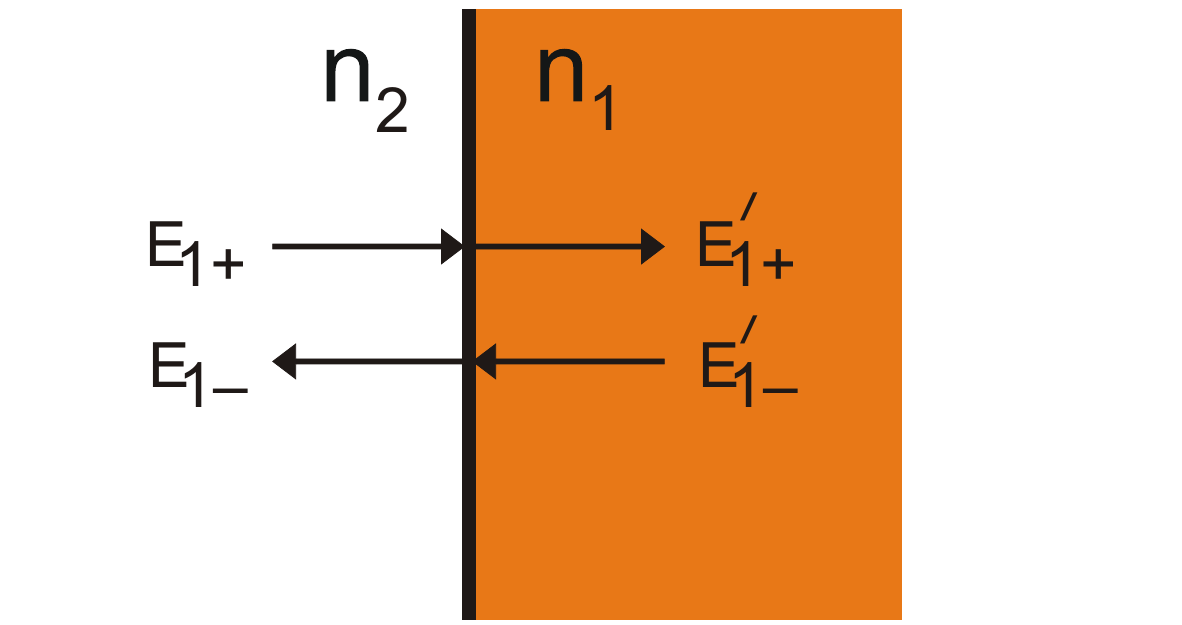}
\end{center}
\vspace{-20pt}
\color{blue}\caption{Electric fields across an interface.}
\label{fig5p1}
\end{figure}

Let us consider a plane interface separating two dielectric media having a refractive indices of $n_{2}$ (left side) and $n_{1}$ (right side) as shown in Figure~\ref{fig5p1}. When a plane wave (amplitude $E_{1+}$) is incident from the left, the electric field gets partially reflected ($E_{1-}$) and partially transmitted ($E'_{1+}$) through the interface. The reflection and the transmission coefficients at the interface are given by

\begin{equation}
r_{12} = \frac{n_1-n_2}{n_1+n_2}, \qquad
t_{12} = \frac{2n_2}{n_1+n_2}
\label{eq5p1}
\end{equation}

These coefficients are called the Fresnel coefficients. The electric field components on the left and right side of the interface can be calculated using the following relationship
\[\left[ {\begin{array}{cc}
E_{1+} \\
E_{1-} \\
\end{array} } \right]=\frac{1}{t_{12}}
\left[ {\begin{array}{cc}
1 & r_{12} \\
r_{12} & 1 \\
\end{array} } \right] \left[ {\begin{array}{cc}
E^{'}_{1+} \\
E{'}_{1-} \\
\end{array} } \right]\]
where `$+$' and `$-$' respectively denote the forward and backward propagating waves, and prime represents the fields on the right side of the interface. The above matrix is known as the scattering matrix. In a lossless homogeneous medium, the fields at any two location (separated by a length $l$) are related by the phase matrix, which is defined as
\[\left[ {\begin{array}{cc}
E_{1+} \\
E_{1-} \\
\end{array} } \right]= \left[ {\begin{array}{cc}
e^{ikl} & 0 \\
0 & e^{-ikl} \\
\end{array} } \right] \left[ {\begin{array}{cc}
E^{'}_{1+} \\
E{'}_{1-} \\
\end{array} } \right]\]
\subsection{\label{sec:level5p1b} A single dielectric layer}
The overall response of a single dielectric layer is determined by multiplying the scattering matrix and the phase matrix from left to right. For example, in a dielectric layer of width $l$, the electric fields at the left hand side ($E_{1+}$, $E_{1-}$) and right hand side ($E'_{1+}$, $E'_{1-}$) are related as follow
\[\left[ {\begin{array}{cc}
E_{1+} \\
E_{1-} \\
\end{array} } \right]= \frac{1}{t_{12}}\left[ {\begin{array}{cc}
1 & r_{12} \\
r_{12} & 1 \\
\end{array} } \right] \left[ {\begin{array}{cc}
e^{ikl} & 0 \\
0 & e^{-ikl} \\
\end{array} } \right] \frac{1}{t_{21}}\left[ {\begin{array}{cc}
1 & r_{21} \\
r_{21} & 1 \\
\end{array} } \right] \left[ {\begin{array}{cc}
E^{'}_{1+} \\
E^{'}_{1-} \\
\end{array} } \right]\]
Note that, the first and the third matrices (right side) are scattering matrices and the matrix between them is the phase matrix. 

\subsection{\label{sec:level5p1c} Multiple dielectric layers}

The transfer matrix of a multilayer structure (say, there are `$n$' layers) can be calculated by multiplying the transfer matrix of every individual layer in the following order
\begin{equation}
M = M_{1}~\times~M_{2}~\times~-----~M_{j}~-----~M_{n}
\label{eq5p1}
\end{equation}

where, $M_{j}$ represents the transfer matrix of $j^{th}$ layer. Using the final matrix `$M$', we can calculate the reflection and the transmission coefficients of the multilayer structure from the matrix elements ($M_{11}$, $M_{12}$, $M_{21}$, $M_{22}$) as
\begin{equation}
r = \frac{M_{21}}{M_{11}}, \qquad
t = \frac{1}{M_{11}}
\label{eq5p3}
\end{equation}

And hence, the reflectance and transmittance of the multilayer structure can be calculated as 
\begin{equation}
R = r^{2}, \qquad
T = t^{2}
\label{eq5p4}
\end{equation}

\begin{figure}[h]
\begin{center}
\includegraphics[scale=0.45]{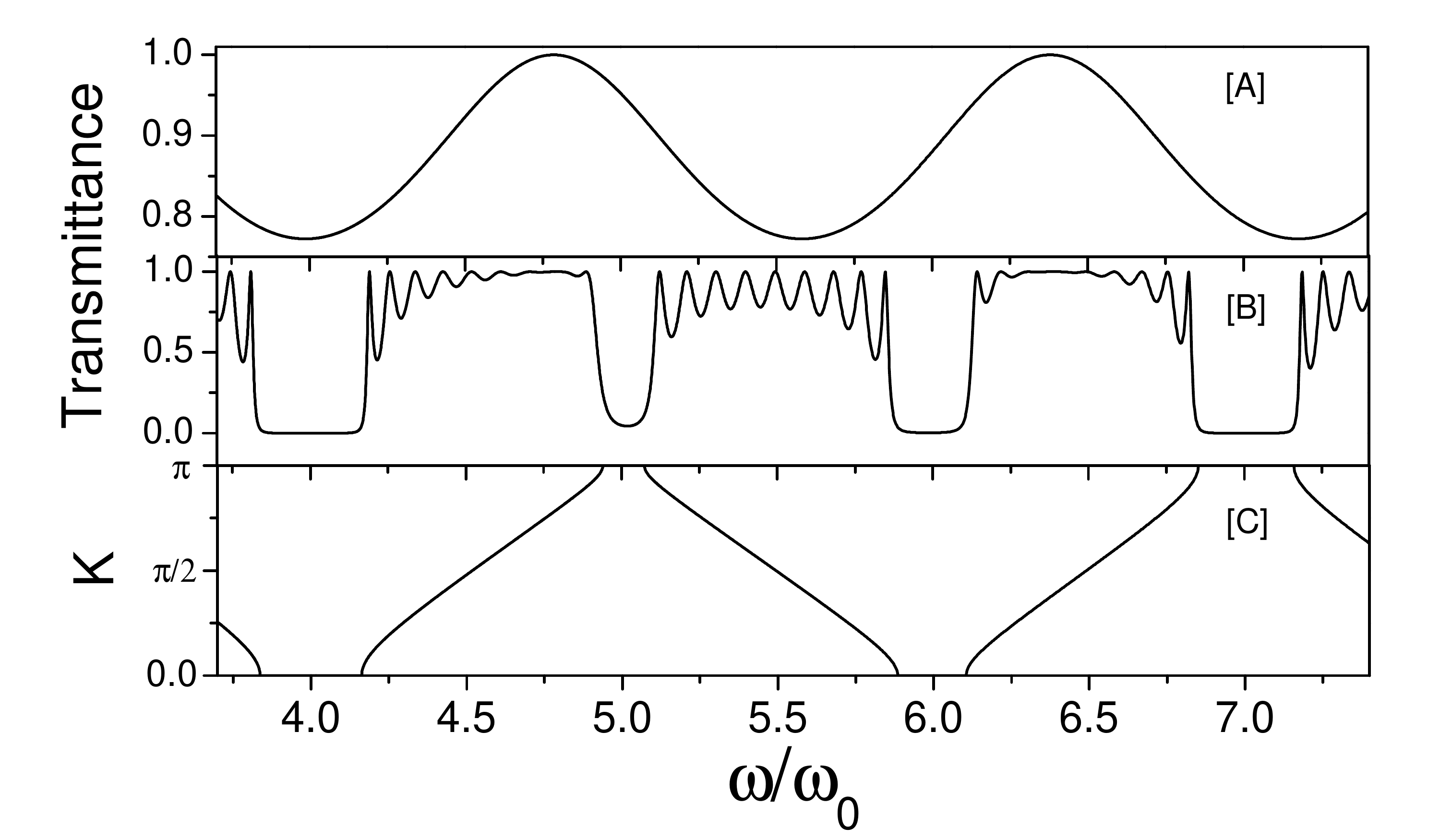}
\end{center}
\vspace{-30pt}
\color{blue}\caption{[A] Transmission spectrum of a single dielectric layer with $A = 400$~nm, and $n_{A} = 2.32$ sandwiched between two layers of $n_{b} = 1.38$. [B] Transmission spectrum from the multilayer photonic structure having $10$ bilayers. [C] Dispersion relation of the multilayer.}
\label{fig5p2}
\end{figure}

Using this formalism, we calculate the transmission response of the multilayer system. We assume that each unit cell (bilayer) consisting of a high ($n_{A} = 2.32$) and a low ($n_{B} = 1.38$) refractive index material with equal thickness ($A = B = 400$~nm). These refractive index corresponding to zinc sulfide and magnesium fluoride materials and are frequently used for the coating of dielectric mirrors. Figure~\ref{fig5p2}[A] plots the transmission spectrum of a single zinc sulfide dielectric layer sandwiched between two layers of magnesium fluoride. Clearly, in the visible wavelength region, the transmission profile has two maxima at $\omega/\omega_{0} = 4.8$ and $6.4$, and three minima at $\omega/\omega_{0} = 4.0, 5.6$ and $7.2$. Figure~\ref{fig5p2}[B] depicts the transmission profile of the multilayer photonic crystal when the number of bilayers is equal to $10$. The transmission spectrum has nearly zero and one transmittance, indicating the stopband and passband of the multilayer structure. Figure~\ref{fig5p2}[C] shows the dispersion curve of the same multilayer photonic structure calculated using the following relationship

\begin{equation}
K = \frac{1}{l_{u}}cos^{-1}[\frac{cos(\frac{\pi\omega}{\omega_{0}})-R~cos(\frac{\pi\zeta\omega}{\omega_{0}})}{T}]
\label{eq5p5}
\end{equation}

where $l_{u}$ is the length of unit cell ($l_{u} = A + B$), $\omega_{0}$ is the normalized frequency given by  $\omega_{0} = 2\pi c /2l_{u}$ and $\zeta = \frac{(n_{A}A)-(n_{B}B)}{(n_{A}A)+(n_{B}B)}$.

The dispersion curve essentially exhibits the band-structure of the one-dimensional photonic crystal formed by the above multilayer.

\section{\label{sec:level5p2}Origin of collective lasing modes}

\begin{figure}[h]
\begin{center}
\includegraphics[scale=1.2]{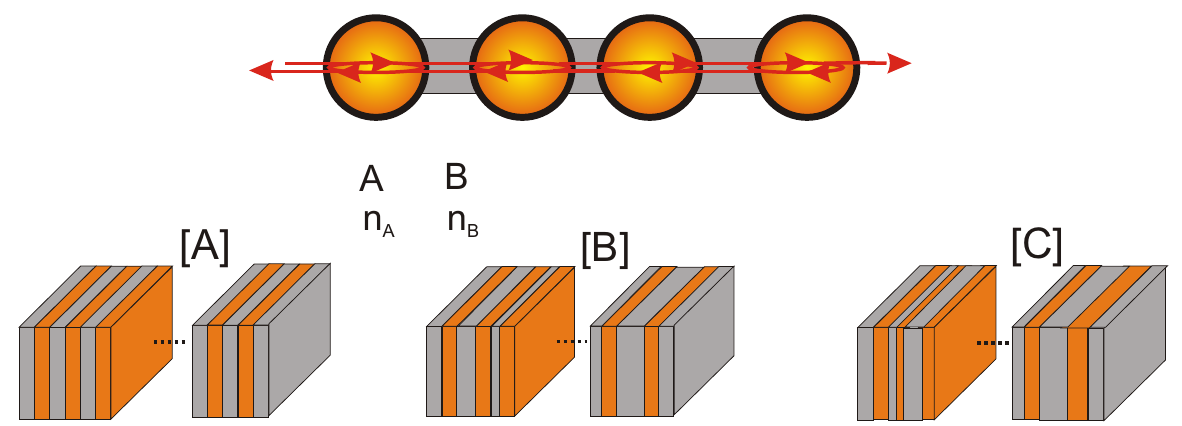}
\end{center}
\vspace{-20pt}
\color{blue}\caption{Schematic representation of the microdroplet array, each droplet is represented by a one-dimensional active (orange) layer. [A] Representation of monodisperse layers with periodic spacing between them. [B] Monodisperse layers with random spacing between them and [C] Polydisperse layers with random spacing between them.}
\label{fig5p3}
\end{figure}

We now apply the above TM calculation to the microdroplet array. Evidently, in this treatment, a three-dimensional system has been approximated as a one-dimensional system. The motivation towards this is that the longitudinal modes only experience the back face and the front face of the droplets, and the droplet sizes are very large, thus offering an effective parallel multilayer to the modes. This approximation is justified via FDTD simulations in the last section of this chapter. The schematic representation of the quasi-one-dimensional microdroplet array is shown in Figure~\ref{fig5p3}. Say, the average diameter of the droplet is `$A$' with a refractive index of $n_{A}$ and the average spacing between two droplets is `$B$' with a refractive index of $n_{B}$. Figure~\ref{fig5p3} [A], [B] and [C] represent three arrangements of the microdroplet array such that
\begin{enumerate}[(A)]
\item Monodisperse microdroplets with periodic spacing between them.
\item Monodisperse microdroplets with random spacing between them.
\item Polydisperse microdroplets with random spacing between them.
\end{enumerate}

As discussed earlier in section \ref{sec:level4.3}, exact monodispersity in the microdroplet array is not possible and thus the first and the second arrangements are not experimentally accessible. In fact, we have an inherent polydispersity of $\sim 80$~nm in the size and $\sim 300$~nm in the spacing. Nonetheless, it is useful to study the spectral behavior of monodisperse system in the simulations to get proper insight on the origin of collective modes.

To simulate the transmission spectra, we consider a multilayer system comprised of $20$ unit cells. Each unit cell consists of two layers, namely a dielectric layer and a spacing layer of air (Figure~\ref{fig5p3}). We assume that, the average thickness of the dielectric layer and the air region is $20~\mu$m and $6~\mu$m respectively. Randomness in the size and the spacing was introduced by adding a random number chosen from a Gaussian distribution. The width in the size distribution and the spacing distribution is quantified by $\delta_{A}$ and $\delta_{B}$ respectively.

\subsection{\label{sec:level5p2a}Passive array}

\begin{figure}[h]
\begin{center}
\includegraphics[scale=0.5]{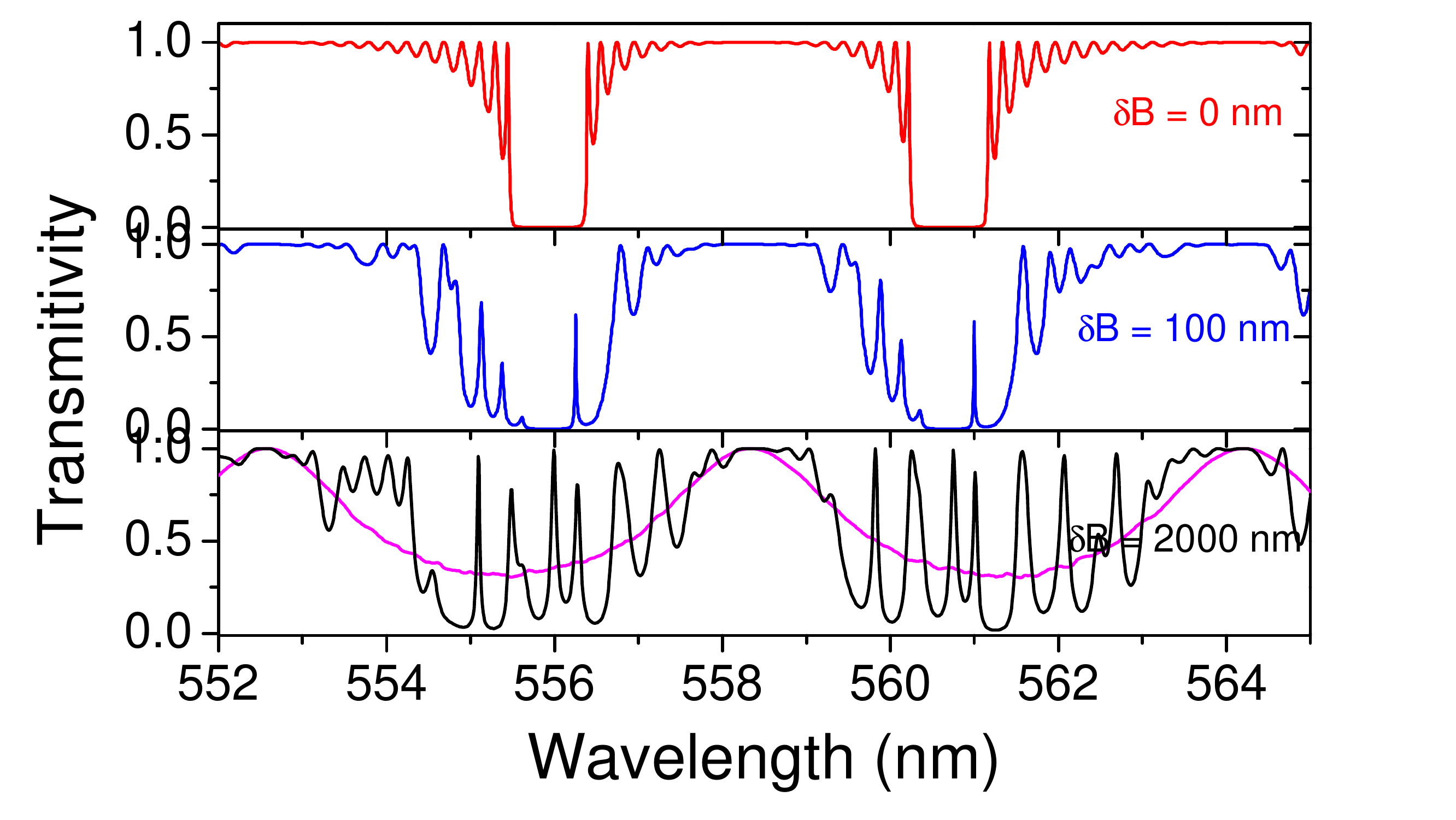}
\end{center}
\vspace{-30pt}
\color{blue}\caption{Transmission spectra from a one-dimensional multilayer system. The red curve: spectrum from a monodisperse configuration with equal spacing. The blue and the black curves respectively represent two PARS systems with weak and strong randomness in the spacing. The magenta curve is the configurationally averaged transmittance generated from $1000$ realizations.}
\label{fig5p4}
\end{figure}

Figure~\ref{fig5p4} shows the transmittance of a periodic system and two passive PARS with different disorder strengths. The red curve depicts the spectrum from a monodisperse multilayer arranged in a perfectly periodic fashion ($A = 20~\mu$m, $B = 6~\mu$m, $\delta_{A} = 0~\mu$m, $\delta_{B} = 0~\mu$m). This arrangement corresponds to a one dimensional photonic crystal superlattice. In the region of interest, the transmission spectrum consists of two stopbands, centered at $\lambda = 556$~nm and $\lambda = 560.7$~nm. In the stopband region, the transmittance goes to zero indicating that the wave cannot transmit through the multilayer. On the other hand, the passband region shows several resonant peaks with transmittance equal to $1$. Discrete peaks are seen, because the sample size is finite. The number of high transmission peaks in the passband is equal to $n-1$, $n$ is the number of layers. Note that the modes at the edge of stopbands, called the band-edge modes, are the highest quality modes.

The blue curve shows a representative spectrum from a monodisperse, weakly aperiodic system ($\delta A = 0$~nm, $\delta B = 100$~nm). Under such weak disorder, the spectrum still has well-defined passbands and stopbands with noticeable differences from the periodic case. The stopbands are widened and the band-edge modes migrate into the stopband region, now known as the gap states. The quality factors of these gap states remain high. With increasing disorder ($\delta B = 2000$~nm, black curve in the bottom panel), many gap states with distributed quality factors are generated in the stopband region. The magenta curve shows the configurationally averaged transmittance ($\delta A = 0$~nm, $\delta B = 2000$~nm) over $1000$ configurations, the parameter that is usually studied in passive systems. The averaged spectrum has equispaced peaks of unit transmittance separated by the Fabry-Perot free spectral spacing of an individual dielectric layer. Here the maxima originate from the serial filtering effect from a collection of independent Fabry-Perot etalons, and are not related to the high-Q gap states.

\subsection{\label{sec:level5p2b}Active array}
\begin{figure}[h]
\begin{center}
\includegraphics[scale=1.35]{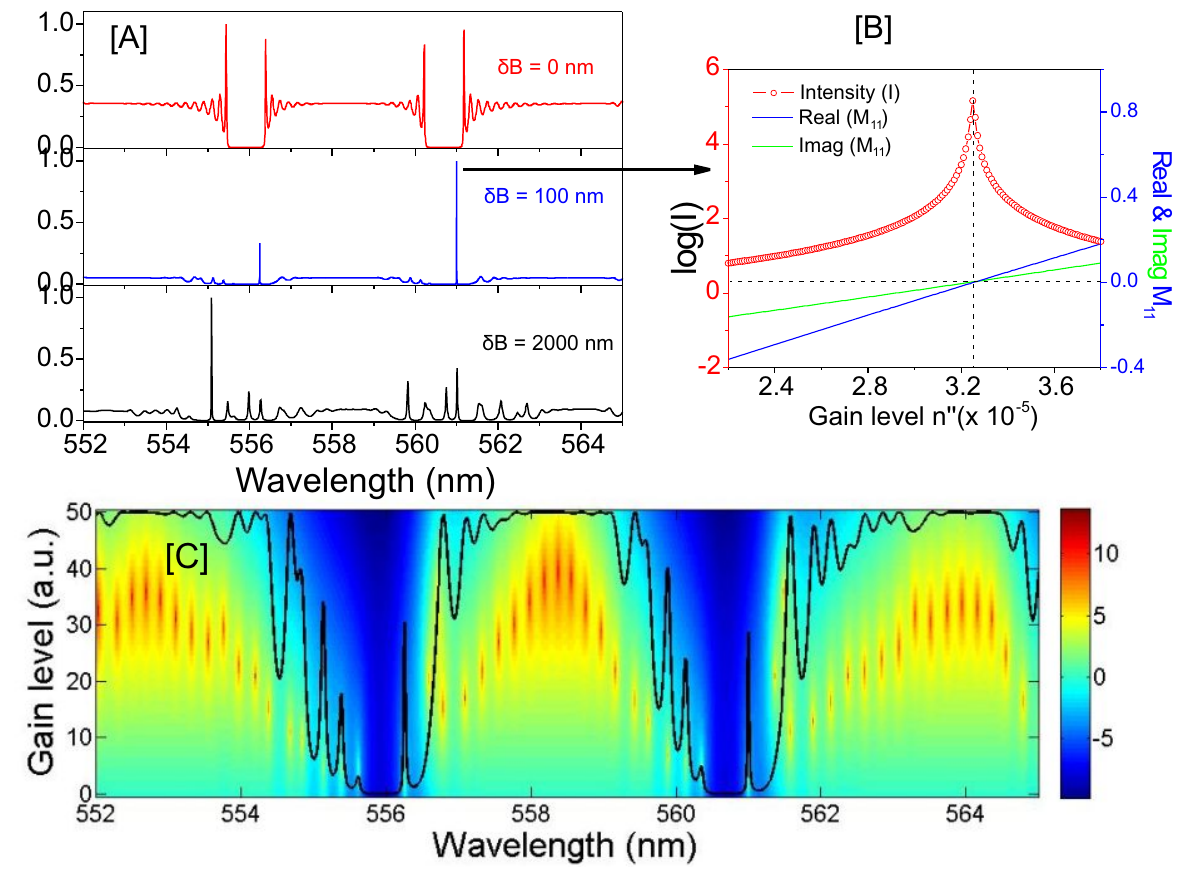}
\end{center}
\vspace{-30pt}
\color{blue}\caption{[A] Transmission spectra under unsaturable gain for the same configuration as in Figure~\ref{fig5p4}. The modes in the stopband region are the highest quality modes and lase first. [B] The red circles: output intensity of the indicated peak as a function of gain level ($n''$). The blue and the green lines are the real and imaginary part of $M_{11}$. [C] Intensity map with varying gain level for a representative transmission spectrum shown by the black line.}
\label{fig5p5}
\end{figure}

We next study the effect of gain on this system. To implement this, we add a negative imaginary part to the refractive index of the dielectric ($n = n'+in''$). This implements unsaturable gain in the system. Although the experimental system has saturable gain, this difference does not affect our conclusions. This is because our calculations only address the frequency of the modes, and not the intensity. Besides, we remain restricted to the near-threshold behavior of the system. The $n''$ (optical gain) was modulated by the gain profile of Rhodamine dye (centered at 563~nm, with a FWHM of $\sim 50$~nm). Figure~\ref{fig5p5}[A] shows the corresponding spectra at $n'' = 2.6~\times~10^{-5}$~\footnote{$n'' = 2.6~\times 10^{-5}$ corresponding to $1$~mM concentration of Rhodamine dye~\cite{penzkofer86}}. In the periodic monodisperse system, the band-edge modes start to gain in intensity (red curve). This is the characteristic band-edge lasing well known in photonic crystals~\cite{rajesh12, rajesh13}. In the weakly random system, the migrated bandedge modes lase first (blue curve), indicating their high quality factor. Interestingly, the intensity of the gap states is higher than the band-edge modes observed in the monodisperse periodic array as seen from non-normalized spectra (not shown here). Finally, in the strongly random system (black curve), multiple lasing modes are initiated in the stopband region.

Figure~\ref{fig5p5}[B] shows the intensity behavior of the mode marked by the arrow. The blue and the green lines are the real and the imaginary parts of the matrix element $M_{11}$. Note that, in equation \ref{eq5p3}, if the matrix element $M_{11}$ tends to zero, the transmission coefficient starts to diverge. This is possible if and only if both the real and imaginary part of $M_{11}$ vanish simultaneously as seen in the figure at $n'' = 3.25~\times~10^{-5}$. The red circles are the peak heights as a function of gain level ($n''$). The output intensity increases and hits a maximum intensity at the same value of $n''$ ($3.25~\times~10^{-5}$). In the transfer matrix method, this value of $n''$ is defined as the lasing threshold. After hitting the lasing threshold, the mode intensity starts falling due to a numerical instability~\cite{andreasen10, jiang99}. In a realistic system, the lasing peak would not have such upper bound. Figure~\ref{fig5p5}[C] shows the intensity map as a function of gain level for a representative spectrum shown by the black curve. Clearly, as the gain level increases, the intensity of every mode also increases. The modes appearing in the stopband region acquire maximum intensity at a lower gain level indicating a lower threshold. The modes in the passband region require a lot of gain to lase due to their low $Q$. Interestingly, the threshold of the gap states originating from the weak random system ($2.77~\times~10^{-5}$) is lesser than that of the band-edge modes ($5.5~\times~10^{-5}$). We present the elaborate discussion of threshold behavior in section~\ref{sec:level5p4}.

\section{\label{sec:level5p3}Frequency distribution of the modes}

\begin{figure}[h]
\begin{center}
\includegraphics[scale=0.5]{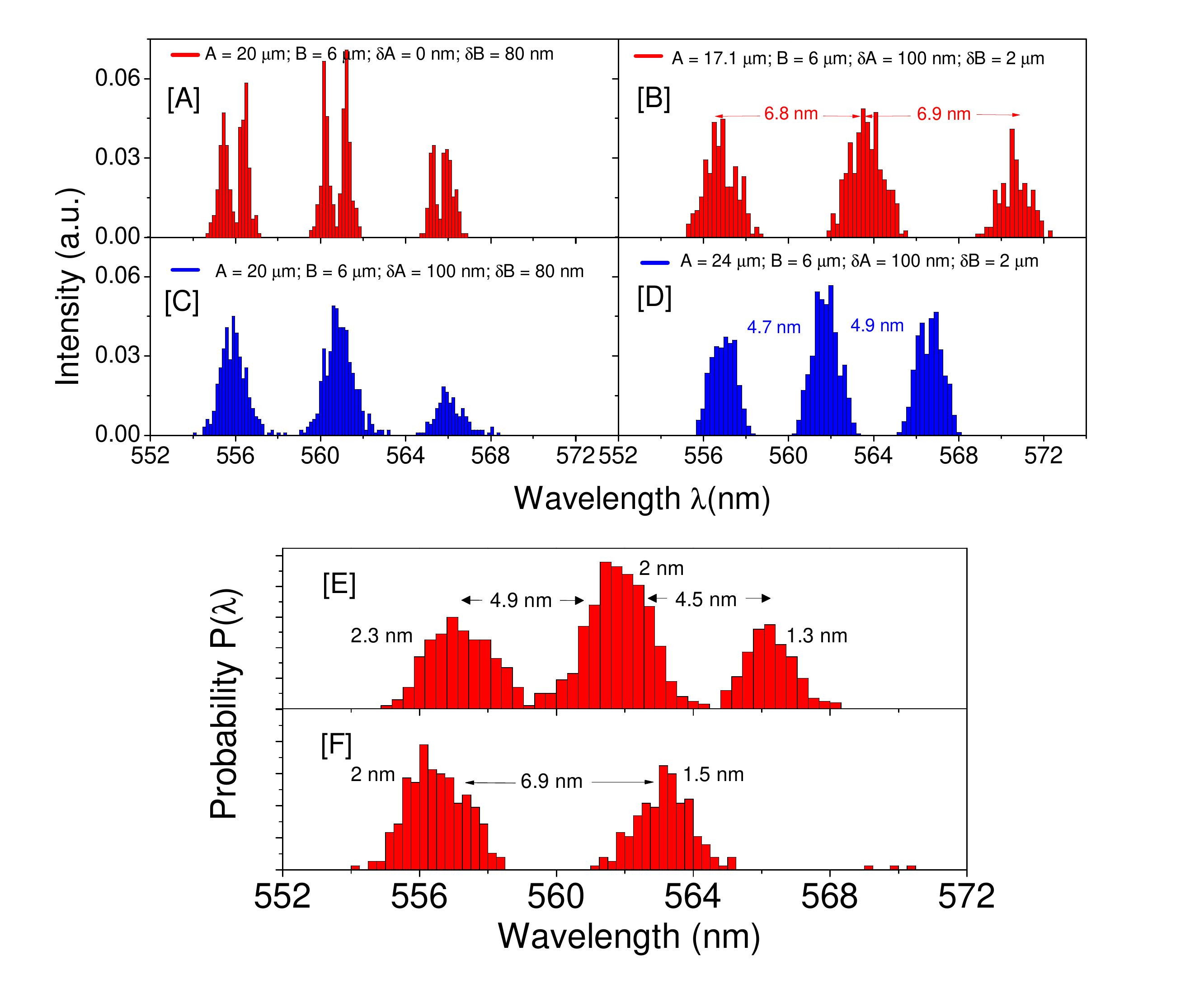}
\end{center}
\vspace{-40pt}
\color{blue}\caption{Distribution of the lasing peaks. [A] For a weakly disordered system with perfect monodispersity, $\delta A = 0$~nm and $\delta B = 80$~nm, the randomness is so weak that the band-edge effects are evident in the form of double-peaked bunches. [B] For a weakly random system with $\delta A = 100$~nm and $\delta B = 80$~nm, the memory of the band-edge effects is washed out. [C, D] For a fairly monodisperse system with a large randomness in spacing, the bunch separation matches the FP free spectral range of an individual layer. [E, F] Experimentally observed probability distribution of lasing wavelengths.}
\label{fig5p6}
\end{figure}

We now focus on the frequency fluctuations of high quality modes. Like the experimental scenario, the frequency fluctuations are quantified by the distribution of the lasing frequencies~\cite{anj12opex}. We choose the first four highest $Q$ modes from every spectrum to construct the frequency distribution of the lasing modes. Figure~\ref{fig5p6} shows various histograms generated over $500$ realizations from systems with varying disorder. 

Panel [A] shows the frequency distribution when the active layers are perfectly monodisperse and the spacing between them is weakly randomized ($A = 20~\mu$m, $\delta A = 0$~nm, $\delta B = 80$~nm). In the region of interest, the distribution has three narrow bunches. For such a weak randomness, each bunch shows a double-peaked profile suggesting that the gap states originate in the vicinity of the band-edges. When the system is made slightly polydisperse ($\delta A = 100$~nm and $\delta B = 80$~nm) the band-edge memory is washed out and the two-peaked bunches collapse into single-peaked bunches (Figure~\ref{fig5p6}[B]). This indicates that the high quality gap states start originating deep in the underlying stopband region. Thus, even a small polydispersity in the active layers can wash out the memory of band-edges. At this disorder strength, we observed that the width of the central bunch is $\sim 1.3$~nm. Figure~\ref{fig5p6}[C, D] represent the numerically calculated frequency distributions with experimentally realistic disorder parameters, where the randomness in the spacing is larger when compared to that in the width of the layers ($\delta A = 100$~nm and $\delta B = 2~\mu$m). The red and the blue histograms correspond to the active layer thickness of $17.1~\mu$m and $24~\mu$m respectively. Interestingly, the spacing between two consecutive two consecutive bunches matches the Fabry-Perot free spectral range of the individual layer. Note that, in a periodic monodisperse system, the spacing between two consecutive stopband centers is not equal, and neither does it match with the Fabry-Perot free spectral range of individual layer. In presence of disorder, the gap states originated in stopband regions are preferentially shifted to make the distribution equispaced. Figure~\ref{fig5p6} [E] and [F] show the experimentally observed histogram bunches for the microdroplet array with $A = 23.9~\mu$m and $A = 17.1~\mu$m respectively. Note that, in [E], the width of the bunches are $1.3$~nm and $2.2$~nm further, the bunches are separated by $4.8$~nm and $4.5$~nm. As shown in [F], the histogram essentially exhibits only two bunches due to the limited gain profile of the dye. The bunch width is $2$~nm and $1.5$~nm and the bunches are separated by $6.9$~nm. Clearly, the experimentally observed mean separation between two bunches shows an excellent agreement with the above calculations. 

\begin{figure}[h]
\begin{center}
\includegraphics[scale=0.5]{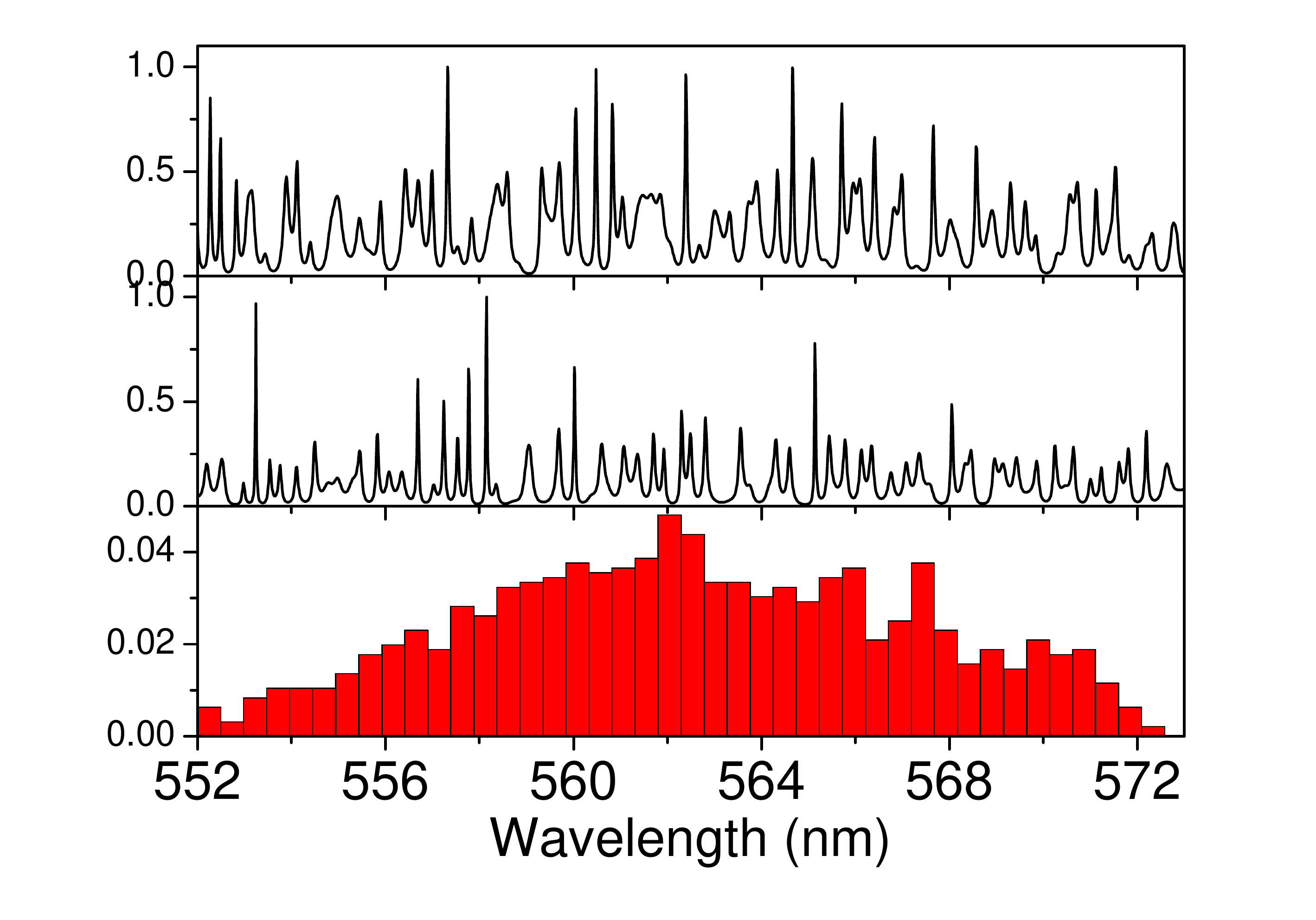}
\end{center}
\vspace{-30pt}
\color{blue}\caption{Lasing in a strongly disordered system with $\delta A = \delta B = 2000$~nm. Black curves: two representative transmission spectra. Red columns: distribution of the lasing modes, which does not show any preference to the lasing wavelength.}
\label{fig5p7}
\end{figure}

Next, we study the systems where both size and spacing are strongly randomized ($\delta A = \delta B = 2000$~nm). Figure~\ref{fig5p7} shows two representative spectra and the distribution of lasing peaks from this system. Note that, the high intensity modes appear at random wavelengths and the distribution of the modes has a width of $\sim 15$~nm which is determined by width of the gain profile. In our calculations, the gain profile was centered at $562$~nm and hence the distribution is also centered at $562$~nm. Clearly, under such a strong disorder, the lasing frequency follows only the underlying gain profile and the system acts as a conventional random laser. These calculations are in excellent agreement with the experimental results presented in the previous chapter. Thus, the TMM calculation also reveal that the microdroplet array behaves as a quasi-one-dimensional aPARS system and offers frequency control of random lasing due to gap states.

\section{\label{sec:level5p4}Dependence of lasing threshold on system size and disorder}

\begin{figure}[h]
\begin{center}
\includegraphics[scale=0.63]{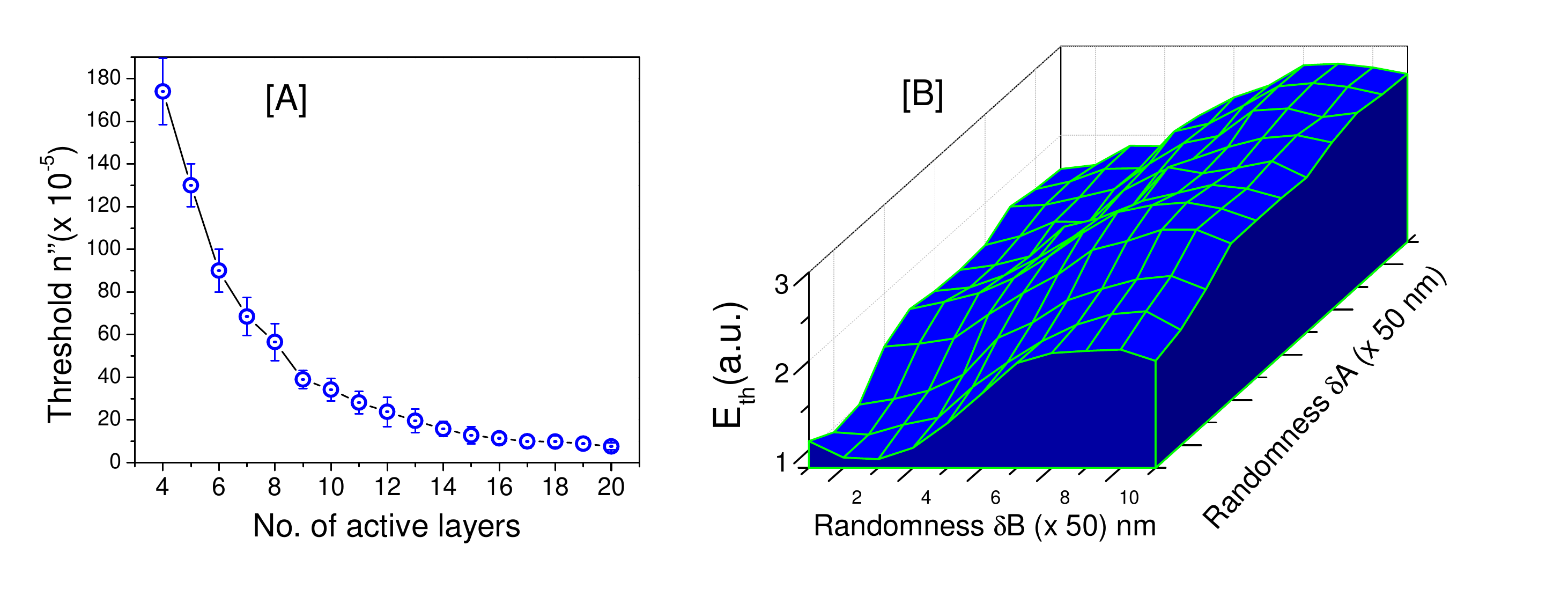}
\end{center}
\vspace{-30pt}
\color{blue}\caption{[A] Lasing threshold versus number of active layers constituting the multilayer system. [B]: Lasing threshold versus disorder strength.}
\label{fig5p8}
\end{figure}

Figure~\ref{fig5p8}[A] plots the variation of lasing threshold versus number of active layers constituting the system. To estimate the lasing threshold, we consider the weakly random systems ($A = 20~\mu$m, $\delta A = 30$~nm and $B = 6~\mu$m, $\delta B = 200$~nm). Here, every data point is averaged over 100 configurations. It can be seen that the lasing threshold monotonically decreases as the number of active layers increases. This can be understood from the following argument. Upon increasing the number of layers, the number of partial reflections increase to create a stronger interference and thus the spectral width of the peaks narrows down. This results in a higher quality factor and hence lower lasing threshold of the modes. Furthermore, a large amount of inverted dye molecules are involved, increasing the gain in the system. The monotonic behavior of lasing threshold shows a good qualitative agreement with the experimental results presented in chapter 4 (see Figure~\ref{fig4p6}[B])~\footnote{For a quantitative agreement, it may perhaps be necessary to invoke the actual curved nature of the
interfaces, which will introduce a coupling coefficient between resonators.}.

Figure~\ref{fig5p8}[B] depicts the dependence of lasing threshold on the disorder strength. The X and the Y axes represent the randomness in spacing and width, respectively, of the active layers. Again, every pixel in the image represents the lasing threshold which is averaged over $100$ configurations. We can see that the lasing threshold has a non-monotonic behavior. The threshold first decreases and then increases as the disorder increases (true for both $\delta A$ and $\delta B$). Importantly, the minimum threshold is observed not in the perfectly ordered system but in the system which has weaker disorder. This confirms that, disorder induced gap states can have larger quality factor compared to the band-edge modes. This counter-intuitive observation, although correct, has to be understood with a bit of caution. The advantage of the lowered threshold in the random system comes at the cost of predictability of frequency. The periodic system supports well-predicted resonances at multiple frequencies. The weakly random sample {\it accidentally} supports only one mode, whose frequency cannot be predicted {\it a-priori}.

\section{\label{sec:level5p5}\it {Verification of the quasi-1D approximation}}

Here, we discuss the validity of the 1-D approximation applied to the resonator array. The large size of microdroplets (size parameter, $kr \sim 120$) in our experiment motivates us to approximate the array as a 1D multilayer system. The collective lasing modes are excited along the axis of the array and are expected to see only a small numerical aperture of the microsphere surface. Over this small aperture, the curvature can be ignored and the droplet array can be approximated as a planar interface. To support this approximation, we carried out finite difference time-domain simulations for the array using a free software, MEEP~\cite{oskooi10}. To this end, we assume the droplet array as a chain of infinite cylinders. These simulations are carried out in a 2D system as 3D computations (involving spheres) are not possible because of the requirement of inordinate computational resources. The simulation of a cylindrical array justifies our approximation. Here, we investigate two systems, where each system has monodisperse cylinders, $11$ in number. The diameters (A) of the cylinders are $18~\mu$m and $20~\mu$m respectively, while they are arranged in a random manner with the average separation (B) of $6~\mu$m. We also study a periodic configuration of the $18~\mu$m cylinders. We introduce optical gain in the system by adding a negative conductivity. A broadband pulse (bandwidth $40$~nm, centered at $560$~nm) excites the array at one end. At the other end, a detector records the temporal evolution of the field over $2000$ optical cycles. Finally, spectra are calculated using a Fast Fourier Transform (FFT) of the temporally evolving fields.

\begin{figure}[h]
\begin{center}
\includegraphics[scale=0.45]{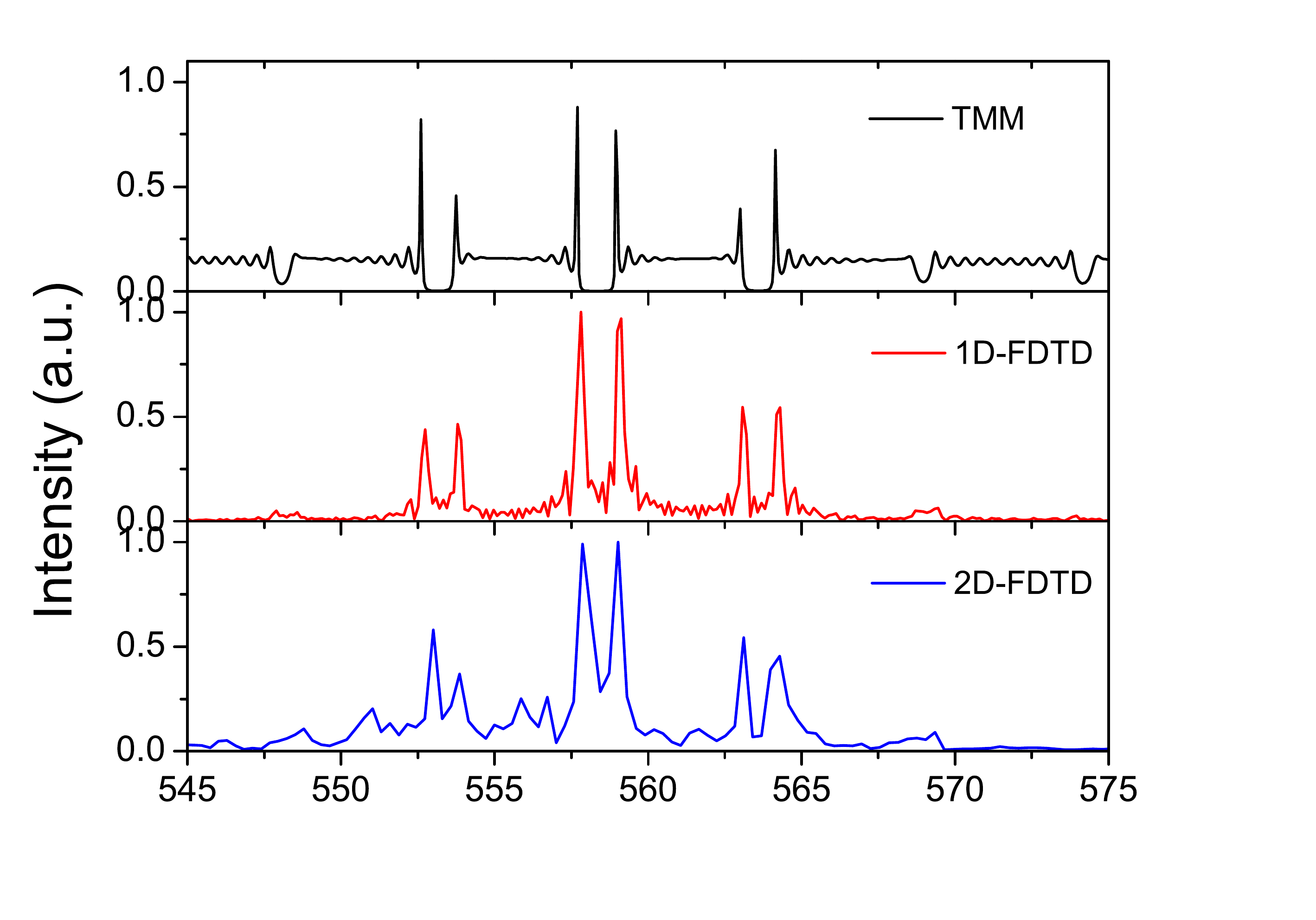}
\end{center}
\vspace{-50pt}
\color{blue}\caption{Transmission spectrum from a monodisperse periodic system. Top panel: Transfer matrix calculations; Middle panel: 1D FDTD calculations; Bottom panel: 2D FDTD calculations.}
\label{fig5p9}
\end{figure}

\begin{figure}[h]
\begin{center}
\includegraphics[scale=0.6]{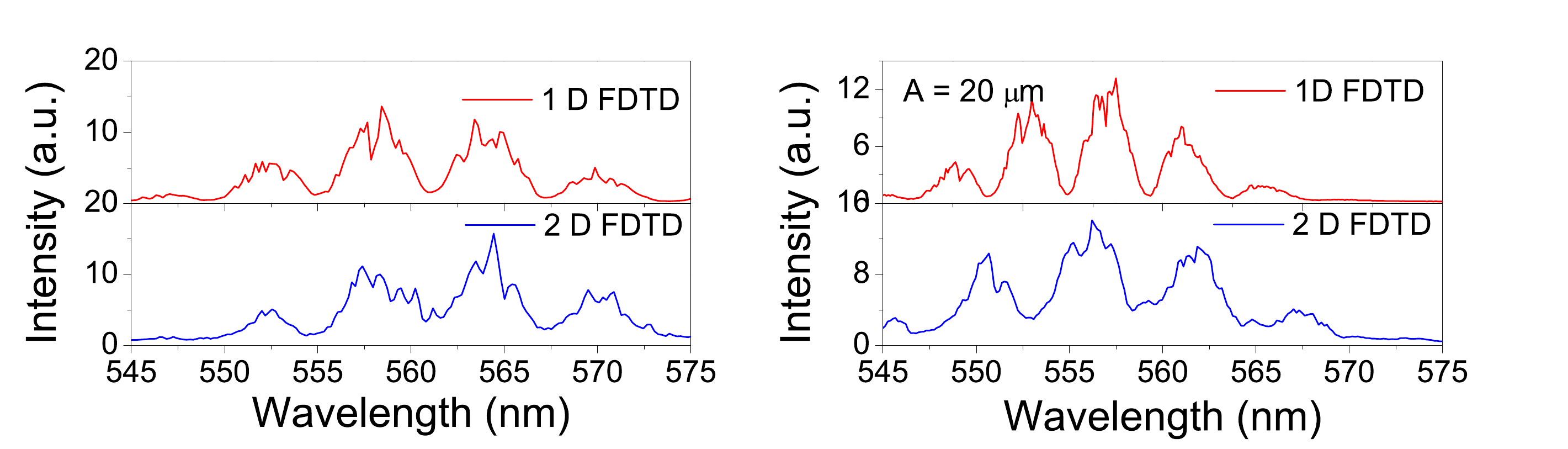}
\end{center}
\vspace{-30pt}
\color{blue}\caption{Ensemble-averaged spectra for two diameters ($18~\mu$m and $20~\mu$m). Red spectra: 1D FDTD calculations, and blue spectra: 2D FDTD calculations.}
\label{fig5p10}
\end{figure}

Figure~\ref{fig5p9} shows the transmission spectrum for the periodic monodisperse system with $A = 18~\mu$m, and $B = 6~\mu$m. The top panel (black spectrum) is calculated using the transfer matrix method. The middle panel shows the result using 1-dimensional FDTD simulations. The bottom panel shows 2D FDTD calculations. To perform the 2D calculations, we used a lower spatio-temporal resolution in order to manage the computational resources. Specifically, for the 2D calculation, we set the spatial resolution, $\Delta x = \Delta y = 33$~nm and temporal resolution, $\Delta t = 0.055$~fs, and for the 1D simulation, $\Delta x = 10$~nm, $\Delta t = 0.017$~fs. The agreement between the three situations is excellent. The band edge modes seen in the TMM calculations are exactly reproduced by the 1D and 2D FDTD simulations. The noisy nature of the background in the 2D simulations originates from the lower resolution and the transient waves in the system. The agreement in the spectra clearly shows that the longitudinal light in the array of cylinders sees the refractive index contrast variation equivalent to a 1D system. This clearly validates our multilayer approximation. Next, we examined the configurationally averaged behavior of the array in the random configuration. To that end, we simulated $30$ spectra for the 2D system with the aforementioned FDTD parameters. Here, the cylinders are monodisperse, while the separations between them are randomized within $1~\mu$m. The spectra exhibited random lasing modes at random locations. We took the sum of these spectra to construct the configurationally averaged spectrum. The same system was also studied in 1D using FDTD. The results are depicted in Figure~\ref{fig5p10}.

The left panel shows the ensemble-averaged spectra for cylinders with diameter $18~\mu$m.The high intensity region in the plots clearly proves that the frequency distribution of the modes in the 2D system is reproduced by the 1D calculations. This behavior is also consistent for $20~\mu$m diameter cylinders. Thus, these calculations confirm the equivalence of the 1D and 2D systems with respect to the frequency behavior of the longitudinal modes. Clearly, the reason is that the curved multilayer has a large radius of curvature at the interface and acts as a planar layer.

{\bf Summary:}\\ In summary, using the transfer matrix method, we model the microdroplet array as an amplifying periodic-on-average random superlattice. We have systematically studied the frequency fluctuations in various PARS systems and explained the origin of frequency controlled coherent random lasing observed in the array. Our calculations show that the experimentally observed lasing modes arise from the states introduced into the stop gap regions of the underlying periodic system. These gap states sustain a high quality factor, and lase under amplification.
	
\chapter{Single-mode, stable coherent random lasing}
In this chapter, we continue our discussion on PARS systems and consolidate our efforts towards frequency control. We experimentally demonstrate single-mode, quasi-stable coherent
random lasing via a mode-matching technique. The chapter is organized as follows: In section~\ref{sec:level6p1}, we consider an amplifying PARS system and discuss the frequency fluctuations of the lasing modes under weak and strong disorder. In section~\ref{sec:level6p2}, we describe the concept of spectral mode-matching in our system. Finally, in section~\ref{sec:level6p3}, we present the experimental results where we successfully implement spectral mode-matching conditions in our system as a result of which the frequency fluctuations get reduced by almost an order of magnitude.

Although the random laser has shown great potential as an optical source, the biggest impediment towards real applications has been the fluctuations in both frequency and intensity,
that result from the inherent randomness of the system~\cite{wiersma07,sushil07,wu08,kumar06,zhu12,uppu12ol,uppu12pra}. Practical applications of random lasers require some control on intensity and frequency fluctuations. In recent years, many attempts have been made to achieve frequency-stabilised lasing emission from such systems~\cite{lagendijk11, gottardo08, sebbah01, sebbah12, sebbah13, shiva12, fujiwara13, leonetti13}. The acknowledged techniques can be broadly categorized as

\begin{enumerate}[(A)]
\item Gain engineering, and
\item Structural engineering
\end{enumerate}

The first technique is based on altering the gain distribution to favor a desirable lasing frequency or a range of frequencies. One such technique involved an active control on the spatial distribution of gain by shaping the pump profile~\cite{sebbah12, sebbah13}. The spatial distribution of the excitation pulse was actively controlled so that only one mode experienced gain and lased at the cost of other modes. The correct pump profile was selected by iterative algorithms using a spatial light modulator (SLM). Thus, any preferential single mode can be extracted from the disordered system. The basic principle behind this technique is based on wavelength {\it selectivity} rather than wavelength {\it control}. Moreover, this method is applicable only for static samples and requires the exact structural information of the sample, which may not be readily available. An alternative method involved the engineering of the spectral gain distribution by the addition of an absorbing dye into the random laser~\cite{lagendijk11}. By varying the concentration of the nonfluorescent dye, the emission wavelength of the diffusive random laser was shifted towards the desired frequency range. 

Controlling a random laser through structural engineering involves the use of resonant scatterers to tune the scattering mean free path of photons inside the gain region. By
varying the physical parameters of resonant scatterers, the interaction time of spontaneously emitted photons can be altered. Photons at the resonant wavelengths experience larger dwell-times which lead to a wavelength-sensitive lasing. Recently, a successful demonstration of wavelength control in diffusive random lasers has been achieved in an ensemble of resonant microspheres~\cite{gottardo08}.

As described earlier, we have implemented the second approach to control the lasing frequency of ultranarrow modes in a system consisting of an amplifying PARS~\cite{anj12opex, tiwari13, anjcleo}. In the previous chapter, we discussed the origin of these random lasing modes in terms of gap states developing in the stopband of the underlying periodic structure. We showed that by tuning the width of the layers, the spectral location of the lasing modes can be controlled. Here, we take a step further and demonstrate that this control can be improved to yield single-mode coherent random lasing with $76~\%$ probability of the modes restricted to an interval of width $\sim 1.2$~nm~\cite{anj14apl}. 

\section{\label{sec:level6p1}aPARS system under weak and strong disorder}

We first numerically study the behavior of an aPARS system wherein we investigate the frequency distribution of lasing modes under weak and strong configurational disorder. We consider that the PARS system consists of $30$ active layers where the width (A) of each layer is $18~\mu$m and the spacing (B) between two consecutive layers is $8~\mu$m.

\begin{figure}[h]
\begin{center}
\includegraphics[scale=0.6]{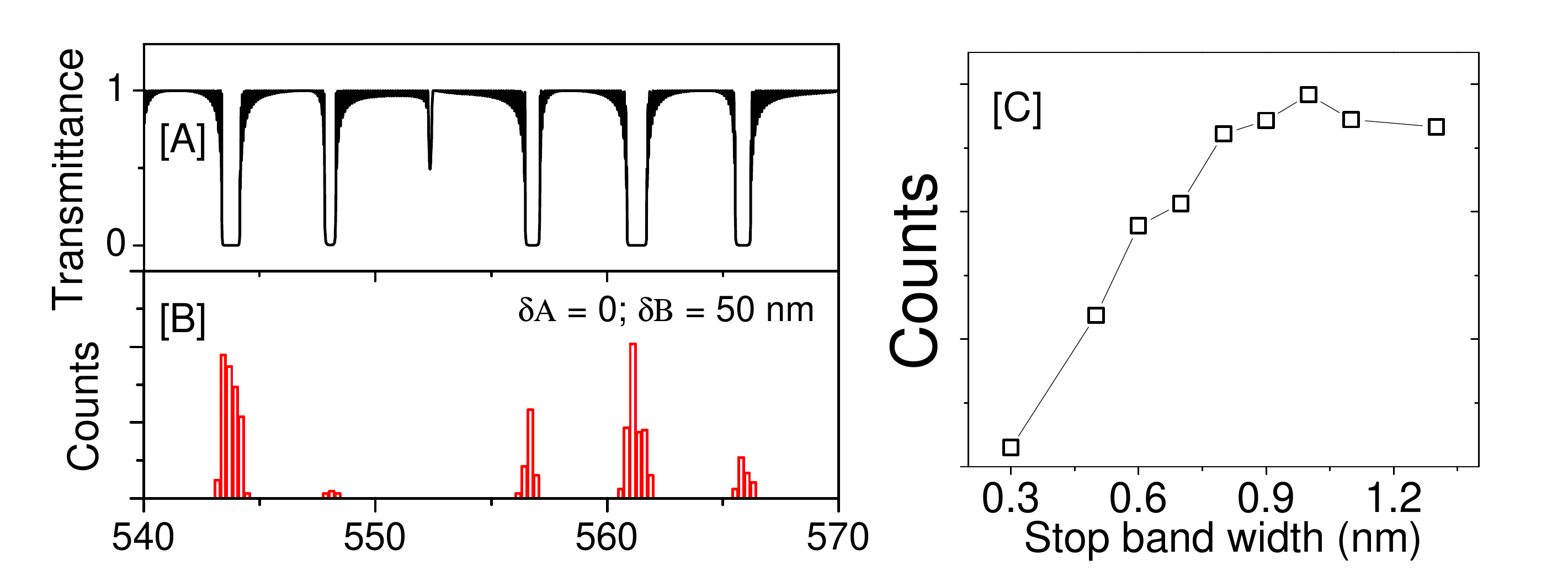}
\end{center}
\vspace{-20pt}
\color{blue}\caption {[A] Transmission spectrum from a multilayer system consisting of $30$ layers ($A = 20~\mu$m, $B = 8~\mu$m, $\delta A = \delta B = 0~\mu$m). [B] Distribution of the random lasing modes under weak randomness in $B$ ($\delta B = 50$~nm). [C] Variation of the mode counts with the width of
the stopband.}
\label{fig6p1}
\end{figure}

The black curve in Figure~\ref{fig6p1}[A] shows the simulated bandstructure (transmittance) from this system, with no randomness ($\delta A = \delta B = 0$~nm). In the wavelength range of interest, the spectrum has five complete stop gaps while one gap is partial. The centers of the complete stopbands lie at $543.7$~nm, $548.0$~nm, $556.8$~nm, $561.1$~nm and $565.9$~nm. Clearly, the spacings between stopbands ($4.3$~nm, $8.8$~nm, $4.3$~nm and $4.8$~nm) are not equal. When weak randomness is introduced in the spacings ($\delta B = 50$~nm), gap modes are created which lase under gain. Figure~\ref{fig6p1}[B] shows the wavelength distribution of these lasing modes. Here, $n=1.34-4\times 10^{-5}i$. The modes are distributed in discrete bunches and the centroid of the bunch matches the stopband position. Thus, the exact bandstructure determines the position of the lasing modes. The same behavior can be observed when the active layer thickness is weakly randomized, while keeping the spacings constant. Thus, the effect of weak disorder in either $A$ or $B$ is same. Figure~\ref{fig6p1}[C] plots the behavior of the number of lasing modes observed in a bunch versus the width of the corresponding stopband. Clearly, the number of lasing modes increases linearly with stopgap width upto a certain width, after which the variation saturates. In these calculations, this saturation width is seen to be $\sim 1 $~nm, but can be expected to change with the level of gain.

\begin{figure}[h]
\begin{center}
\includegraphics[scale=0.55]{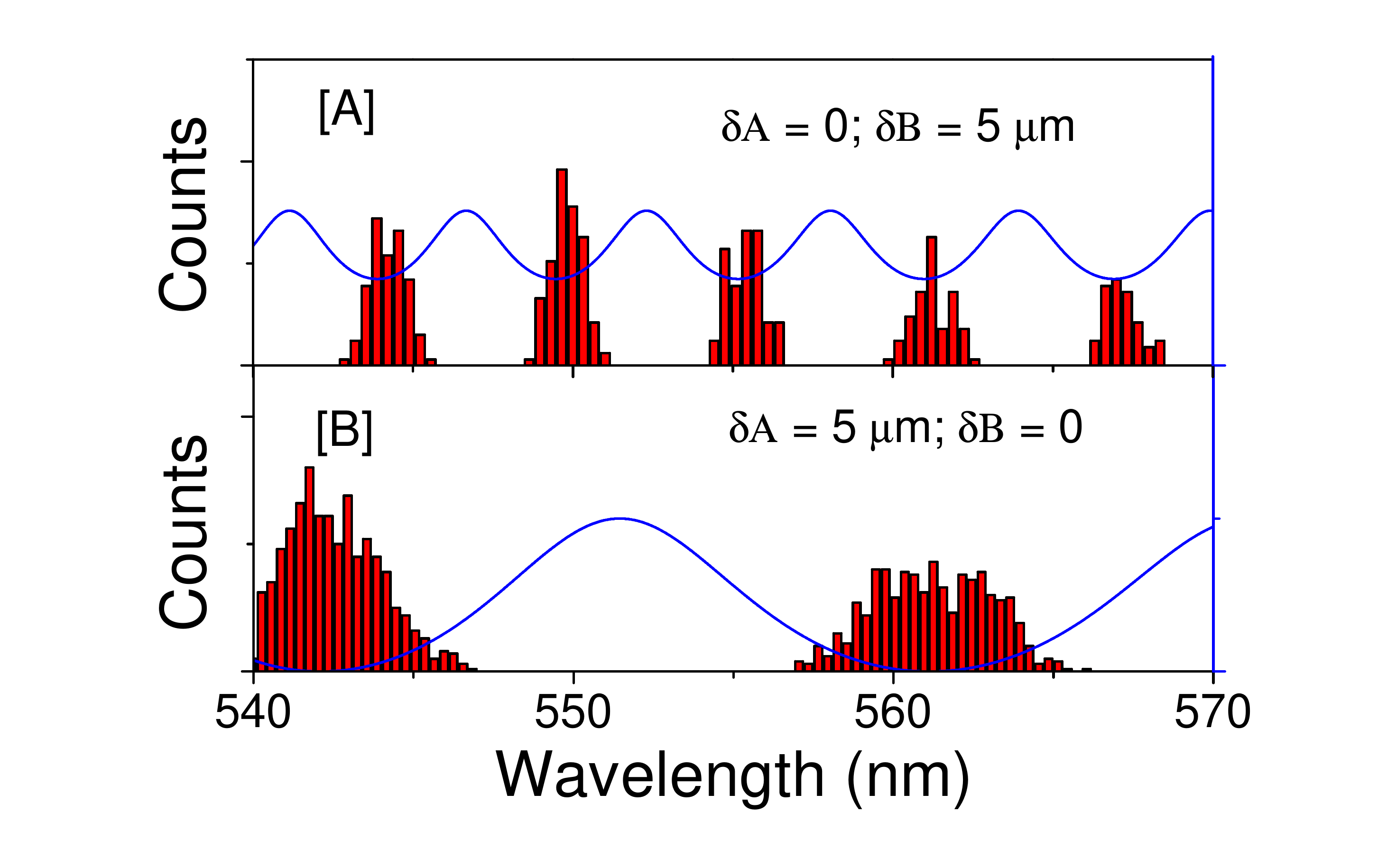}
\end{center}
\vspace{-30pt}
\color{blue}\caption {[A] Distribution of random lasing modes for stronger disorder in layer spacing. [B] Same, for the stronger disorder in layer width.}
\label{fig6p2}
\end{figure} 

Figure~\ref{fig6p2} shows the effect of stronger disorder in the system. [A] shows the effect when $\delta B = 5~\mu$m. The blue curve in [A] is the FP profile of a single dielectric layer of $18~\mu$m. It is important to note now that, the lasing modes occupy regular wavelength intervals. The high Q lasing modes occur at the FP minima of an individual active layer with almost equal probability in each bunch. In fact, once the dielectric layers are fairly monodisperse, the distribution becomes independent of the spacing B and the bunch position is always determined by the width of the dielectric layer. Figure~\ref{fig6p2}[B] shows the situation when the width of the dielectric is strongly randomized, whilst keeping the spacings monodisperse. The bunches now occur at the FP minima of the individual spacing layer. Here, the FP profile is broader due to the smaller spacing between the layers. Clearly, this situation is symmetric with respect to the Figure~\ref{fig6p2}[A]. Thus, we see an interesting behavior such that, when one parameter is monodisperse, the actual value of the other parameter is not relevant.

\section{\label{sec:level6p2}The concept of spectral mode-matching}

In a conventional laser, the mode frequencies are determined by the resonator, while the gain is the fundamental property of the active medium. To maximize the emission, the resonator geometry is chosen such that one mode lies at the wavelength of the gain maximum, thereby achieving spectral mode-matching. In disordered systems, the best one can do is to match the gain maximum with one interval of lasing frequencies.

\begin{figure}[h]
\begin{center}
\includegraphics[scale=0.59]{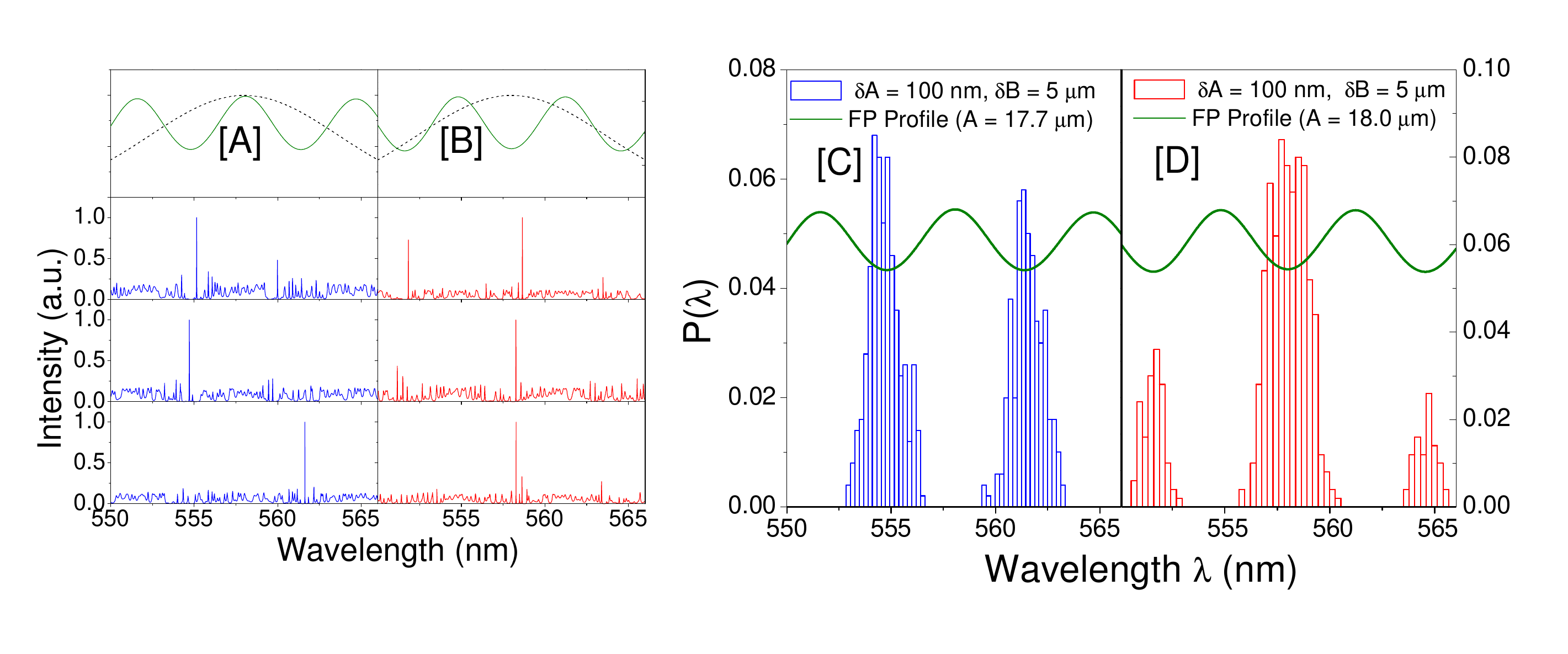}
\end{center}
\vspace{-40pt}
\color{blue}\caption{Top panel shows the gain profile of the amplifying medium (dotted black curve) along with the Fabry-Perot profile (green curve) of an individual resonator. [A] For a resonator of diameter $A = 17.7~\mu$m. [B] For a resonator of diameter $A = 18.0~\mu$m. Bottom panels show three calculated spectra from the same system under strong disorder ($\delta A = 100$~nm, $\delta B = 5000$~nm). [C] Frequency distribution of lasing modes in mode-mismatched configuration along with the FP profile. [D] Same for the mode-matched configuration.}
\label{fig6p3}
\end{figure}

From the analysis of Figure~\ref{fig6p1}, it can be seen that mode-matching under conditions of weak disorder is a challenging task. This is because, the bandstructure of the underlying periodic structure depends on both the layer thickness and the inter-layer spacing, requiring both the parameters to be simultaneously tuned to match the gain maximum to the desired stopband. Moreover, it would be required to know {\it apriori} which bunch is more populated, so that the corresponding stopband is matched to the gain maximum. Under stronger configurational disorder, however, the condition is quite relaxed because all the bunches are comparably populated and overlap the Fabry-Perot minima of the monodisperse parameter. Figure~\ref{fig6p3} illustrates the idea of spectral mode-matching in our system. Figure~\ref{fig6p3}[A] and [B] represent the spectral properties of random multilayer systems with active layer thickness of $17.7~\mu$m and $18.0~\mu$m respectively. These parameters are motivated from the experimental observations as described in the next section. The dotted black curve in the top panels is a numerically generated gain profile centered at $\lambda_{0} = 558$~nm. This $\lambda_{0}$ is consistent with the experimentally measured $\lambda_{0}$. The shape of the gain curve, here, was maintained to be gaussian. The green curves are the respective FP profiles of a single dielectric layer of the said thickness. In Figure~\ref{fig6p3}[A], one FP maximum is matched with the gain maximum. On the other hand, in Figure~\ref{fig6p3}[B], the gain maximum and the center of one FP minimum is matched. We refer to this latter situation as the spectral mode-matched condition.

Next, we simulate the random multilayer systems with a weak disorder in the size and relatively stronger disorder in the spacing ($\delta A = 100$~nm, $\delta B = 5000$~nm). Three representative transmission spectra from this system are shown underneath. As expected, the spectra consist of ultranarrow lasing modes which fluctuate from configuration-to-configuration. In the Figure~\ref{fig6p3}[B], we consistently observe a mode in the vicinity of the central Fabry-Perot minimum. The corresponding histogram of the lasing wavelength is depicted in Figure~\ref{fig6p3}[C] and [D]. The blue and the red bars are the respective distributions in the mode-mismatched ($17.7~\mu$m) and mode-matched ($18.0~\mu$m) configurations. It is clear that the lasing probability in the mode-mismatched situation is equally distributed into two bunches. This implies that the lasing modes can appear in any of the two intervals. On the other hand, in the mode-matched configuration, there are three bunches of unequal heights. Importantly, the height of the central bunch is maximum. The two side bunches indicate modes which have a high enough Q to compensate for the low gain experienced at their spectral positions. The histogram clearly shows a large central bunch where $76\%$ of the lasing peaks appear. Thus, the mode-matching technique can significantly enhance the lasing emission in a desired wavelength range.

\section[Experimental observations: mode-matched and mode-mismatched coherent~.~]{\label{sec:level6p3}Experimental observations: mode-matched and mode-mismatched coherent random lasing}

\begin{figure}[h]
\begin{center}
\includegraphics[scale=0.8]{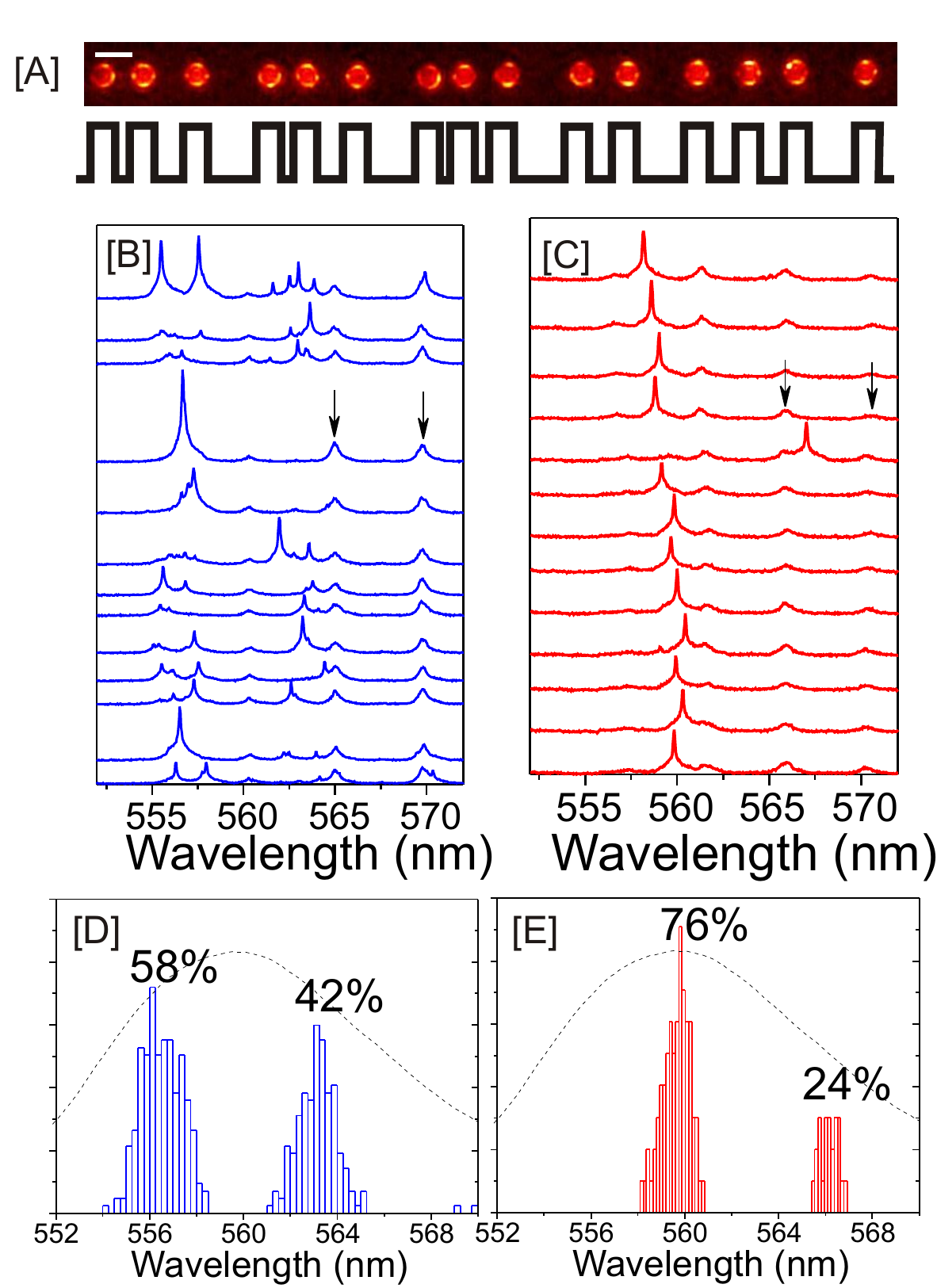}
\end{center}
\vspace{-30pt}
\color{blue}\caption{[A] Image of the array, for the mode-matched case. White bar $= 25~\mu$m. [B] Experimentally observed spectra from the mode-mismatched case, the arrows indicate the
WGM modes. [C] Same, for the mode-matched case. [D] Frequency distribution of the lasing modes for the mode-mismatched case. [E] Same for the mode-matched case. Dotted line: the emission profile of Rhodamine dye solution.}
\label{fig6p4}
\end{figure}

In this section, we discuss the experimental results wherein we implement the modematching technique in our aPARS system. Figure~\ref{fig6p4}[A] shows a representative image of the microdroplet array, and its axial refractive index profile underneath. Figure~\ref{fig6p4}[B] depicts a series of experimentally captured longitudinal spectra taken at excitation energy $E_{p} = 2.2~\mu$J, when the droplet diameter was $17.7~\mu$m. The estimation of the size was made from the whispering gallery modes which can be identified from the fact that they occupy the same spectral position in every spectrum (see the black arrows). This confirms that the microdroplets in every embodiment are of the same diameter and maintain their monodispersity. In this system, multiple lasing peaks are seen in the spectra which fluctuate from pulse-to-pulse. The frequency distribution of lasing modes over $200$ spectra is shown in Figure~\ref{fig6p4}[D]. Clearly, the modes are distributed into two bunches with the distribution of the modes being $58~\%$ and $42~\%$. Thus, this system represents a mode-mismatched situation.

To achieve the mode-matched condition in the array, we vary the microdroplet diameter keeping them monodisperse. This was done by scanning the piezo gate frequency at a fixed chamber pressure. We observed that the scan of $15$ kHz resulted in a shift of about $5$~nm in the WGMs and about $6.5$~nm in FP modes. This ensured that the FP minimum overlapped the gain maximum and the array achieved mode-matched condition. Figure~\ref{fig6p4}[C] shows the lasing spectra for the PARS system where the diameter of the droplet was $18.0~\mu$m as measured from the WGM modes. Here, this system emits a single mode in every spectrum. Further, these modes are generated at very comparable wavelengths. It can be seen that, from this particular series of $13$ consecutive spectra, only one mode hopped to a band away from the gain maximum. The frequency distribution of the modes over $200$ spectra is shown in Figure~\ref{fig6p4}[E] which shows $76~\%$ population in the first bunch. This implies that $3$ out of every $4$ pulses gave a single mode in this favored frequency interval. Further, the width of the favored bunch is only $\sim 1.2$~nm, as compared with the $10$~nm band seen in coherent random lasers. The black curve in Figure~\ref{fig6p4}[D] and [E] is the slightly shifted emission profile of the bulk solution of Rhodamine dye dissolved in methanol. We shifted the gain curve to the right by $1.7$~nm. This shift is justified by the fact that the emission was measured from bulk dye solution contained in a cuvette at $E_{p} = 4.6~\mu$J. Pumped at different energy, in the microdroplet form, it can be expected to have a minor modification in the emission profile. With this shifted gain curve, the mode-matching and mode-mismatching conditions are clearly evident.

\begin{figure}[h]
\begin{center}
\includegraphics[scale=0.4]{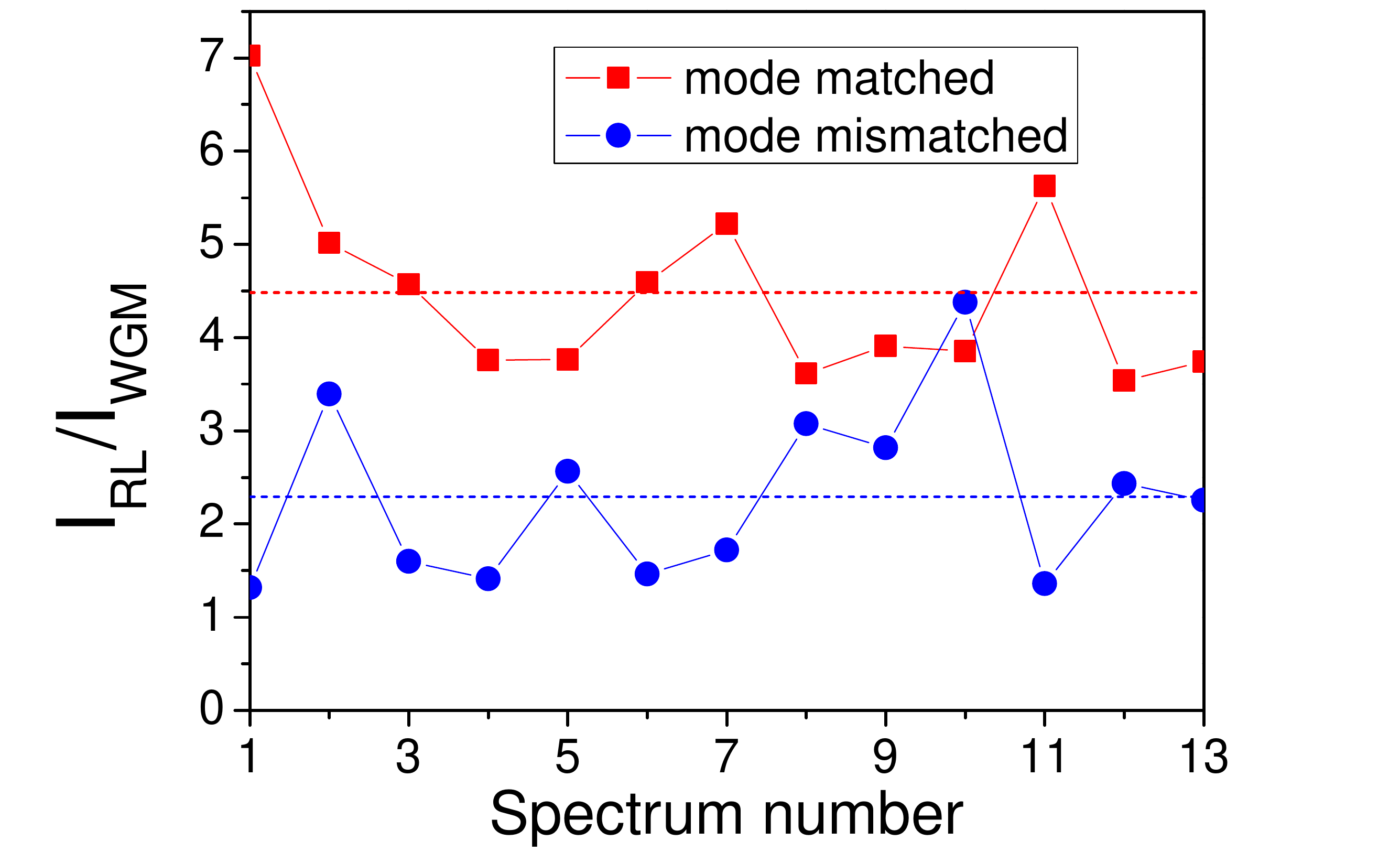}
\end{center}
\vspace{-35pt}
\color{blue}\caption{Ratio of the highest intensity random lasing peak to that of a WGM peak for mode-mismatched case (blue curve, 	$\bullet$ markers) and mode-matched case (red curve, $\blacksquare$ markers).}
\label{fig6p5}
\end{figure}

We, next, quantified the intensity which was coupled into the random lasing modes. To this end, we measure the ratio of the highest intensity random lasing mode to the WGM mode existing in the same spectrum. The WGM modes do not fluctuate in intensity, and hence this normalization is valid. Figure~\ref{fig6p5} displays this ratio. The intensity of the random lasing peak in the mode-matched case is always larger (barring one point at spectrum no. $10$) than the mode-mismatched case. Note that intensity fluctuations exist in both the cases, however, the average intensity in the mode-matched case is larger by a factor of $\sim 2$ (see the respective horizontal lines). Thus using a mode-matched aPARS system, we have experimentally achieved highly efficient, quasistable, single-mode coherent random lasing within a narrow interval of $1.2~$nm regardless of the instantaneous randomness in its configuration.

{\bf Summary:}\\ In summary, we have discussed the frequency behavior of a PARS system under weak and strong configurational disorder. We showed that, under strong randomness, the high Q modes appear with equal probability at each FP minimum of the dielectric layer. We used the approach of spectral mode-matching to restrict the frequency fluctuations of coherent random laser. Using this technique, in our aPARS system, we have successfully demonstrated single-mode, quasi-stable coherent random lasing where the fluctuations of the coherent lasing modes were reduced to a narrow band of $1.2$~nm as compared to a $10$~nm band seen in conventional coherent random lasers.
	
\chapter{Anderson localization-induced random lasing}

In this chapter, we show that the observed lasing originates from Anderson localized states. We carry out systematic simulations to estimate various quantities such as localization
length, field distributions of the modes and the distribution of their Inverse Participation Ratio (IPR). This is followed by the experimental results observed in the microdroplet array. The chapter is organized as follows: In section~\ref{sec:level7p1}, we analyze the passive PARS system with varying disorder and study the variation of ensemble averaged transmittance and localization length across the spectral band of interest. In section~\ref{sec:level7p2}, we calculate the field profile of various gap states and conclude that the field profiles do have an exponentially decaying behavior induced by disorder, a hallmark of Anderson localization. In section~\ref{sec:level7p3}, we present the experimentally observed mode profiles and show that the array indeed supports Anderson localization and the lasing modes are the consequence of the localization. Finally, in section~\ref{sec:level7p4}, we study the spatial extent of various high quality modes in terms of their Inverse Participation Ratio (IPR).

\section{\label{sec:level7p1}Ensemble-averaged transmittance and localization length}

\begin{figure}[h]
\begin{center}
\includegraphics[scale=0.57]{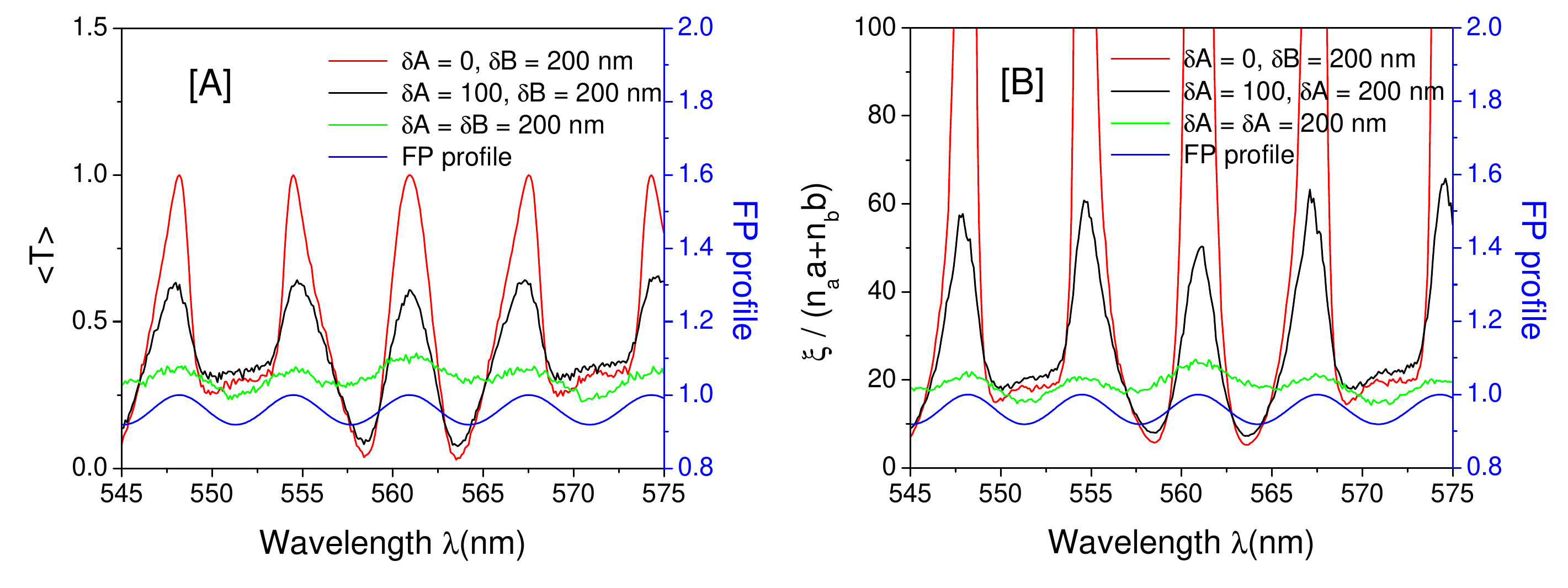}
\end{center}
\vspace{-20pt}
\color{blue}\caption {[A] Variation of ensemble-averaged transmittance ($<T>$) versus wavelength.The red line, the black line and the green line are the numerically calculated ensembleaveraged transmittance at the said disorder strength. The blue line is the FP profile of a single dielectric layer. [B] Variation of localization length $\xi$ (in units of no. of unit cells) as a function of wavelength at the same disorder. The horizontal dotted line represents the system size.}
\label{fig7p1}
\end{figure}

In Chapter 5, we showed how the interferences in the forward and backward propagating waves led to a modification of transmission in the random multilayer. Disorder-induced interference can lead to Anderson localized states within the system when the interference happens in a manner which forces the wavefunction to decay exponentially from a location of maximum intensity~\cite{anj14photonics}. In these conditions, transmittance already shows an exponential decay with system length, the decay constant indicating the localization length. To understand the general characteristics of the transmittance of the PARS system, we first calculate the ensemble averaged transmittance ($<T>$) spectra at various disorder strengths. These spectra are averaged over $1000$ configurations. We consider the multilayer structure to consist of $40$ unit cells where each unit cell contains a dielectric layer of $18~\mu$m ($n = 1.34$) and a gap layer of 8~$\mu$m. Figure~\ref{fig7p1}[A] shows the dependence of $<T>$ on wavelength at different degrees of randomness. The blue curve (right Y axis) is the Fabry-Perot profile of a single dielectric slab. The red curve represents a weakly disordered system with the randomness only in the spacing layers ($\delta A = 0$~nm, $\delta B = 200$~nm). The spectrum shows alternating high and low transmission regions similar to the FP profile of a dielectric layer.

The black curve shows the spectral behavior of $<T>$, when a small randomness in the size is introduced along with a weak disorder in spacing ($\delta A = 100$~nm, $\delta B = 200$~nm). A marginal enhancement of $<T>$ in the spectral region close to the FP minima (stopband of the underlying periodic structure) is seen in this case (black curve) as compared to the red curve. Further, $<T>$ reduces strongly at the wavelength of the FP maxima (passband of the underlying periodic structure). At even stronger disorder ($\delta A = \delta B = 500$~nm), the ensemble-averaged transmittance loses the wavelength sensitivity, resulting in a nearly flat curve across the spectral band (see the green curve). In this situation, the disorder is so strong that the system cannot be treated as PARS anymore.

The length scale of interest, here, is the localization length ($\xi$) which is given by

\begin{equation}
\xi = - \frac{L}{<\ln T>}
\label{eq7p1}
\end{equation}
where, $L$ is the total length of the multilayer system, and $<$...$>$ denotes the ensemble averaging over various random configurations. We calculate the localization length in the
PARS system at the aforementioned disorder strengths to study the contribution of localization to the random lasing modes. Figure~\ref{fig7p1}[B] shows the wavelength dependence of localization length calculated using Eq.~\ref{eq7p1} The horizontal dotted line represents the system size.

In the weakly random systems (red curve), the localization length diverges at the passband region indicating that all modes appearing in these regions are extended and the system is not able to localize them. In comparison, modes originating in the stopband region (the previously discussed gap states) have a smaller localization length and hence a propensity to be strongly localized. A reduction in contrast is seen in the localization length across the spectrum in the experimentally feasible disorder configurations (black curve). Assuming a system size of $40$ unit cells, $\xi < L$ in the stopband region at this disorder strength. Next, in systems with strong disorder (green curve), this contrast is washed out and modes across the entire spectral range can localize light. It can be seen that $\xi$ drops to very small values in PARS at the stopband region, and even small system sizes can achieve Anderson localization.

\begin{figure}[h]
\begin{center}
\includegraphics[scale=0.6]{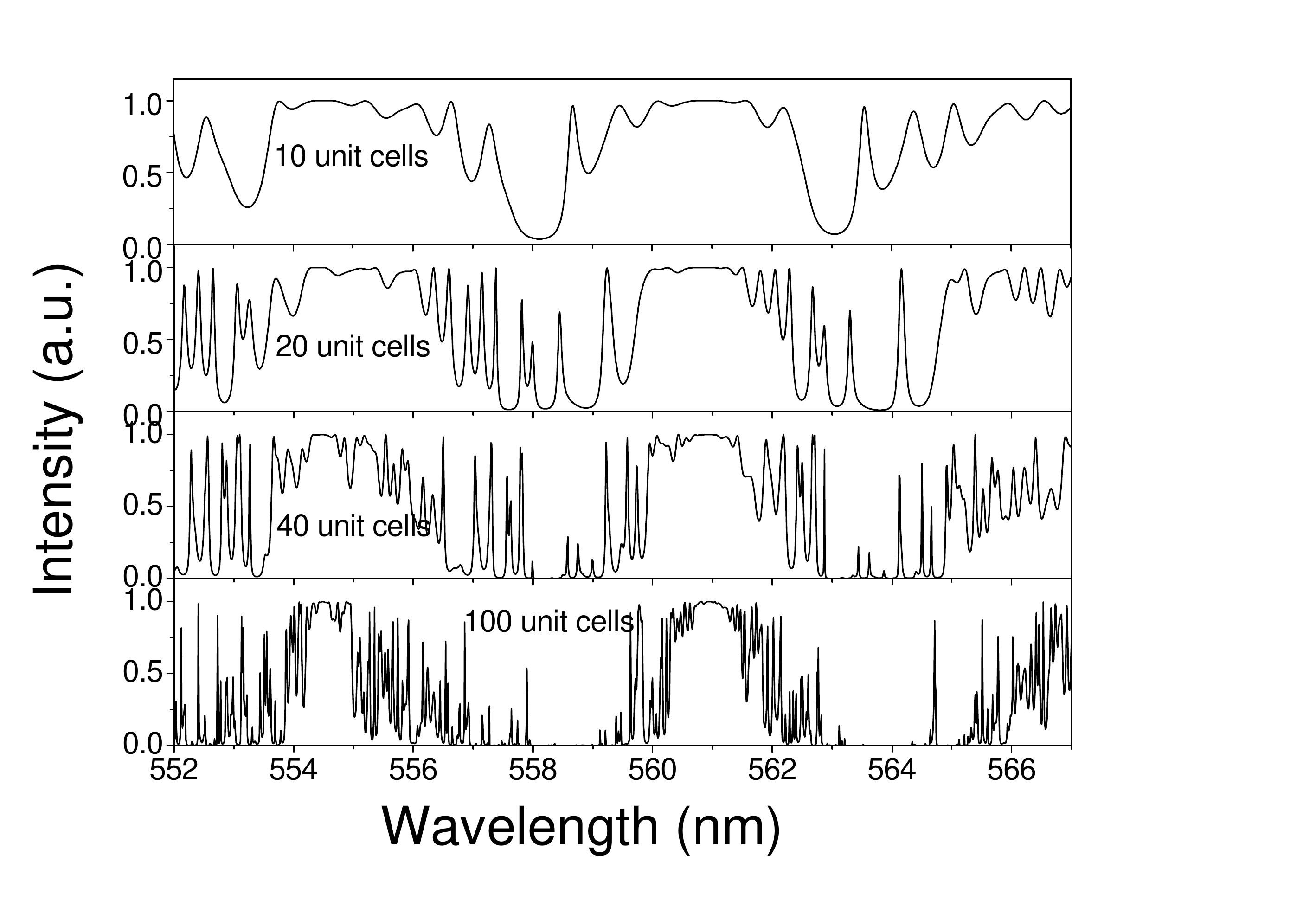}
\end{center}
\vspace{-50pt}
\color{blue}\caption {Simulated transmitted spectrum from four systems ($\delta A = 0$~nm and $\delta B = 200$~nm) consisting of $10$, $20$, $40$ and $100$ unit cells.}
\label{fig7p2}
\end{figure} 

To study the effect of system size on localization length in a PARS system, we consider a disorder configuration with $A = 18~\mu$m, $B = 8~\mu$m, $\delta A = 0$~nm and $\delta B = 200$~nm. Figure~\ref{fig7p2} depicts a representative spectrum for the multilayer system consisting of $10$, $20$, $40$ and $100$ unit cells (from top to bottom). It can be seen that, with increasing system size, the number of modes increases and the bandwidth of the modes narrows (quality factor increases). The narrow bandwidths are the result of larger number of multiple reflections from the multilayer structure.

\begin{figure}[h]
\begin{center}
\includegraphics[scale=0.45]{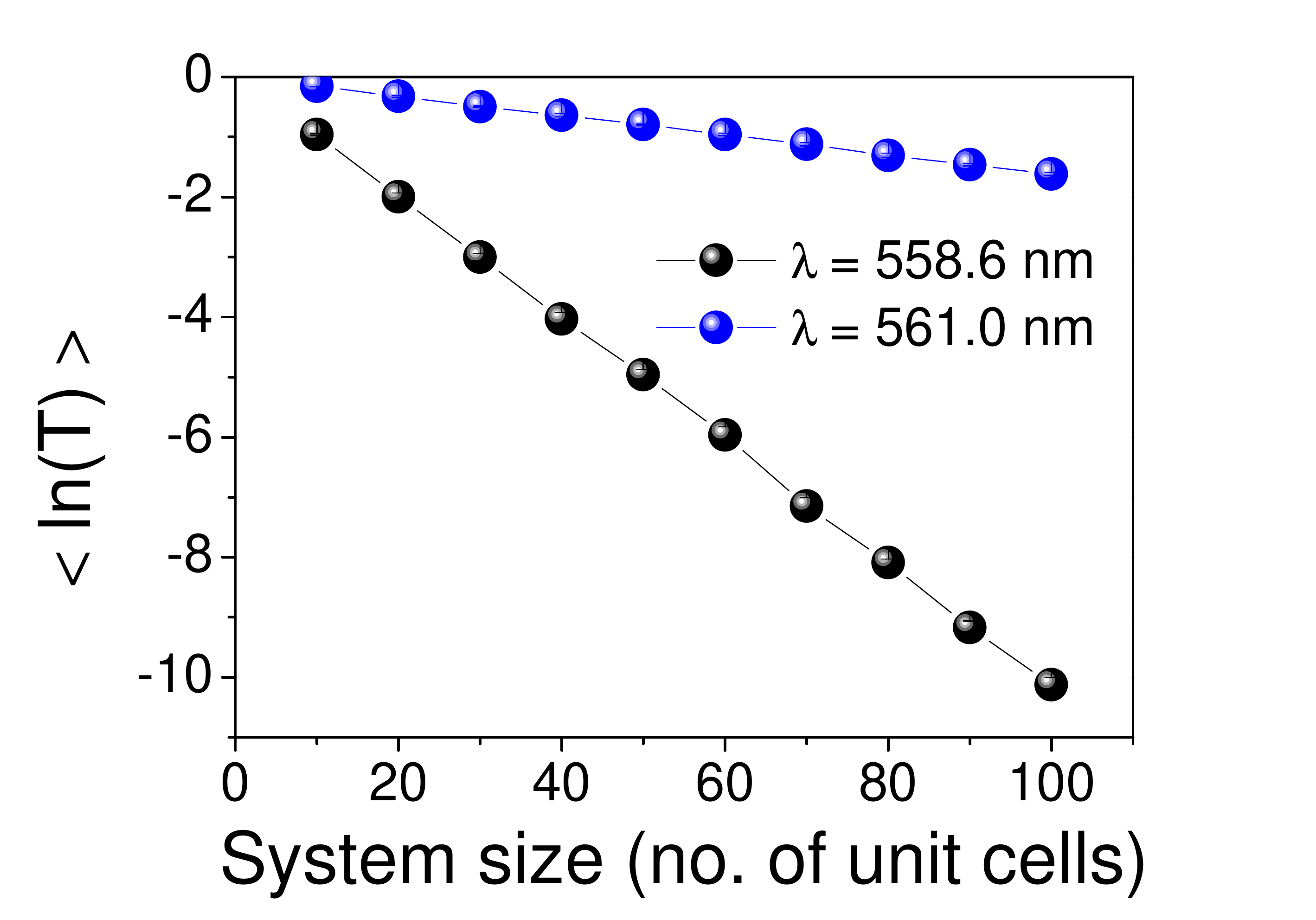}
\end{center}
\vspace{-20pt}
\color{blue}\caption{Variation of ensemble average of `logarithm of transmission' for the systems with $\delta A = 100$~nm and $\delta B = 200$~nm at two wavelengths, namely $558.6$~nm and $561$~nm. These wavelengths represent the center of the stop band and the pass band of the multilayer structure respectively. Clearly, the transmission is always larger in the pass band region.}
\label{fig7p3}
\end{figure}

Figure~\ref{fig7p3} plots the variation of ensemble average of `logarithm of the transmission' ($<ln(T)>$) versus system size. Again, each data point is averaged over $1000$ random configurations. To calculate $<ln(T)>$, we select two wavelengths namely $\lambda = 558.6$~nm (black curve) and $\lambda = 561.0$~nm (blue curve). These wavelengths are selected since they correspond to the underlying stopband and passband centers of the multilayer structure respectively. It is clear that at both wavelengths, $<ln(T)>$ decreases linearly as the system size increases. The slope of $<ln(T)>$ curve is the inverse of the localization length. An asymptotic convergence to zero transmission can be seen at both wavelengths as the system size approaches infinity. However, finiteness of every practical system results in light leakage from its boundaries and hence a broadening in the width of the modes (due to finite lifetimes). According to the Thouless criterion, when Anderson localization sets in, the Thouless parameter ($\delta \omega/\Delta\omega) < 1$, where, $\delta \omega$ is the width of the mode and $\Delta\omega$ is the average spacing between the modes~\cite{thouless77}. Clearly, in the passband region, the modes have larger $\delta \omega$ as compared to the modes lying in the stopband region (see Figure~\ref{fig7p3}). This again confirms that the localized modes preferentially lie in the stopband region, even in sample sizes as small as $10$ layers. 

\section{\label{sec:level7p2}Field distribution inside the PARS system}

To compute the field profiles of the high quality modes inside the multilayer structure, we first find the transfer matrix $M$ of the entire multilayer structure, which is defined by
\[
\left[ {\begin{array}{cc}
E_{1+} \\
E_{1-} \\
\end{array} } \right]= M
\left[ {\begin{array}{cc}
E^{'}_{1+} \\
E{'}_{1-} \\
\end{array} } \right]
\]
Note that, as there are no waves traveling in the backward direction from outside the last layer, E$^{'}_{1-} = 0$. This information allows us to calculate the field amplitudes E$^{'}_{1+}$ and E$_{1-}$ as a function of the incident field E$_{1+}$. For instance, if  E$_{1+} = 1$, then E$_{1-} = r$ and E$^{'}_{1+} = t$, where $r$ and $t$ are the reflection and transmission coefficients of the entire structure. Once we know the fields at the input plane ($Z = 0$), we introduce an artificial plane, say at $Z = Z_{p}$ and compute the transfer matrix M$^{'}_p$ from the input plane $Z = 0$ to $Z = Z_{p}$. The electric fields at the arbitrary plane $Z_{p}$ are related by its transfer matrix. By shifting the position of the arbitrary plane, we calculate the fields across the entire multilayer structure. Thus, the TM method can be first used to calculate the transmission spectrum, and thereafter, for a given frequency, the spatial distribution can be computed.

\begin{figure}[h]
\begin{center}
\includegraphics[scale=0.57]{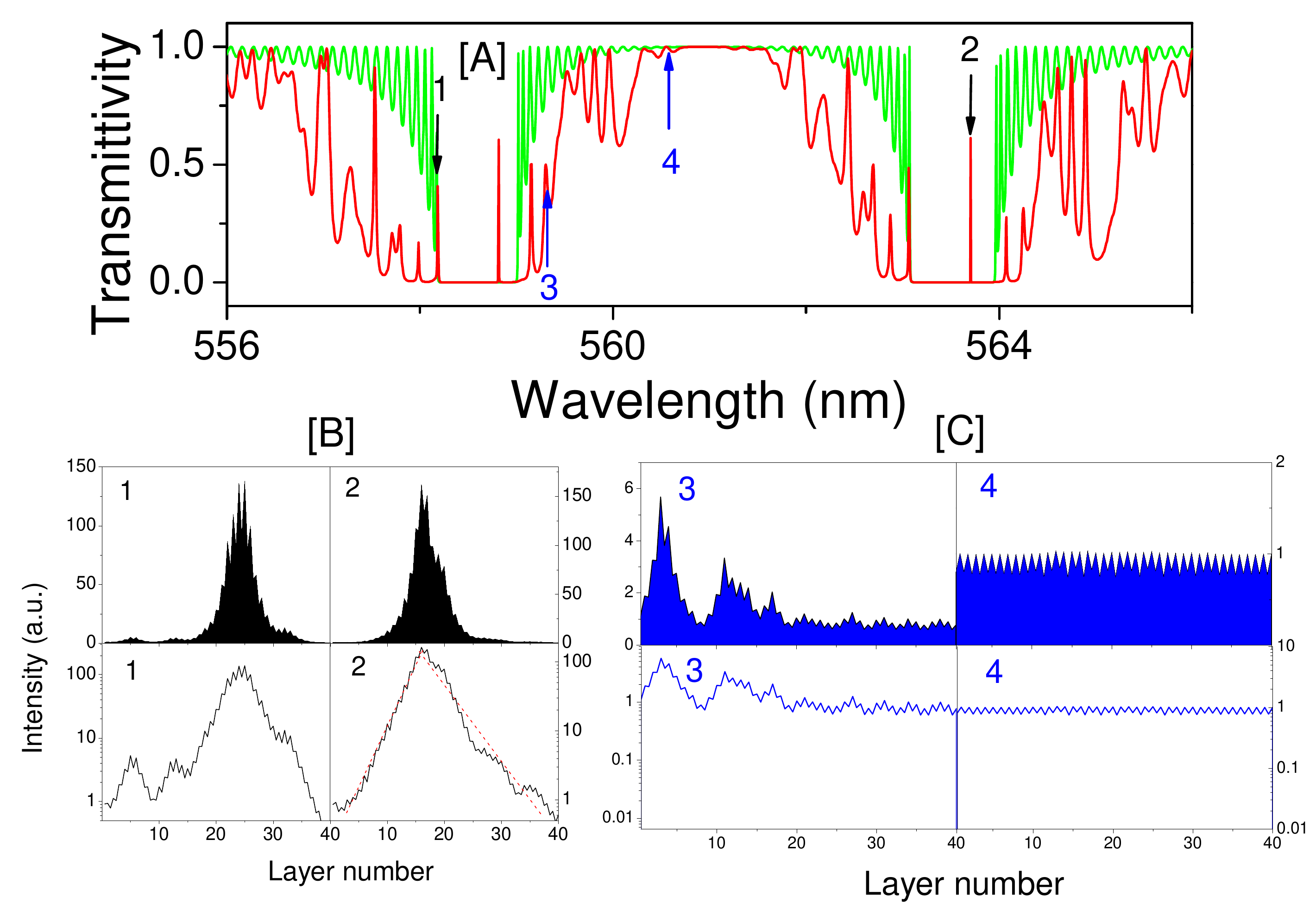}
\end{center}
\vspace{-40pt}
\color{blue}\caption{[A] Red curve: A representative spectrum from a passive PARS system with $\delta A = 0$~nm and $\delta B = 100$~nm. Green curve: The underlying band-structure of the periodic monodisperse system. The black arrows mark two gap states, the blue arrows mark two modes appearing in the passband region. [B and C] Intensity distribution of the respective modes inside the multilayer structure, the bottom panels is the same intensity distribution on the semi-log scale.}
\label{fig7p4}
\end{figure}

Figure~\ref{fig7p4} describes the spectrum and spatial distribution of the modes together. The green curve in Figure~\ref{fig7p4} [A] depicts the transmission spectrum of a monodisperse periodic system. The red curve is the transmission spectrum of one representative configuration of a passive PARS with $\delta A = 0$~nm and $\delta B = 100$~nm. The spectrum from the PARS system has several modes with distributed quality factors, out of which we choose two high-$Q$ gap states, appearing at $\lambda = 558.19$~nm and $563.71$~nm (marked by the black arrows). We can clearly associate these modes to the states which appear inside the stopband region i.e. the gap states. We also select two other modes which appear at $\lambda = 539.3$~nm and $560.5$~nm (marked by the blue arrows). The quality factor of these modes is relatively lower and they clearly appear in the passband region.

Figure~\ref{fig7p4}[B] shows the calculated intensity distributions of both the high $Q$ modes along the structure. The bottom panel is the same intensity envelope plotted on a semi-log scale. Clearly, the intensity distribution of the mode does follow an exponentially decaying profile as shown by the dotted red line. Further, the mode is confined within $\sim 8$ unit cells which is consistent with the simulations presented in section~\ref{sec:level7p1}. Thus, the field distribution shows Anderson localization in this one-dimensional system, when the states are in the gap. Figure~\ref{fig7p4}[C] depicts the intensity profiles of the modes which appear in the passband regions (marked by the blue arrows). These modes are extended over the entire multilayer sample and the system of $40$ unit cells is not capable to localize these modes. It is interesting to note that modes with strongly varying spatial extents can appear at different spectral positions for a single given configuration.

\section{\label{sec:level7p3}Experimental observation, intensity profile of the lasing modes}

\begin{figure}[h]
\begin{center}
\includegraphics[scale=1.45]{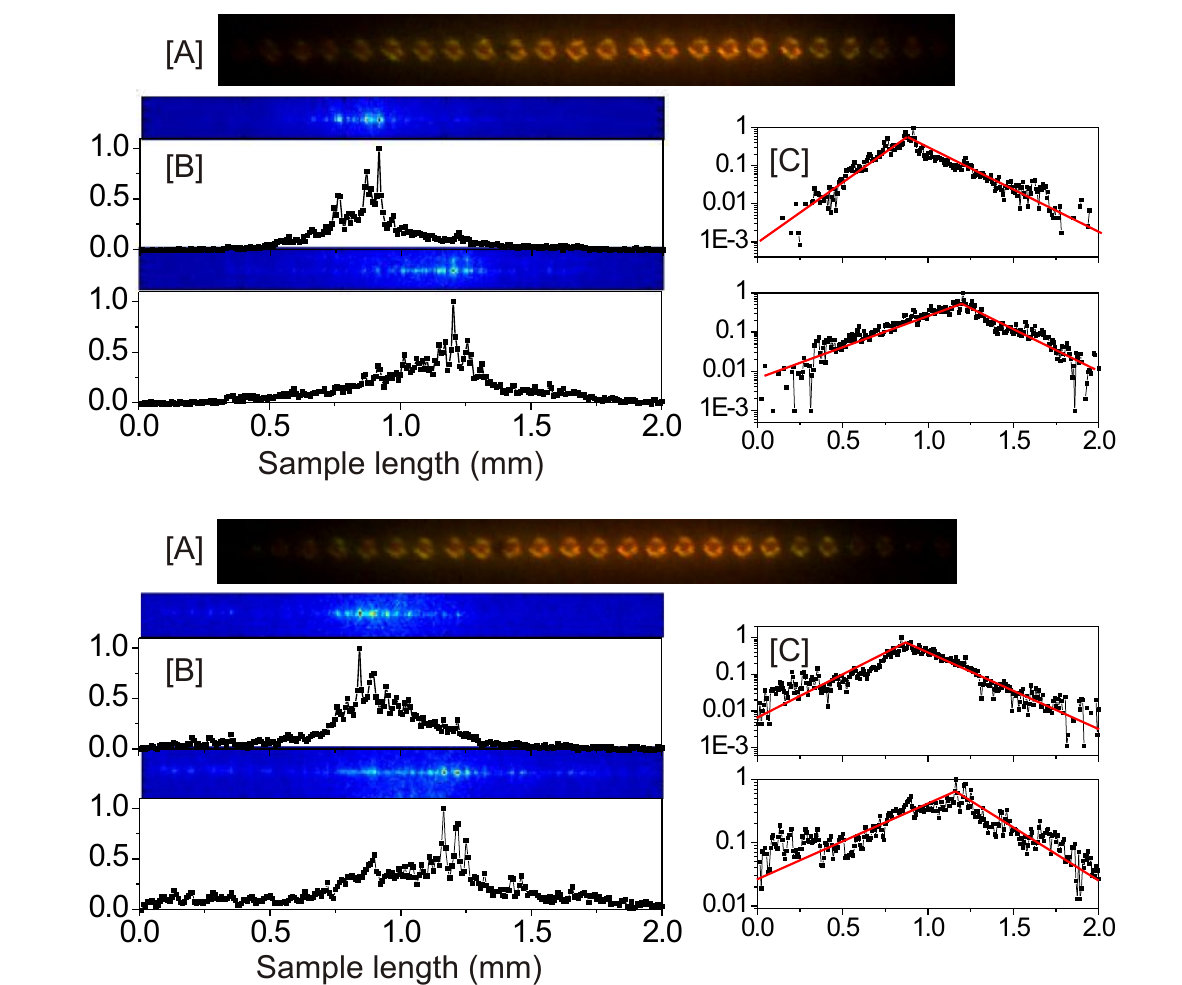}
\end{center}
\vspace{-30pt}
\color{blue}\caption{[A] Image of the monodisperse microdroplet array. [B] Simultaneously captured intensity profile of lasing modes. [C] The same intensity profile on the semi-log plot.}
\label{fig7p5}
\end{figure}

Although transmittance cannot be measured in our system, the transverse emission gives the direct signature of the intensity distribution across the array. Figure~\ref{fig7p5} depicts the intensity profile of the lasing modes for two excitation pulses. Figure~\ref{fig7p5}[A] shows a part of typical lattice with monodisperse resonators. While the monodispersity is good, the inter-resonator separations are seen to vary mildly. The sizes were characterized accurately from the whispering gallery modes which were observed in the transverse emission, while the separations were estimated from the images by a CCD.

Figure~\ref{fig7p5}[B] depicts the spectral image of the random lasing modes which is directly obtained by the spectrographic imaging system. Clearly, the lasing modes have the maximum intensity at random positions in the array despite the fact that they are observed from the same configuration (the top and the bottom panel of Figure~\ref{fig7p5}[B]). Furthermore, the modes are strongly localized in the multilayer and have varying spatial extent which covers several microdroplets. Figure~\ref{fig7p5}[C] depicts the same mode profiles on the semi-log scale. Clearly, the modes display an exponentially decaying profile as seen from the red line. The localization length of these modes (from top to bottom) are $\sim 405~\mu$m ($15$ unit cells), $\sim 380~\mu$m ($14$ unit cells), $\sim 405~\mu$m ( $15$ unit cells) and $\sim 515~\mu$m ($20$ unit cells) respectively. The estimation of the localization length shows a good agreement with the simulation presented in section~\ref{sec:level7p1}. Direct measurement of the intensity profile confirms the Anderson localization of light in this system. Thus, the lasing modes observed from the arrays essentially constitute lasing over Anderson-localized states.

{\bf \$ Comment on the angular profile of the random lasing modes}\\

\begin{figure}[h]
\begin{center}
\includegraphics[scale=0.6]{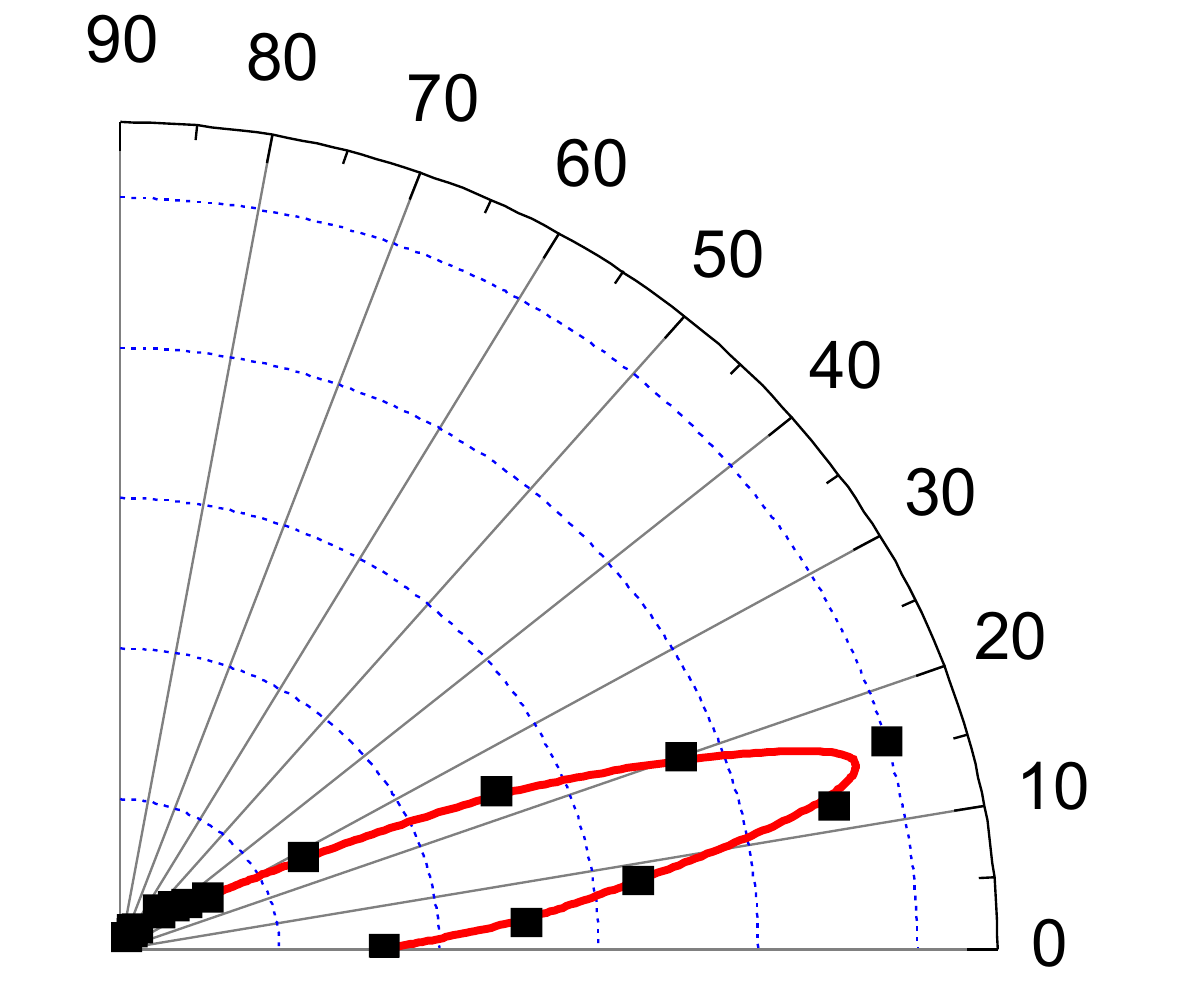}
\end{center}
\vspace{-20pt}
\color{blue}\caption{Angular distribution of the random lasing modes observed from the droplet array.}
\label{fig7p6}
\end{figure}

We now elucidate the angular distribution of the random lasing modes observed from the microdroplet array. Figure~\ref{fig7p6} shows the intensity distribution of random lasing modes, as reproduced from inset of Figure~\ref{fig4p7}. As mentioned earlier, the average intensity of the random lasing modes shows a minimum in the longitudinal direction, and in this case, has a maximum at $\sim 16^{\circ}$. The longitudinal minimum can now be understood as follows. We have shown that random lasing modes originate from the gap states. These gap states are localized due multiple scattering from the surface of the microdroplet array. Localization, by definition, forces the light to remain in the interior of the sample, and realizes very low transmittance outwards along the axial direction. This forbids the intensity distribution of the modes to maximize along the axis of the array. Therefore, the maximum intensity is observed away from the axial direction. It should be noted that the maximum at $\sim 16^{\circ}$ is not for granted and depends on the size and spacing of the microdroplets.

\section{\label{sec:level7p4} Mode extent and IPR distribution} 

The exponentially decaying profile of the random lasing modes confirms that the modes originate due to disorder induced localization. As discussed previously, our aPARS system
provides the possibility to measure several modes and various configurationally-averaged parameters which is of great interest in the field of disorder system. One such parameter is the Inverse Participation Ratio (IPR)~\cite{schwartz07}. It should be noted that, in our experiments, the system has inherent optical gain and hence, under optical pumping, the probability of observing the high quality modes is high. Hence, in the simulations, we deliberately choose the high quality modes of the PARS system and discuss their IPR.

\subsection{\label{sec:7p4a}IPR distribution of the gap states}

To quantify the spreading of the field and degree of localization, we calculate the probability distribution of the IPR, which is defined as follows

\begin{equation}
IPR = \frac{\sum\limits_{i} |E_i|^4}{(\sum\limits_{i} |E_i|^2)^2}
\label{eq7p2}
\end{equation}

where E$_i$ is the amplitude of the field at the site $i$. In our experiment, each pixel of the CCD represent a site. Note that, the spectrometer detects the intensity of the modes
$I_{i} = |E_{i}|^2$. Hence, the above equation can be translated into

\begin{equation}
IPR = \frac{\sum\limits_{i} I_{i^{2}}}{(\sum\limits_{i} I_{i})^{2}}
\label{eq7p3}
\end{equation}

\begin{figure}[h]
\begin{center}
\includegraphics[scale=0.60]{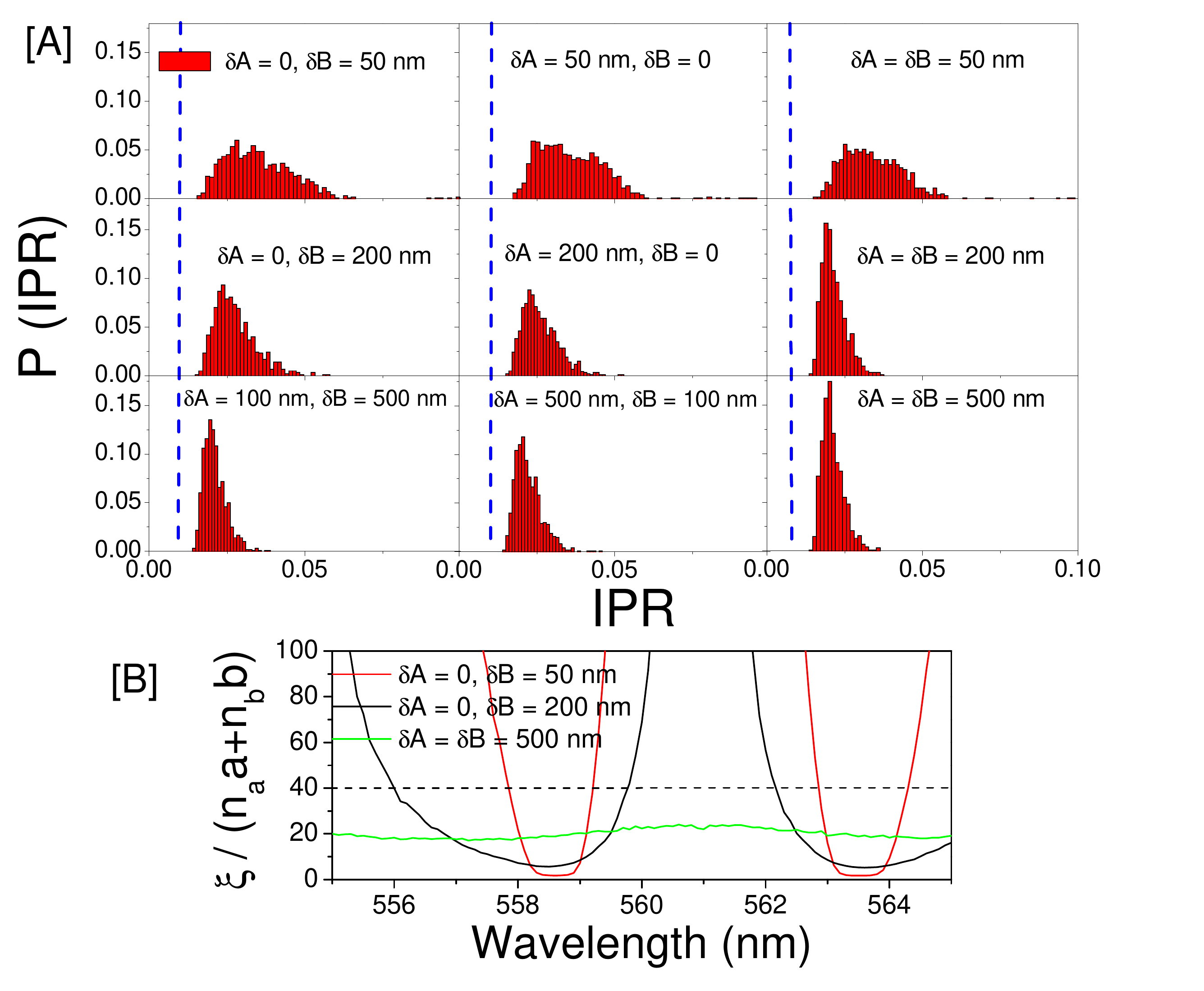}
\end{center}
\vspace{-30pt}
\color{blue}\caption{[A] IPR distributions at nine disorder strengths. In the weakly random system, the centroid of the distribution is shifted towards higher IPR which is the signature of mode localization. Furthermore, the distribution is wider which implies that the fluctuations in the IPR are higher in the PARS. [B] Variation of localization length at three disorder strengths.}
\label{fig7p7}
\end{figure}

This definition implies that if the field is localized at a single lattice site (ideal localized state), then the IPR $= 1$ and if the field is completely extended (flat distribution), then the IPR = 1/N, where N is the total number of lattice sites. Clearly, a larger IPR is indicative of light localization. To study the IPR distribution at a given strength of disorder, we generate $250$ PARS configurations and choose four highest quality modes from each PARS spectrum. To pick the high quality modes, we monitor their respective thresholds. The IPR of these high Q modes is calculated from the field profiles using Eq.~\ref{eq7p3}. Figure~\ref{fig7p7}[A] depicts the distribution at nine different disorder strengths. The blue dashed line marks the limit, which is the IPR of a completely extended mode. Since there are $40$ unit cells with alternating dielectric and gap layers, the IPR of a completely extended mode corresponds to $1/80 = 0.0125$.

Clearly, the shape of the IPR distribution is asymmetric and sensitive to the randomness ($\delta A$ and $\delta B$) of the multilayer structure. In weak random systems, the IPR distribution is relatively broader. As the randomness increases, the width of the distribution narrows down and the centroid of the distribution shifts towards left. This variation in the distribution width can be described using Figure~\ref{fig7p7}[B]. Note that, in the weak random systems, the localization length has a huge variation from $6$ unit cells upto $40$ unit cells. This implies that largely varying distribution in the spatial mode profile is expected from weakly random system. The mode corresponding to smaller localization length gets tightly localized by the multilayer system and hence has a higher IPR and vice versa. It should be noted that, since the gap modes maintain their high $Q$ and have a smaller localization length, the centroid of the IPR distribution has higher value. On the other hand, as the randomness increases, the variation in the localization length decreases resulting in a narrower distribution. Further, as compared to the weak disorder system, these systems have relatively low quality factor modes (see Figure~\ref{fig5p8}[B]). The intensity distribution of these low quality modes are extended and the distribution gets shifted towards lower IPR. Finally, the asymmetry in the distribution can be explained from Figure~\ref{fig7p2}. It can be seen that, the modes situated closer to the bandedges are always more in number than those generated deep inside the gap region. The former have a lower IPR, while the latter are tightly localized and have higher IPR. Hence, the higher IPR modes are fewer in occurrence, leading to the asymmetry.

\subsection{\label{sec:7p4b}Experimentally measured IPR distribution of the random lasing modes}

\begin{figure}[h]
\begin{center}
\includegraphics[scale=0.50]{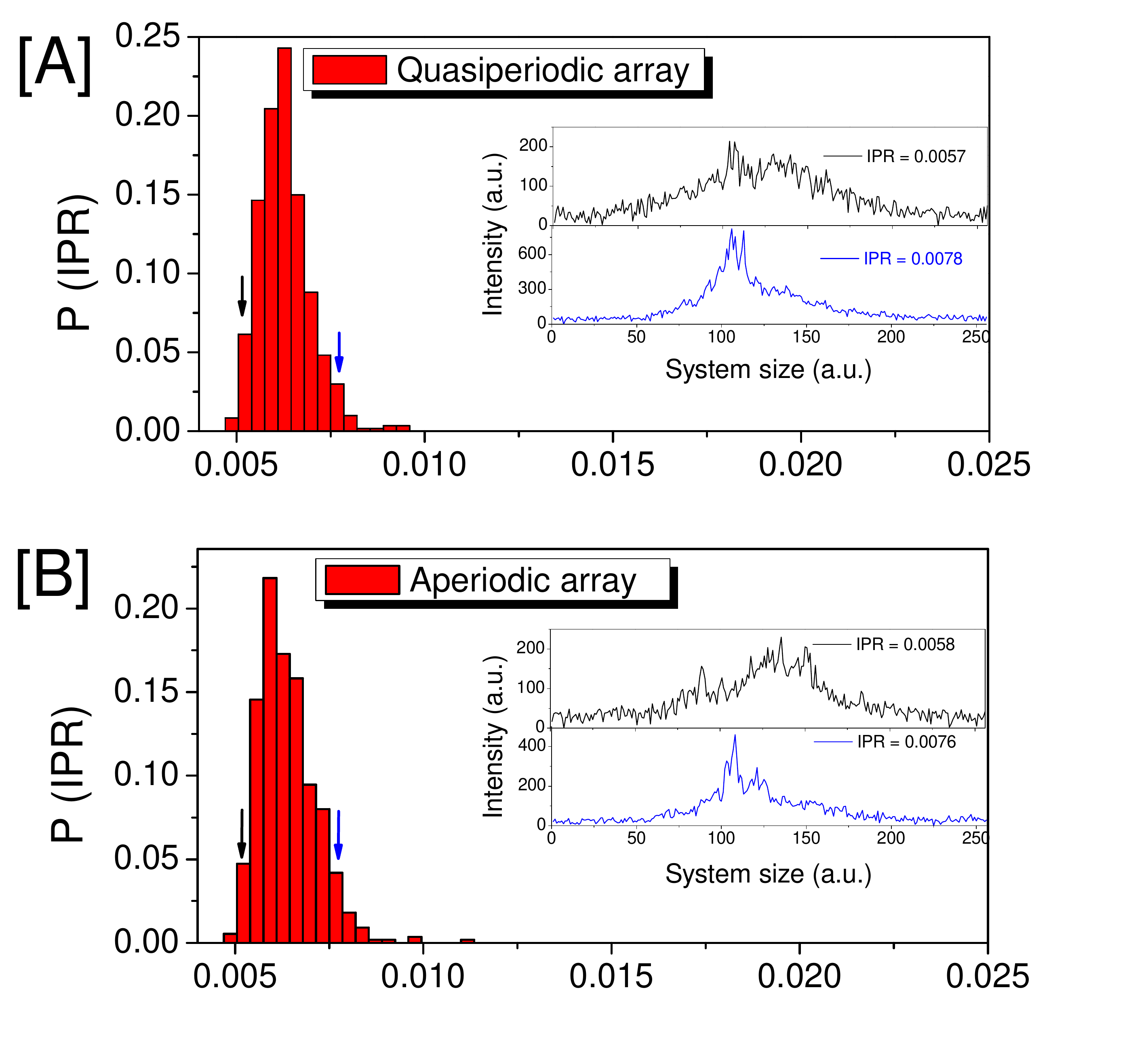}
\end{center}
\vspace{-30pt}
\color{blue}\caption{Experimentally observed IPR distribution and field profile of the random lasing modes. [A] From a quasi-periodic array with $\delta A = 100$~nm and $\delta B = 1000$~nm. The inset curve shows the representative mode profiles, corresponding to an IPR marked by the same color arrows. [B] IPR distribution from an aperiodic array with $\delta A = 100$~nm and $\delta B = 5000$~nm.}
\label{fig7p8}
\end{figure}

Figure~\ref{fig7p8} shows the experimentally observed IPR distributions from two different configurations, namely, quasi-periodic and aperiodic array. These distributions are created over $2000$ spectra. First, the asymmetry in the distribution is clearly evident. The inset shows the field distribution at two IPR values marked by the black and the blue arrows. Clearly, in both situations, the mode corresponding to the lower IPR has extended behavior and is weakly localized. On the other hand, the modes corresponding to the higher IPR are strongly localized in the small region. We note that, the quasi-periodic and aperiodic configurations do not show any significant difference in the distribution. The distribution is centered at $\sim 0.0065$ and is asymmetric. We discussed in section~\ref{sec:level7p3} that, if the inherent randomness is sufficiently large ($> 200$~nm), the IPR distribution does not show any significant overall change. As the droplet array always has such small inherent randomness, these observations are consistent with our calculations. It appears that a better control on the periodicity and monodispersity needs to be obtained in the microdroplet array. Then, it would be possible to trace the complete behavior seen in section~\ref{sec:level7p3}. Nonetheless, these results still form the first measurements of IPR distribution from an Anderson-localized random laser.

{\bf Summary:}\\ In this chapter, we provide evidence of Anderson localization in the quasi-one-dimensional system. First, by using transfer matrix method, we have analyzed various quantities like localization length, ensemble-averaged transmittance, etc. of a PARS system. Particularly, we calculated the field distribution of high quality modes in various random multilayer systems, and characterized the exponential decays. We confirmed the exponential decays in the experimental arrays. Next, we showed that the IPR distributions of the modes are asymmetric and sensitive to the randomness of PARS system. We also observed that the distribution are shifted towards lower IPR for higher randomness in the multilayer. Experimentally, we have studied two situations namely quasi-periodic and aperiodic array. We analyzed their field profiles and IPR distributions and found an excellent agreement with our simulations.
	
\chapter[Collective modes in direct-coupled, bent chains of active spherical~.~.~.~.~.]{Collective modes in direct-coupled, bent chains of active spherical microcavities}%

In this final chapter, we present the existence of collective modes in the arrays which are bent at small angles. So far, we have focused on the randomness of the system, and have shown that the random system sustains Anderson localized modes, the study of which has a paramount significance in fundamental physics. This chapter is slightly tilted towards a practical application of our system. Importantly, here we focus on the applicability of a periodic array, and the consequences of the inherent randomness are mentioned from the vantage point of ``inherent randomness in an otherwise periodic system''. Indeed, fundamental studies of Anderson localization in bent systems also evoke a lot interest, but they are out of the scope of this thesis.

Here, we discuss the experimental results in the context of coupled resonator optical waveguides (CROW's) that are coupled via nanojet-induced modes (NIM's). Next, using the FDTD technique, we theoretically examine the ability of gain to overcome the bending and other radiative losses. This chapter is organized as follows: section~\ref{sec:level8p1} discusses the experimental setup to observe collective modes in microdroplet arrays which are bent at small angles. In section~\ref{sec:level8p2}, we discuss the mode probability and modal density at various bending angles. Finally, in \ref{sec:level8p3}, we numerically study the mode propagation using two-dimensional FDTD calculations, and show that the transport of light is mediated by the filamentation of nanojet-induced modes. This chapter essentially recasts our resonator array as a CROW, a photonic device that exploits the collective behavior of optical resonators.

The main motivation to study the collective behavior of any system is to obtain a better functionality of the system compared to its individual components~\cite{boriskina11, boriskina06}. Collective phenomena in the optical domain, especially in microcavities, have been a subject of intense
scrutiny~\cite{boriskina06, deych06, deych08, hara05, nakagawa05, hara03}. In the recent past, such collective phenomena have been studied in various arrangements of microdiscs, microspheres and photonic crystal microcavities~\cite{scheuer05, astratov10, yang08, morichetti12, chen06, darafsheh12, kapitonov07, astratov04, yariv99}, creating coupled resonator optical waveguides (CROWs). In general, a CROW comprises of a series of identical resonators, typically placed in contact with each other. When light is coupled at one end of the waveguide, light propagates due to the evanescent coupling of whispering gallery modes (WGMs).

Several important discoveries like generation of slow or fast light states~\cite{heebner02}, all-optical processes using nonlinearities~\cite{melloni03}, or higher-order photonic structures such as delay lines~\cite{poon04} have been demonstrated in a CROW. Although such waveguides mainly used evanescent coupling using WGMs, recent experiments have demonstrated the ability of direct-coupling of light in the domain of straight chains~\cite{chen06}. In this scenario, the propagation of light is realized by directly-coupled nanojet-induced modes (NIMs), mediated by optical jets which form at the edge of a microsphere. A prime advantage with the NIM modes is that they are less sensitive to size disorder compared to the WGMs, and hence they are shown to dominate the transport of light in long straight chains of spherical resonators. Recently, many interesting features of NIM, such as periodic focussing, beam profile narrowing, etc. have been revealed using ray analysis~\cite{darafsheh12}.

A desirable feature of a waveguide is the possibility to steer light in a bent arrangement, either in sharp angles or over gradual turns. In this regard, evanescently coupled WGMs have a natural advantage due to their ability of omni-directional coupling. Significant studies of bent arrays and CROW branches have been reported in various systems using evanescent coupling~\cite{mitsui08, pishko04, boriskina07}. However, the WGMs are very sensitive to size disorder, hence a slight mismatch between the cavities affects the coupling strength of the modes in the array. In comparison, the significant concern about the nanojet-induced modes is that they have inherent radiative losses. Hence, extra losses due to bending are undesirable. One accomplished way of overcoming the losses is by incorporating optical gain in the system~\cite{smotrova06, mookherjea04, moeller05, dumige08, poon07}. In the previous chapters, we have studied the emission properties of a linear chain comprising amplifying microdroplets and discussed their collective behavior. Here, we further demonstrate the existence of collective modes in the chains which are bent at small angles~\cite{anj13apl}.

\section{\label{sec:level8p1}Experimental setup}

\begin{figure}[h]
\begin{center}
\includegraphics[scale=1.4]{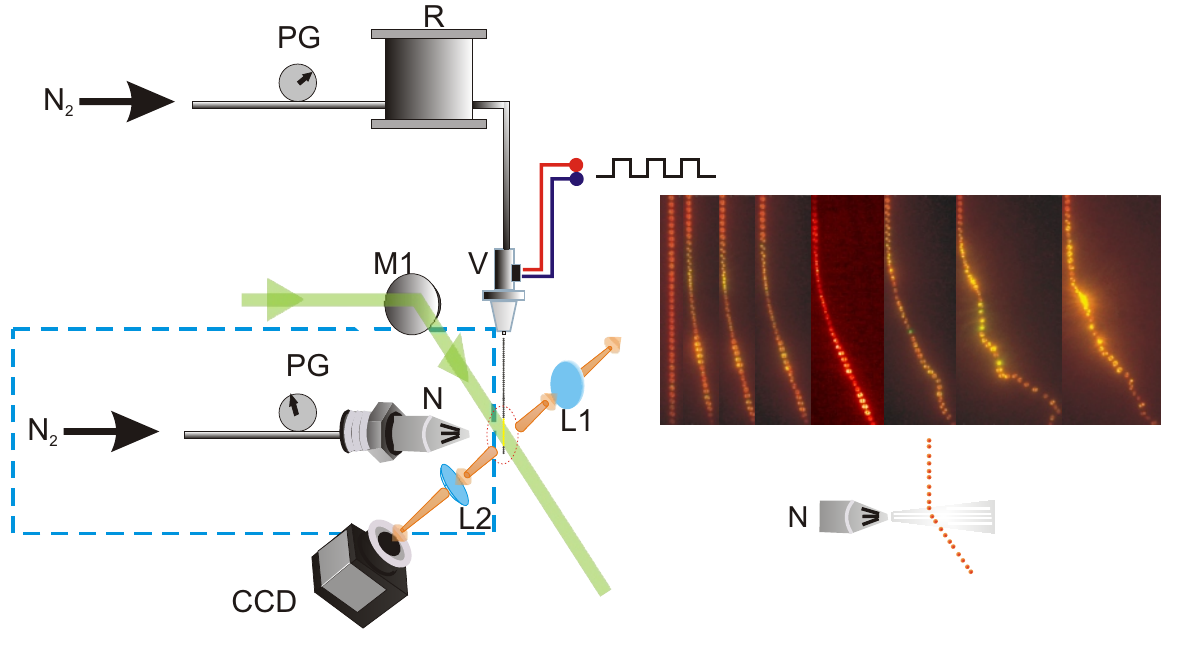}
\end{center}
\vspace{-30pt}
\color{blue}\caption {Experimental setup to generate and observe transverse emission from a bent chain of microdroplets. PG: pressure gauge, R: reservoir, V: vibrating orifice aerosol generator, M1: mirror, L1 and L2: lens, CCD: imaging CCD, N: tapered nozzle, which blows N2 gas onto the droplets to achieve desired bending. Right panel shows various images of the bent
arrays upto $24^{\circ}$. Thereafter, the array undergoes distortion.}
\label{fig8p1}
\end{figure}

The schematic setup to generate the bent CROW is shown in Figure~\ref{fig8p1}. The details of the experimental setup are already discussed in chapter 3. To achieve desirable bending in the array, we employed a tapered nozzle N into the vicinity of microdroplets (highlighted by the dashed rectangle) and N2 gas was blown at a mild pressure ($\sim 15$ atmospheres). By adjusting the gas pressure and the distance between the nozzle and the array, we bent the array at various angles. Using this technique, we could bend the array in a controlled fashion upto $\sim 24^{\circ}$, without severely distorting the chain. The right panel in Figure~\ref{fig8p1} shows the representative images of the bent microdroplet array. While the first image shows the
straight array, the rest of the images show bending at $\sim 10^{\circ}$, $\sim 14^{\circ}$, $\sim 18^{\circ}$ and $\sim 24^{\circ}$ respectively. Beyond $\sim 24^{\circ}$, the microdroplets randomly deviate from their original path, which results in a chaotic configuration as depicted by the other three images. Note that, in the interaction region (bend part), the shape and size of the droplets get perturbed by the gas jet. However, due to the surface tension force, the droplets immediately regain their spherical shape and attain monodispersity. Clearly, in each configuration, the droplets are separated by several micrometers, thus preventing the possibility of evanescent coupling.

\section{\label{sec:level8p2}Collective modes in bent arrays}

\begin{figure}[h]
\begin{center}
\includegraphics[scale=0.95]{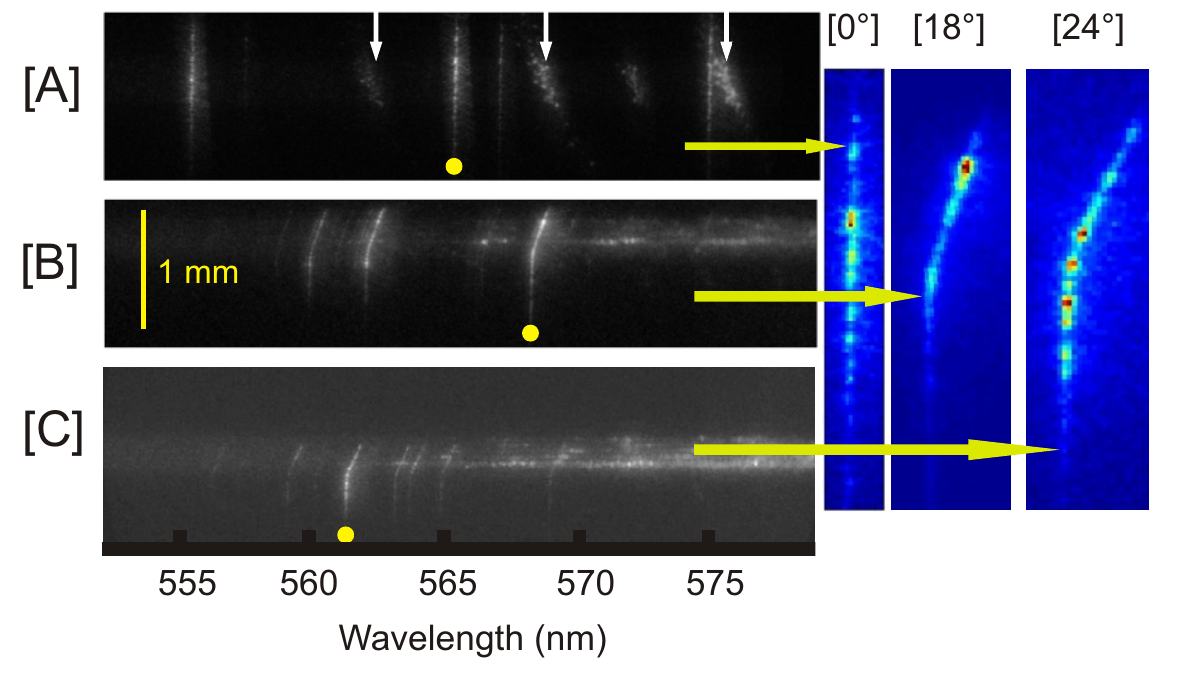}
\end{center}
\vspace{-40pt}
\color{blue}\caption {Spatio-spectral images of the array. [A] Spectral image from the straight array. Vertical bright lines are the carrier modes, tilted patches represent the whispering gallery modes (indicated by white arrows). [B, C] Spectral images from the arrays bent through $18^{\circ}$ and $24^{\circ}$ respectively. Right panel shows the magnified images of the modes marked by the
yellow dots.}
\label{fig8p2}
\end{figure}

Figure~\ref{fig8p2} depicts the spatio-spectral image of the array in the transverse direction. Top panel [A] shows the emission properties from a straight array consisting of highly uniform microdroplets. In this system, the spectrum has well-defined WGMs and collective carrier modes. Equispaced broad patches in the background (labelled by the white arrows) are the WGMs. On the other hand, vertical bright streaks depict the collective carrier modes. These carrier modes are extended over the entire illuminated region and manifested by the direct coupling of various Fabry-Perot resonances, as discussed in section~\ref{sec:level4.3} in this thesis.

Figure~\ref{fig8p2}[B] and [C] show the spectral images when the array was bent at $18^{\circ}$ and $24^{\circ}$ respectively. Note that, as we blow the N2 jet, the monodispersity of the array gets disturbed, and as a result the WGMs are washed out from the background. However, the multiple carrier modes still exist in the bent configuration despite the polydispersity and bending losses. This clearly confirms that the carrier modes are less sensitive to the size disorder compared to the WGMs and follow the same spatial trail of the array. The right panel shows the magnified image of three carrier modes labelled by the yellow dots. Each carrier mode has a non-uniform intensity distribution and some intense hot-spots are seen along the array at random positions. These hot-spots do not necessarily happen in the bent region.

\begin{figure}[h]
\begin{center}
\includegraphics[scale=0.55]{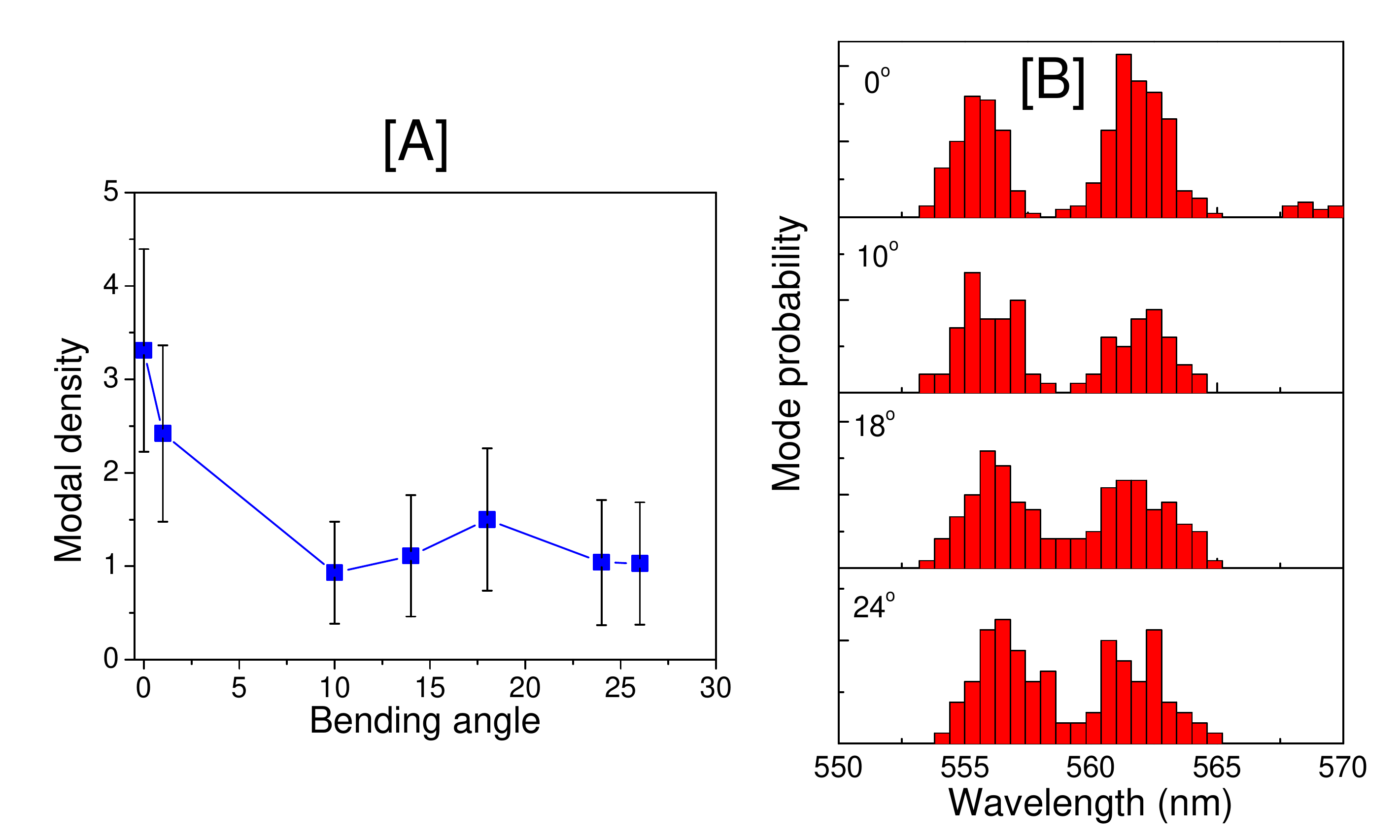}
\end{center}
\vspace{-30pt}
\color{blue}\caption{[A] Modal density as a function of $\theta_{bend}$. At small angles, modal density drops sharply and subsequently shows a gradual and non-monotonic variation. [B] Frequency distribution of the carrier modes at various $\theta_{bend}$. Well-defined bunching is seen in the straight array. Upon bending, the modes start to populate the gap region, though the bunching is still evident.}
\label{fig8p3}
\end{figure}

We next measure the frequency distribution of carrier modes. To that end, one hundred spectra were grabbed at a given bending angle ($\theta_{bend}$). This also allows us to quantify the dependence of carrier mode buildup on the bending angles. Figure~\ref{fig8p3}[A] depicts the variation of modal density as a function of $\theta_{bend}$, at $E_{p} = 1.17~\mu$J. Modal density is defined by the average number of carrier modes observed in the entire spectral range. On an average, the
straight array has more than $3$ modes per pulse. Upon bending, the modal density rapidly decays up to $30~\%$ within few degrees of bending ($\sim 10^{\circ}$). Thereafter, the density does not seem to vary significantly.

Figure~\ref{fig8p3}[B] shows the frequency distribution of the modes. In the straight array, as expected, the modes appear in a well-defined region which is separated by the FP free spectral range of the individual resonator. In the context of the CROW, a similar system of linear chains of microspheres has been studied by Yang and Astratov, where the microspheres were in physical contact with each other~\cite{yang08}. They observed a similar signature of Fabry-Perot resonances in the transmittance. Interestingly though, the spectral spacing between the Fabry-Perot maxima corresponded to an effective cavity length of two microspheres. This behavior was explained by simulating the intensity pattern inside the chain which indeed showed a period equal to two sphere diameters. In our experiment, we observed that upon bending, the bunches start to merge together due to the polydispersity introduced by the N2 jet. Nonetheless, the bunches can still be identified at all bending angles and carrier modes obey the frequency behavior similar to the straight array. This implies that the carrier modes are still ruled by the monodisperse droplets and show the signature of FP resonances of a single droplet.

\section{\label{sec:level8p3}2D FDTD simulations}

To study the mode propagation in the droplet chains, we first simulate an ideal situation of a CROW. In general, studies on CROW's are performed on samples with long static chains where the surfaces of the resonators touch each other. To estimate the waveguide losses, a source at a fixed wavelength excites one end of the chain and the transmission is measured at the other end. Note that, in our experiment, the dynamical nature of the microdroplets prohibits us from measuring quantities like transmittance, bending losses etc. Therefore, we studied these parameters by carrying out FDTD simulations through the well-acknowledged, free, open-source software package MEEP (MIT Electromagnetic Equation Propagation)~\cite{oskooi10}. The large size of the system restricted our calculations to two-dimensions. Nonetheless, the important features of the study, like the bending of the modes and the effects of curved surfaces, etc. are well served by these simulations. Note that, we focus our study only on the spatial characteristics of the modes and not on their spectral behavior.

We assume that the array consists of $15$ monodisperse microdroplets where each droplet is represented by an infinite cylinder of diameter $d = 18~\mu$m. These identical cylinders are arranged in a non-touching periodic fashion with a surface to surface separation of $6~\mu$m to form the CROW. The large separation forbids evanescent coupling. To implement the
bend, we choose three central cylinders and then shift their centers on a circular locus. The curvature of the locus defines the bending angle of the waveguide. The spatial and temporal resolution were maintained at $dx = dy = 1/15~\mu$m and $dt = 0.11$~fs (in real units) respectively. Altogether, the size of the computational cell was $420~\mu$m $\times~23~\mu$m wide in the straight waveguide and $390~\mu$m $\times~125~\mu$m for the largest bend angle.

We studied the propagation of the carrier mode at a single wavelength chosen as follows. It is known that the propagation band of light in a periodic CROW is centered on the central wavelength of an individual resonator~\cite{scheuer05}. Accordingly, we excite the CROW at $\lambda = 561.4$~nm. It is important to note that, the wavelength that we chose lies in the passband region of the array (propagation band). Furthermore, we make sure that the source wavelength does not excite WGM resonances. A single line source was used to illuminate the array from one end and the transmitted signal was measured by placing a detector at the other end of the waveguide. We simulated eight configurations with bending angles of $5^{\circ}$, $7.5^{\circ}$, $10^{\circ}$, $12.5^{\circ}$, $15^{\circ}$, $20^{\circ}$, $25^{\circ}$ and $30^{\circ}$. While the results presented here describe the frequency of TM polarization (E-field in the plane of the paper), we also studied the TE polarization, which yielded consistent results.

\begin{figure}[h]
\begin{center}
\includegraphics[scale=1]{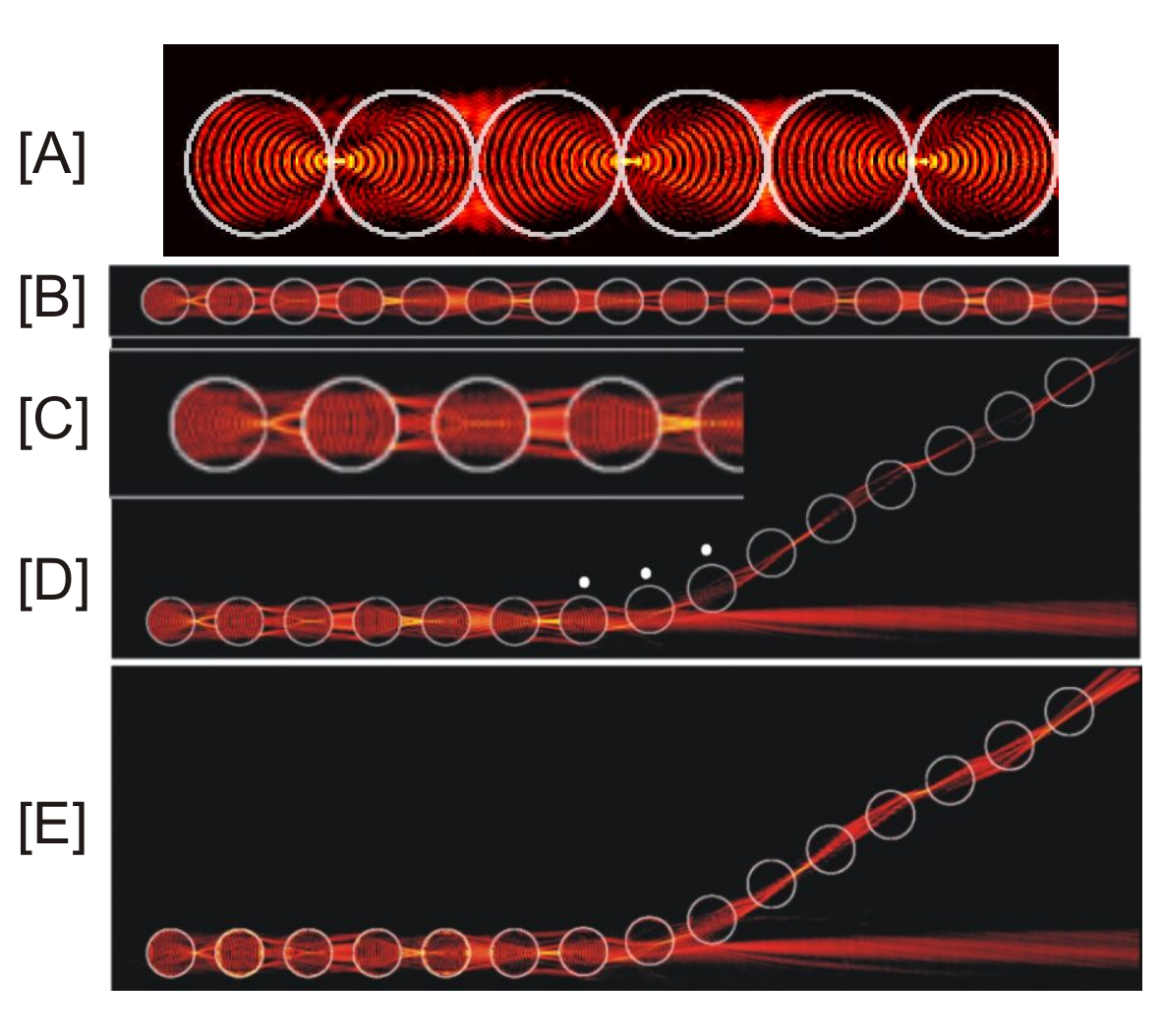}
\end{center}
\vspace{-25pt}
\color{blue}\caption{FDTD simulations of the carrier modes [A] Formation of nanojet-induced mode in a chain of touching spheres [B] Field profile in a straight chain consisting of $15$ nontouching cylinders. The nanojet-induced mode is formed at the edge of first cylinder which breaks up into various filaments, clearly depicted in the magnified image [C]. [D] Details
of mode propagation at $\theta_{bend} = 30^{\circ}$ which depicts the bending losses and self-correcting propagation mechanism. [E] Effect of gain in the array of [D]. In an active array the gain compensates for the bending and radiative losses.}
\label{fig8p4}
\end{figure}

Figure~\ref{fig8p4}[A-E] illustrate the detailed analysis of the mode propagation in various situations, as revealed by the FDTD simulations. As mentioned in the previous section, in a waveguide consisting of touching spheres, the mode propagates through the nanojet formation which is formed on the surface of alternating spheres~\cite{yang08}. Figure~\ref{fig8p4}[A] depicts this behavior clearly. Furthermore, in the straight waveguide, the mode travels along the axis of the array which is a usual property of the NIM's. Figure~\ref{fig8p4}[B] depicts the case of non-touching spheres. It is clear that the nanojet that is formed at the edge of the first cylinder breaks up into a series of transverse maxima and minima in the air region. These maxima propagate in the form of filaments and are coupled to the subsequent cylinders (see the magnified inset Figure~\ref{fig8p4}[C]). The spherical surface of the cylinder is responsible to refocus these maxima into the nanojet. In this particular configuration, the next nanojet is formed at the sixth cylinder. Figure~\ref{fig8p4}[D] shows the mode profile at $\theta_{bend} = 300$, the white dots marking the bending region. The image clearly illustrates significant radiative losses due to bending, seen as a trail of forward propagating light. Since the transverse intensity profile of the nanojet filaments is not uniform, the coupling efficiency does not vary monotonically with bending angle. Note that, the mode becomes off-axis after the bend. However, the lensing effect induced by the spherical surface redirects the mode towards the axis of the array. Thus, a self-correcting mechanism is realized in the array. This is clearly an advantage of using a spherical resonator over a plane-parallel FP resonator. Figure~\ref{fig8p4}[E] shows the effect of gain on the same system as in Figure~\ref{fig8p4}[D]. The unsaturable optical gain was implemented by adding a negative imaginary component to the refractive index ($n' + in'' = 1.34 - 0.002i$), where $n''$ was maintained uniform across the array. Clearly, the transported mode is seen enhanced in intensity, confirming that
the gain compensates for the bending and radiative losses.

\begin{figure}[h]
\begin{center}
\includegraphics[scale=0.55]{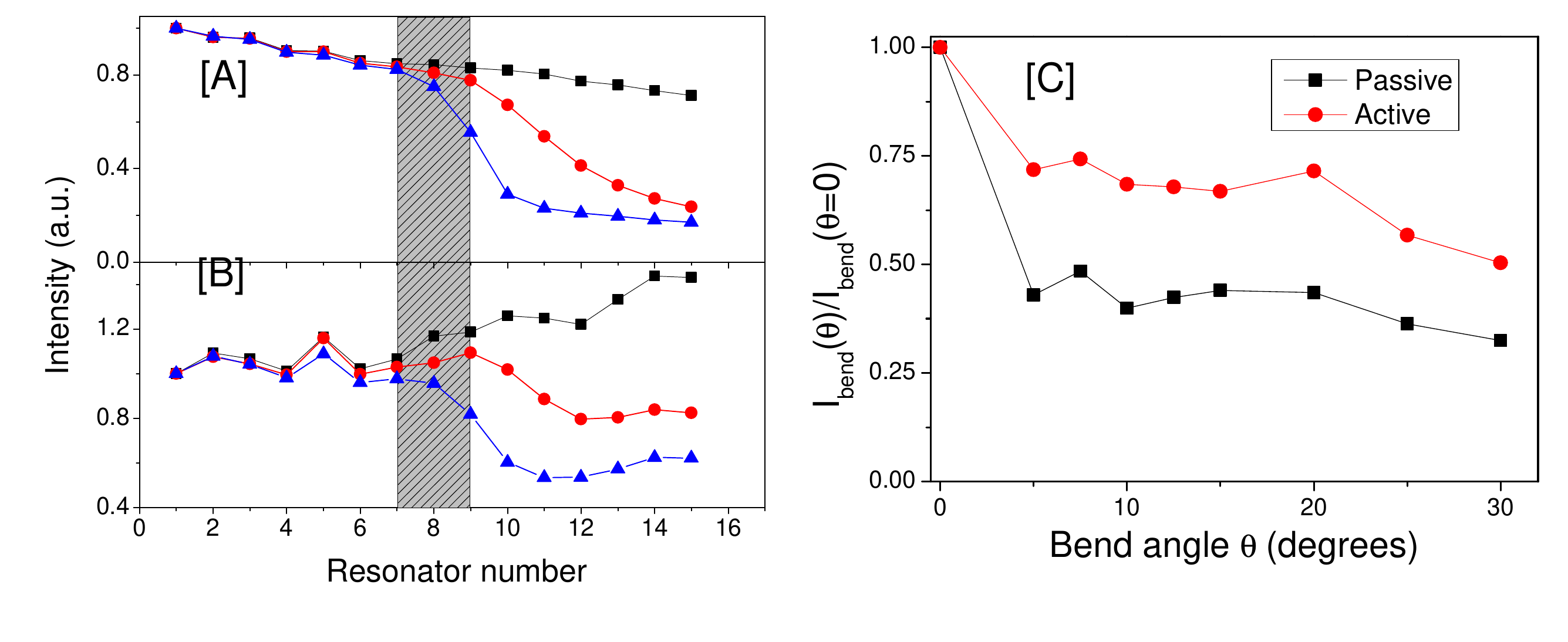}
\end{center}
\vspace{-45pt}
\color{blue}\caption{A] Intensity distribution in each cylinder along the passive array at $\theta_{bend} = 0^{\circ}, 10^{\circ}$ and $15^{\circ}$. [B] Same array under gain. [C] Intensity at the termination end as a function of $\theta_{bend}$, amplification leads to the compensation of loss by about $2$~dB.}
\label{fig8p5}
\end{figure}

To estimate the intensity along the array, we calculate the average intensity of light inside each cylinder. To that end, we draw a concentric circle of $12~\mu$m diameter on each cylinder and estimate the average intensity inside this region. Figure~\ref{fig8p5} [A] shows the intensity distributions along the passive array and the shaded region labels the bending part. The black line is the intensity distribution in the straight array which shows a gradual decay indicating normal radiative loss in direct-coupling. The red and the blue curves represent the array bent at $10^{\circ}$ and $15^{\circ}$ respectively. It is evident from the plots that the intensity drops over about $5$ to $6$ spheres due to bending and afterwards the loss is determined by the normal radiative losses. Figure~\ref{fig8p5}[B] describes the effect of gain. Around sphere no. $12$, the intensity starts to grow again and overcomes the bending and far field losses. The net intensity transport at the termination point as a function of bending angles is quantified in Figure~\ref{fig8p5}[C]. The red and the black curves respectively represent the transmission intensity in the active and passive array. The Y axis of the plot is normalized by the transmitted intensity of the straight waveguide $\theta_{bend} = 0^{\circ}$. As can be seen, both the active and passive systems show an initial rapid drop in intensity which is followed by a gradual non-monotonic decay. By comparing the intensity at the termination point in the active and passive systems, we observed that the gain compensates for the far field and bending losses and improves the transmission intensity by about $2$~dB.

\begin{figure}[h]
\begin{center}
\includegraphics[scale=1.3]{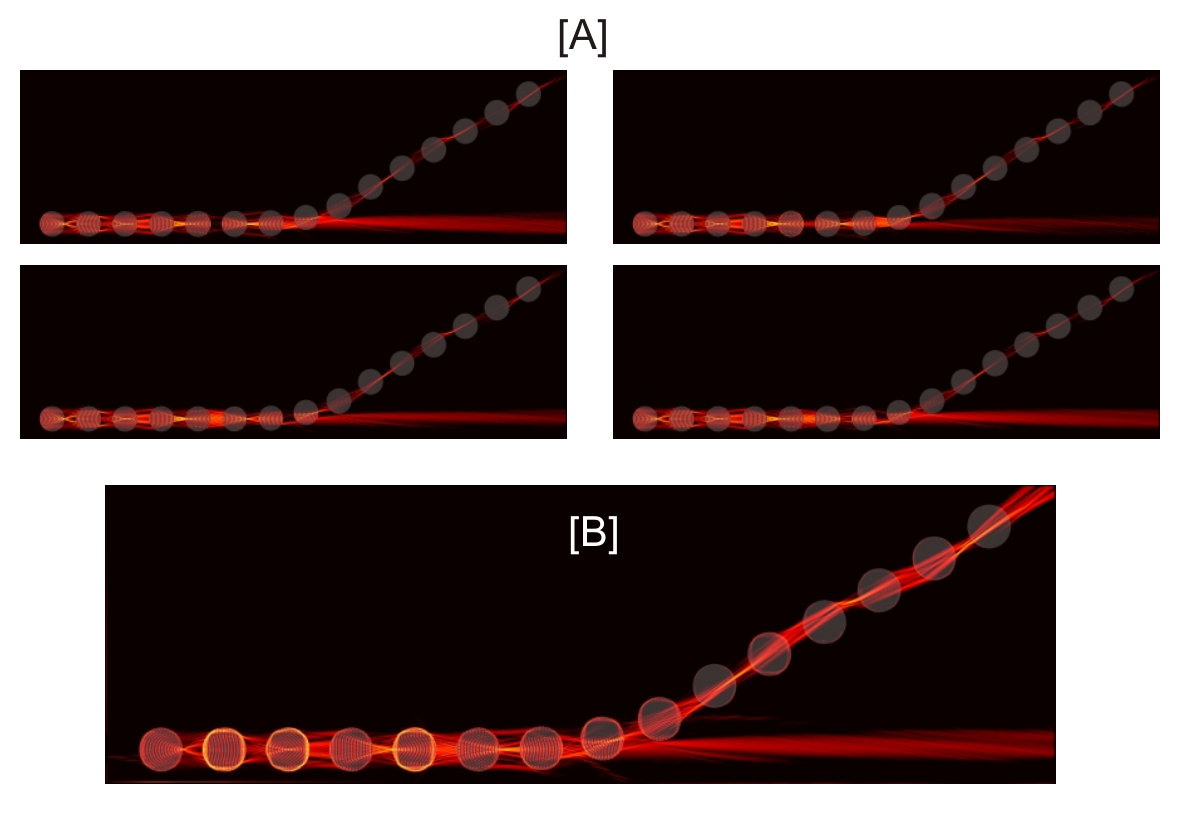}
\end{center}
\vspace{-30pt}
\color{blue}\caption{[A] Mode propagation in four disordered situations. [B] Transport of TE light.}
\label{fig8p6}
\end{figure}

To confirm that the above-described features are generic, we carried out various different simulations. Figure~\ref{fig8p6}[A] depicts the field distribution in four disordered arrays bent at $\theta_{bend} = 30^{\circ}$. Note that, in every configuration, the coupling between two spheres weakens at random positions which may result in a smaller throughput. However, other generic features mentioned earlier do not change. Figure~\ref{fig8p6}[B] shows the transport of light of the other polarization (TE) in a periodic array. Again, the generic features are maintained for both polarizations. The only difference is that, at the simulated wavelength ($\lambda = 561.4$~nm), for TE light, there is a larger probability of hitting the WGM in certain spheres.

{\bf Summary:}\\ In this chapter, we have experimentally demonstrated the existence of carrier modes in a chain of microspheres bent at small angles. The spatial separation between the resonators rules out the possibility of WGM mode coupling and the modes couple through direct Fabry-Perot resonances. We measured the frequency response and variation in modal density with respect to the bending angle. Via FDTD calculations, we showed in an ideal CROW that the filamentation in nanojet-induced modes determines mode coupling. The spherical shape of the resonators self-corrects the propagation of the mode and redirects it towards the axis of array. This is the extra advantage over the plane FP arrangement existing in literature. This study reveals the possible implementation of our system in making aCROW's (active CROW's). Furthermore, the fluidic nature of the system is best suited towards the application in optofluidic devices.	
\chapter{Epilogue}

Random lasers have come a long way since their conception in the sixties. Several new observed phenomena, several novel experimental systems and versatile theoretical treatments have made this intricate subject an inseparable part of complex, mesoscopic physics. Over the years, the holy grail for research in random lasers remained the goal of resonant lasing via Anderson localization of light. With experimental achievement of Anderson localization, random lasing thereof was only going to be a matter of time. In this thesis, we have made a significant step forward towards that goal.

We have explored random lasing from a novel system consisting of amplifying aerosol in the form of an array of tiny microdroplets. The size and spacing between the droplets were controlled to create several aPARS systems, a promising candidate to study disorder-induced light localization and lasing. Further, this system allowed us to find out various configurationally-averaged quantities that are of prime interest in the field of disordered systems. We have shown that, a complete random manifestation of the array acts as a conventional coherent random laser. However, by reducing one degree of randomness (i.e. by making them monodisperse), many interesting properties can be obtained. For example, we established that the monodisperse embodiments have lower lasing thresholds and better lasing properties compared to the completely random configuration of the array. Further, we have successfully demonstrated that, by using a novel technique of spectral mode-matching, the frequency fluctuations in the monodisperse system can eventually be reduced by an order of magnitude ($\sim 1.2$~nm) compared to a conventional random laser. To our knowledge, this is the first demonstration of an amplifying PARS system with conclusive experimental evidence of frequency control in coherent random lasers.

Our experimental results were well supported by various simulations techniques like the transfer matrix method, finite difference time domain calculations and Monte Carlo simulations. Particularly, we adopted the transfer matrix method to reveal that the coherent random lasing modes that were observed in the experiments arise due to the migrated bandedge states of the underlying periodic system. Since the migrated states (also known as the gap states) always appears in the underlying stopband region, they lead to a significant reduction in the frequency fluctuations. We have also studied the spatial extent and field profiles of the lasing modes and conclusively attributed those to the Anderson-localized modes, which have exponentially decaying wings due to the disorder. Thus, lasing due to localized states was achieved in our system.

Furthermore, it is known that linear arrays of spherical scatterers have been vastly used in coupling, guiding, slowing and storage of light. We, too, have examined the mode propagation and its transmittance, especially in bent coupled-resonator waveguides. Using FDTD, we showed that the scattering and the bending losses get compensated during the propagation due to gain. We showed that the nanojet filamentation formed in the air region influence the coupling in the bend region.

{\bf Future directions:}  

These results have opened up a plethora new ideas that can be experimentally implemented. First and foremost, this system offers the possibility of studying the transition from bandedge lasing to random lasing (localization-induced lasing) by creating a periodic array first and then introducing controlled disorder. This seems attainable by techniques like creating the array in vacuum, using high-viscosity fluids etc which essentially reduced fluctuations. An interesting problem in the field relates to the interaction, or the lack thereof, of multiple localized modes existing in the system. This situation is readily available in our system. Further, the phenomena of mode-locking of periodic modes and that of random modes can also be systematically addressed by a controlled periodic-to-disordered transition. Level statistics, level-spacing statistics, intensity statistics of localized modes etc are crucial studies of the field which can be carried out.

Apart from gain systems, the linear chain of passive microdroplets can help measure Lifshitz tails, which arise in the density of states in the gaps when randomness is introduced. The relation between the localization length in the passive and active system can be examined. The behavior of the Lyapunov exponent, inverse of the localization length, can be investigated. Importantly, the fluids used for creating the microdroplets can have a high nonlinear coefficient. Optics of nonlinear disordered systems, which is a rapidly growing field currently, is thus accessible with our technique. Further, due to the fluidic nature of our system, the micro-droplet array promises immediate applications in optofluidic random lasers.

In conclusion, this system promises to be an ideal testbed for various fundamental studies in disordered systems. I hope the results presented in this thesis trigger more activity in these directions.	

\singlespacing
\bibliographystyle{unsrt}
\bibliography{}

\providecommand{\href}[2]{#2}\begingroup\raggedright\begin{thebibliography}{10%
0}
\bibitem {ishimaru78}
A. Ishimaru, ``{Wave propagation and scattering in random media,}'' {\em Academic Press} New York, (1978).

\bibitem {hulst81}
H. C. van de Hulst, ``{Light scattering by small particles,}'' (1981).

\bibitem {ping90}
P. Sheng, ``{Scattering and localization of classical waves in random media,}'' {\em World Scientific} Singapore, (1990).

\bibitem {anderson58}
P. W. Anderson, ``{Absence of diffusion in certain random lattices,}'' {\bf 109}, 1492 (1958).

\bibitem {lagendijk09}
A. Lagendijk, Bart Van Tiggelen, and D. S. Wiersma, ``{Fifty years of Anderson localization,}'' {\em Phys. Today} {\bf 62}, 24-29 (2009).

\bibitem {john83}
S. John, H. Sompolinksy and M. J. Stephen, ``{Localization in a disordered elastic medium near two dimensions,}'' {\em Phys. Rev. B} {\bf 27}, 5592-5603 (1983).

\bibitem {graham90}
I. S. Graham, L. Piche and M. Grant, ``{Experimental evidence for localization of acoustic waves in three dimensions,}'' {\em Phys. Rev. Lett.} {\bf 64}, 3135 (1990).

\bibitem {sajeev84}
S. John, ``{Electromagnetic absorption in a disordered medium near a photon mobility edge,}''  {\em Phys. Rev. Lett.} {\bf 53}, 2169 (1984).

\bibitem {albada85}
M. Van Albada and A. Lagendijk, ``{Observation of weak localization of light in a random medium,}'' {\em Phys. Rev. Lett.} {\bf 55}, 2692-2695 (1985).

\bibitem {wolf85}
P. E. Wolf and G. Maret, ``{Weak localization and coherent backscattering of photons in disorder media,}'' {\em Phys. Rev. Lett.} {\bf 55}, 2696-2699 (1985).

\bibitem {sajeev87}
Sajeev John, ``{Strong localization of photons in certain disordered dielectric superlattices,}'' {\em Phys. Rev. Lett.} {\bf 58}, 2486 (1987).

\bibitem {sajeev91}
S. John, ``{Localization of light,}'' {\em Phys. Today} {\bf 44}, 32-40 (1991).

\bibitem {siegman99a}
A. E. Siegman, ``{Laser beams and resonators: the 1960s,}'' (1999).

\bibitem {siegman99b}
A. E. Siegman, ``{Laser beams and resonators: beyond the 1960s,}'' (1999).

\bibitem {Ambartsumyan66}
R.V. Ambartsumyan, N.G. Basov, P.G. Kryukov, and V.S. Letokhov, ``{A laser with a non-resonant feedback,}'' {\em IEEE J. Quantum Electron.} {\bf QE-2}, 442-446 (1966).

\bibitem {letokhov68}
V. S. Letokhov, ``{Light generation by a scattering medium with a negative resonant absorption,}'' {\em Sov. Phys. JETP} {\bf 26}, 835 (1968).

\bibitem {lawandy94}
N. M. Lawandy, R. M. Balachandran, A. S. L. Gomes and E. Sauvain,``{Laser action in strongly scattering media,}'' {\em Nature} {\bf 368}, 436 (1994).

\bibitem {gouedard93}
C. Gouedard, D. Husson, C. Sauteret, F. Auzel, and A. Migus, ``{Generation of spatially incoherent short pulses in laser-pumped neodymium stoichiometric crystals and powders,}'' {\em J. Opt. Soc. Am. B} {\bf 10}, 2358 (1993).

\bibitem {noginov95}
M. A. Noginov, H. J. Caulfield, N. E. Noginova, and P. Venkateswarlu, ``{Line narrowing in the dye solution with scattering centers,}'' {\em Opt. Commun.} {\bf 118}, 430 (1995).

\bibitem {siddique96}
M. Siddique, R. R. Alfano, G. A. Berger, M. Kempe, and A. Z. Genack, ``{Time-resolved studies of stimulated emission from colloidal dye solutions,}'' {\em Opt. Lett.} {\bf 21}, 450 (1996).

\bibitem {siegman86}
A. E. Siegman, ``{Lasers,}'' Mill Valley, CA: Univ. Science,(1986).

\bibitem {wiersma96}
D. S. Wiersma, and A. Lagendijk, ``{Light diffusion with gain and random lasers,}'' {\em Phys. Rev. E} {\bf 54}, 4256 (1996).

\bibitem {varsanyi71}
F. Varsanyi, ``{Surface lasers,}'' {\em Appl. Phys. Lett.} {\bf 19}, 169-171 (1971).

\bibitem {fork74}
R. L. Fork, D. W. Taylor, K. R. German, A. Kiel, and E. Buehler, ``{Unusual luminescence from crystals containing Eu2+,}'' {\em Phys. Rev. Lett.} {\bf 32}, 781-783 (1974).

\bibitem {fork79}
R. L. Fork and D. W. Taylor, ``{Unusual optical emission from microcrystals containing Eu2+ : Experiment,}'' {\em Phys. Rev. B} {\bf 19}, 3365-3398 (1979).

\bibitem {cao99}
H. Cao, Y. G. Zhao, S. T. Ho, E. W. Seelig, Q. H. Wang and R. P. H. Chang,``{Random laser action in semiconductor powder,}'' {\em Phys. Rev. Lett.} {\bf 82}, 2278 (1999).

\bibitem {cao02}
H. Cao, ``{Random lasers with coherent feedback},'' {\em in Optical properties of nanostructured random media, V.M. Shalaev, ed.} {\bf 82}, {\em Topics in Applied Physics Springer-Verlag} 303-330 (2002).

\bibitem {cao00}
H. Cao, J. Y. Xu, S.-H. Chang, and S. T. Ho, ``{Transition from amplified spontaneous emission to laser action in strongly scattering media,}'' {\em Phys. Rev. E} {\bf 61}, 1985 (2000).

\bibitem {sushil04}
S. Mujumdar, M. Ricci, R. Torre and D. S. Wiersma, ``{Amplified extended modes in random lasers,}'' {\em Phys. Rev. Lett.} {\bf 93}, 053903 (2004).

\bibitem {wu06}
X. Wu, W. Fang, A. Yamilov, A.A. Chabanov, A.A. Asatryan, L.C. Botten, H. Cao, ``{Random lasing in weakly scattering systems,}'' {\em Physical Review A} 053812 (2006).

\bibitem {sebbah01}
C. Vanneste and P. Sebbah, ``{Selective excitation of localized modes in active random media,}'' {\em Phys. Rev. Lett.} {\bf 87}, 183903 (2001).

\bibitem{soukoulis00}
X. Jiang, C.M. Soukoulis,``{Time dependent theory for random lasers,}'' {\em Phys. Rev. Lett.} {\bf 85}, 70 (2000).

\bibitem {soukoulis99}
C. M. Soukoulis and E. N. Economou ``{Electronic localization in disordered systems,}'' {\em Waves in Random Media} {\bf 9}, 255 (1999).

\bibitem {soukoulis02}
X. Jiang and C. M. Soukoulis, ``{Localized random lasing modes and a path for observing localization,}'' {\em Phys. Rev. E} {\bf 65}, 025601 (2002).

\bibitem {polson04}
R.C. Polson, Z.V. Vardeny, ``{Random lasing in human tissues,}'' {\em Applied Physics Letters} {\bf 85}, 1289-1291 (2004).

\bibitem {tulek10}
A. Tulek, R.C. Polson, Z.V. Vardeny, ``{Naturally occurring resonators in random lasing of $\pi$-conjugated polymer films,}'' {\em Nature Physics} {\bf 6}, 303-310 (2010).

\bibitem {strangi06}
G. Strangi, S. Ferjani, V. Barna, A. D. Luca, C. Versace, N. Scaramuzza, and R. Bartolino, ``{Random lasing and weak localization of light in dye-doped nematic liquid crystals,}'' {\em Opt. Express} {\bf 14}, 7737 (2006).

\bibitem {turistsyn10}
S.K. Turitsyn, S.A. Babin, A.E. El-Taher, P. Harper, D.V. Churkin, S.I. Kablukov, J.D. Ania-Castanon, V. Karalekas, E.V. Podivilov, ``{Random distributed feedback fibre laser,}'' {\em Nature Photonics} {\bf 4}, 231-235 (2010).

\bibitem {noginov05}
M. Noginov, ``{Solid State Random Lasers,}'' {\em Springer Series in Optical Sciences}, (2005).

\bibitem {beenakker96}
C. W. J. Beenakker, J. C. J. Paasschens, and P. W. Brouwer, ``{Probability of reflection by a random laser,}'' {\em Phys. Rev. Lett.} {\bf 76}, 1368 (1996).

\bibitem {hcao01}
H. Cao, Y. Ling, J. Y. Xu, C. Q. Cao, and P. Kumar, ``{Photon statistics of random lasers with resonant feedback,}'' {\em Phys. Rev. Lett.} {\bf 86}, 4524 (2001).

\bibitem {hackenbroich05}
G. Hackenbroich, ``{Statistical theory of multimode random lasers,}'' {\em J. Phys. A} {\bf 38}, 10537 (2005).

\bibitem {zaitsev06}
O. Zaitsev, ``{Mode statistics in random lasers,}'' {\em Phys. Rev. A} {\bf 74}, 063803 (2006).

\bibitem {zaitsev07}
O. Zaitsev, ``{Spacing statistics in two-mode random lasing,}'' {\em Phys. Rev. A} {\bf 76}, 043842 (2007).

\bibitem {uppu10}
R. Uppu, S. Mujumdar, ``{Statistical fluctuations of coherent and incoherent intensity in random lasers with nonresonant feedback,}'' {\em Optics Letters} 2831-2833 (2010).

\bibitem {wiersma07}
S. Lepri, S. Cavalieri, G-L. Oppo and D. S. Wiersma ``{Statistical regimes of random laser fluctuations,}'' {\em Phys. Rev. A} {\bf 75}, 063820 (2007).

\bibitem {sushil07}
S. Mujumdar, V. T{\"u}erck, R. Torre and D. S. Wiersma, ``{Chaotic behavior of a random laser with static disorder,}'' {\em Phys. Rev. A} {\bf 76}, 033807 (2007).

\bibitem {lagendijk07}
K. L. van der Molen, R. W. Tjerkstra, A. P. Mosk, and A. Lagendijk, ``{Spatial extent of random laser modes,}'' {\em Phys. Rev. Lett.} {\bf 98}, 143901 (2007).

\bibitem {lagendijk11}
R.G.S. El-Dardiry, Ad Lagendijk, ``{Tuning random lasers by engineered absorption,}'' {\em Applied Physics Letters} {\bf 98}, 161106 (2011).

\bibitem {gottardo08}
S. Gottardo, R. Sapienza, P. D. Garcia, A. Blanco, D. S. Wiersma, and C. Lopez, ``{Resonance-driven random lasing,}'' {\em Nat. Photon.} {\bf 2}, 429 (2008).

\bibitem {sebbah12}
N. Bachelard, J. Andreasen, S. Gigan and P. Sebbah, ``{Taming random lasers through active spatial control of the pump,}'' {\em Phys. Rev. Lett.} {\bf 109}, 033903 (2012).

\bibitem {sebbah13}
N. Bachelard, S. Gigan, X. Noblin, and P. Sebbah, ``{Turning a random laser into a tunable singlemode laser by active pump shaping,}''
\href{http://arXiv.org/abs/1303.1398}{{\tt arXiv:1303.1398} [physics.optics]}, (2013).

\bibitem {shiva12}
B. N. S. Bhaktha, N. Bachelard, X. Noblin, and P. Sebbah, ``{Optofluidic random laser,}'' {\em Appl. Phys. Lett.} {\bf 101}, 151101 (2012).

\bibitem {leonetti13}
M. Leonetti, C. Lopez, ``{Active subnanometer spectral control of a random laser,}'' {\em Appl. Phys. Lett.} {\bf 102}, 071105 (2013).

\bibitem {fujiwara13}
H. Fujiwara, R. Niyuki, Y. Ishikawa, N. Koshizaki, T. Tsuji, and K. Sasak, ``{Low-threshold and quasi-single-mode random laser within a submicrometer-sized ZnO spherical particle film,}'' {\em Appl. Phys. Lett.} {\bf 102}, 061110 (2013).

\bibitem {abrahams79}
E. Abrahams, P. W. Anderson, D. C. Licciardello, and T. V. Ramakrishnan, ``{Scaling theory of localization: Absence of quantum diffusion in two dimensions,}'' {\em Phys. Rev. Lett.} {\bf 42}, 673-676 (1979).

\bibitem {iofee60}
A. F. Ioffe and A. R. Regal, {\em Prog. Semicond.} {\bf 4}, 237 (1960).

\bibitem {wiersma97}
D. S. Wiersma, Paolo Bartolini, Ad Lagendijk and Roberto Righini, ``{Localization of light in a disordered medium,}'' {\em Nature} {\bf 390}, 671-673 (1997).

\bibitem {maret99}
F. Scheffold, R. Lenke, R. Tweer, and G. Maret, ``{Localization or classical diffusion of light?,}'' {\em Nature} {\bf 398},206-207 (1999).

\bibitem {chabanov00}
A. A. Chabanov, M. Stoytchev, and A. Z. Genack, ``{Statistical signatures of photon localization,}'' {\em Nature} {\bf 404}, 850-853 (2000).

\bibitem {garcia91}
N. Garcia, and A. Z. Genack, ``{Anomalous photon diffusion at the threshold of the Anderson localization transition,}'' {\em Phys. Rev. Lett.} {\bf 66}, 1850-1853 (1991).

\bibitem {genack91}
A. Z. Genack, and N. Garcia, ``{Observation of photon localization in a three-dimensional disordered system,}'' {\em Phys. Rev. Lett.} {\bf 66}, 2064-2067 (1991).

\bibitem{sperling13}
T. Sperling, W. Buhrer, C. M. Aegerter, and G. Maret, ``{Direct determination of the transition to localization in three dimensions,}'' {\em Nature Photon.} {\bf 7}, 48-52 (2013).

\bibitem{schffold13}
F. Scheffold and D. S. Wiersma, ``{Inelastic scattering puts in question recent claims of Anderson localization of light,}'' {\em Nature Photon.} {\bf 7}, 934 (2013).

\bibitem {rachida91}
R. Dalichaouch, J. P. Armstrong, S. Schultz, P. M. Platzman and S. L. McCall, ``{Microwave localization by two-dimensional random scattering,}'' {\em Nature} {\bf540}, 53-55 (1991).

\bibitem{vollmer07}
J. Topolancik, B. Ilic, and F. Vollmer, ``{Experimental observation of strong photon localization in disordered photonic crystal waveguides,}'' {\em Phys. Rev. Lett.} {\bf99}, 253901 (2007).

\bibitem{lodahl10}
L. Sapienza, H. Thyrrestrup, S. Stobbe, P. D. Garcia, S. Smolka, P. Lodahl, ``{Cavity quantum electrodynamics with Anderson-localized modes,}'' {\em Science} {\bf 327}, 1352-1355 (2010).

\bibitem {lagendijk89}
H. De Raedt, Ad Lagendijk, and P. de Vriest, ``{Transverse localization of light,}'' {\em Phys. Rev. Lett.}  {\bf 62}, 47-50 (1989).

\bibitem {lahini08}
Y. Lahini, A. Avidan,F. Pozzi, M. Sorel, R. Morandotti, D. N. Christodoulides, and Y. Silberberg, ``{Anderson localization and nonlinearity in one-dimensional disordered photonic lattices,}'' {\em Phys. Rev. Lett.}  {\bf 100}, 013906 (2008).

\bibitem {segev07}
T. Schwartz, G. Bartal, S. Fishman and M. Segev, ``{Transport and Anderson localization in disordered two-dimensional photonic lattices,}'' {\em Nature} {\bf446}, 52-55 (2007).

\bibitem {pradhan94}
P. Pradhan and N. Kumar, ``{Localization of light in coherently amplifying random media,}'' {\em Phys. Rev. B} {\bf 50}, 9644-9647 (1994).

\bibitem {genack05}
V. Milner and A. Z. Genack, ``{Photon localization laser: low-threshold lasing in a random amplifying layered medium via wave localization,}'' {\em Phys. Rev. Lett.} {\bf 94}, 073901 (2005).

\bibitem {mujumdar00}
S. Mujumdar and H. Ramachandran, ``{Spectral features of emissions from random amplifying media,}'' {\em Opt. Commun.} {\bf 176}, 31-41 (2000).

\bibitem {mujumdar10}
S. Mujumdar, R. Torre, H. Ramachandran, D.S. Wiersma, ''{Monte Carlo calculations of spectral features in random lasing,}'' {\em Journal of Nanophotonics} 041550 (2010).

\bibitem {ruppuaip}
R. Uppu, A. K. Tiwari and S. Mujumdar, ``{Coherent random lasing in diffusive resonant media,}'' {\em AIP Conf. Proc.} 1398, 103-105 (2011).

\bibitem {ruppu11}
R. Uppu, S. Mujumdar, ``{Persistent coherent random lasing using resonant scatterers,}'' {\em Optics Express} 23523-23531 (2011).

\bibitem {anjpnfa}
A. K. Tiwari, R. Uppu, S. Mujumdar, ``{Frequency behavior of coherent random lasing in diffusive resonant media,}'' {\em Photonics and Nanostructures - Fundamentals and Applications} {\bf 10}, 416-422 (2012). 

\bibitem {matzler02}
C. M{\"a}tzler, ``{Matlab functions for Mie scattering and absorption Version 2,}'' {\em Tech. Rep. 2002-11, Institut f{\"u}r Angewandte Physik} (2002).

\bibitem {fallert09}
J. Fallert, R. J. B. Dietz, J. Sartor, D. Schneider, C. Klingshirn, and H. Kalt, ``{Co-existence of strongly and weakly localized random laser modes,}'' {\em Nat. Photon.} {\bf 3}, 279 (2009).

\bibitem {lin90}
H. B. Lin, J. D. Eversole, and A. J. Campillo, ``{Vibrating orifice droplet generator for precision optical studies,}'' {\em Rev. Sci. Instrum.} {\bf 61}, 1018-1023 (1990).

\bibitem {berguland73}
R. N. Berguland and Benjamin Y. H. Liu, ``{Generation of monodisperse aerosol standards,}'' {\em Environmental Science and Technology} {\bf 7}, 147 (1973).

\bibitem {qian86}
S. X. Qian, J. B. Snow, H. M. Tzeng, and R. K. Chang, ``{Lasing droplets: highlighting the liquid-air interface by laser emission,}'' {\em Science} {\bf 231}, 486-488 (1986).

\bibitem {chang96}
R. K. Chang and A. J. Campillo, ``{Optical processes in microcavities,}'' eds. (World Scientific, 1996). 

\bibitem {campillo88}
A. J. Campillo and H. B. Lin, ``{Absorption and fluorescence spectroscopy of aerosols,}'' in {\em Optical Effects Associated with Small Particles}, P. W. Barber and R. K. Chang, eds. (World Scientific, 1988), pp. 141-202.

\bibitem {bohren83}
C. F. Bohren and D. R. Huffman, ``{Absorption and Scattering of Light by Small Particles,}'' {\em Wiley, New York} (1983).

\bibitem {anj12ol}
A. K. Tiwari, R. Uppu and S. Mujumdar, ``{Aerosol-based coherent random laser,}'' {\em Optics Letters} {\bf 37}, 1053 (2012).

\bibitem {fleming76}
G. R. Fleming, J. M. Morris and G. W. Robinson, ``{Direct observation of rotational diffusion by picosecond spectroscopy,}'' {\em Chemical Physics} {\bf 17}, 91-100 (1976).

\bibitem {eichler79}
H. J. Eichler, U. Klein and D. Langhans, ``{Measurement of orientational relaxation times of Rhodamine 6G with a streak camera,}'' {\em Chem. Phys. Lett.} {\bf 67}, 21 (1979). 

\bibitem {porter77}
C. Porter, P. J. Sadkowski and C. J. Tredwell, ``{Picosecond rotational diffusion in kinetic and steady state fluorescence spectroscoply,}'' {\em Chem. Phys. Lett.} {\bf 49}, 416 (1977).


\bibitem {mcgurn93}
A. R. McGurn, K. T. Christensen, F. M. Mueller, and A. A. Maradudin, ``{Anderson localization in one-dimensional randomly disordered optical systems that are periodic on average,}'' {\em Phys. Rev. B} {\bf 47}, 13120 (1993).

\bibitem {freilikhr95}
V. D. Freilikher, B. A. Liansky, I. V. Yurkevich, A. A. Maradudin, and A. R. McGurn, ``{Enhanced transmission due to disorder,}'' {\em Phys. Rev. E} {\bf 51}, 6301 (1995).

\bibitem {deych98}
L. I. Deych, D. Zaslavsky, and A. A. Lisyansky, ``{Statistics of the Lyapunovexponent in 1D eandom periodic-on-everage systems,}'' {\em Phys. Rev. Lett.} {\bf 81}, 5390 (1998).

\bibitem {chang03}
S. H. Chang, H. Cao, and S. T. Ho, ``{Cavity formation and light propagation in partially ordered and completely random one-dimensional systems,}'' {\em IEEE J. Quant. Elect.} {\bf 39}, 364-374 (2003).

\bibitem {born80}
M. Born and E. Wolf, ``{Principles of optics,}'' {\em Pergamon, New York} (1980).

\bibitem {penzkofer86}
Y. Lu and A. Penzkofer, ``{Absorption behaviour of methanolic Rhodamine 6G solutions at high concentration,}'' {\em Chemical Physics} {\bf 107}, 175-184 (1986).

\bibitem {rajesh12}
R. V. Nair, A. K. Tiwari, S. Mujumdar, and B. N. Jagatap,
``{Photonic-band-edge-induced lasing in self-assembled dye-activated photonic crystals,}'' {\em Phys. Rev. A} {\bf 85}, 023844 (2012).

\bibitem {rajesh13}
R. V. Nair, A. K. Tiwari, S. Mujumdar, and B. N. Jagatap, ``{Inhibition and enhancement of spontaneous emission using photonic band gap structures,}'' Advanced Material Letters {\bf 4}, Issue 6, pp. 497-501 (2013).

\bibitem {andreasen10}
J. Andreasen, A. A. Asatryan, L. C. Botten, M. A. Byrne, H. Cao, L. Ge, L. Labont\'{e}, P. Sebbah, A. D. Stone, H. E. T\..{u}reci and C. Vanneste, ``{Modes of random lasers,}'' {\em Advances in Optics and Photonics} {\bf 3}, 88 (2010).

\bibitem {jiang99}
X. Jiang, Q. Li, and C. M. Soukoulis, ``{Symmetry between absorption and amplification in disordered media,}'' {\em Phys. Rev. B} {\bf 59}, R9007 (1999).

\bibitem {anj12opex}
A. K. Tiwari, B. Chandra, R. Uppu and S. Mujumdar, ``{Collective lasing from a linear array of dielectric microspheres with gain,}'' {\em Optics Express} {\bf 20}, 6598 (2012).

\bibitem {oskooi10}
A. F. Oskooi, D. Roundy, M. Ibanescu, P. Bermel, J. D. Joannopoulos, and S. G. Johnson, ``{Meep: A flexible free-software package for electromagnetic simulations by the FDTD method,}'' {\em Computer Physics Communications} {\bf 181}, 687-702 (2010).


\bibitem {wu08} 
X. Wu and H. Cao, ``{Statistical studies of random-lasing modes and amplified spontaneous-emission spikes in weakly scattering systems,}'' {\em Phys. Rev. A} {\bf 77}, 013832 (2008).

\bibitem {kumar06}
D. Sharma, H. Ramachandran, and N. Kumar, ``{Levy statistical fluctuations from a random amplifying medium,}'' {\em Fluctuat. Noise Lett.} {\bf 6}, L95 (2006).

\bibitem {zhu12}
G. Zhu, Lei Gu, and M. A. Noginov, ``{Experimental study of instability in a random laser with immobile scatterers,}'' {\em Phys. Rev. A} {\bf 85}, 043801 (2012).

\bibitem {uppu12ol}
R. Uppu, A. K. Tiwari, and S. Mujumdar, ``{Identification of statistical regimes and crossovers in coherent random laser emission,}'' {\em Opt. Lett.} {\bf 37}, 662 (2012).

\bibitem {uppu12pra}
R. Uppu and S. Mujumdar, ``{Dependence of the Gaussian-L'evy transition on the disorder strength in random lasers,}'' {\em Phys. Rev. A} {\bf 87}, 013822 (2013).

\bibitem {tiwari13}
A. K. Tiwari and S. Mujumdar, ``{Random lasing over gap states from a quasi-one-dimensional amplifying periodic-on-average random superlattice,}'' {\em Phys. Rev. Lett.} {\bf 111}, 233903
(2013).

\bibitem {anjcleo}
A. K. Tiwari and S. Mujumdar, ``{Photon-localization induced random lasing from an amplifying periodic-on-average random system,}'' {\em The European Conference on Lasers and Electro-Optics} CK-P-34 (2013).

\bibitem {anj14apl}
A. K. Tiwari, K. S. Alee, R. Uppu and S. Mujumdar, ``{Single-mode, quasi-stable coherent random lasing in an amplifying periodic-on-average random system,}'' {\em Appl. Phys. Lett.} {\bf 104}, 131112 (2014).

\bibitem {anj14photonics} 
S. Mujumdar, A. K. Tiwari, K. S. Alee and R. Uppu, ``{Amplifying periodic-on-average random systems: Route to Anderson-localization random lasers,}'' {\em 12th International Conference on Fiber Optics and Photonics} M3D.2 (2014).


\bibitem {thouless77}
D. J. Thouless, ``{Maximum metallic resistance in thin wires,}'' {\em Phys. Rev. Lett.} {\bf 39}, 1167-1169 (1977).

\bibitem {schwartz07}
T. Schwartz, G. Bartal, S. Fishman and M. Segev, ``{Transport and Anderson localization in disordered two-dimensional photonic lattices,}'' {\em Nature} {\bf 446}, 52-55 (2007).


\bibitem{boriskina11}
S. V. Boriskina, M. Povinelli, V. N. Astratov, A. V. Zayats, and V. A. Podolskiy, ``{Collective phenomena in photonic, plasmonic and hybrid structures,}'' {\em Opt. Express} {\bf 19}, 22024-22028 (2011).

\bibitem{boriskina06}
S. V. Boriskina, ``{Spectrally engineered photonic molecules as optical sensors with enhanced sensitivity: a proposal and numerical analysis,}'' {\em J. Opt. Soc. Am. B} {\bf 23}, 1565-1573 (2006).

\bibitem{deych06}
L. I. Deych and O. Roslyak, ``{Photonic band mixing in linear chains of optically coupled microspheres,}'' {\em Phys. Rev. E} {\bf 73}, 036606 (2006).

\bibitem {deych08}
L. I. Deych, C. Schmidt, A. Chipouline, T. Pertsch, and A. T\"{u}nnermann, ``{Propagation of the fundamental whispering gallery modes in a linear chain of microspheres,}'' {\em App. Phys. B} {\bf 93}, 21-30 (2008).

\bibitem {nakagawa05}
A. Nakagawa, S. Ishii, and T. Baba, ``{Photonic molecule laser composed of GaInAsP microdisks,}'' {\em Appl. Phys. Lett.} {\bf 86}, 041112 (2005).

\bibitem {hara05}
Y. Hara, T. Mukaiyama, K. Takeda, and M. Kuwata-Gonokami, ``{Heavy photon states in photonic chains of resonantly coupled cavities with supermonodispersive microspheres,}'' {\em Phys. Rev. Lett.} {\bf 94}, 203905 (2005).

\bibitem {hara03}
Y. Hara, T. Mukaiyama, K. Takeda, and M. Kuwata-Gonokami, ``{Photonic molecule lasing,}'' {\em Opt. Lett.} {\bf 28}, 2437-2439 (2003).

\bibitem {scheuer05}
J. Scheuer, G. T. Paloczi, J. K. S. Poon, and A. Yariv, ``{Coupled resonator optical waveguides: toward the slowing and storage of light,}'' {\em Opt. Photonics News} {\bf 16}, 36-40 (2005).

\bibitem {astratov10}
V. N. Astratov, ``{Photonic microresonator research and applications,}'' {\em Series in Optical Sciences} edited by I. Chremmos, O. Schwelb, and N. Uzunoglu {\bf 156}, 423-457  (Springer, 2010).

\bibitem {yang08}
S. Yang and V. N. Astratov, ``{Photonic nanojet-induced modes in chains of size-disordered microspheres with an attenuation of only 0.08 dB per sphere,}'' {\em Appl. Phys. Lett.} {\bf 92}, 261111 (2008).

\bibitem {morichetti12}
F. Morichetti, C. Ferrari, A. Canciamilla, and A. Melloni, ``{The first decade of coupled resonator optical waveguides: bringing slow light to applications,}'' {\em Laser Photonics Rev.} {\bf 6}, 74-96 (2012).

\bibitem {chen06}
Z. Chen, A. Taflove, and V. Backman, ``{Highly efficient optical coupling and transport phenomena in chains of dielectric microspheres,}'' {\em Opt. Lett.} {\bf 31}, 389-391 (2006).

\bibitem {darafsheh12}
A. Darafsheh and V. N. Astratov, ``{Periodically focused modes in chains of dielectric spheres,}'' {\em Appl. Phys. Lett.} {\bf 100}, 061123 (2012).

\bibitem {kapitonov07}
A. M. Kapitonov and V. N. Astratov, ``{Observation of nanojet-induced modes with small propagation losses in chains of coupled spherical cavities,}'' {\em Opt. Lett.} {\bf 32}, 409-411 (2007).

\bibitem {astratov04}
V. N. Astratov, J. P. Franchak, and S. P. Ashili, ``{Optical coupling and transport phenomena in chains of spherical dielectric microresonators with size disorder,}'' {\em App. Phys. Lett.} {\bf 85}, 5508-5510 (2004).

\bibitem {yariv99}
A. Yariv, Y. Xu, R. K. Lee, and A. Scherer, ``{Coupled-resonator optical waveguide: a proposal and analysis,}'' {\em Opt. Lett.} {\bf 24}, 711-713 (1999).

\bibitem {heebner02}
J. E. Heebner and R. W. Boyd, ``{`Slow' and `fast' light in resonator-coupled waveguides,}'' {\em J. Mod. Opt.} {\bf 49}, 2629-2636 (2002).

\bibitem {melloni03}
A. Melloni, F. Morichetti, and M. Martinelli, ``{Linear and nonlinear pulse propagation in coupled resonator slow-wave optical structures,}'' {\em Opt. Quantum Electron.} {\bf 35}, 365-379 (2003).

\bibitem {poon04}
J. K. S. Poon, J. Scheuer, Y. Xu, and A. Yariv, ``{Designing coupled-resonator optical waveguide delay lines,}'' {\em J. Opt. Soc. Am. B} {\bf 21}, 1665-1673 (2004).

\bibitem {mitsui08}
T. Mitsui, Y. Wakayama, T. Onodera, Y. Takaya, and H. Oikawa, ``{Observation of light propagation across a $90^{o}$ corner in chains of microspheres on a patterned substrate,}'' {\em Opt. Lett.} {\bf 33(11)}, 1189-1191 (2008).

\bibitem {pishko04}
S. V. Pishko, P. Sewell, T. M. Benson, and S. V. Boriskina, ``{Efficient analysis and design of low-loss whispering-gallery-mode coupled resonator optical waveguide bends,}'' {\em J. Lightwave Technol.} {\bf 25(9)}, 2487-2494 (2007).

\bibitem {boriskina07}
S. V. Boriskina, ``{Spectral engineering of bends and branches in microdisk coupled-resonator optical waveguides,}'' {\em Opt. Express} {\bf 15(25)}, 17371-17379 (2007).

\bibitem {smotrova06}
E. I. Smotrova, A. I. Nosich, T. M. Benson, and P. Sewell, ``{Threshold reduction in a cyclic photonic molecule laser composed of identical microdisks with whispering-gallery modes,}'' {\em Opt. Lett.} {\bf 31}, 921-923 (2006).

\bibitem {mookherjea04}
S. Mookherjea, ``{Semiconductor coupled-resonator optical waveguide laser,}''{\em Appl. Phys. Lett.} {\bf 84}, 3265-3267 (2004).

\bibitem {moeller05}
B. M. Moeller, U. Woggon, and M. V. Artemyev, ``{Coupled-resonator optical waveguides doped with nanocrystals,}'' {\em Opt. Lett.} {\bf 30}, 2116-2118 (2005).

\bibitem {dumige08}
Y. Dumeige, T. K. N. Nguyen, L. Ghisa, S. Trebaol, and P. Feron, ``{Measurement of the dispersion induced by a slow-light system based on coupled active-resonator-induced transparency,}'' {\em Phys. Rev. A} {\bf 78}, 013818 (2008).

\bibitem {poon07}
J. K. S. Poon and A. Yariv, ``{Active coupled-resonator optical waveguides. I. Gain enhancement and noise,}'' {\em J. Opt. Soc. Am. B} {\bf 24}, 2378-2388 (2007).

\bibitem {anj13apl}
A. K. Tiwari, R. Uppu, and S. Mujumdar ``{Experimental demonstration of small-angle bending in an active direct-coupled chain of spherical microcavities,}'' {\em Appl. Phys. Lett.} {\bf 103}, 171108 (2013).

\end{thebibliography}\endgroup


\begin{thebibliography}{99}
\bibitem {syn_siegman86} A. E. Siegman, Lasers. Mill Valley, CA: Univ. Science (1986).
\bibitem {syn_letokhov68} V. S. Letokhov, ``Light generation by a scattering medium with a negative resonant absorption'' Sov. Phys. JETP {\bf 26}, 835 (1968).
\bibitem {syn_lawandy94} N. M. Lawandy, R. M. Balachandran, A. S. L. Gomes and E. Sauvain,``Laser action in strongly scattering media'' Nature {\bf 368}, 436 (1994).
\bibitem {syn_wiersma08} D. S. Wiersma, ``The physics and applications of random lasers'' Nature Phys. {\bf 4}, 359 (2008).
\bibitem {syn_noginov05} M. Noginov, Solid State Random Lasers, Springer Series in Optical Sciences (2005).
\bibitem {syn_cao99} H. Cao, Y. G. Zhao, S. T. Ho, E. W. Seelig, Q. H. Wang and R. P. H. Chang,``Random Laser Action in Semiconductor Powder'' Phys. Rev. Lett. {\bf 82}, 2278 (1999).
\bibitem {syn_fallert09} J. Fallert, R. J. B. Dietz, J. Sartor, D. Schneider, C. Klingshirn and H. Kalt, ``Co-existence of strongly and weakly localized random laser modes'' Nat. Photon. {\bf 3}, 279 (2009).
\bibitem {syn_sushil04} S. Mujumdar, M. Ricci, R. Torre and D. S. Wiersma, ``Amplified Extended Modes in Random Lasers '' Phys. Rev. Lett. {\bf 93}, 053903 (2004).
\bibitem {syn_gottardo08} S. Gottardo, R. Sapienza, P. D. Garcia, A. Blanco, D. S. Wiersma, and C. Lopez, ``Resonance-driven random lasing'' Nat. Photon. {\bf 2}, 429 (2008).
\bibitem{syn_turitsyn10} S. K. Turitsyn, S. A. Babin, A. E. El-Taher, P. Harper, D. V. Churkin, S. I. Kablukov, J. D. Ania-Castanon, V. Karalekas, and E. V. Podivilov, ``Random distributed feedback fibre laser'' Nat. Photon. {\bf 4}, 231 (2010).
\bibitem {syn_cao02} H. Cao, ``Random lasers with coherent feedback'' in Optical Properties of Nanostructured Random Media, V. M. Shalaev, ed., Vol. 82 of Topics in Applied Physics (Springer-Verlag, 2002), pp. 303--330.
\bibitem {syn_wiersma96} D. S. Wiersma, and A. Lagendijk, ``Light diffusion with gain and random lasers'' Phys. Rev. E {\bf 54}, 4256 (1996).
\bibitem {syn_sajeevjohn96} S. John, G. Pang, ``Theory of lasing in a multiple-scattering medium'' Phys. Rev. A {\bf 54}, 3642 (1996).
\bibitem{syn_carminati07} R. Pierrat and R. Carminati,``Threshold of random lasers in the incoherent transport regime'' Phys. Rev. A {\bf 76}, 023821 (2007).
\bibitem {syn_genack97} G. A. Berger, M. Kempe, and A. Z. Genack, ``Dynamics of stimulated emission from random media'' Phys. Rev. E {\bf 56}, 6118 (1997).
\bibitem {syn_sebbah01} C. Vanneste and P. Sebbah, ``Selective Excitation of Localized Modes in Active Random Media'' Phys. Rev. Lett. {\bf 87}, 183903 (2001).
\bibitem{syn_soukoulis00} X. Jiang, C.M. Soukoulis,``Time Dependent Theory for Random Lasers'' Phys. Rev. Lett.  {\bf 85}, 70 (2000).
\bibitem {syn_soukoulis99} Xunya Jiang and C. M. Soukoulis, ``Transmission and reflection studies of periodic and random systems with gain'' Phys. Rev. B {\bf 59}, 6159 (1999).
\bibitem {syn_soukoulis02} X. Jiang and C. M. Soukoulis, ``Localized random lasing modes and a path for observing localization'' Phys. Rev. E {\bf 65}, 025601 (2002).
\bibitem {syn_wiersma07} S. Lepri, S. Cavalieri, G-L. Oppo and D. S. Wiersma ``Statistical regimes of random laser fluctuations'' Phys. Rev. A {\bf 75}, 063820 (2007).
\bibitem {syn_sushil07} S. Mujumdar, V. Tuerck and D. S. Wiersma, ``Chaotic behavior of a random laser with static disorder'' Phys. Rev. A {\bf 76}, 033807 (2007).
\bibitem {syn_lagendijk07} K. L. van der Molen, R. W. Tjerkstra, A. P. Mosk, and A. Lagendijk, ``Spatial Extent of Random Laser Modes'' Phys. Rev. Lett. {\bf 98}, 143901 (2007).
\bibitem {syn_sebbah12} N. Bachelard, J. Andreasen, S. Gigan and P. Sebbah, ``Taming Random Lasers through Active Spatial Control of the Pump'' Phys. Rev. Lett. {\bf 109}, 033903 (2012).
\bibitem {syn_berglund73} Richard N. Berglund and Benjamin Y. H. Liu,``Generation of monodisperse aerosol standards'' Environmental Science and Technology, {\bf 7(2)} 147 (1973).
\bibitem {syn_benner80} R. E. Benner, P. W. Barber, J. F. Owen, and R. K. Chang, ``Observation of Structure Resonances in the Fluorescence Spectra from Microspheres'' Phys. Rev. Lett. {\bf 44}, 475 (1980).
\bibitem{syn_yariv99} A. Yariv, Y. Xu, R. K. Lee, and A. Scherer, ``Coupled-resonator optical waveguide: a proposal and analysis"  Opt. Lett. {\bf 24,} 711-713 (1999).
\bibitem{syn_kapitonov07} A. M. Kapitonov and V. N. Astratov, ``Observation of nanojet-induced modes with small propagation losses in chains of coupled spherical cavities"  Opt. Lett. {\bf 32,} 409-411 (2007).
\bibitem{syn_chen06} Z. Chen, A. Taflove and V. Backman, ``Highly efficient optical coupling and transport phenomena in chains of dielectric microspheres"  Opt. Lett. {\bf 31,} 389-391 (2006).
\bibitem{syn_freilikher95} V.~D.~Freilikher, B.~A.~Liansky, I.~V.~Yurkevich, A.~A.~Maradudin, and A.~R.~McGurn, ``Enhanced transmission due to disorder" Phys. Rev. E {\bf 51,} 6301 (1995).
\bibitem {syn_deych98} Lev I. Deych, D. Zaslavsky, and A. A. Lisyansky, ``Statistics of the Lyapunov Exponent in 1D Random Periodic-on-Average Systems'' Phys. Rev. Lett. {\bf 81}, 5390 (1998).
\bibitem{syn_daozhong94} Z.~Daozhong, H.~Wei, Z.~Youlong, L.~Zhaolin, C.~Bingying, and Y.~Guozhen, ``Experimental verification of light localization for disordered multilayers in the visible-infrared spectrum"  Phys. Rev. B {\bf 50,} 9810 (1994).



\end{thebibliography}

\end{document}